\documentclass[11pt,a4paper]{article}
\usepackage{jcappub,cancel}
\usepackage{graphicx,verbatim}
\usepackage{multirow}
\usepackage{morefloats}
\usepackage[section,subsection,subsubsection]{extraplaceins}

\usepackage{dcolumn}% Align table columns on decimal point
\usepackage{bm}% bold math
\usepackage{color}
\usepackage{ulem}

\newcommand{\neut}{{\tilde{\chi}^0_1}}

\definecolor{orange}{rgb}{1,0.5,0}
\definecolor{purple}{rgb}{0.627451,0.125490, 0.941176}

\newcommand{\loc}{\mathrm{loc}}

\newcommand{\be}{\begin{equation}}
\newcommand{\ee}{\end{equation}}
\newcommand{\sigmaSI}{\sigma_{\neut-p}^\text{SI}}
\newcommand{\sigmaSDp}{\sigma_{\neut-p}^\text{SD}}
\newcommand{\sigmaSDn}{\sigma_{\neut-n}^\text{SD}}

\newcommand{\lsim}{\lower.7ex\hbox{$\;\stackrel{\textstyle<}{\sim}\;$}}
\newcommand{\gsim}{\lower.7ex\hbox{$\;\stackrel{\textstyle>}{\sim}\;$}}

\def\neut{{\tilde\chi_{1}^0}}
\def\charg{{\tilde\chi_{1}^{\pm}}}
\def\relic{\Omega_{\neut}}

\newcommand\mchi{m_\chi^0}

\newcommand{\BR}{BR}
\newcommand\RBtaunu{\frac{\BR(B_u \to \tau \nu)}{\BR(B_u \to \tau \nu)_{SM}}}
\newcommand\DeltaO{\Delta_{0-}}
\newcommand\RBDtaunuBDenu{\frac{\BR(B \to D \tau \nu)}{\BR(B \to D e \nu)}}
\newcommand\Rl{R_{l23}}

\newcommand\Dstaunu{\BR(D_s \to \tau \nu)}
\newcommand\Dsmunu{\BR(D_s\to \mu \nu)} 
\newcommand\Dmunu{\BR(D \to \mu \nu)} 
\newcommand\brbsmumu{\BR(\overline{B}_s\to\mu^+\mu^-)}
\newcommand\afb{A_{FB} (B \to K^* \mu^+ \mu^-)}

\newcommand\mhl{m_h}

\newcommand{\brbsgamma}{BR(\bar{B} \rightarrow X_s\gamma) }

\newcommand{\OhLSP}{\Omega_{\chi} h^2}
\newcommand{\siW}{\sigma_{\text{Planck}}}
\newcommand{\muW}{\mu_\text{Planck}}

\newcommand{\mugg}{\mu_{\gamma \gamma}}
\newcommand{\muww}{\mu_{W^+ W^-}}
\newcommand{\muzz}{\mu_{Z Z}}
\newcommand{\mubb}{\mu_{b \bar{b}}}
\newcommand{\mutautau}{\mu_{\tau^+ \tau^-}}

\newcommand{\gev}{\mbox{ GeV}}
\newcommand{\tev}{\mbox{ TeV}}

\newcommand{\pb}{\mbox{ pb}}
\newcommand{\fb}{\mbox{ fb}}

\newcommand{\cl}{\text{C.L.}}

\newcommand{\gmt}{$g-2$}
\newcommand{\like}{\mathcal{L}}
\newcommand{\order}{\mathcal{O}}

\def\beq{\begin{eqnarray}}
\def\eeq{\end{eqnarray}}
\def\bea{\begin{eqnarray*}}
\def\eea{\end{eqnarray*}}
\def\pmatrix{\begin{pmatrix}}

\newcommand{\pMSSM}{MSSM-15}
\newcommand{\sgn}[1]{\text{sgn}(#1)}
\newcommand{\multinest}{\texttt{MultiNest}}
\newcommand{\SB}{\texttt{SuperBayeS}}
\newcommand{\likeJ}{\like_\text{Joint}}
\newcommand{\Ohsq}{\Omega_\chi h^2}

\newcommand{\lumifb}{fb$^{-1}$}

\newcommand{\lumitext}{integrated luminosity}
\newcommand{\gevklam}{[GeV]}
\newcommand{\etmis}{$E_{\text{T}}^{\text{miss}}$}

\newcommand{\meff}{m$_{\text{eff}}$}
\newcommand{\pt}{p$_{\text{T}}$}

\newcommand{\effbruch}{$\frac{\Delta \varepsilon}{\varepsilon}$}

\newcommand{\tanb}{tan $\beta$}

\usepackage{booktabs}
\usepackage{ifthen}

\usepackage{rotating}
\usepackage{longtable}
\usepackage{pdflscape}

\newcommand{\forloop}[5][1]{%
\setcounter{#2}{#3}%
\ifthenelse{#4}{#5\addtocounter{#2}{#1}%
\forloop[#1]{#2}{\value{#2}}{#4}{#5}}%
}

\newcounter{crcounter}

\newcommand{\compensaterule}[1]{%
\forloop{crcounter}{1}{\value{crcounter} < #1}%
{\vspace*{-\aboverulesep}\vspace*{-\belowrulesep}}}

\newcommand{\multirowbt}[4][\anzlines]{%
\def\anzlines{#2}%
\multirow{#2}{#3}%
{\compensaterule{#1}#4}}

\begin{document}

\title{Profile likelihood maps of a 15-dimensional MSSM } 

\author{ C. Strege\,$^{1}$, G.~Bertone\,$^{2}$, G.J. Besjes\,$^{3,4}$, S.~Caron\,$^{3,4}$, R.~Ruiz de Austri\,$^{5}$, A.~Strubig\,$^{3,4}$, R.~Trotta\,$^{1}$}
\affiliation{${^1}$ Astrophysics Group, Imperial Centre for Inference and Cosmology, Imperial College London, Blackett Laboratory, Prince Consort Road, London SW7 2AZ, UK} 
\affiliation{${^2}$ GRAPPA Center of Excellence, University of Amsterdam, Science Park 904, 1090 GL Amsterdam, The Netherlands} 
\affiliation{${^3}$ Experimental High Energy Physics, IMAPP, Faculty of Science, Radboud University Nijmegen, Mailbox 79, P.O. Box 9010, 6500 GL Nijmegen, The Netherlands}
\affiliation{${^4}$ Nikhef, Science Park 105, 1098 XG Amsterdam, The Netherlands}
\affiliation{${^5}$ Instituto de F\'isica Corpuscular, IFIC-UV/CSIC, Valencia, Spain}

\abstract{
We present statistically convergent profile likelihood maps obtained via global fits of a phenomenological Minimal Supersymmetric Standard Model with 15 free parameters (the MSSM-15), based on over 250M points. We derive constraints on the model parameters from direct detection limits on dark matter, the Planck relic density measurement and data from accelerator searches. We provide a detailed analysis of the rich phenomenology of this model, and determine the SUSY mass spectrum and dark matter properties that are preferred by current experimental constraints. We evaluate the impact of the measurement of the anomalous magnetic moment of the muon ($g-2$) on our results, and provide an analysis of scenarios in which the lightest neutralino is a subdominant component of the dark matter. The MSSM-15 parameters are relatively weakly constrained by current data sets, with the exception of the parameters related to dark matter phenomenology ($M_1$, $M_2$, $\mu$), which are restricted to the sub-TeV regime, mainly due to the relic density constraint. The mass of the lightest neutralino is found to be $< 1.5$ TeV at 99\% C.L., but can extend up to $3$ TeV when excluding the $g - 2$ constraint from the analysis. Low-mass bino-like neutralinos are strongly favoured, with spin-independent scattering cross-sections extending to very small values, $\sim 10^{-20}$ pb. ATLAS SUSY null searches strongly impact on this mass range, and thus rule out a region of parameter space that is outside the reach of any current or future direct detection experiment. The best-fit point obtained after inclusion of all data corresponds to a squark mass of $2.3$ TeV, a gluino mass of $2.1$ TeV and a $130$ GeV neutralino with a spin-independent cross-section of $2.4 \times 10^{-10}$ pb, which is within the reach of future multi-ton scale direct detection experiments and of the upcoming LHC run at increased centre-of-mass energy.
}

\maketitle

%%%%%%%%%%%%%%%%%%%%%%%%%%%%%%%%%%%%%%%%%%%%%%%%%%%%%%%%%%%%%%%%%%%%%%%%%%%%
\section{Introduction}
\label{sec:introduction}
%%%%%%%%%%%%%%%%%%%%%%%%%%%%%%%%%%%%%%%%%%%%%%%%%%%%%%%%%%%%%%%%%%%%%%%%%%%%%

The Large Hadron Collider (LHC) has delivered $\sim 20 \fb^{-1}$ of integrated luminosity at $\sqrt{s}=8  \tev$, but evidence for new physics beyond the Standard Model (SM) is still lacking. In particular, the data contain no signature of Supersymmetry (SUSY), which is the most widely studied theory of physics beyond the SM, as it may offer a solution to the hierarchy problem and to the dark matter problem of the universe. In light of the lack of a signal in direct searches for SUSY, the ATLAS and CMS collaborations have placed strong bounds on gluinos and squarks with masses $\lsim 1 \tev$. On the other hand, the recent discovery of a SM-like Higgs boson with a mass $m_h \sim 125$ GeV requires very large top squark sector masses, generally of several TeV. This excludes generic supersymmetric theories in which all the superpartners have masses below  $\sim 1 \tev$.

The simplest SUSY realization, the Minimal Supersymmetric Standard Model (MSSM), has 126 Lagrangian parameters, including complex phases, which makes its phenomenological study impractical.
If one applies  a concrete mechanism that mediates SUSY breaking to the observable sector, then the number of parameters can be reduced significantly. This is for instance the case for models like the constrained MSSM (cMSSM) in which one demands universal scalar masses, gaugino masses and the trilinear couplings at a high energy scale. The cMSSM is certainly the most studied model in the literature and its viability with respect to the relevant available data has been assessed using several methods; from grid and random scans to -- more recently -- statistically convergent methods (both Bayesian and profile likelihood). 

The LHC data have severely constrained this model, so much that it is in tension with the naturalness of the electroweak breaking at the correct scale since the SUSY-breaking parameters are pushed to large values. One exception is the focus point region where the weak scale is insensitive to variations in these parameters. However, this region is becoming increasingly constrained by direct dark matter searches, such as the XENON100 and LUX experiments. This conclusion also applies to less constrained models such as the non-universal Higgs mass model (NUHM)~\cite{Cabrera:2012vu, Buchmueller:2013rsa}, the non-universal gaugino mass model (NUGM)~\cite{Chakrabortty:2013voa} and the non-universal gaugino and Higgs mass model (NUGHM)~\cite{Cabrera:2013jya}.

One approach to address this issue is to avoid explicitly assuming a SUSY breaking mechanism. Instead, one can reduce the 126 MSSM parameters to $19$ parameters by using phenomenological constraints, that define the so-called phenomenological MSSM (pMSSM)~\cite{Djouadi:1998di}. In this scheme, one assumes that: (i) All the soft SUSY-breaking parameters are real, therefore the only source of CP-violation is the CKM matrix. (ii) The matrices of the sfermion masses and the trilinear couplings are diagonal, in order to avoid FCNCs at the tree-level. (iii) First and second sfermion generation universality to avoid severe constraints, for instance, from  $K^0-\bar{K}^0$ mixing.

The pMSSM has been studied in the past using random scans~\cite{Berger:2008cq, Arbey:2011un,Cahill-Rowley:2013dpa}, as well as Bayesian methods~\cite{Baltz:2006fm,AbdusSalam:2009qd, AbdusSalam:2012ir, Boehm:2013qva}. Both approaches have limitations. While appearing uniformly distributed in 1D and 2D projections, random scans in large-dimensional parameter spaces are actually highly concentrated in a thin shell of the hypersphere inscribed in the scan box (the ``concentration of measure'' phenomenon). This means that only a negligible fraction of the pMSSM parameter space is explored by random scans. Furthermore, such scans typically only retain points within e.g.~$2\sigma$ cuts of the observed experimental constraints. Without the explicit use of a likelihood function, random scans have no way of directing the exploration towards more interesting regions of parameter space, i.e. regions where the likelihood is larger.  The Bayesian approach is much more efficient, but the prior dependence of the posterior distribution can be very strong, especially for high-dimensional models such as the pMSSM with a large number of effectively unconstrained parameters. 
 
In this paper, we adopt a Bayesian approach to scanning (using a full likelihood function and an algorithm that generates samples from the posterior distribution), but then derive profile likelihood maps ---which are in principle prior-independent--- for a more robust statistical interpretation. We perform a profile likelihood analysis of a simplified version of the pMSSM with 15 parameters, which we refer to as the MSSM-15. The number of model parameters is reduced by some reasonable assumptions which retain the most relevant phenomenological aspects of the pMSSM in terms of collider and dark matter searches. Our likelihood includes all available accelerator constraints and a newly developed technique to approximate joint constraints from inclusive searches at the LHC. We also adopt cosmological (from Planck) and astro-particle physics constraints (from direct detection experiments) that apply to the lightest neutralino, discussing both the case where it constitutes the entirety or just part of the dark matter in the universe (see e.g. Refs. \cite{Jungman:1995df, Munoz:2003gx, Bertone:2004pz} and references therein).

This paper is organised as follows. We introduce our theoretical model and statistical approach in Section~\ref{sec:theory}. In Section \ref{sec:results} we present the profile likelihood maps from our scans, both with and without LHC constraints. Section~\ref{sec:conclusions} contains our conclusions. In the Appendix, we describe our approach to approximating the likelihood for ATLAS 0-lepton and 3-lepton inclusive searches. 

%%%%%%%%%%%%%%%%%%%%%%%%%%%%%%%%%%%%%%%%%%%%%%%%%%%%%%%%%%%
\section{Theoretical and statistical framework}
\label{sec:theory}
%%%%%%%%%%%%%%%%%%%%%%%%%%%%%%%%%%%%%%%%%%%%%%%%%%%%%%%%%%% 

%%%%%%%%%%%%%%%%%%%%%%%%%%%%%%%%%%%%%%%%%%%%%%%%%%%%%%%
\subsection{Theoretical model}
%%%%%%%%%%%%%%%%%%%%%%%%%%%%%%%%%%%%%%%%%%%%%%%%%%%%%

If one is mainly interested in the phenomenology of the MSSM the number of parameters can be significantly reduced using a number of reasonable simplifying assumptions. 
In this paper, we study such a phenomenological version of the MSSM that is described by 15 model parameters, which we call the \pMSSM.  This is motivated by the present lack of experimental evidence for SUSY: while highly constrained models as the cMSSM are under pressure in the light of the recent negative sparticle searches at the LHC, there is no experimental indication that one requires the full freedom of the 19-dimensional pMSSM at present.

The sfermion soft-masses are defined as in the pMSSM. Namely, the sfermion mass sector is completely described by the first and second generation squark mass $m_Q$, the third generation squark masses $m_{Q_3}$, $m_{U_3}$ and $m_{D_3}$, the first and second generation slepton mass $m_L$ and the third generation slepton masses $m_{L_3}$ and $m_{E_3}$ (where $m_{U_3}$, $m_{D_3}$ and $m_{E_3}$ are the superpartners of the right-handed third-generation quarks and leptons, respectively). 

The trilinear couplings of the sfermions enter in the off-diagonal parts of the sfermion mass matrices. Since these entries are proportional to the Yukawa couplings of the respective fermions, we can approximate the trilinear couplings associated with the first and second generation fermions to be zero. Furthermore, due to the large top Yukawa coupling, the trilinear coupling of the top $A_t$ is in general more relevant than the trilinear couplings of the other third generation couplings. Therefore, we assume unification of the bottom and tau trilinear couplings at the GUT scale, so that both are described by the same parameter $A_0 \equiv A_b = A_\tau$\footnote{This is equivalent to the assumption of bottom-tau Yukawa unification, as motivated for example by SU(5) models \cite{su5}.}. 

After the application of the electroweak symmetry breaking conditions, the Higgs sector can be fully described by the ratio of the Higgs vacuum expectation values $\tan \beta$ and the Higgs masses $m_{H_i}^2$. Instead of the Higgs masses, we choose to use the higgsino mass parameter $\mu$ and the mass of the pseudoscalar Higgs $m_A$ as input parameters, as they are more directly related to the phenomenology of the model. The final ingredient of our model are the three gaugino masses: the bino mass $M_1$, the wino mass $M_2$ and the gluino mass $M_3$.

The above parameters describe a 15-dimensional realisation of the pMSSM which encapsulates all phenomenologically relevant features of the full model that are of interest for dark matter and collider experiments. The model parameters are displayed in Table~\ref{pMSSM_priors}, along with their prior ranges (see next section). All of the input parameters are defined at the SUSY scale $\sqrt{m_{\tilde{t_1} } m_{\tilde{t_2} } }$, with the exception of $A_0$, which is defined at $10^{16}$ GeV and run to the SUSY scale using the RGEs.

In this scenario, in principle, there are five arbitrary phases embedded in the parameters $M_i (i=1,2,3)$, $\mu$ and the one corresponding to the trilinear couplings provided we assume that the trilinear matrices are flavour diagonal. However, one may perform a $U(1)$-$R$ rotation  on the gaugino fields to remove one of the phases of $M_i$. For consistency with the literature we choose the phase of $M_2$ to be zero. Note that this $U(1)_R$ transformation affects neither the phase of the trilinear couplings, since the Yukawa matrices being real fixes the phases of the same fields that couple to the trilinear couplings, nor the phase of $\mu$. Therefore in the CP-conservation case $M_1$, $M_3$, $\mu$ and the trilinear couplings can be chosen both positive and negative.

\begin{table}
\begin{center}
\begin{tabular}{l l | l l}
\hline
\hline
\multicolumn{4}{c}{\pMSSM\ parameters and priors} \\
\multicolumn{2}{c}{Flat priors} & \multicolumn{2}{c}{Log priors} \\
\hline
$M_1$ [TeV] & (-5, 5) & $\sgn{M_1} \log |M_1|/\text{GeV}$ & $(-3.7, 3.7)$ \\
$M_2 $ [TeV] & (0.1, 5) & $\log M_2/\text{GeV}$ & $(2,3.7)$ \\
$M_3 $ [TeV] & (-5, 5) & $\sgn{M_3} \log |M_3|/\text{GeV}$ & $(-3.7, 3.7)$ \\
$m_{L} $ [TeV] & (0.1,10) & $\log m_L/\text{GeV}$ & $(2,4)$ \\
$m_{L_3} $ [TeV] & (0.1,10) & $\log m_{L_3}/\text{GeV}$ & $(2,4)$ \\
$m_{E_3} $ [TeV] & (0.1,10) & $\log m_{E_3}/\text{GeV}$ & $(2,4)$ \\
$m_{Q} $ [TeV] & (0.1,10) & $\log m_{Q}/\text{GeV}$ & $(2,4)$ \\
$m_{Q_3} $ [TeV] & (0.1,10) &$\log m_{Q_3}/\text{GeV}$ & $(2,4)$\\
$m_{U_3} $ [TeV] & (0.1,10) & $\log m_{U_3}/\text{GeV}$ & $(2,4)$ \\
$m_{D_3} $ [TeV] & (0.1,10) & $\log m_{D_3}/\text{GeV}$ & $(2,4)$ \\
$A_t $ [TeV] & (-10, 10) & $\sgn{A_t} \log |A_t|/\text{GeV}$ & $(-4, 4)$  \\
$A_0 $ [TeV] & (-10,10) & $\sgn{A_0} \log |A_0|/\text{GeV}$ & $(-4, 4)$ \\
$\mu $ [TeV] & (-5,5) & $\sgn{\mu} \log |\mu|/\text{GeV}$ & $(-3.7, 3.7)$ \\
$m_A $ [TeV] & (0.01, 5) & $\log m_A/\text{GeV}$ & $(1, 3.7)$ \\
$\tan\beta$ & $(2, 62)$ & $\tan\beta$ & $(2, 62)$ \\
\hline
%& Gaussian prior  & Range scanned (170.6, 175.8) \\
$M_t$ [GeV] & \multicolumn{3}{c}{$173.2 \pm 0.87$~\cite{CDF:2013jga} (Gaussian prior)}   \\
\hline
\end{tabular}
\end{center}
\caption{\fontsize{9}{9}\selectfont \pMSSM\ parameters and top mass value used in this paper and prior ranges for the two prior choices adopted in our scans. ``Flat priors'' are 
uniform on the parameter itself (within the ranges indicated), while ``Log priors'' are uniform in the log of the parameter (within the ranges indicated).}\label{pMSSM_priors}
\end{table}

 %%%%%%%%%%%%%%%%%%%%%%%%%%%%%%%%%%%%%%%%%%%%%%%%%%%%%%%%%%%
 \subsection{Scanning algorithm and profile likelihood maps}
%%%%%%%%%%%%%%%%%%%%%%%%%%%%%%%%%%%%%%%%%%%%%%%%%%%%%%%%%%% 

We adopt a Bayesian approach to sample the \pMSSM\ parameter space, and then use the resulting posterior samples to produce profile likelihood maps. This is because the large dimensionality of the \pMSSM\ and the relatively weak constraints imposed by experimental data result in a (Bayesian) posterior distribution suffering from severe prior-dependent volume effects. These would make the interpretation of the Bayesian posterior problematic.  

We therefore focus on the profile likelihood (PL) for one or two parameters at the time. The profile likelihood is obtained by maximising the likelihood function over the parameters that are not displayed. For example, for a single parameter of interest $\theta_i$ the other parameters $\Psi = \{\theta_1,...,\theta_{i-1},\theta_{i+1},...,\theta_{n}\}$ are eliminated from the 1D profile likelihood by maximising over them: 
\begin{equation}
{\mathcal L}(\theta_i)= \max_{\Psi}\mathcal{L}(\theta_i, \Psi) = \mathcal{L}(\theta_i, \hat{\hat{\Psi}}),
\end{equation}
where $\mathcal{L}(\theta_i, \Psi)$ is the full likelihood function. 
Our samples of the \pMSSM\ parameter space are distributed according to the posterior pdf, but we simply ignore their density in producing profile likelihood maps by maximising over the hidden variables.  Confidence intervals/regions from the resulting 1D/2D profile likelihood maps are determined by adopting the usual Neyman construction with the profile likelihood ratio $\lambda(\theta_i)$ as test statistics:
\be
\lambda(\theta_i) = \frac{\mathcal L (\theta_i, \hat{\hat{\Psi}})}{\mathcal L (\hat{\theta_i}, \hat{\Psi})},
\ee
where $\hat{\hat{\Psi}}$ is the conditional maximum likelihood estimate (MLE) of $\Psi$ with $\theta_i$ fixed and $\hat{\theta_i}, \hat{\Psi}$ are the unconditional MLEs. Values of the $\Delta\chi^2 = -2 \ln \lambda(\theta_i)$ corresponding to 68\%, 95\% and 99\% confidence intervals are obtained from Wilks' theorem. The generalisation to 2D PL maps is straightforward. 

We have upgraded the publicly available \SB-v1.5 package~\cite{deAustri:2006pe, Roszkowski:2007fd, Trotta:2008bp, Bertone:2011nj} to a new version, \SB-v2.0, which will shortly be released to the public\footnote{Visit the webpage \texttt{superbayes.org} to download the new version.}. This latest version of \SB\ is interfaced with SoftSUSY 3.3.10~\cite{SoftSUSY, Allanach:2001kg} as SUSY spectrum calculator, MicrOMEGAs 2.4~\cite{MicrOMEGAs, Belanger:2006is} to compute the abundance of dark matter, DarkSUSY 5.0.5~\cite{DarkSUSY, Gondolo:2004sc} for the computation of $\sigmaSI$ and $\sigmaSDp$, SuperIso 3.0~\cite{SuperIso, Mahmoudi:2008tp} to compute $\delta a_\mu^{\mathrm{SUSY}}$ and B(D) physics observables, SusyBSG 1.5 for the determination of $\brbsgamma$\cite{SusyBSG,Degrassi:2007kj} and FeynHiggs 1.9 \cite{feynhiggs} to compute the Higgs production cross-sections and decay amplitudes. 
For the computation of the electro-weak observables described in Section~\ref{sec:constraints} we have implemented the complete one-loop corrections and the available MSSM two-loop corrections as well as the full Standard Model results~\cite{Heinemeyer:2007bw}.

\SB-v2.0 is interfaced with the publicly available code \multinest\ v2.18~\cite{Feroz:2007kg,Feroz:2008xx}, which we use to obtain samples from the posterior distribution. As a multi-modal implementation of the nested sampling algorithm~\cite{skilling}, \multinest\ is an extremely efficient scanning algorithm that can reduce the number of likelihood evaluations required for an accurate mapping of the Bayesian posterior probability distribution function by up to two orders of magnitude with respect to conventional MCMC methods. This Bayesian algorithm, originally designed to compute the model likelihood and to accurately map out the posterior,  is also able to reliably evaluate the profile likelihood, given appropriate settings, as demonstrated in~\cite{Feroz:2011bj}.  

As motivated above, we use the posterior samples to extract 1D and 2D (prior-independent) profile likelihood maps by maximising the likelihood over all other parameter dimensions. This however requires a much larger number of samples than marginalization of the posterior~\cite{Feroz:2011bj}, as well as dedicated settings of the \multinest\ code. We adopt the recommendations of Ref.~\cite{Feroz:2011bj}, and use a tolerance parameter $\text{tol} =  10^{-4}$ and a number of live points $N_\text{live} = 2\times10^4$. To further increase the resolution of our profile likelihood maps, we store the likelihood and parameter values of all likelihood evaluations performed by \multinest. This includes all samples that would usually be discarded because they do not lie above the iso-likelihood contour in the replacement step in the nested sampling algorithm.
This increases the number of likelihood values by a factor $>20$, and allows for a higher-resolution profile likelihood mapping, especially in the tails of the profile likelihood, at no additional computational cost.

%%%%%%%%%%%%%%%%%%%%%%%%%%%%%%%%%%%%%%%%%%%%
\subsection{Prior choices and ranges} 
%%%%%%%%%%%%%%%%%%%%%%%%%%%%%%%%%%%%%%%%%%%%

\begin{figure}[tb]
\begin{center}
\includegraphics[width=0.24\linewidth, trim = 0.7cm 0cm 0.7cm 0cm]{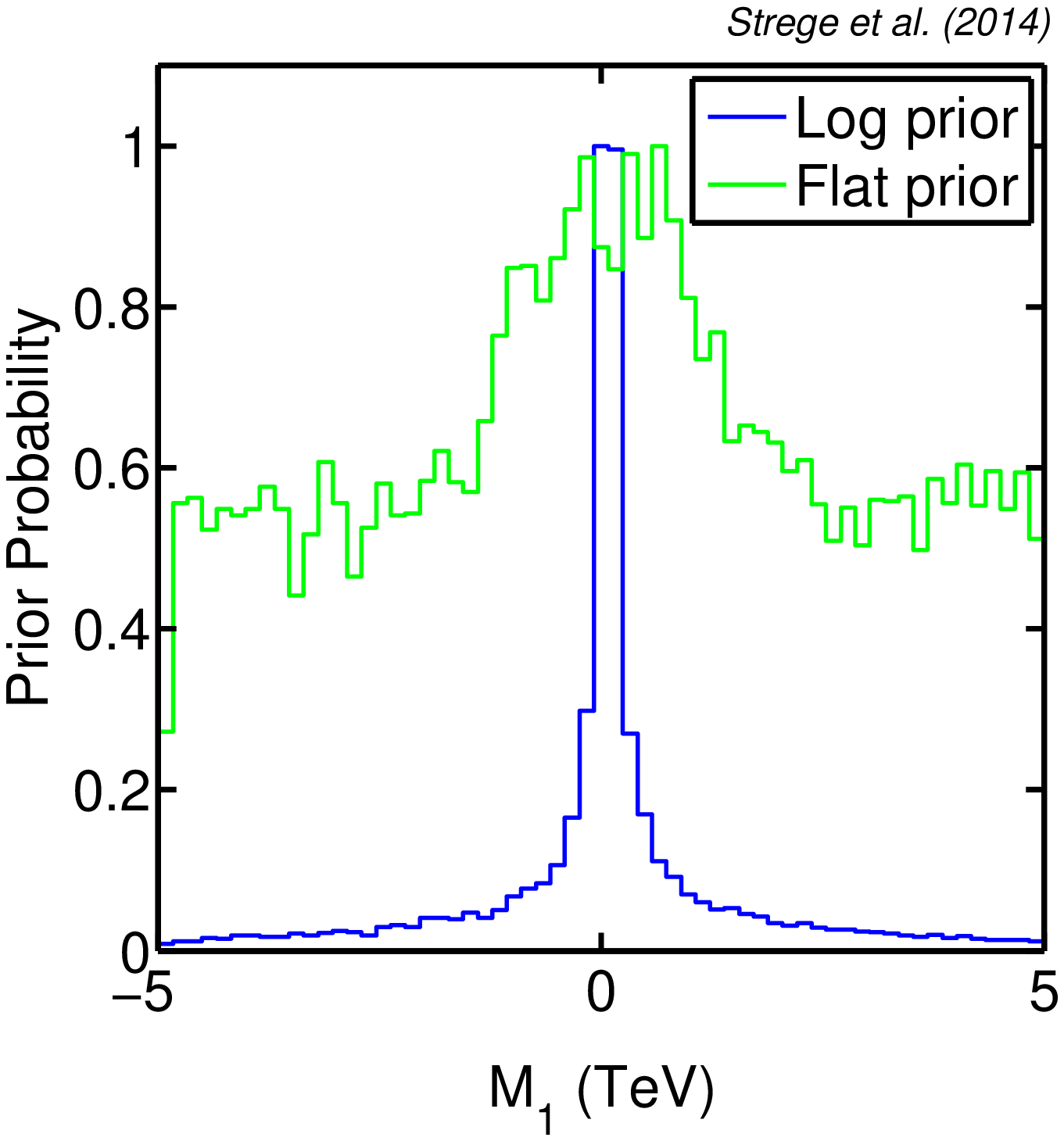}
\includegraphics[width=0.24\linewidth, trim = 0.7cm 0cm 0.7cm 0cm]{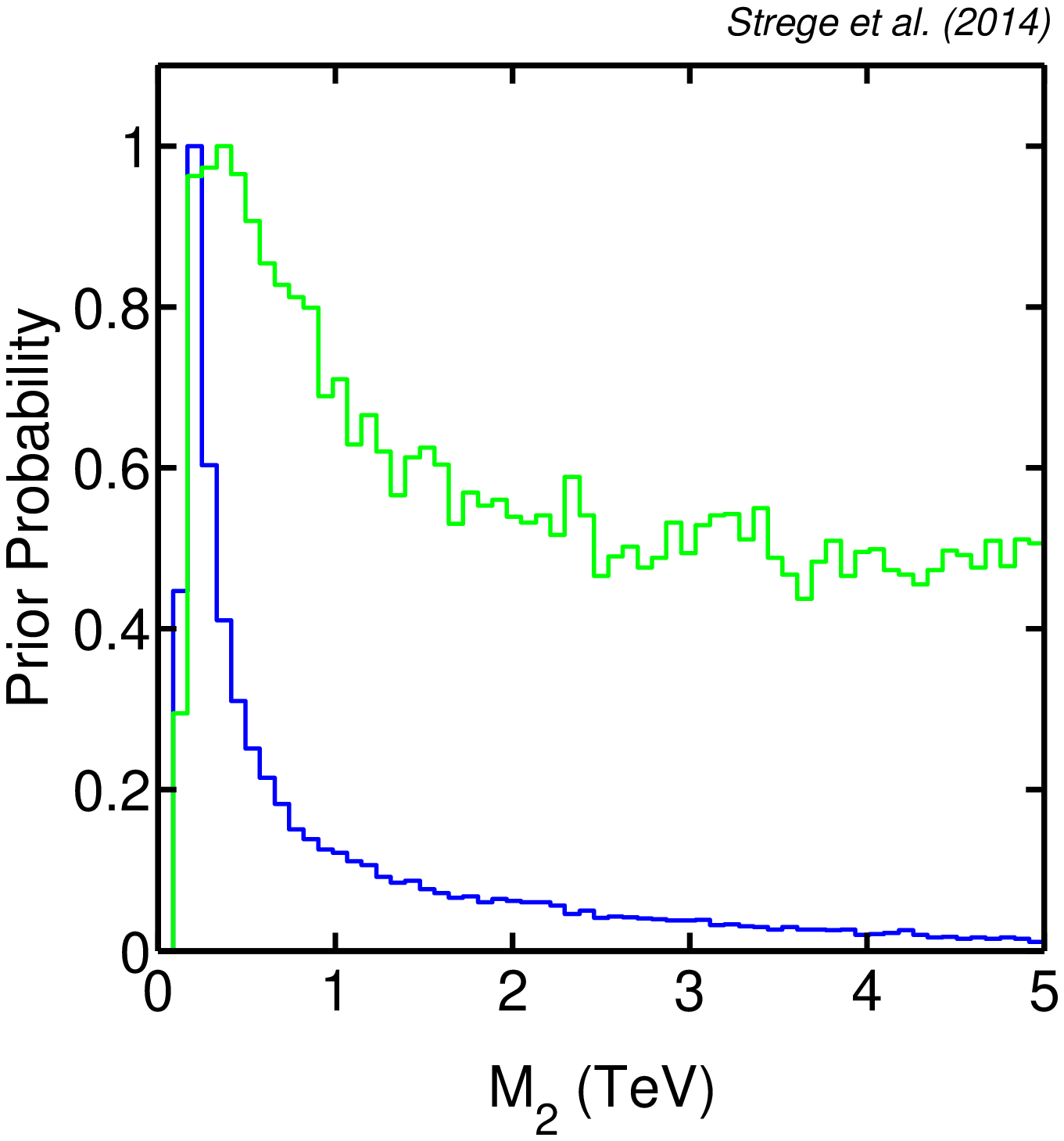}
\includegraphics[width=0.24\linewidth, trim = 0.7cm 0cm 0.7cm 0cm]{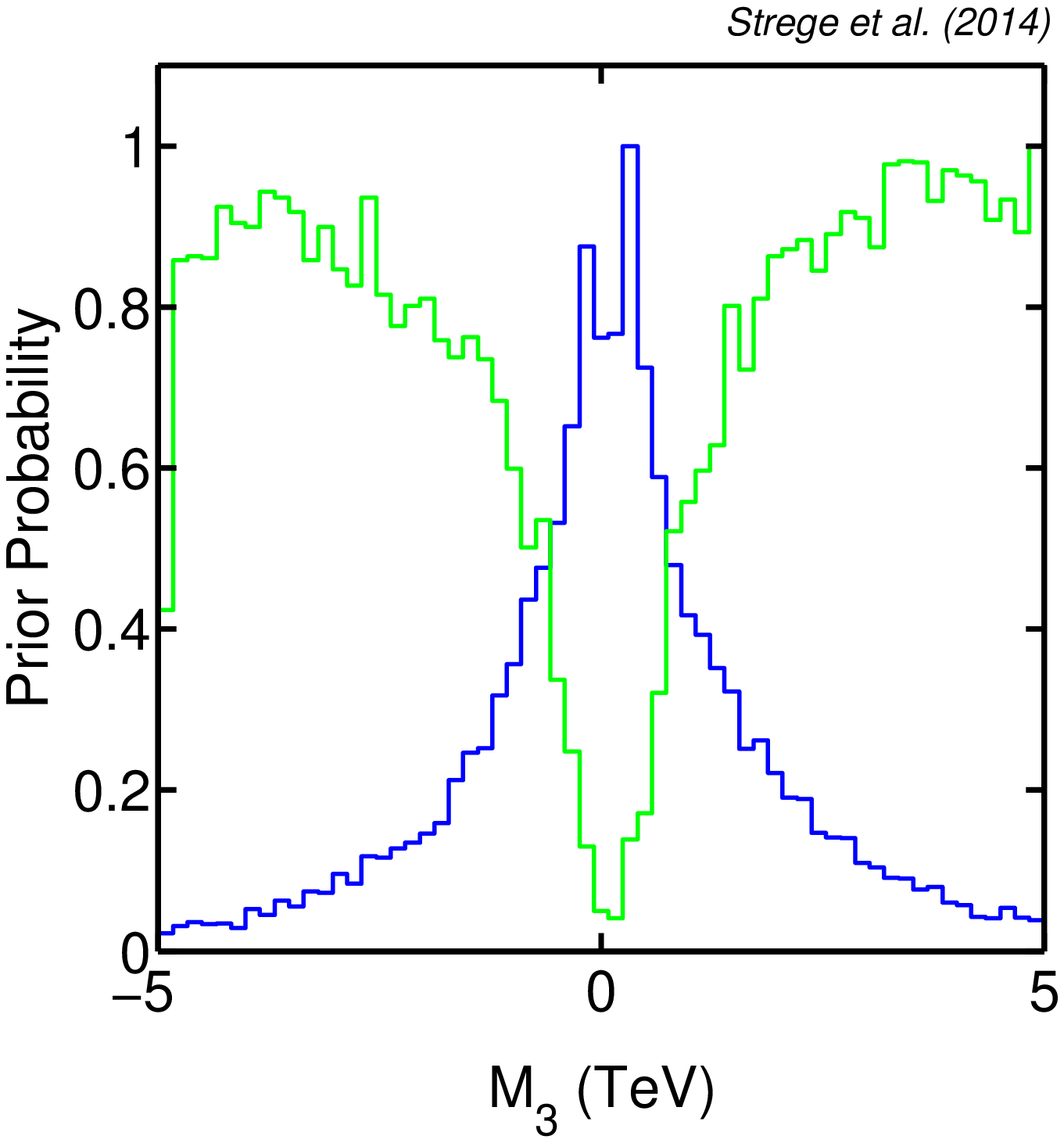}
\includegraphics[width=0.24\linewidth, trim = 0.7cm 0cm 0.7cm 0cm]{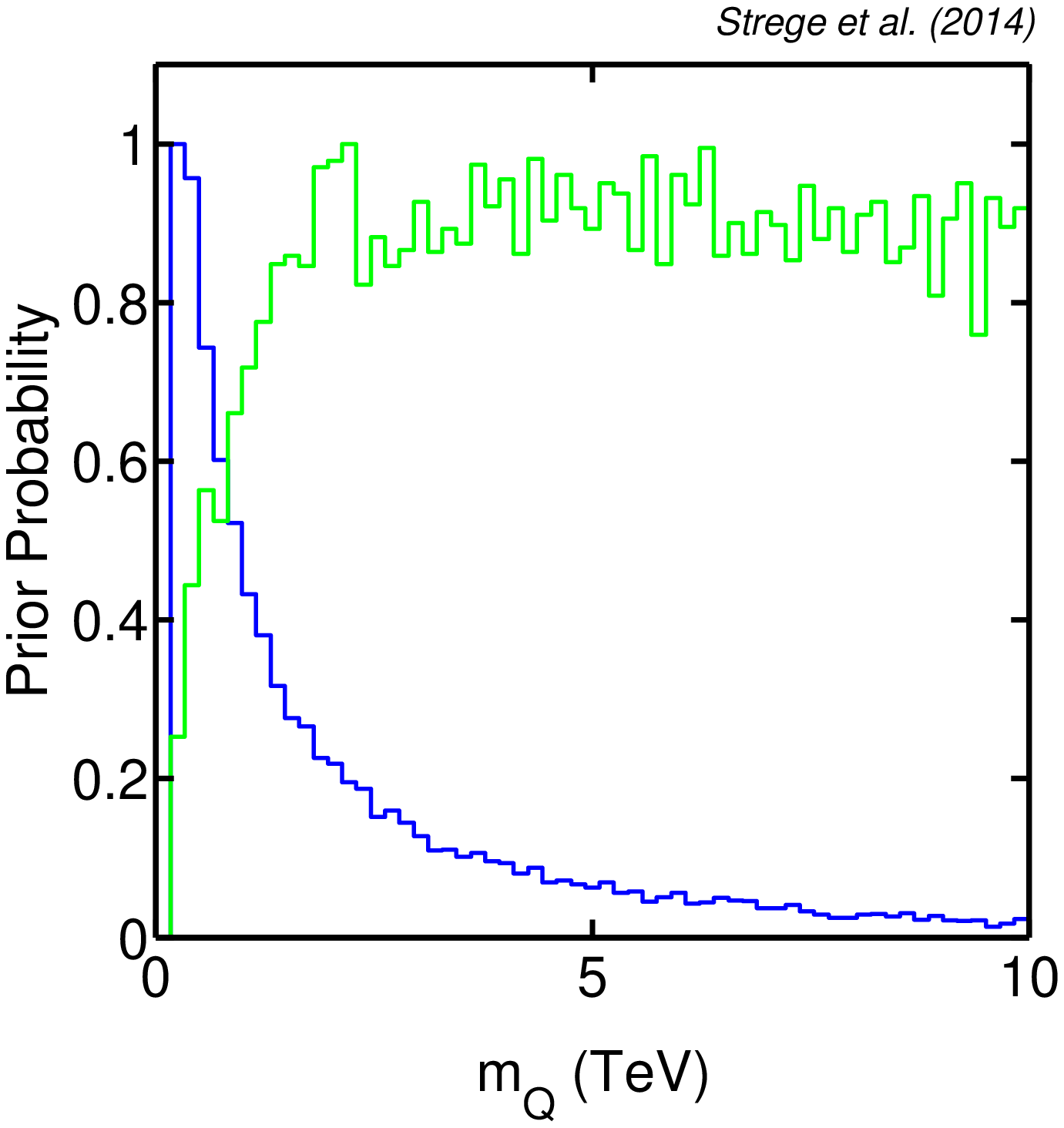} \\
\includegraphics[width=0.24\linewidth, trim = 0.7cm 0cm 0.7cm 0cm]{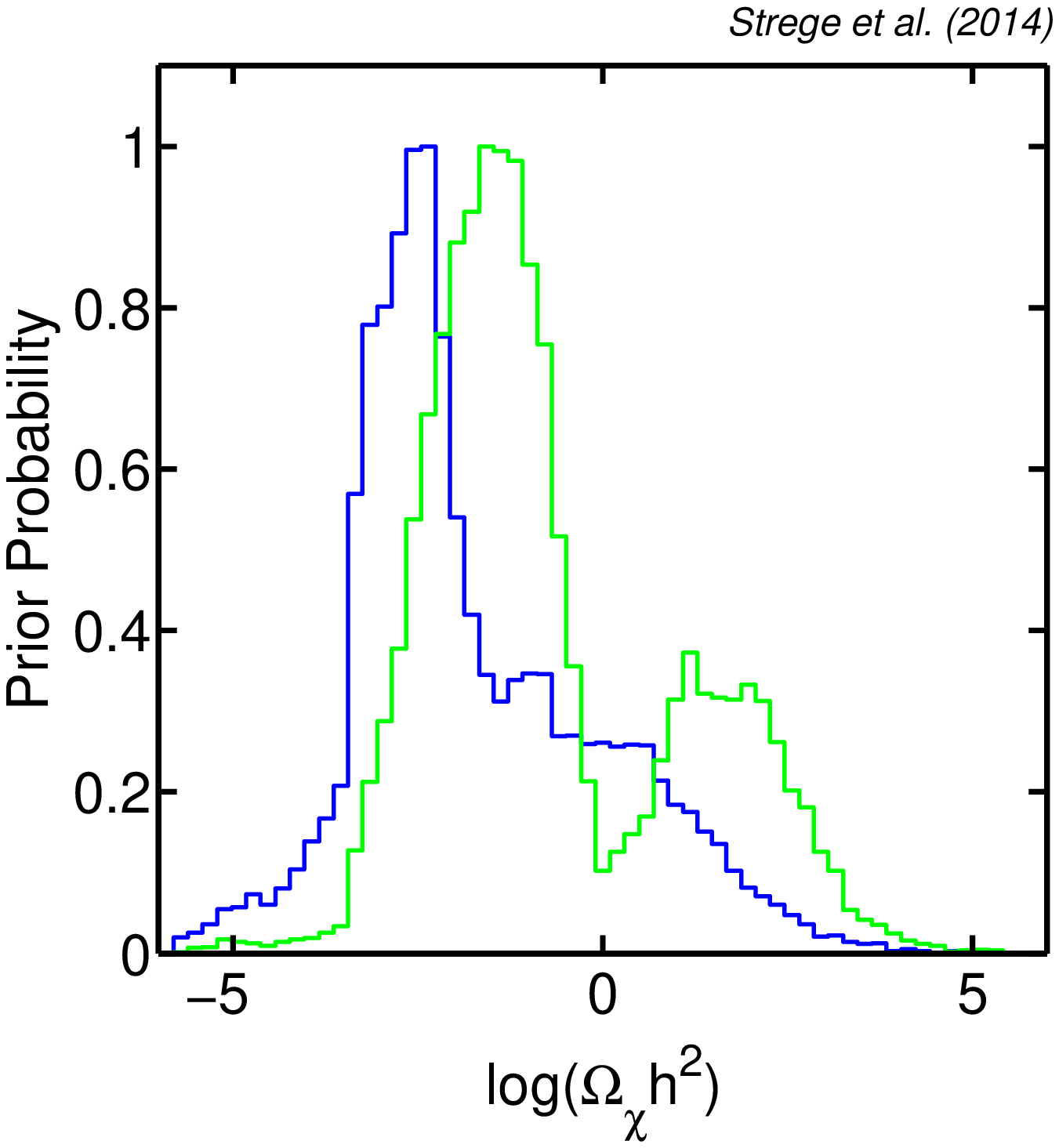}
\includegraphics[width=0.24\linewidth, trim = 0.7cm 0cm 0.7cm 0cm]{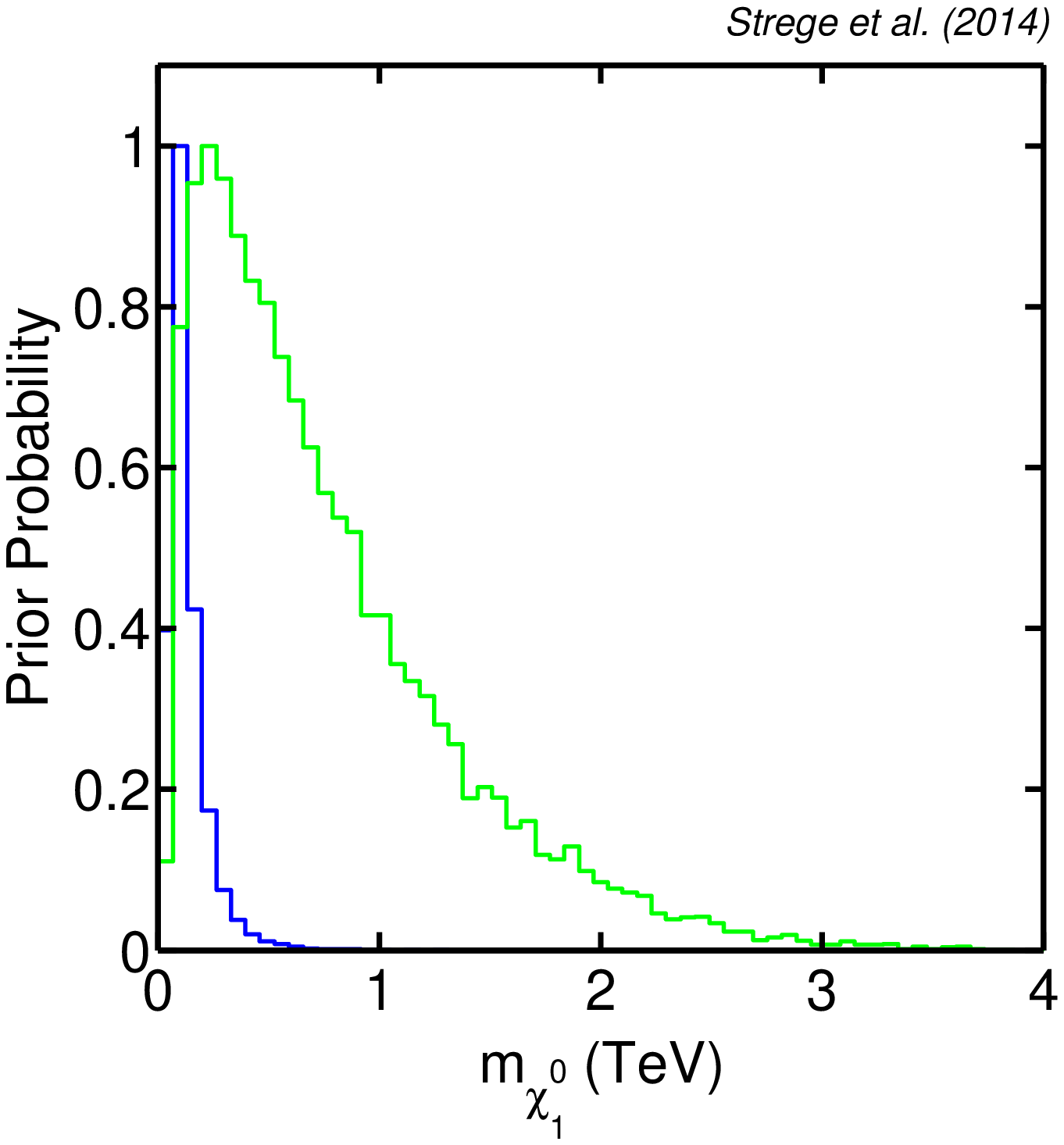}  
\includegraphics[width=0.24\linewidth, trim = 0.7cm 0cm 0.7cm 0cm]{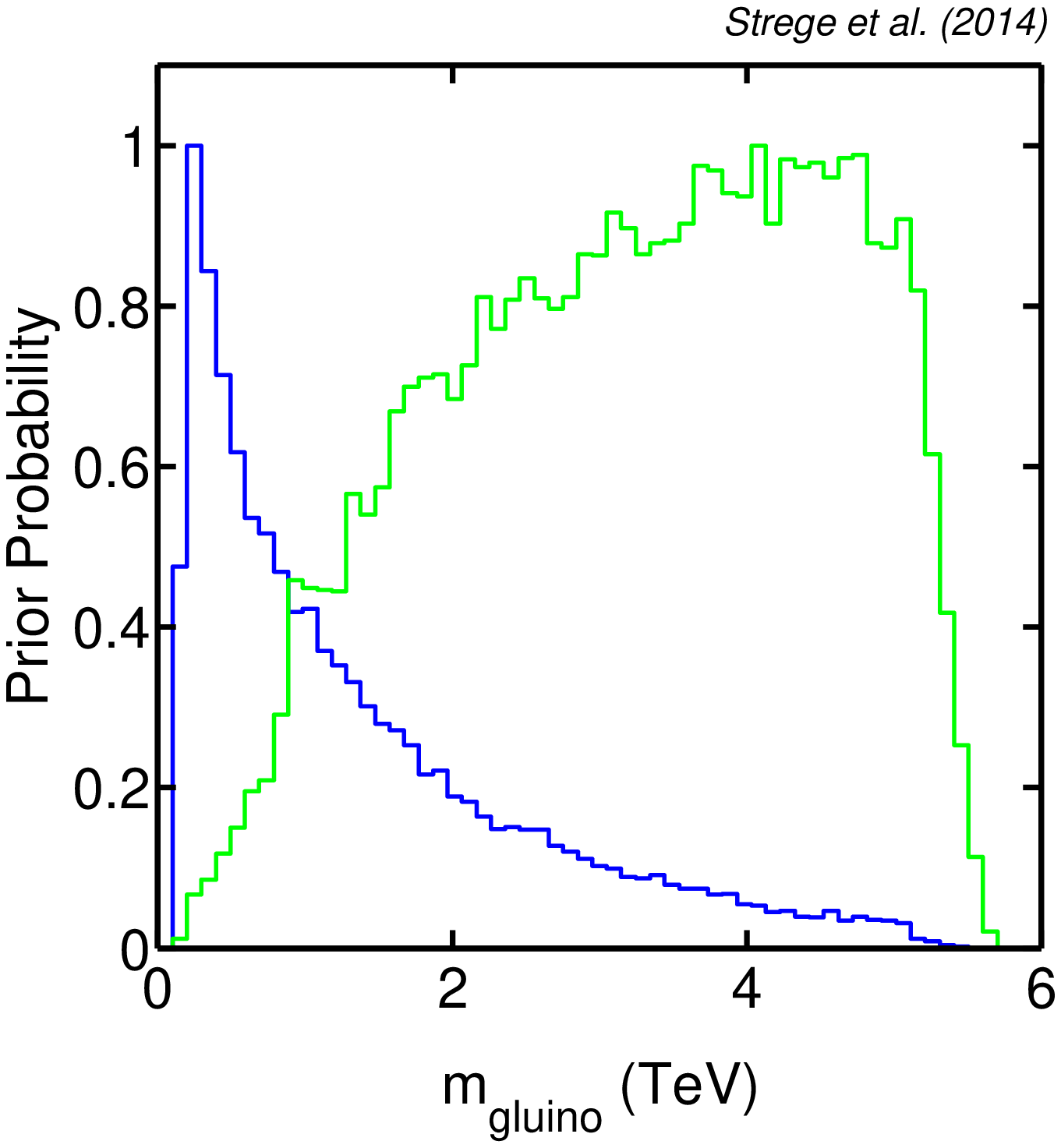} 
\includegraphics[width=0.24\linewidth, trim = 0.7cm 0cm 0.7cm 0cm]{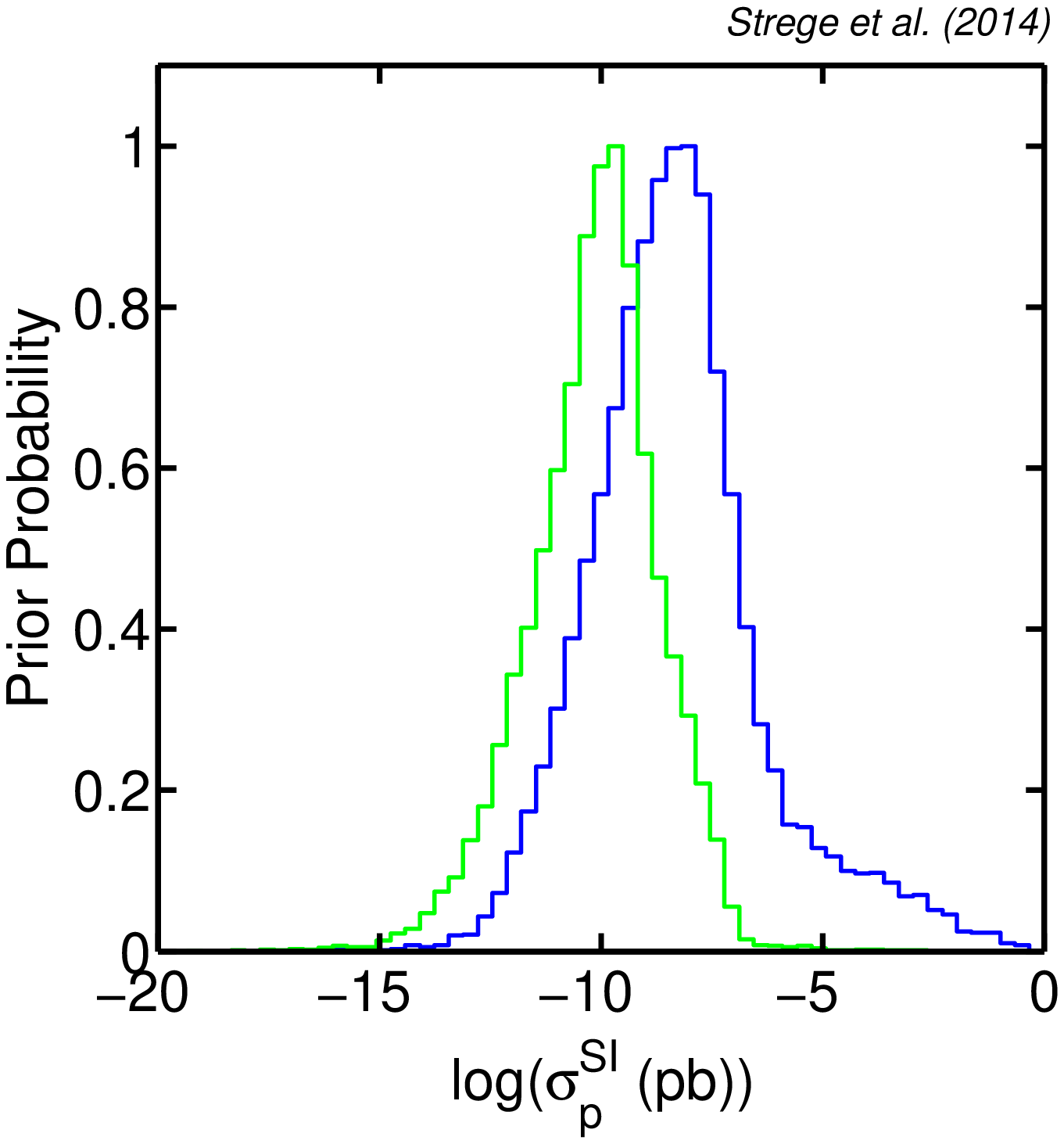} 
\caption{1D prior distributions for our two choices of priors for the three gaugino masses and the first and second generation squarks (top row).  The bottom row depicts the implied distribution for some observable of interest, namely the relic abundance, the neutralino mass, the gluino mass and the spin-independent neutralino-proton scattering cross-section. When samples from both priors are merged in our profile likelihood analysis, we obtain a detailed sampling of the entirety of the parameter space. }
\label{fig:priors}
\end{center}
\end{figure}

In our approach, the prior becomes a device to concentrate the scan in certain regions of parameter space. We adopt two very different prior distributions: ``Flat priors'' are uniform on all model parameters, while ``Log priors'' are uniform in the log of all model parameters, except for $\tan \beta$, on which a uniform prior is chosen (see Table ~\ref{pMSSM_priors}). Flat priors tend to concentrate sampling towards large values of the parameters (as most of the prior volume lies there), while log priors concentrate the scan in the low-mass region (as every decade in the parameter values is given the same a priori probability under this metric). We then merge the chains resulting from the flat and log prior scans to achieve a reliable mapping of the (prior-independent) profile likelihood function, as advocated in Ref.~\cite{Feroz:2011bj}. 

Our profile likelihood maps, which we obtain from merging the samples gathered with both priors, explore in detail both the low-mass and the high-mass region, for a more thorough scanning of the entire parameter space. This is demonstrated in Fig.~\ref{fig:priors}, which shows the 1D prior distributions (marginalised) for a few representative quantities.  

In terms of prior ranges,  we set the upper limit for the gaugino masses, $\mu$ and $m_A$ to $5$ TeV. For the squark and slepton masses and trilinear couplings we choose an upper prior boundary of $10$ TeV, to allow for large stop masses as favoured by the Higgs mass measurement. For consistency, the same upper boundary is applied to the trilinear couplings $A_t$ and $A_0$. All of the above choices for the upper boundary can be justified by considering that the profile likelihood becomes approximately flat below the boundaries, which implies that further increasing the range would have no qualitative impact on our results. For the ratio of the Higgs vacuum expectation values we chose a prior range $\tan \beta = [2, 62]$, ensuring that the Yukawa couplings do not become non-perturbatively large.

In accordance with previous analyses (see Ref.~\cite{Strege:2012bt}), we run 10 different \multinest\ scans for the ``All data" case and 5 scans for each of the ``without $g-2$" and ``Planck upper limit" cases.  We compared the best-fit points and profile likelihood function resulting from the different scans and found consistent results (within numerical noise). This verifies that a reliable exploration of the \pMSSM\ parameter space is achieved and confirms the robustness of our profile likelihood results.

The profile likelihood maps presented in this work are obtained from a combined total of 261M (all data), 124M (excluding the $g - 2$ constraint) and 91M (relaxing the requirement that the neutralino is the only dark matter component) likelihood values. We estimate that the total computational effort expended for these analyses is approximately 105 CPU years.

%%%%%%%%%%%%%%%%%%%%%%%%%%%%%%%%%%%%%%%%%%%%%%%%%%%%%%%%%%%
\subsection{Nuisance parameters and astrophysical quantities}
%%%%%%%%%%%%%%%%%%%%%%%%%%%%%%%%%%%%%%%%%%%%%%%%%%%%%%%%%%%

Residual uncertainties on the measured value of the top mass\footnote{As this paper was being finalised, a new top mass determination was presented~\cite{MTopCombo}, stemming from a joint analysis of ATLAS, CMS, CDF and D0 data, giving $M_t = 173.34 \pm 0.76 \gev$, which is compatible with the value used here.}, $M_t = 173.2 \pm 0.87 \gev$~\cite{CDF:2013jga}, can have a significant impact on the results of SUSY analyses~\cite{Roszkowski:2007fd}. Therefore, in addition to the model parameters described above we include $M_t$ as a nuisance parameter in our scans. We adopt an informative Gaussian prior for this quantity, with mean and standard deviation chosen according to recent experimental measurement above. Uncertainties in other SM parameters, namely the bottom mass $m_b(m_b)^{\bar{MS}} = 4.18\pm 0.03$~\cite{SMnuis}, the electroweak coupling constant $[\alpha_{em}(M_Z)^{\bar{MS}}]^{-1} = 127.944 \pm 0.014$~\cite{SMnuis} and the strong coupling constant $\alpha_s(M_Z)^{\bar{MS}} = 0.1184 \pm  0.0007$~\cite{SMnuis}, can also have an impact on the results of SUSY analyses~\cite{Roszkowski:2007fd}. However, this effect is subdominant compared to the impact of the top mass. Therefore, in order to keep the dimensionality of the scanned parameter space as small as possible to ensure statistical convergence of our results, we fix these three SM quantities to their experimentally measured values. 

In previous analyses of lower dimensional SUSY models, we included additional nuisance parameters in the analysis to account for residual (potentially large) uncertainties in astrophysical and nuclear physics quantities entering the likelihood for direct detection searches. A detailed discussion of the relevant uncertainties, and our parameterisation of the local astrophysical and nuclear physics was given in Ref.~\cite{Bertone:2011nj}. As shown explicitly in Ref.~\cite{Bertone:2011nj,Strege:2011pk}, the effect of marginalizing or maximising over these uncertainties is relatively small, and the main conclusions remain qualitatively unchanged when excluding the corresponding nuisance parameters from the scans. Therefore, we choose to fix these quantities in this analysis, again for the sake of limiting the dimensionality of our parameter space. 
 
The relevant astrophysical quantities are the local dark matter density, $\rho_\loc$, and three quantities entering the weakly interactive massive particle (WIMP) velocity distribution. Following our previous work~\cite{Bertone:2011nj,Strege:2011pk,Strege:2012bt}, for the WIMP velocity distribution we use the parameterisation given in Eq.(3.3) of Ref.~\cite{Bertone:2011nj}. The three velocities entering this equation are the escape velocity $v_{\textnormal{esc}} = 544$ km/s, the local circular velocity $v_{\textnormal{lsr}}=30$ km/s and the velocity dispersion $v_{\textnormal{d}}=282$ km/s. In this work, we fix these velocities to the above values, as well as the local dark matter density to $\rho_\loc = 0.4$ GeV/cm$^3$, following Refs.~\cite{Bertone:2011nj, Pato:2010zk}. 

The most important hadronic uncertainties arise in the computation of the WIMP-proton scattering cross-sections from the SUSY input parameters. The cross-section for spin-independent elastic scattering of neutralinos off atomic nuclei depends on the hadronic matrix elements $f_{T_u}$, $f_{T_d}$ and $f_{T_s}$, which parameterise the contributions of the light quarks to the proton composition $f_{T_q} \propto \langle N|\bar{q}q|N \rangle$. These matrix elements can not directly be measured, but instead there are two different approaches to calculate the values of these quantities: they can either be calculated directly using lattice QCD calculations, or derived from experimental measurements of the pion-nucleon sigma term, that can be extrapolated to zero momentum exchange, as required for the calculation of $\sigmaSI$, using chiral perturbation theory.

For $f_{T_u}$ and $f_{T_d}$ estimates from the two approaches are in reasonably good agreement, so that we use the recent results presented in Ref.~\cite{Ren:2012aj} and fix them to the experimental central values: $f_{Tu} =0.0457 \pm 0.0065$~\cite{Ren:2012aj}, $f_{Td} =0.0457 \pm 0.0065$~\cite{Ren:2012aj}. The strange content of the nucleon is much more uncertain, and different groups have found very different results for the scalar strange-quark matrix element $f_{T_s}$. While there still exist strong differences in the results of different groups extracting $f_{T_s}$ from $\pi-N$ scattering data using chiral perturbation theory, recent results from various lattice QCD computations of $f_{T_s}$ tend to be in good agreement both with each other, and with a recent analysis of pion-nucleon scattering data from the CHAOS group~\cite{Stahov:2012ca}. Therefore, in this work we use a recently determined average of various lattice QCD calculations $f_{T_s} = 0.043 \pm 0.011$~\cite{Junnarkar:2013ac}. 

The spin-dependent neutralino-proton scattering cross-section depends on the contribution of the light quarks to the total proton spin $\Delta_{u}$, $\Delta_{d}$ and $\Delta_{s}$. For these quantities, we use results from a lattice QCD computation presented in~\cite{QCDSF:2011aa}, namely $\Delta_{u} = 0.787 \pm 0.158$, $\Delta_{d} = -0.319 \pm 0.066$, $\Delta_{s}=-0.02 \pm 0.011$~\cite{QCDSF:2011aa}. As above, we fix all quantities to their central values. These results are in agreement with experimental measurements, with the possible exception of $\Delta_{s}$, which however gives a sub-dominant contribution to the total cross-section. For a recent discussion of the discrepancy between the two approaches and the impact of the resulting uncertainties on predictions for the detectability of SUSY with direct and indirect detection experiments, see Ref.~\cite{deAustri:2013saa}.

%%%%%%%%%%%%%%%%%%%%%%%%%%%%%%%%%%%%%%%%%%%%%%%%%%%%%%%%%%%
\subsection{Experimental constraints}
\label{sec:constraints}
%%%%%%%%%%%%%%%%%%%%%%%%%%%%%%%%%%%%%%%%%%%%%%%%%%%%%%%%%%%

We implement experimental constraints with a joint likelihood function,  whose logarithm takes the following form:
\begin{equation}
\ln \likeJ =  \ln \like_\text{EW}+ \ln \like_\text{B(D)} + \ln \like_{g-2} + \ln \like_{\Ohsq}  + \ln \like_\text{DD} + \ln \like_\text{Higgs} + \ln \like_\text{SUSY},     
\end{equation}
where $\like_\text{EW}$ represents electroweak precision observables, $\like_\text{B(D)}$ B and D physics constraints, $\like_{g-2}$ measurements of the anomalous magnetic moment of the muon, $\like_{\Ohsq}$ measurements of the cosmological dark matter relic density, $\like_\text{DD}$ direct dark matter detection constraints, $\like_\text{Higgs}$ LHC measurements of the properties of the Higgs boson and $\like_\text{SUSY}$ ATLAS sparticles searches. We discuss each component in turn. The values used are also summarised in Table~\ref{tab:exp_constraints}. 

%%%%%%%%%%%%%%%%%%%%%%%%%%%%%%%%%%%%%%%%%%%%%%%%%%%%%%%%%%
\subsubsection{Electroweak precision observables}
%%%%%%%%%%%%%%%%%%%%%%%%%%%%%%%%%%%%%%%%%%%%%%%%%%%%%%%%%
 
Constraints on several observables obtained from Z-pole measurements at LEP~\cite{lepwwg} are included: the constraint on the effective electroweak mixing angle for leptons $\sin^2\theta_\text{eff}$, the total width of the Z-boson $\Gamma_{Z}$, the hadronic pole cross-section $\sigma^0_{had}$, as well as the decay width ratios $R^0_l$, $R^0_b$, $R^0_c$. We do not include the constraints on the asymmetry parameters $A_l$, $A_b$, $A_c$ and $A^{0,l}_{FB}$, $A^{0,b}_{FB}$, $A^{0,c}_{FB}$ in the analysis, since we found that supersymmetric contributions to their value were very small and well below the experimental error. Due to the strong correlation between these parameters and $\sin^2\theta_\text{eff}$, the inclusion of these constraints would qualitatively not change the profile likelihood contours.
In addition, we also use the measurement of the mass of the W boson $m_W$ from the LEP experiment~\cite{lepwwg}. We apply a Gaussian likelihood for all of these quantities, with mean and standard deviation as in Table~\ref{tab:exp_constraints}. 

%%%%%%%%%%%%%%%%%%%%%%%%%%%%%%%%%%%%%%%%%%%%%%%%%%%%%%%%%
\subsubsection{B and D physics constraints}
%%%%%%%%%%%%%%%%%%%%%%%%%%%%%%%%%%%%%%%%%%%%%%%%%%%%%%%%

Several B and D physics constraints are applied with a Gaussian likelihood, as summarised in Table~\ref{tab:exp_constraints}. We include a number of results obtained by the Heavy Flavor Averaging Group, including the measurement of the branching fraction of the decay $\brbsgamma$, the ratio of the measured branching fraction of the decay $B_u \to \tau \nu$ to its branching fraction predicted in the SM, and the decay branching fraction $\Dstaunu$~\cite{hfag}. 
Additionally, we include the ratio of the measurement of the $B_s^0-\bar{B_s^0}$ oscillation frequency to its SM value $R_{\Delta M_{B_s}} = 1.04 \pm 0.11$~\cite{deltambs}.
We also include the constraint on the integrated forward-backward asymmetry $\afb$ in the bin  $q^2  \in [1, 6] \gev^2$, which has been shown to have a powerful impact on simple SUSY models~\cite{Mahmoudi:2012un}.

Finally, we include the latest measurement of the rare decay $\brbsmumu$ from the LHCb experiment at the LHC. Using a combination of 1.0  fb$^{-1}$ data at $\sqrt{s}$ = 7 TeV collision energy and 1.1 fb$^{-1}$ data at $\sqrt{s}$ = 8 TeV collision energy, collected in 2011 and 2012, the LHCb collaboration reported an excess of decay events with respect to the background expectation, leading to the value $\brbsmumu = (3.2^{+1.5}_{-1.2}) \times 10^{-9}$ with a 3.5$\sigma$ signal significance~\cite{:2012ct}\footnote{This constraint is in good agreement with the CMS measurement of this quantity, $\brbsmumu = (3.0^{+1.0}_{-0.9}) \times 10^{-9}$~\cite{Chatrchyan:2013bka}, which became available at a later date and is thus not included in the analysis.}. We apply this constraint as a Gaussian likelihood function with a conservative (symmetric) experimental uncertainty of $1.5 \times 10^{-9}$ and a theoretical error of $0.38 \times 10^{-9}$~\cite{Arbey:2012ax}.

We include a measurement of the isospin asymmetry between $B^0$ and $B^+$ decay widths from the radiative decay $B \rightarrow K^* \gamma$. We combine results from three different groups to obtain the constraint $\DeltaO = (3.1 \pm 2.3) \times 10^{-2}$~\cite{delta0} and  we adopt a theoretical error of $1.75 \times 10^{-2}$~\cite{Mahmoudi:2008tp}.

%%%%%%%%%%%%%%%%%%%%%%%%%%%%%%%%%%%%%%%%%%%%%%%%%%%%%%%
\subsubsection{Cosmological relic abundance}
%%%%%%%%%%%%%%%%%%%%%%%%%%%%%%%%%%%%%%%%%%%%%%%%%%%%%%

We include the Planck constraint on the dark matter relic abundance in our analysis. When assuming that the neutralino makes up all of the dark matter in the universe, we apply the result from Planck temperature and lensing data $\Ohsq = 0.1186 \pm 0.0031$ as a Gaussian likelihood in the analysis~\cite{Ade:2013zuv}. We also add a (fixed) theoretical uncertainty, $\tau = 0.012$, in quadrature, in order to account for the numerical uncertainties entering in the calculation of the relic density from the SUSY parameters. 

When we allow for the possibility that neutralinos are a sub-dominant dark matter component,  then the Planck relic density measurement is applied as an upper limit.  As shown in the Appendix of~\cite{Bertone:2010ww}, the effective likelihood for this case is given by the expression 
\be \label{eq:upperbound}
\like_{\Ohsq} = \like_0 \int_{\OhLSP/\siW}^\infty  
e^{-\frac{1}{2}(x-r_\star)^2} x^{-1}{\rm d}x ,
\ee
where $ \like_0$ is an irrelevant normalization constant,  $r_\star \equiv \muW/\siW$ and $\OhLSP$ is the predicted relic density of neutralinos as a function of the model parameters.

When neutralinos are not the only constituent of dark matter, the rate of events in a direct detection experiment is proportionally smaller as the local neutralino density,  $\rho_\chi$,  is now smaller than the total local dark matter density, $\rho_\text{DM}$. The suppression is given by the factor $\xi \equiv \rho_\chi / \rho_{\rm DM}$. Following~\cite{Bertone:2010rv}, we assume that ratio of local neutralino and total dark matter densities is equal to that for the cosmic abundances, thus we adopt the scaling Ansatz 
\be \label{eq:scaling_ansatz}
\xi \equiv \rho_\chi / \rho_{\rm DM}= \Omega_\chi / \Omega_{\rm DM}. 
\ee
For $\Omega_{\rm DM}$ we adopt the central value measured by Planck, $\Omega_{\rm DM} = 0.1186$~\cite{Ade:2013zuv}. 

%%%%%%%%%%%%%%%%%%%%%%%%%%%%%%%%%%%%%%%%%%%%%%%%%%%%%
\subsubsection{Direct detection constraints}
%%%%%%%%%%%%%%%%%%%%%%%%%%%%%%%%%%%%%%%%%%%%%%%%%%%%%

We include the latest constraints from the XENON100 direct detection experiment, obtained from 224.6 live days and 34 kg fiducial volume~\cite{Aprile:2012nq}. The data set contained two candidate WIMP scattering events inside the signal region, compatible with the expected number of background events $b = 1.0 \pm 0.2$. The resulting XENON100 exclusion limits currently places tight limits in the plane of WIMP mass $m_\chi$ vs. spin-independent cross-section $\sigmaSI$, as well as in the ($m_\chi$, $\sigma_{\neut-n}^\text{SD}$) plane, and also places competitive constraints on $\sigmaSDp$ as a function of WIMP mass~\cite{Aprile:2012nq,Aprile:2013doa}. We use an approximate XENON100 likelihood function to incorporate these data in our analysis. For a detailed description of our approximate XENON100 likelihood function we refer the reader to Ref.~\cite{Bertone:2011nj,Strege:2012bt}.

In previous studies of the cMSSM~\cite{Strege:2012bt,Strege:2011pk} and the NUHM~\cite{Strege:2012bt} we neglected the contribution of spin-dependent neutralino-nucleon scattering to the total number of events, since in these constrained models this contribution was always subdominant compared to the spin-independent event rate, and in the favoured regions the number of events from spin-dependent scattering was much smaller than 1. However, in models like the \pMSSM\ considered here there are regions of parameter space in which the spin-dependent scattering event rate can exceed the spin-independent contribution. Therefore, we now include the spin-dependent contribution to the event rate, and calculate the constraints by using $R^{\textnormal{tot}} = R^{\textnormal{SD}}+R^{\textnormal{SI}}$. For the axial-vector structure functions entering into the spin-dependent differential WIMP-nucleus cross-section we use the results by Ref.~\cite{Menendez:2012tm}, as 
recommended by the XENON100 collaboration~\cite{Aprile:2013doa}. 

While this work was being finalised, the LUX collaboration reported results from a search for WIMPs based on 85.3 live days of data and 118 kg fiducial volume~\cite{Akerib:2013tjd}. No significant excess above the background expectation was observed, and new limits on the WIMP properties were derived. The resulting limit on the spin-independent WIMP-proton interaction improved on the XENON100 limits used here, by a factor of $\sim 2$ for WIMP masses $m_\chi \gsim 50 \gev$ and by a larger factor for lighter WIMPs. We do not implement the LUX results in this paper, but notice that their impact is comparatively small given the many orders of magnitude spanned by the predictions of the \pMSSM\ in the relevant spin-independent vs mass plane (see Fig.~\ref{2D_plots}, upper left panel).  

%%%%%%%%%%%%%%%%%%%%%%%%%%%%%%%%%%%%%%%%%%%%%%%%
\subsubsection{Anomalous magnetic moment of the muon}
%%%%%%%%%%%%%%%%%%%%%%%%%%%%%%%%%%%%%%%%%%%%%%%%

The experimentally measured value of the anomalous magnetic moment of the muon~\cite{Davier:2010nc} $a_{\mu} \equiv (g-2)/2$ shows a $3.6\sigma$ discrepancy with the value predicted in the SM. Therefore, a strong supersymmetric contribution is required in order to explain the discrepancy $\delta a_{\mu}^{\text{SUSY}} = (28.7 \pm 8.2) \times 10^{-9}$, where experimental and theoretical errors have been added in quadrature. However, there remain strong theoretical uncertainties in the computation of the SM value of the muon anomalous magnetic moment, most importantly in the computation of the hadronic loop contributions. Additionally, the discrepancy between the experimental measurement and the SM value is reduced to $2.4\sigma$ when relying on $\tau$ data instead of $e^+e^-$ data~\cite{Davier:2010nc}.

In previous global fits analyses of constrained MSSM scenarios, such as the cMSSM~\cite{Strege:2012bt,Strege:2011pk} and the NUHM~\cite{Strege:2012bt}, it was found that the constraint on $g-2$ can play a dominant role in driving the profile likelihood results. In order to evaluate the dependence of the our results on this somewhat controversial constraint, we repeat our analysis after excluding the $g-2$ constraint from the likelihood.

%%%%%%%%%%%%%%%%%%%%%%%%%%%%%%%%%%%%%%%%%%%%%%%
\subsubsection{Higgs properties}
%%%%%%%%%%%%%%%%%%%%%%%%%%%%%%%%%%%%%%%%%%%%%%

Both ATLAS and CMS have recently reported new results for the measurement of the mass of the Higgs boson. CMS reported a measured value of $\mhl = 125.8 \pm 0.4 \pm 0.4$ GeV, where the first error is statistical and the second error is systematic~\cite{CMS_Higgs}. This result was derived from a combination of 5.1 fb$^{-1}$ data at $\sqrt{s}$ = 7 TeV collision energy, and 12.2 fb$^{-1}$ data at $\sqrt{s}$ = 8 TeV collision energy. The ATLAS collaboration found a value $\mhl = 125.5 \pm 0.2^{+0.5}_{-0.6}$ GeV, derived from a combination of 4.8 fb$^{-1}$ $\sqrt{s}$ = 7 TeV data and 20.7 fb$^{-1}$ $\sqrt{s}$ = 8 TeV data~\cite{ATLAS_Higgs}. We use a weighed average of the ATLAS and CMS measurements, resulting in $\mhl = 125.66 \pm 0.41$ GeV. We add a theoretical error of 2 GeV~\cite{Allanach:2004rh} in quadrature. 

The observation of the Higgs boson in several channels has allowed the measurement of some of its couplings with relatively good accuracy. The standard way to infer the couplings of the produced Higgs boson is to consider their deviation from the SM expectation. For a given channel this is parametrized through the signal strength parameter $\mu$. For the $h \rightarrow X  X$ channel one has
\begin{equation}
 \mu_{XX} = \frac{\sigma(pp \rightarrow h) \times BR(h \rightarrow X X)}{\sigma(pp \rightarrow h)_{SM} \times BR(h \rightarrow X X)_{SM}}.  
\end{equation}
This quantity is compared directly with experimental measurements. Note that from here on what we call ``Higgs" in the MSSM context, we are referring to the lightest CP-even Higgs.

The channels considered in the likelihood function are listed in Table~\ref{tab:exp_constraints}. We apply the experimental constraints obtained by the CMS collaboration. For the $\gamma \gamma$~\cite{Higgsgg}, $W^+ W^-$~\cite{Higgsww}, $ZZ$~\cite{Higgszz} and $\tau^+ \tau^-$~\cite{Higgstautau} decay modes the constraints were derived from datasets corresponding to an integrated luminosity of $\sim5$ fb$^{-1}$ at $\sqrt{s} = 7$ TeV collision energy and $\sim19$ fb$^{-1}$ at $\sqrt{s} = 8$ TeV collision energy. The constraint on the $h \rightarrow b \bar{b}$~\cite{Higgsbb} decay channel was derived from $\sim5$ fb$^{-1}$ integrated luminosity at $\sqrt{s} = 7$ TeV and $\sim12$ fb$^{-1}$ integrated luminosity at $\sqrt{s} = 8$ TeV collision energy.

%%%%%%%%%%%%%%%%%%%%%%%%%%%%%%%%%%%%%%%%%%%%%%%%%%%%%%%%%%%%%%%%
\subsubsection{ATLAS SUSY searches}
%%%%%%%%%%%%%%%%%%%%%%%%%%%%%%%%%%%%%%%%%%%%%%%%%%%%%%%%%%%%%%%

The SUSY searches constraints applied come from bounds on SUSY masses from LEPII and Tevatron for which we apply the likelihood as outlined in~\cite{deAustri:2006pe} and from LHC searches looking for 0-lepton and multi-jets with missing transverse energy and 3-leptons with missing transverse energy in the ATLAS experiment, both with data recorded at $\sqrt{7}$ TeV and a total integrated luminosity of 4.7 fb$^{-1}$~\cite{ATLAS-CONF-2012-033-5fb,3-lepton-ana}. Details about the construction and validation of the LHC likelihood function associated with these two channels are given in Appendix~\ref{LHC_likelihood} and Appendix~\ref{LHC_validation}, respectively. The likelihood implementation has been done in the ROOT framework through the RooFit/RooStats packages.

For each likelihood evaluation, we simulate the SUSY kinematical distributions of  $10^4$ events with PYTHIA 6.4~\cite{pythia} using the ATLAS MC09 tune~\cite{ATLAS:2010zyu}. 
The parton distribution functions are obtained from the CTEQ6L1 set~\cite{Pumplin:2002vw}. The SUSY cross-sections for gluino and squarks production are normalized by 
NLO K-factors in the strong coupling constant, including the resummation of soft gluon emission at next-to-leading-logarithmic (NLO+NLL) accuracy with NLL-fast 1.2~\cite{nllfast} and 
outside the available NLL-fast grid by PROSPINO2~\cite{prospino} at NLO. For the electroweakino production we use PROSPINO2 which provides a NLO calculation. 
The detector simulation employed is DELPHES3~\cite{delphes}. Details about the efficiencies validation are provided in Appendix~\ref{LHC_validation}.

%%%%%%%%%%%%%%%%%%%%%%%%%%%%%%%%%%%%%%%%%%%%%%%%%%%%%%%%%%%%%
%\section{Data}
%%%%%%%%%%%%%%%%%%%%%%%%%%%%%%%%%%%%%%%%%%%%%%%%%%%%%%%%%%%%%

\begin{table*}
\begin{center}
\begin{tabular}{|l | l l l | l|}
\hline
\hline
Observable & Mean value & \multicolumn{2}{c|}{Standard deviation} & Ref. \\
 &   $\mu$      & ${\sigma}$ (exper.)  & $\tau$ (theor.) & \\\hline
$M_W$ [GeV] & 80.385 & 0.015 & 0.01 & \cite{lepwwg} \\
$\sin^2\theta_\text{eff}$ & 0.23153 & 0.00016 & 0.00010 & \cite{lepwwg} \\
$\Gamma_{Z}$ [GeV] & 2.4952 & 0.0023 & 0.001 & \cite{lepwwg} \\
$\sigma^0_{had}$ [nb] & 41.540 & 0.037 & - & \cite{lepwwg} \\
$R^0_l$ & 20.767 & 0.025 & - & \cite{lepwwg} \\
$R^0_b$ & 0.21629 & 0.00066 & - & \cite{lepwwg} \\
$R^0_c$ & 0.1721 & 0.003 & - & \cite{lepwwg} \\
$^\#$$A^{0,l}_{FB}$ & 0.0171 & 0.001 & - & \cite{lepwwg} \\
$^\#$$A^{0,b}_{FB}$ & 0.0992 & 0.0016 & - & \cite{lepwwg} \\
$^\#$$A^{0,c}_{FB}$ & 0.0707 & 0.0035 & - & \cite{lepwwg} \\
$^\#$$A_{l} (SLD)$ & 0.1513 & 0.0021 & - & \cite{lepwwg} \\
$^\#$$A_{b}$ & 0.923 & 0.02 & - & \cite{lepwwg} \\
$^\#$$A_{c}$ & 0.670 & 0.027 & - & \cite{lepwwg} \\
$\delta a_\mu^{\mathrm{SUSY}} \times 10^{10}$ & 28.7 & 8.0 & 2.0 & \cite{Davier:2010nc} \\
$\brbsgamma \times 10^4$ & 3.55 & 0.26 & 0.30 & \cite{hfag}\\
$R_{\Delta M_{B_s}}$ & 1.04 & 0.11 & - & \cite{deltambs} \\
$\RBtaunu$   &  1.63  & 0.54  & - & \cite{hfag}  \\
$\DeltaO  \times 10^{2}$   &  3.1 & 2.3  & 1.75 & \cite{delta0}  \\
$^\#$$\RBDtaunuBDenu \times 10^{2}$ & 41.6 & 12.8 & 3.5  & \cite{Aubert:2007dsa}  \\
$^\#$$\Rl$ & 0.999 & 0.007 & -  &  \cite{Antonelli:2008jg}  \\
$\afb$ & -0.18 & 0.063 & 0.05 &  \cite{Mahmoudi:2012un}  \\
$\Dstaunu \times 10^{2}$ & 5.44 & 0.22 & 0.1  & \cite{hfag}  \\
$^\#$$\Dsmunu  \times 10^{3}$ & 5.54 & 0.24 & 0.2  & \cite{hfag}  \\
$^\#$$\Dmunu \times 10^{4}$  & 3.82 & 0.33 & 0.2  & \cite{hfag} \\
$\brbsmumu \times 10^{9} $ & 3.2 & 1.5 & 0.38 & \cite{:2012ct}\\
$\relic h^2$ & 0.1186 & 0.0031 & 0.012 & \cite{Ade:2013zuv} \\
$\mhl$ [GeV] & 125.66  & 0.41  & 2.0 & \cite{ATLAS_Higgs,CMS_Higgs} \\
$^\dagger$$\mugg$  &  0.78 & 0.27  & 15\% & \cite{Higgsgg}  \\
$^\dagger$$\muww$ &  0.76 & 0.21  & 15\% & \cite{Higgsww}  \\
$^\dagger$$\muzz$ &  0.91 & 0.27  & 15\% & \cite{Higgszz}  \\
$^\dagger$$\mubb$ &  1.3 & 0.65  & 15\% & \cite{Higgsbb}  \\
$^\dagger$$\mutautau$ &  1.1 & 0.4  & 15\% & \cite{Higgstautau}  \\
\hline\hline
   &  Limit (95\%~$\cl$)  & \multicolumn{2}{r|}{$\tau$ (theor.)} & Ref. \\ \hline
Sparticle masses  &  \multicolumn{3}{c|}{LEP, Tevatron. As in Table~4 of
  Ref.~\cite{deAustri:2006pe}.}  & \cite{deAustri:2006pe}\\
 $^\dagger$0-lepton SUSY search & \multicolumn{3}{c|}{ATLAS, $\sqrt{s} = 7$ TeV, $4.7$ fb$^{-1}$} & \cite{ATLAS-CONF-2012-033-5fb} \\  
 $^\dagger$3-lepton SUSY search &  \multicolumn{3}{c|}{ATLAS, $\sqrt{s} = 7$ TeV, $4.7$ fb$^{-1}$} & \cite{3-lepton-ana} \\
$m_\chi - \sigmaSI$ & \multicolumn{3}{l|}{XENON100 2012 limits ($224.6 \times 34$ kg days)} & \cite{Aprile:2012nq} \\
$m_\chi - \sigmaSDp$ & \multicolumn{3}{l|}{XENON100 2012 limits ($224.6 \times 34$ kg days)} & \cite{Aprile:2013doa} \\
\hline
\end{tabular}
\end{center}
\caption{\fontsize{9}{9} \selectfont Summary of experimental constraints used in the likelihood. Upper part: measured observables, modelled with a Gaussian likelihood of mean $\mu$, and standard deviation $(\sigma^2+\tau^2)^{1/2}$, where $\sigma$ is the experimental and $\tau$ the theoretical uncertainty. Lower part: observables for which only limits currently exist. See text for further information on the explicit form of the likelihood function. Experimental constraints tagged with $^\#$ have been found to contribute an approximately constant value to the log-likelihood and hence have been omitted. Observables tagged with $^\dagger$ are applied via post-processing of the samples. \label{tab:exp_constraints}}
\end{table*}

%%%%%%%%%%%%%%%%%%%%%%%%%%%%%%%%%%%%%%%%%%%%%%%%%%%%%%%%%%%
\section{Results}
\label{sec:results}
%%%%%%%%%%%%%%%%%%%%%%%%%%%%%%%%%%%%%%%%%%%%%%%%%%%%%%%%%

In the following sections we present the combined impact of all present day constraints shown in Table~\ref{tab:exp_constraints} on the \pMSSM. In addition to the analysis including all available data, we also show results for two other cases. In particlar, we present results for an analysis excluding the $g - 2$ constraint, in order to evaluate the impact of this controversial measurement on our profile likelihood results. In a third analysis we relax the requirement that the lightest neutralino is the only dark matter component, by adopting the Planck measurement of the dark matter relic density as an upper limit.

We start by presenting 1D and 2D profile likelihood results for all three cases excluding LHC constraints on the sparticle masses and the Higgs production cross-sections (note however that the LHC measurement of the Higgs mass, $m_h$, is included in all of the results). Since this paper presents the first high-resolution profile likelihood analysis of the \pMSSM, we discuss in detail the favoured model phenomenology, in particular the different neutralino compositions that can be achieved throughout the parameter space and the dark matter detection prospects. In the final section we present the impact of constraints from LHC SUSY searches and Higgs signal strengths measurements on this parameter space, obtained with a simplified statistical treatment.  A full profile likelihood analysis of the \pMSSM\ including all LHC constraints is beyond the scope of this paper, and is the focus of a dedicated work~\cite{LHC_PP_Future}.

%%%%%%%%%%%%%%%%%%%%%%%%%%%%%%%%%%%%%%%%%%%%%%%%%%%%%%%%%%%%%
\subsection{Global fits from all data and excluding $g-2$}
\label{sec:1DPL_wog2}
%%%%%%%%%%%%%%%%%%%%%%%%%%%%%%%%%%%%%%%%%%%%%%%%%%%%%%%%%%%%%

We begin by showing in Fig.~\ref{fig:1D_wog2_1}--\ref{fig:1D_wog2_3} the combined impact of the present day constraints shown in Table~\ref{tab:exp_constraints}, with the exception of the LHC constraints on SUSY and the Higgs production cross-sections (which are discussed separately below). We compare results for the analysis including all data (red), and for the analysis excluding the $g - 2$ constraint (purple); the encircled crosses show the corresponding best-fit points. For observable quantities, the applied likelihood function is shown in black. Fig.~\ref{fig:1D_wog2_1} shows the 1D profile likelihood results for the 15 input model parameters and the top mass, Fig.~\ref{fig:1D_wog2_2} shows results for the relevant observables and Fig.~\ref{fig:1D_wog2_3} shows results for some SUSY quantities of interest. From here on we will refer to the lightest neutralino as ``neutralino'', for brevity. 

\subsubsection{Profile likelihood for the \pMSSM\ parameters}

\begin{figure}
\begin{center}
\includegraphics[width=0.24\linewidth, trim = 0.7cm 0cm 0.7cm 0cm]{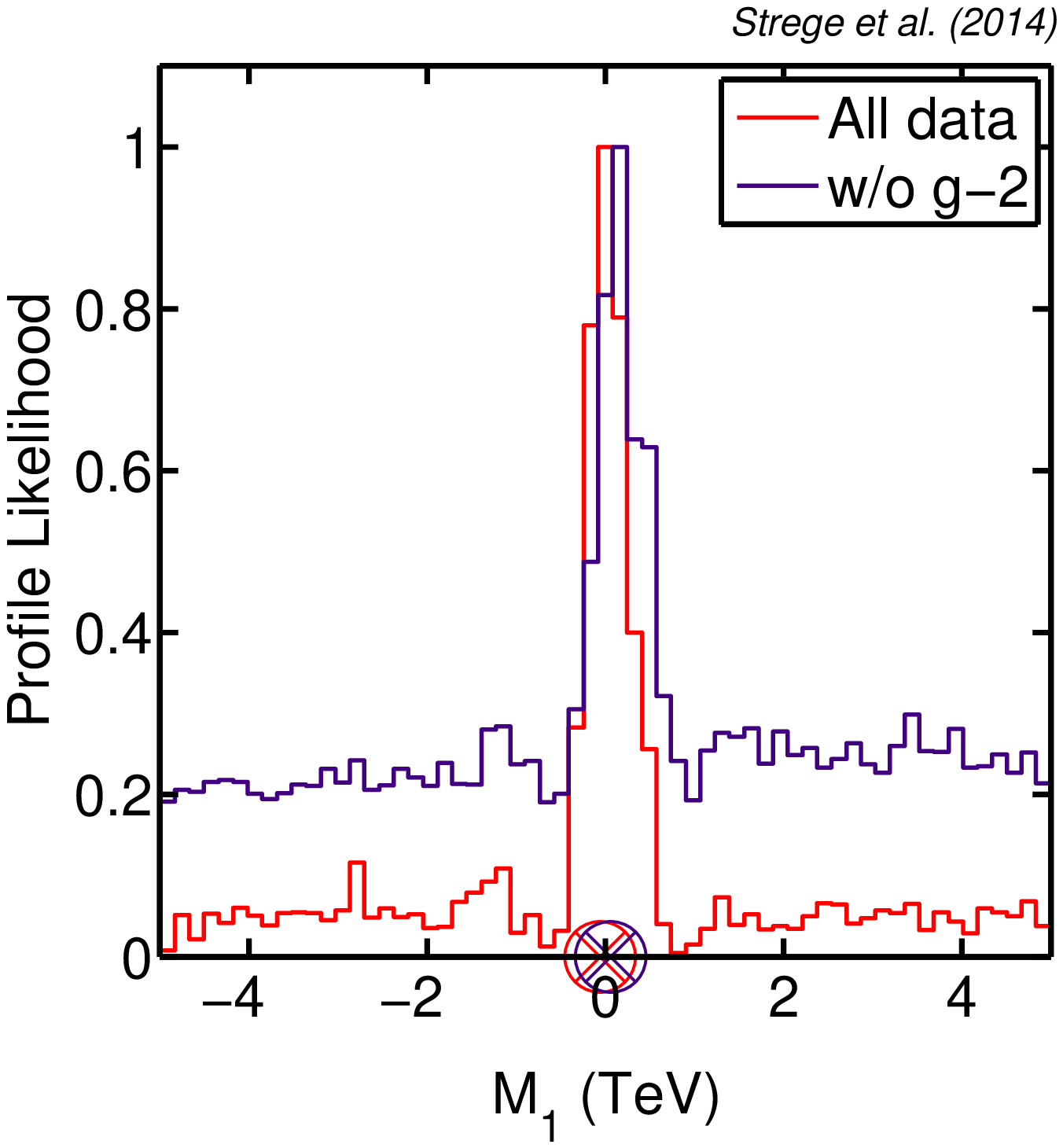} \hfill
\includegraphics[width=0.24\linewidth, trim = 0.7cm 0cm 0.7cm 0cm]{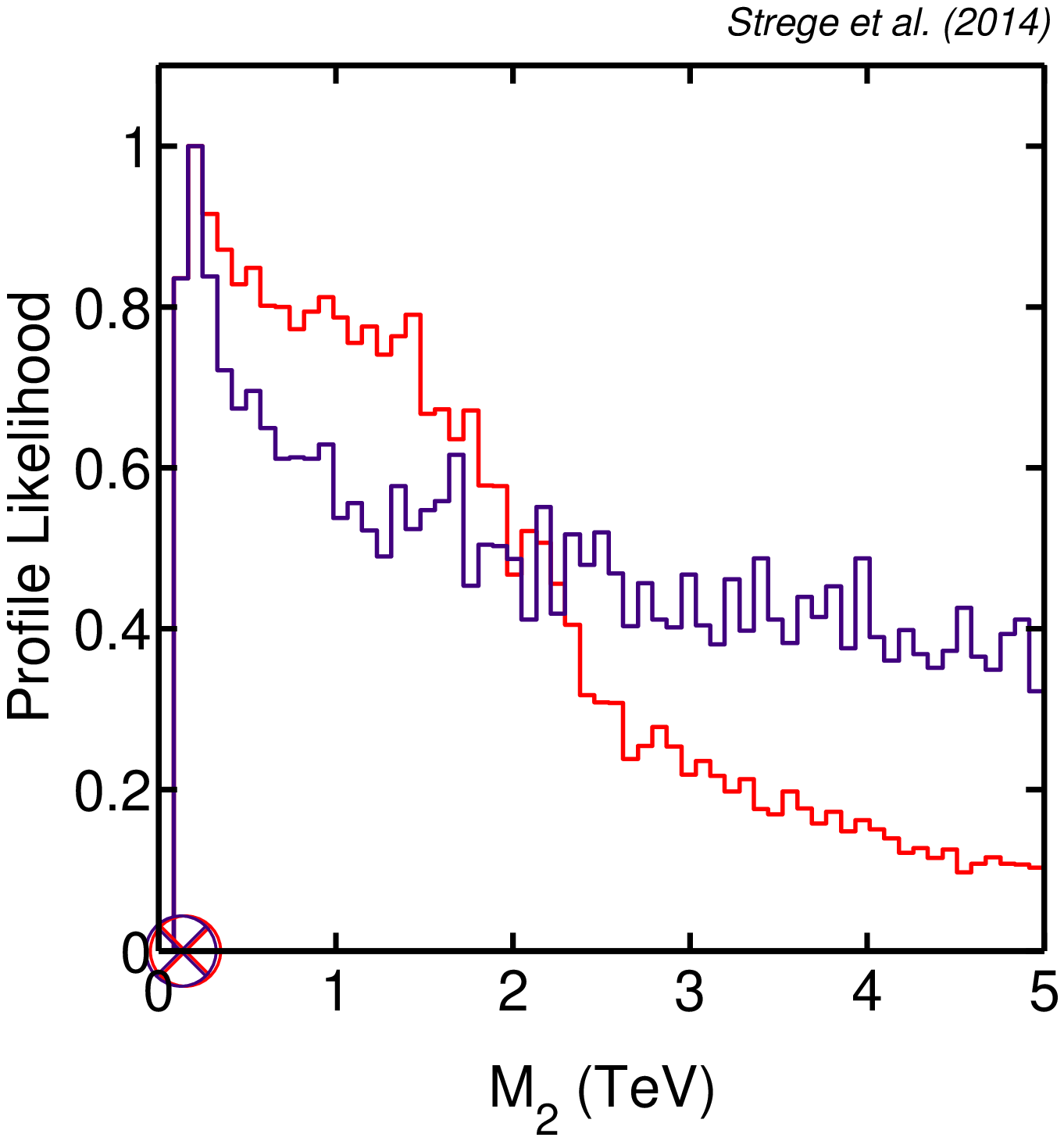}\hfill
\includegraphics[width=0.24\linewidth, trim = 0.7cm 0cm 0.7cm 0cm]{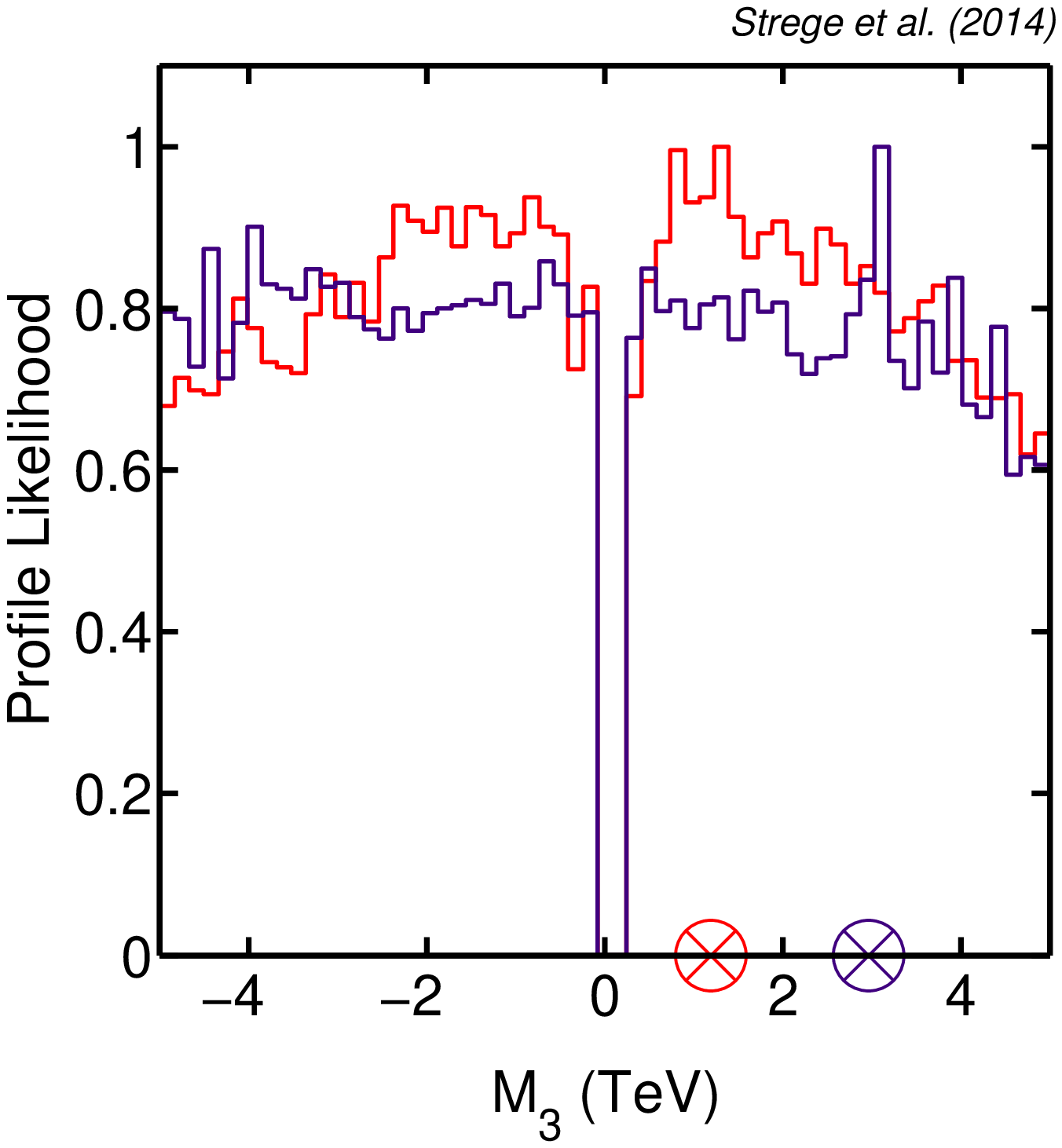}\hfill
\includegraphics[width=0.24\linewidth, trim = 0.7cm 0cm 0.7cm 0cm]{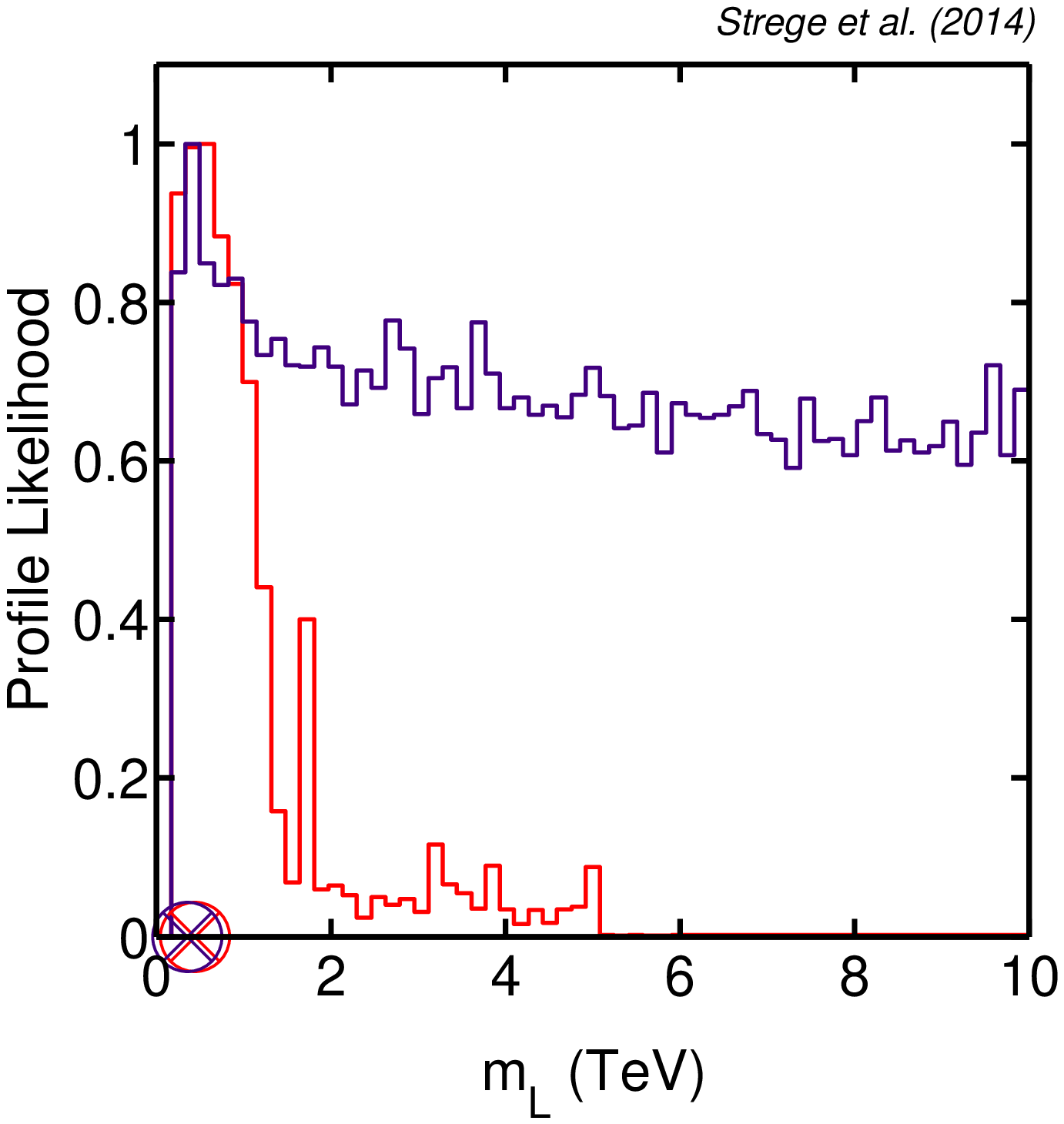} \\
\includegraphics[width=0.24\linewidth, trim = 0.7cm 0cm 0.7cm 0cm]{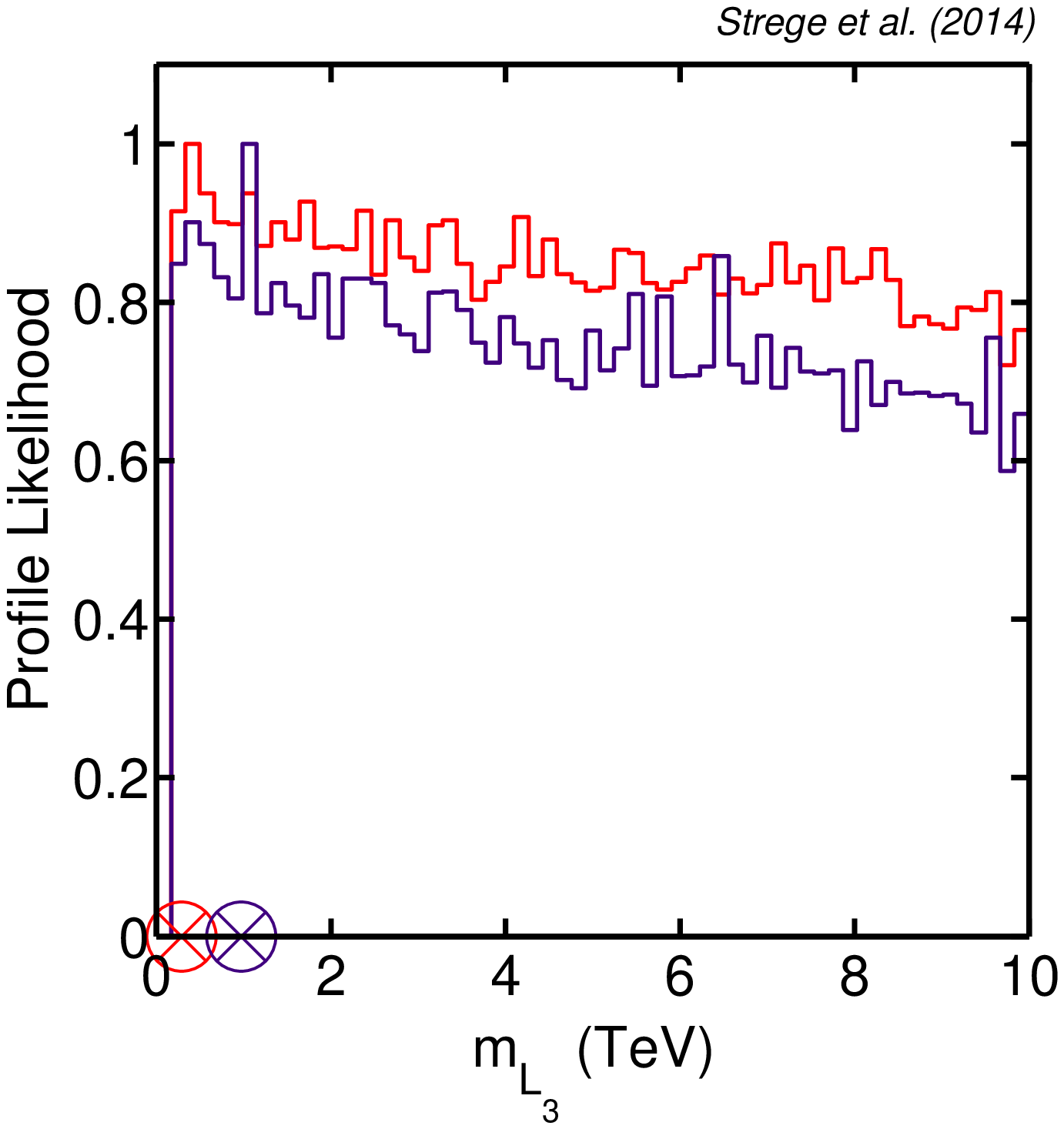}\hfill
\includegraphics[width=0.24\linewidth, trim = 0.7cm 0cm 0.7cm 0cm]{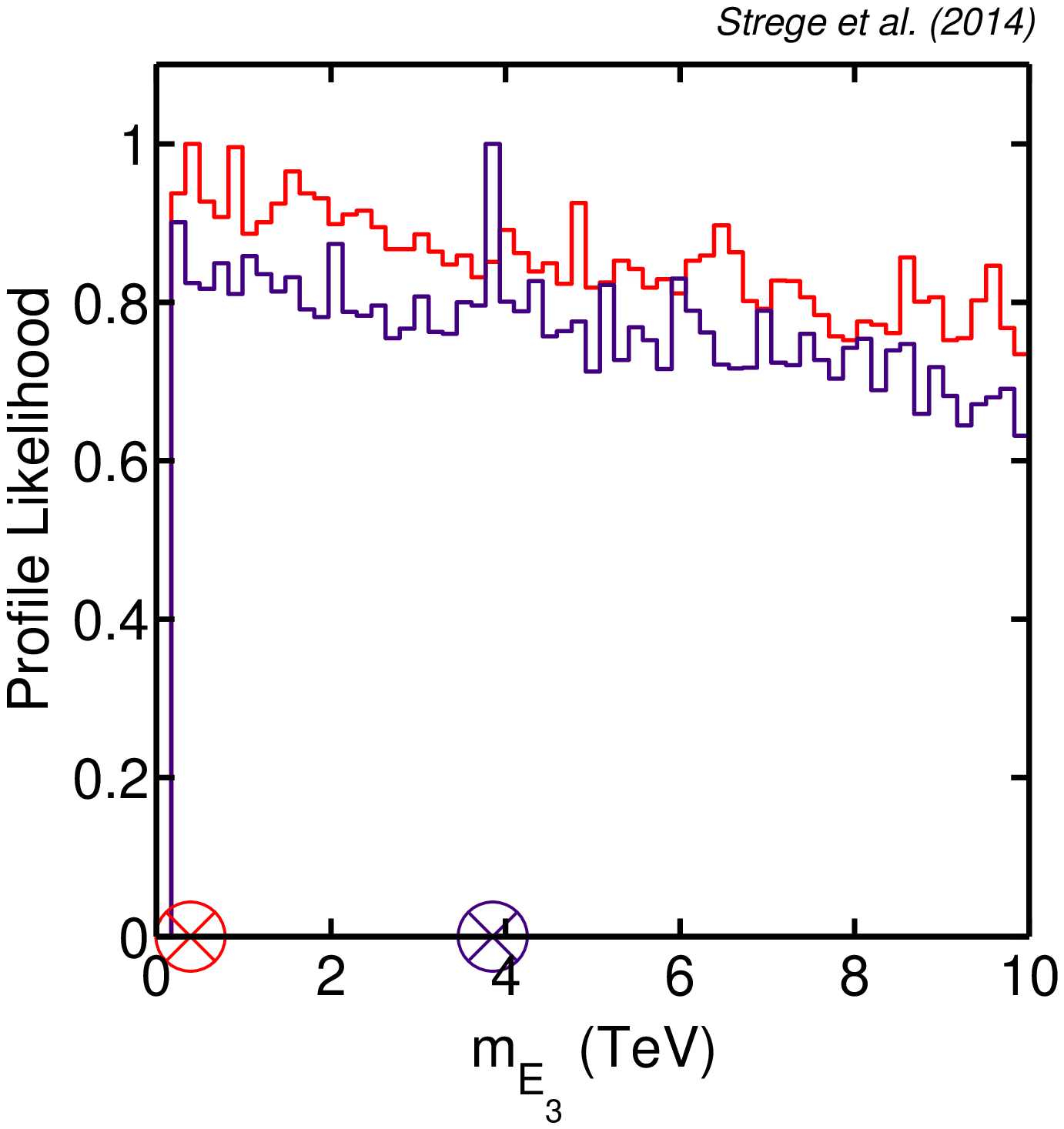}\hfill
\includegraphics[width=0.24\linewidth, trim = 0.7cm 0cm 0.7cm 0cm]{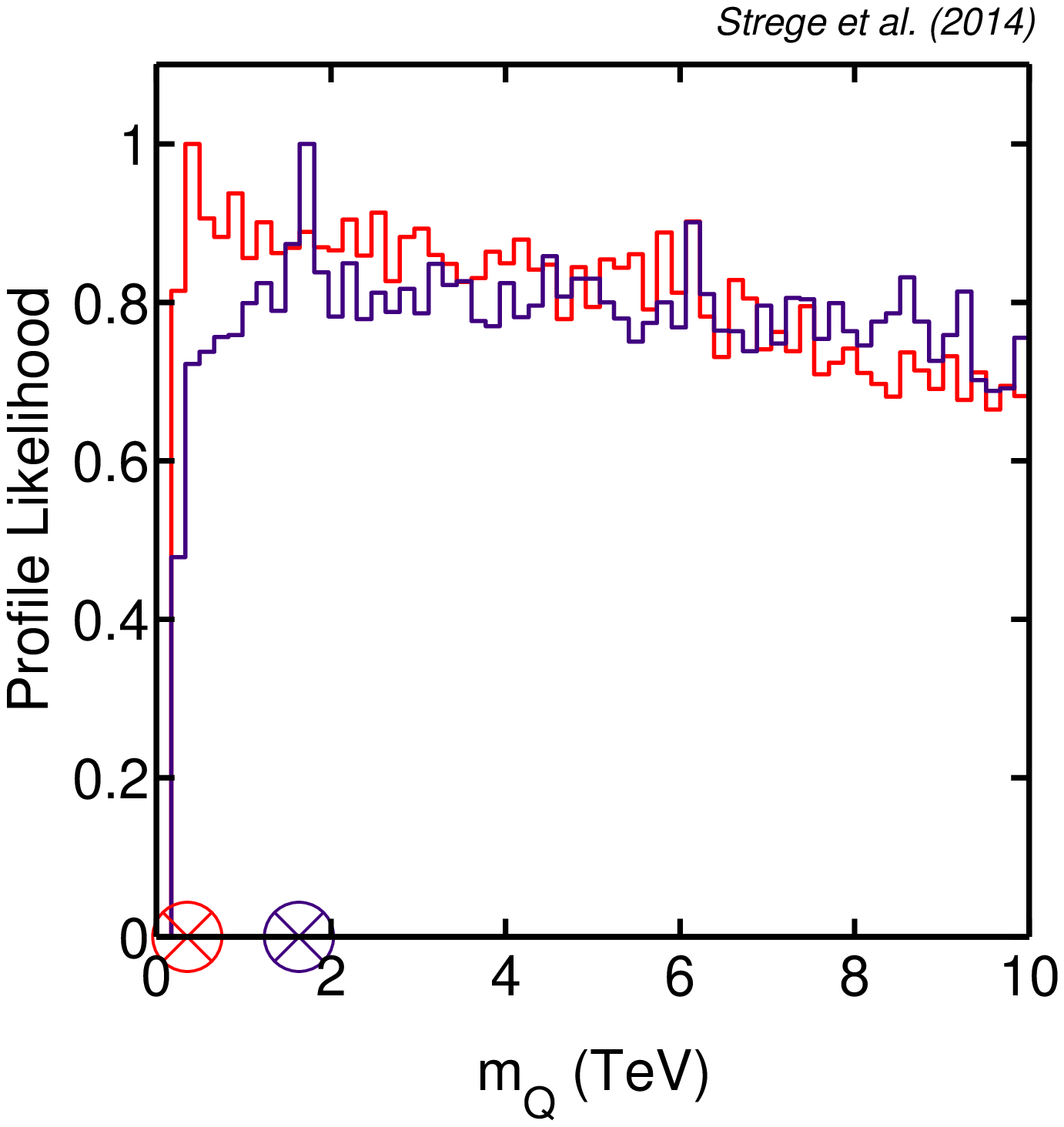}\hfill
\includegraphics[width=0.24\linewidth, trim = 0.7cm 0cm 0.7cm 0cm]{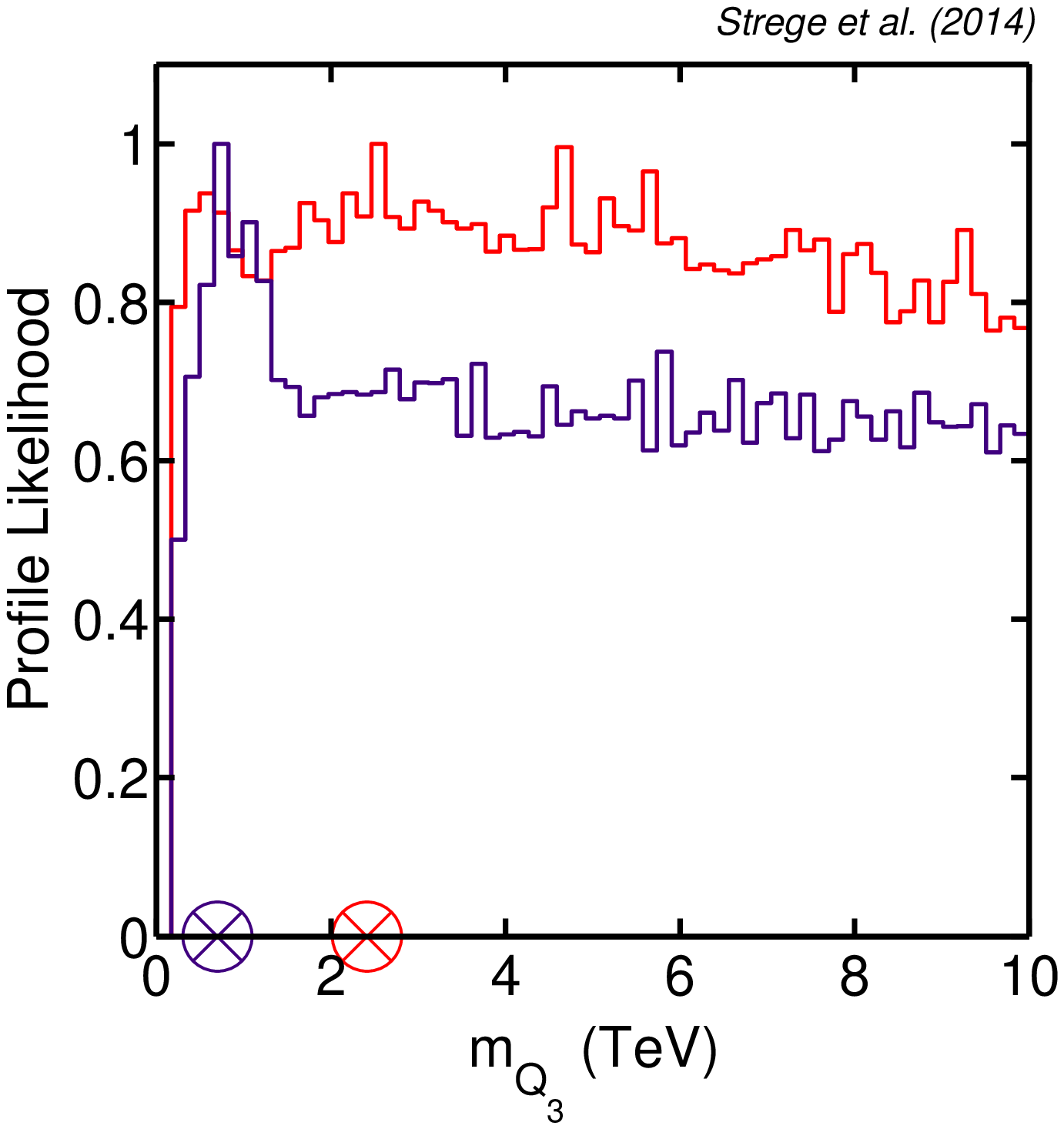} \\
\includegraphics[width=0.24\linewidth, trim = 0.7cm 0cm 0.7cm 0cm]{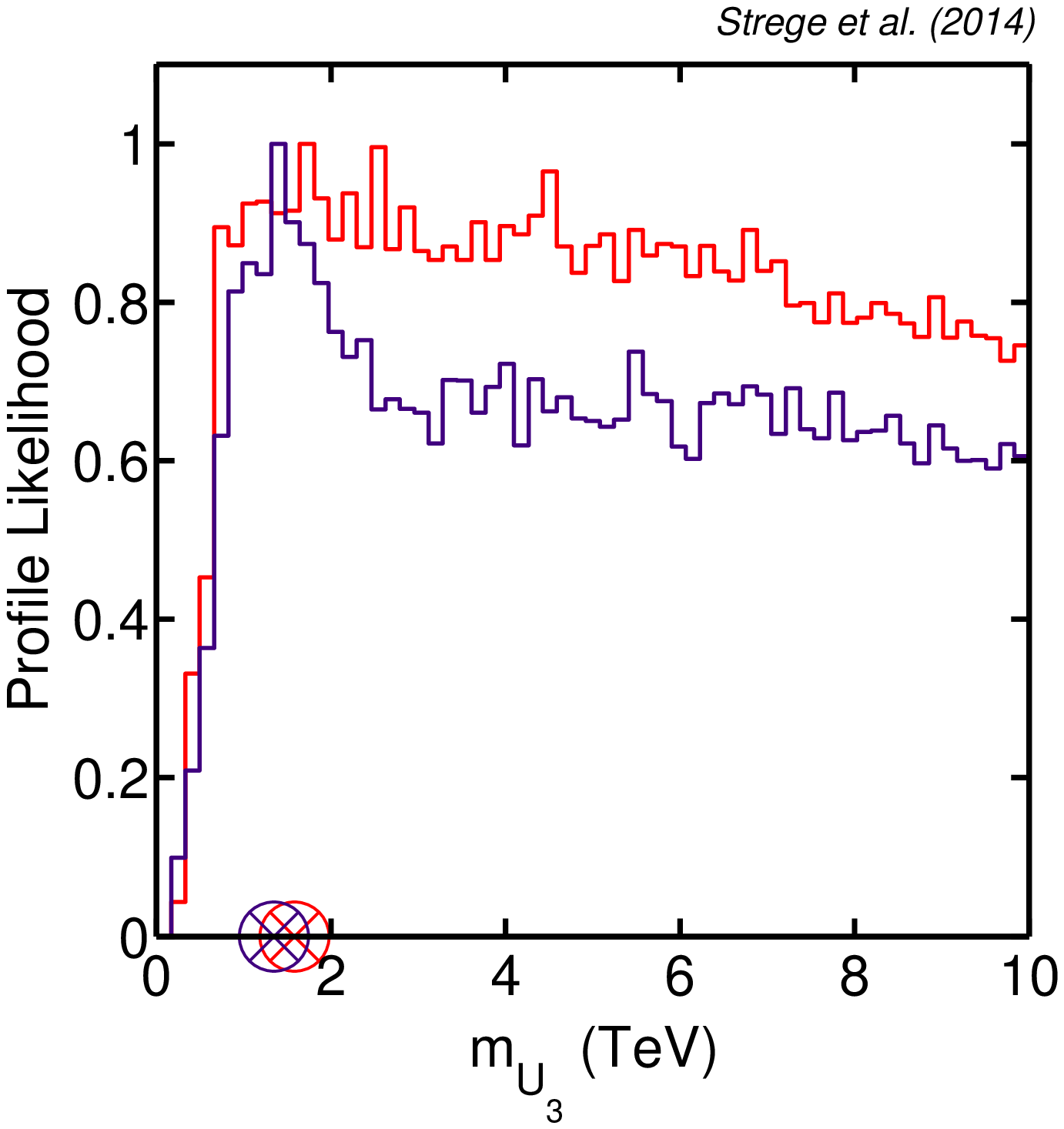}\hfill
\includegraphics[width=0.24\linewidth, trim = 0.7cm 0cm 0.7cm 0cm]{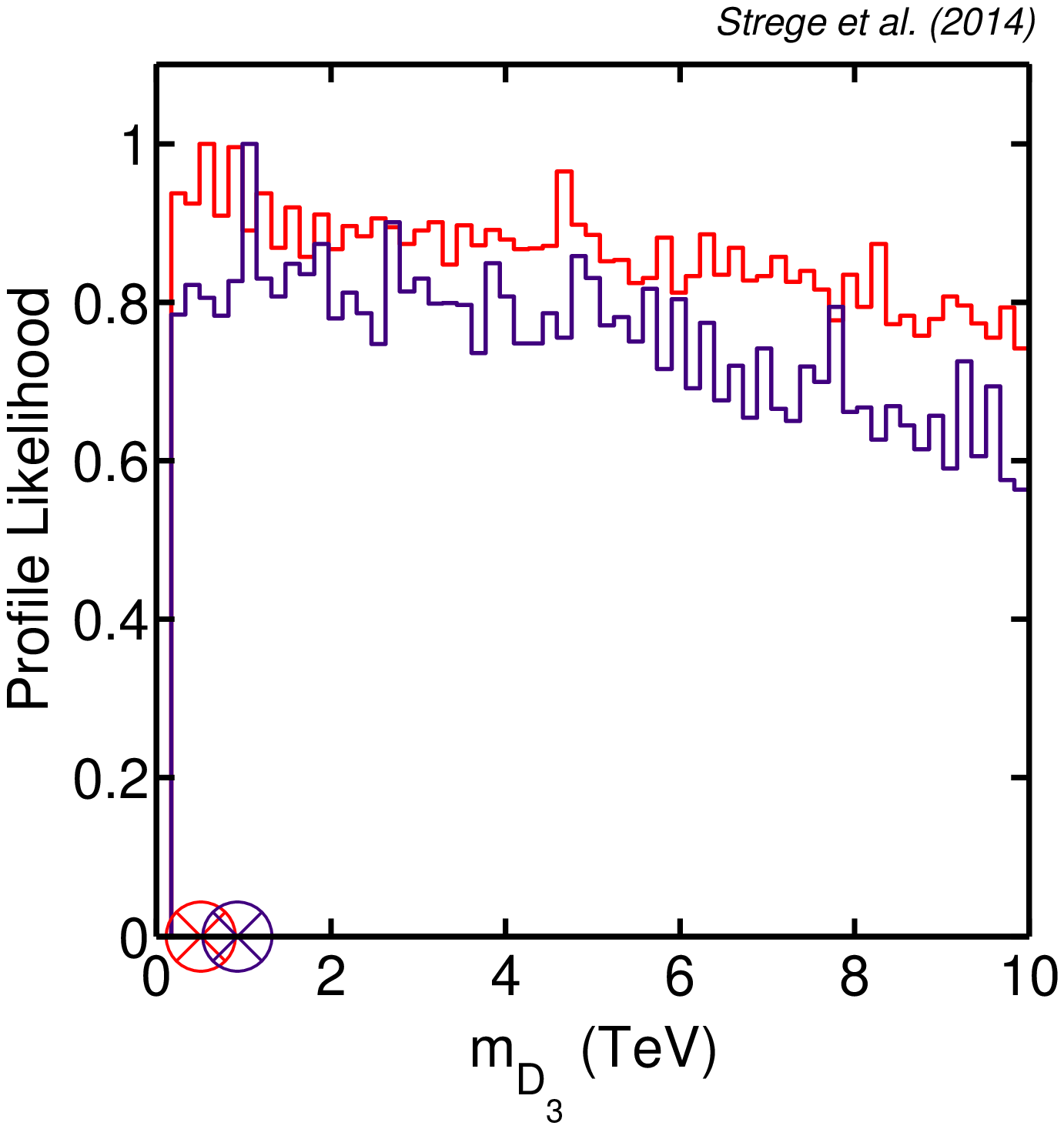}\hfill
\includegraphics[width=0.24\linewidth, trim = 0.7cm 0cm 0.7cm 0cm]{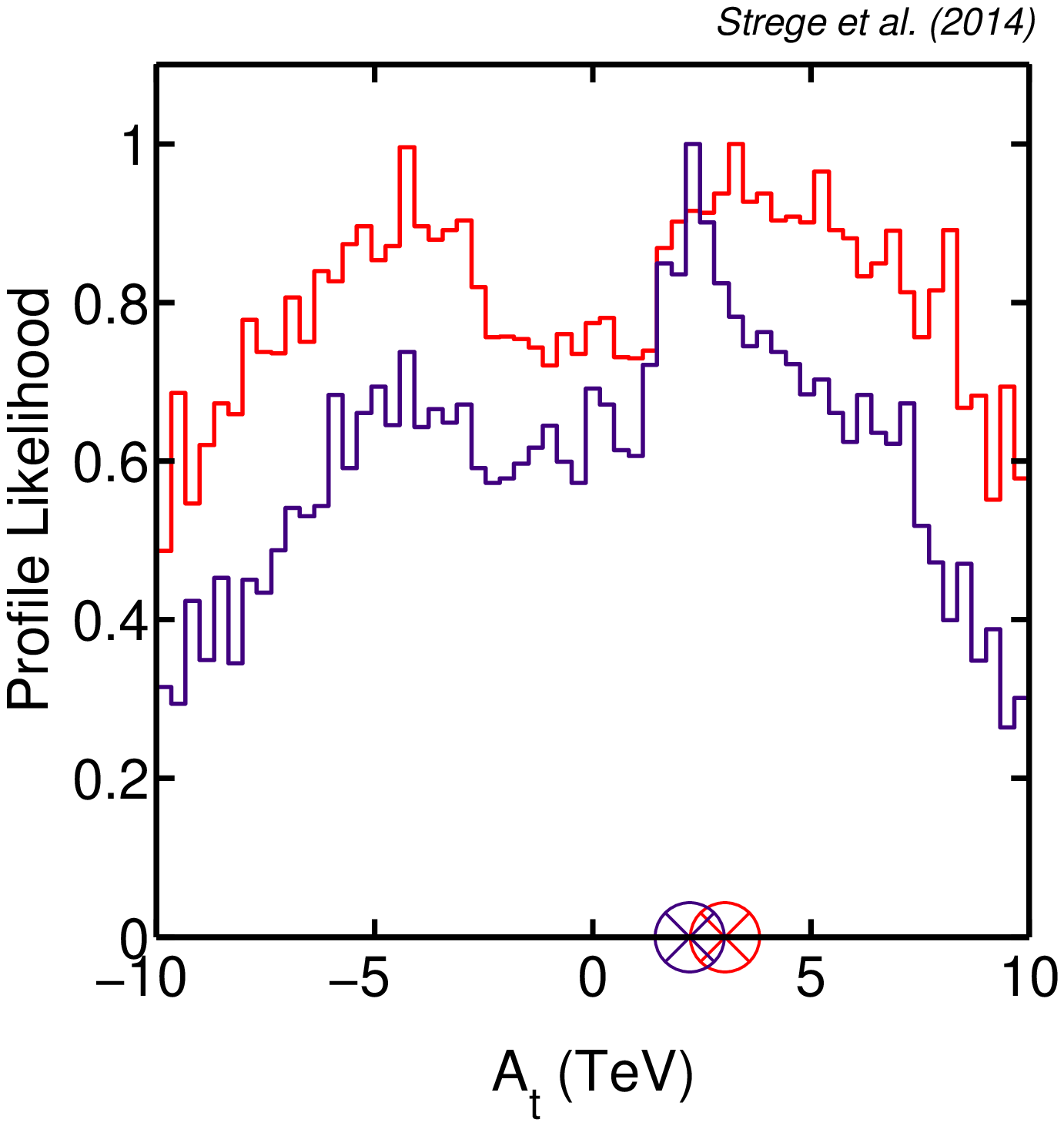}\hfill
\includegraphics[width=0.24\linewidth, trim = 0.7cm 0cm 0.7cm 0cm]{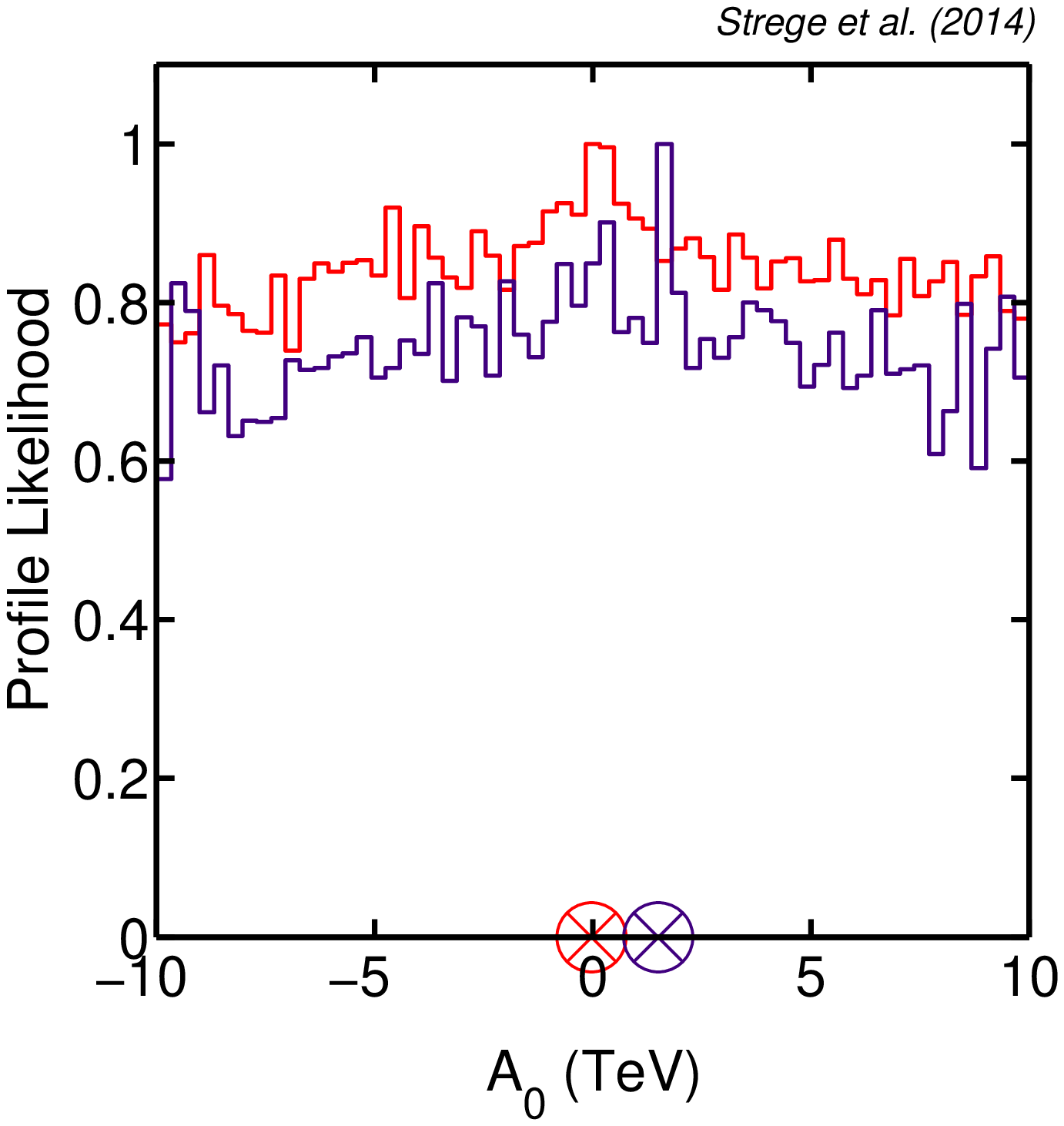}\\
\includegraphics[width=0.24\linewidth, trim = 0.7cm 0cm 0.7cm 0cm]{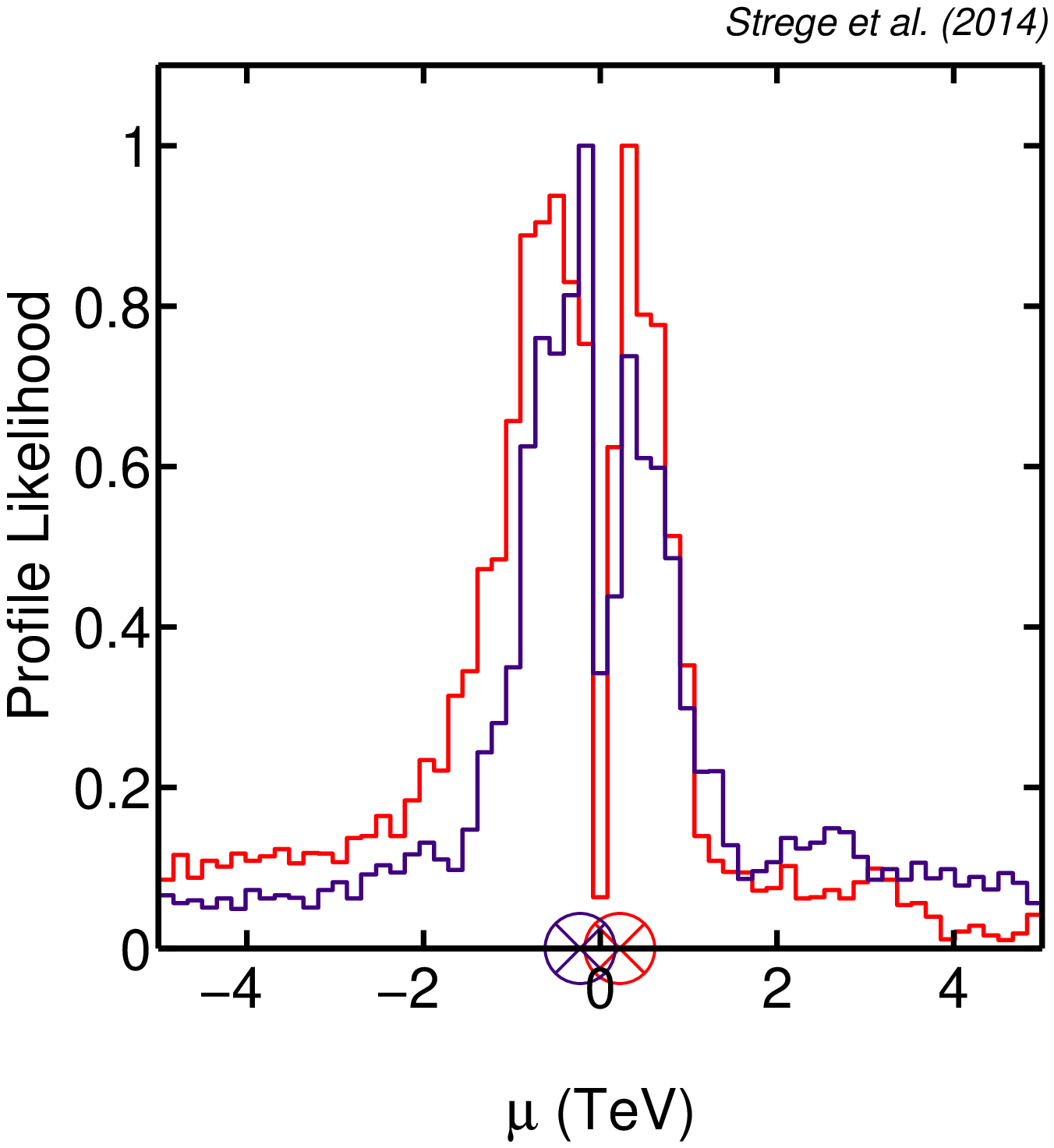}\hfill
\includegraphics[width=0.24\linewidth, trim = 0.7cm 0cm 0.7cm 0cm]{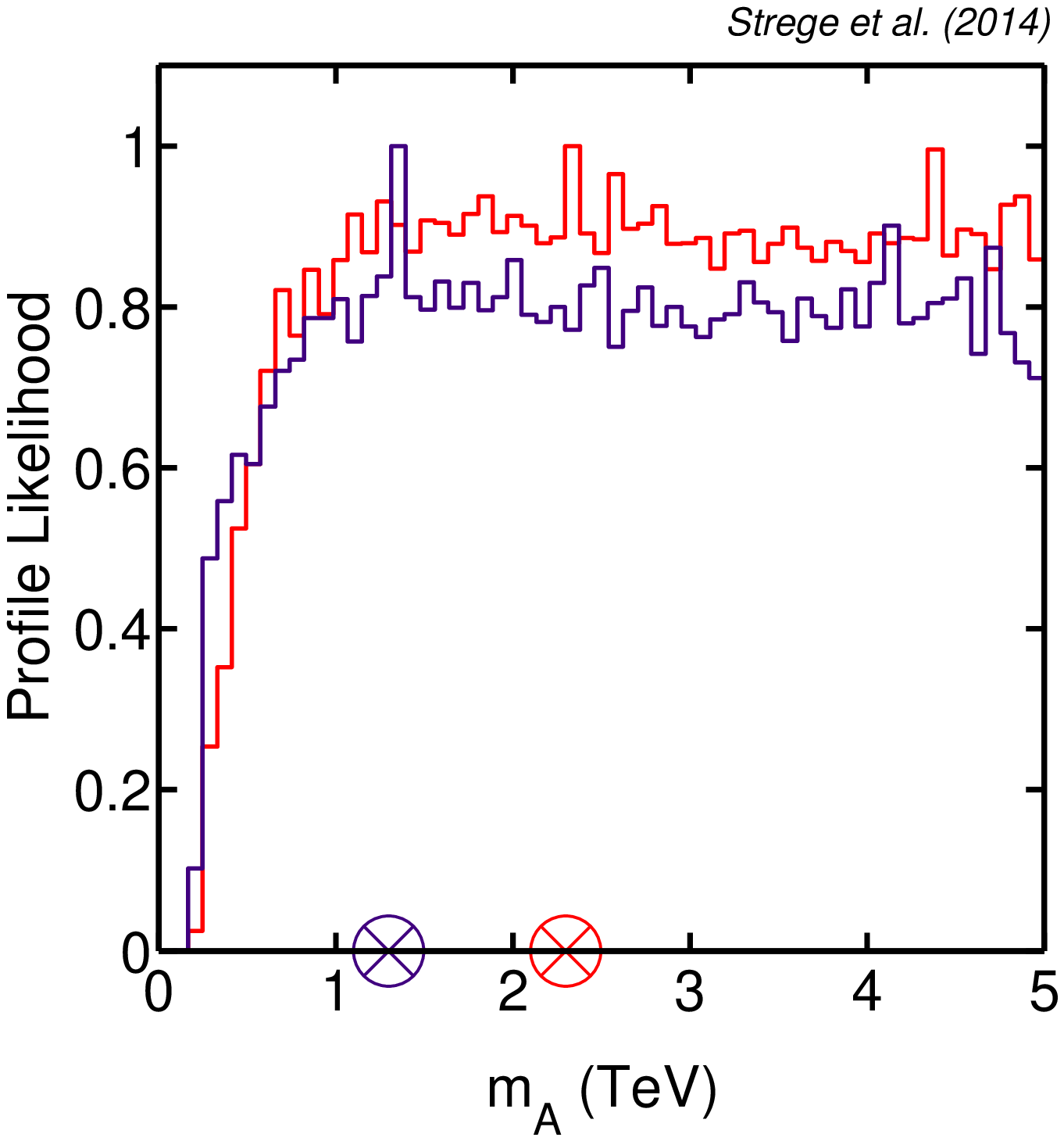}\hfill
\includegraphics[width=0.24\linewidth, trim = 0.7cm 0cm 0.7cm 0cm]{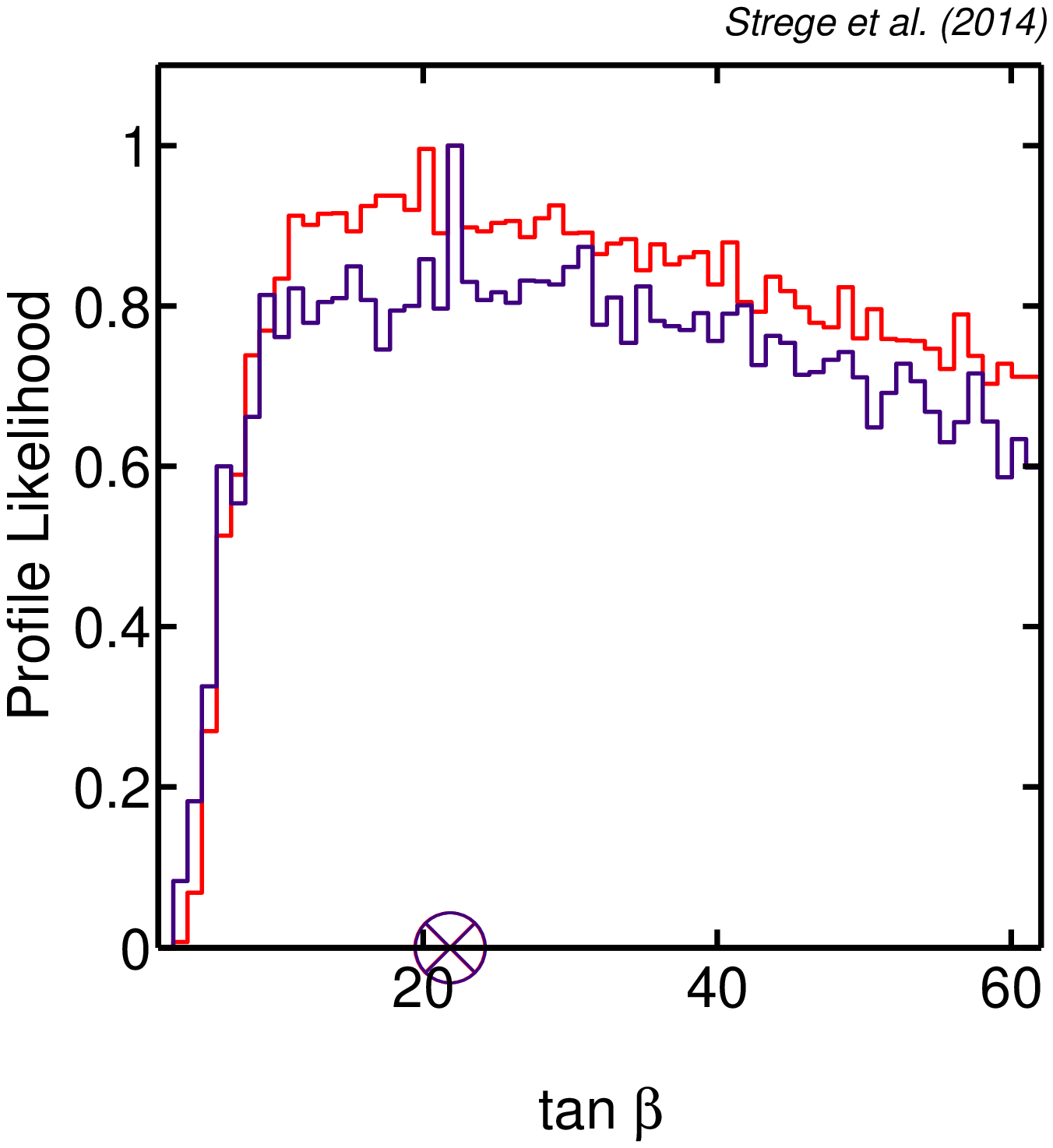}\hfill
\includegraphics[width=0.24\linewidth, trim = 0.7cm 0cm 0.7cm 0cm]{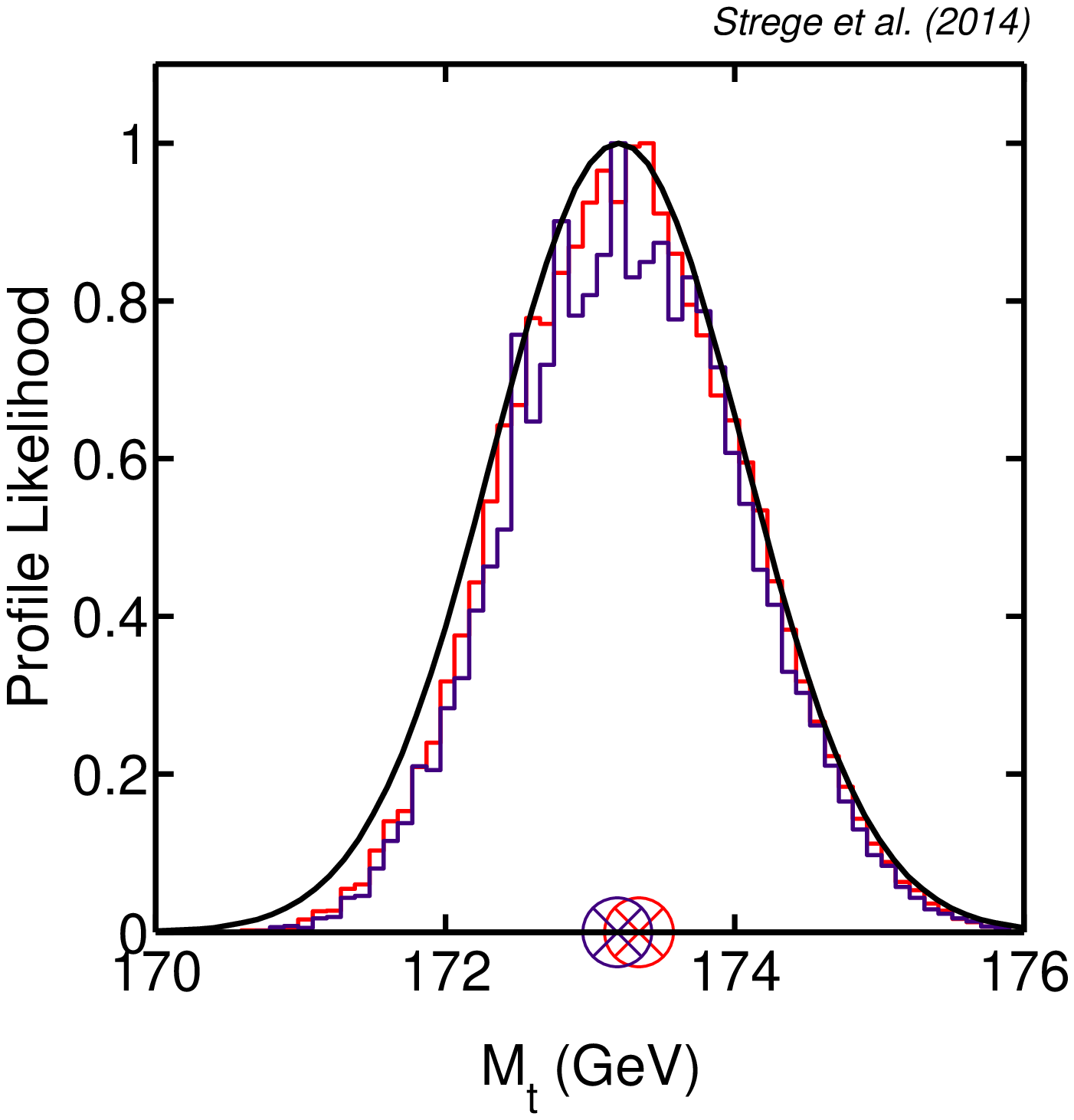}\hfill
\caption{1-D profile likelihood global fits results including all data except LHC SUSY searches and Higgs couplings (red) and further excluding the $g - 2$ constraint (purple), for the input \pMSSM\ parameters. Encircled crosses represent the best-fit points. The black curve in the top mass panel is the applied prior distribution.}
\label{fig:1D_wog2_1}
\end{center}
\end{figure}

Fig.~\ref{fig:1D_wog2_1} shows that most of the \pMSSM\ model parameters are relatively weakly constrained. 
We start by discussing the 1D profile likelihood (PL) functions for the parameters entering at tree level into the electro-weakino (EWK) sector, namely $M_1$, $M_2$, $\mu$ and \tanb. 

The bino mass $|M_1|$ shows a clear preference for relatively low values, peaking at $\sim 50 \gev$. In this region, the neutralino is bino-like and annihilates mainly into a pair of fermions through  $Z/h$-funnels in the early universe. When $|M_1|$ takes values of $\order (100 \gev)$ the neutralino gets some mixing with higgsinos, so that its relic density is reduced to the experimentally measured value by co-annihilations with the second lightest neutralino and the lightest chargino. Notice that at low $m_\neut$ the degree of mixing can be at most of a few percent, otherwise the relic density would fall below the Planck measurement and the neutralino nucleon spin-independent cross-section would be in tension with the XENON100 limit. Additionally, the relic density can be reduced by efficient annihilation to a pair of fermions via the exchange of relatively light sleptons and squarks (the so-called bulk region), and co-annihilations with sleptons of the first and second generation. Note that the $A$-funnel region is suppressed in this mass range due to the preference for $m_A > 1$ TeV (see below).
For heavier binos the 1D PL drops abruptly due to the fact that it can not mix sizeably and therefore the wino mass $M_2$ and the higgsino mass $\mu$ are pushed to large values, in tension with the muon \gmt\ constraint and, to a lesser extent, with several flavor physics observables, such as $\brbsgamma$, $\DeltaO$ and $\afb$.  

The wino mass $M_2$ is only mildly constrained from below by the LEP constraint on the chargino mass~\cite{LEP_SUSY}, and it peaks at values around a few hundred \gev.  Above this value, we observe a moderate decrease of the PL towards the boundary of the prior. The shape of the 1D PL for $M_2$ is mostly driven by the muon \gmt\ constraint, whose  MSSM contributions are dominated by chargino-sneutrino and neutralino-smuon loop diagrams. One can write the chargino-sneutrino contribution, which often is the dominant one, as follows~\cite{Endo:2013bba}
\begin{equation}
\delta a_\mu(\tilde{W}, \tilde{H}, \tilde{\nu}_\mu) \sim 15 \times 10^{-9} \left( \frac{ \tan \beta}{10} \right) \left( \frac{(100 \gev)^2}{M_2 \mu} \right) \left(  \frac{f_C}{1/2} \right),
\label{eq:gm2w}
\end{equation}
where $f_C$ is a loop function with maximum value $f_C=1/2$ when the masses in the loop are degenerate. When $M_2$ and $\mu$ are of $\order (100 \gev)$ and \tanb\ of $\order (10)$, the contribution becomes $\order (10^{-9})$, which can explain the muon \gmt\ ``anomaly" provided that $\text{sgn}(M_2, \mu) > 0$ and the smuon/sneutrino soft-masses, which we assume to be universal, are $\order (100 \gev)$. The degree of decoupling allowed depends on the value of \tanb\ (note that values above $\gtrsim 60$ are forbidden by imposing the perturbativity of the bottom Yukawa coupling). While in general winos with masses $\gtrsim 1 \tev$ are decoupled, the neutralino-smuon contribution can still give a large contribution to \gmt\ and thus give a good fit. Since we have set $M_2 > 0$,  $\mu > 0$ is favoured. Notice that the best-fit point fulfils this condition. It is worth mentioning that the wino mass can also play an important role in new physics contributions to the Wilson coefficient $C_7$ which is a fundamental quantity in the most relevant flavor observables entering into our analysis~\cite{DescotesGenon:2011yn} \footnote{The Wilson coefficient $C_9$ is also potentially important, though within the MSSM its role is diluted~\cite{Altmannshofer:2013foa}.}.

The 1D PL for the higgsino mass $|\mu|$ shows an almost symmetrical distribution about zero.  Like the wino-mass, $|\mu|$ is constrained from below by the LEP constraint on the chargino mass~\cite{LEP_SUSY}. Relatively small values $|\mu| \lsim 1$ TeV are strongly favoured, while large values of $|\mu|$ are disfavoured. This is because the SM predictions for flavour physics observables such as $\DeltaO$ and $\afb$ are discrepant with the experimental measurement at $\sim 1 - 2 \sigma$ level, and thus require rather large new physics contributions to $C_7$. Note that, in contrast, for $\brbsgamma$ the discrepancy is below $1 \sigma$. For intermediate/large \tanb\ values the leading SUSY corrections scale as $1/m^2_\chi$, so that light higgsinos are preferred. It is worth noticing that the PL drops much faster for larger values of the higgsino mass than for the wino mass. This is due to the fact that the higgsinos mass enters in both the chargino-sneutrino and the neutralino-smuon diagrams that contribute to $\delta a_\mu$.
Moreover, one can see that there is not a full decoupling at $\order (1 \tev)$ higgsino masses for the negative branch, due to an enhancement of the pure-bino contribution for large higgsino masses as a result of a large left-right mixing in the smuon mass matrix. One can write this contribution as follows
\begin{equation}
\delta a_\mu (\tilde{\mu}_L, \tilde{\mu}_R,\tilde{B}) \sim 1.5 \times 10^{-9} 
\left( \frac{\tan \beta}{10} \right) \left( \frac{(100 \gev)^2}{m_{\tilde{\mu}_L}^2 m_{\tilde{\mu}_R}^2/(M_1 \mu)} \right) \left(  \frac{f_N}{1/6} \right),
\end{equation}
where $f_N$ is a loop function with maximum value $f_N=1/6$ when the masses in the loop are degenerated. A sizeable contribution can be achieved provided that $\text{sgn}(M_1\mu ) > 0$. While the chargino-sneutrino contribution to \gmt\ in Eq.~\eqref{eq:gm2w} prefers $\mu > 0$, large positive $\mu$ tend to lead  to tachyons in the Higgs sector due to the fact that the sign of $\mu$ enters into the RGE of the bilinear soft breaking term, resulting in a subtle preference for negative values of $\mu$. Recall that the bounday conditions are applied at  the scale $\sqrt{m_{\tilde{t}_1} m_{\tilde{t}_2}}$ and there is a running to $m_Z$ which is where the SUSY thresholds corrections are applied. The sign of the associated $\beta$ function tends to be positive  and large unless the pseudoscalar mass $m_A$ is not too large and/or the gaugino mass $M_3 > 0$ is large. Note that the 1D PL for negative $\mu$ is shifted to slightly larger values compared to the positive branch, as very small values of $|\mu|$ for $\mu < 0$ would lead to a sizeable negative contribution to \gmt\ from Eq.~\eqref{eq:gm2w}. 

The 1D PL for \tanb\ is suppressed below $\sim 10$, mainly due to the Higgs mass measurement, since at tree level $m_h \le m_Z |cos 2 \beta|$. Values close to the upper prior boundary are also slightly disfavoured as they are close to the non-perturbativity limit of the bottom Yukawa coupling. This effect is stronger when the SUSY threshold corrections to the bottom mass are large, which occurs mainly for a low-mass SUSY spectrum.  

The 1D PL for the pseudoscalar Higgs mass, $m_A$, is severely suppressed for values $\lesssim 1 \tev$, 
mainly due to the Higgs mass measurement, but also due to 
the LHCb measurement of $\brbsmumu$, which is in good agreement with the SM expectation. 
Since SUSY contributions enter as $\propto \tan^6 \beta/m_A^2$ and larger values of $\tan\beta$ are favoured, heavier 
pseudoscalars masses are preferred. 
It implies that $m_A \gg m_Z$ and the SUSY decoupling regime~\cite{Gunion:2002zf} is favoured in the model which leads to $m_H \simeq m_{H^\pm} \simeq m_A$, up to corrections of $\order{(m_Z^2/m_A^2)}$. 

We now turn to the discussion of the sfermion sector. The 1D PL for the first and second generation slepton mass, $m_L$, shows a clear preference for relatively low values when \gmt\ is included, as follows from the discussion above.  Such a preference is almost entirely driven by the \gmt\ constraint, as is clear from the comparison with the corresponding PL for the analysis excluding the constraint on the anomalous magnetic moment of the muon. Both of the third generation slepton soft-masses $m_{L_3}$ and $m_{E_3}$ remain essentially unconstrained, as their PL functions are almost flat. The slight preference for low values is due to the impact of relatively light staus in the electroweak precision observables (EWPOs). 
Similarly, relatively small $m_{Q}$, $m_{U_3}$ and $m_{D_3}$ are somewhat preferred, as low values of these quantities lead to greater freedom to satisfy the constraints on several constrained flavour observables. In contrast, the 1D PL for $m_{Q_3}$ is almost flat up to $\sim 6$ TeV, as in general TeV-scale values of $m_{Q_3}$ are required to achieve a good fit to the constraint on the lightest Higgs mass.

The top trilinear coupling $A_t$ shows a symmetric PL around 0. We have checked that the peaks correspond to the maximal mixing scenario where $|X_t/M_S | = \sqrt{6}$ with $X_t \equiv (A_t - \mu cot \beta) $ and $M^2_S \equiv 1/2 (m_{\tilde{t}_1}^2 + m_{\tilde{t}_2}^2)$. In the maximal mixing region, $m_h \sim 125$ GeV can be achieved even for relatively small stop masses, which in general are preferred by the SM precision observables.

Finally, the gluino mass $|M_3|$ is constrained from below due to the Tevatron lower limit $m_{\rm gluino} > 289$ GeV~\cite{Shamim:2007yy}.  Above this, the distribution is nearly flat with a slight suppression near the prior boundary. Very large gluino masses are expected to be disfavoured because they tend to induce the presence of tachyons in the staus and sbottoms, due to the RGE running of the trilinear coupling $A_0$. Notice that the gluino plays a role both in higher loop corrections to flavor physics observables and at the level of the RGEs of SUSY parameters.

Exclusion of the muon \gmt\ constraint has a strong impact on the electroweakino sector, in particular on the 1D PL for the wino mass $M_2$ and the universal slepton mass $m_L$, both of which enter in the contribution from the chargino-sneutrino loop in Eq.~\eqref{eq:gm2w}. The 1D PL for the  bino and higgsino masses $M_1$ and $\mu$ are also affected, albeit to a lesser extent. Dropping the \gmt\ constraint makes the data more compatible with heavier winos, and hence heavy binos are less disfavoured in this scenario. Additionally, upon exclusion of the \gmt\ constraint the positive branch of $\mu$ is no longer favoured, and indeed negative values of $\mu$ are preferred.

The most remarkable difference in the PL for the sfermion soft-masses occurs for the mass parameters related to the stop spectrum. Namely, the 1D PL for $m_{Q_3}$ and $m_{U_3}$ peak at small values, while large masses are significantly suppressed. As mentioned above, the Wilson coefficient $C_7$ plays a fundamental role in the most relevant flavor observables entering into our analysis, most importantly the isospin asymmetry $\DeltaO$. As can be seen in Fig.~\ref{fig:1D_wog2_2} below, after exclusion of the \gmt\ constraint the 1D PL for $\DeltaO$ peaks much closer to the experimentally measured value. Removing the \gmt\ constraint from the analysis leads to greater freedom to satisfy the experimental constraint on $\DeltaO$. In particular, Higgsino-stop loops can lead to a sizeable contribution to $C_7$~\cite{Altmannshofer:2012ks} 
\begin{equation}
\delta C_7 \propto \frac{M_t^2 \mu A_t}{2 m_{Q_3}^4} \tan \beta f_7 \left( \frac{\mu^2}{m_{Q_3}^2},  \frac{\mu^2}{m_{U_3}^2} \right),
\end{equation}
where $f_7$ is a loop function. For small $m_{Q_3}$, medium $m_{U_3}$ and sizeable $\tan \beta$, $\delta C_7$ becomes large. Additionally, for ${\rm sgn}(\mu A_t) < 0$, the sign of this loop contribution is opposite to the SM contribution~\cite{Altmannshofer:2012ks}, and values of $\DeltaO$ in good agreement with the experimental constraint can be achieved. The requirement that ${\rm sgn}(\mu A_t) < 0$ also explains the preference for the peak in the positive branch of $A_t$, which is clearly favoured with respect to negative values. We point out that $C_7$ also enters in a range of other flavour observables, in particular $\brbsgamma$. In contrast to the isospin asymmetry, the measurement of this quantity is in excellent agreement with the SM predictions, so that large SUSY contributions to $C_7$ are disfavoured by this constraint. Note that we use the {\it SusyBSG} code for the computation of $\brbsgamma$, while {\it SuperIso} is used to computed $\DeltaO$. We caution that, for some fine-tuned points, the simultaneous achievement of a good fit to $\brbsgamma$ and $\DeltaO$ (and other flavour observables) can be a numerical effect, related to differences in the numerical implementation of the $C_7$ calculation in these codes\footnote{In particular, {\it SusyBSG} implements the full gluino two-loop contributions to the Wilson coefficient $C_7$ within the minimal flavour violation scenario~\cite{Degrassi:2006eh}, while {\it SuperIso} implements those in the heavy gluino limit~\cite{Bobeth:2004jz}, which does not necessarily hold in all regions of the MSSM-15 parameter space.}.

%%%%%%%%%%%%%%%%%%%%%%%%%%%%%%%%%%%%%%%%%%%%%%%%%%%%%%%%%%
\subsubsection{Profile likelihood for observable quantities}
%%%%%%%%%%%%%%%%%%%%%%%%%%%%%%%%%%%%%%%%%%%%%%%%%%%%%%%%%

\begin{figure}
\begin{center}
\includegraphics[width=0.24\linewidth, trim = 0.7cm 0cm 0.7cm 0cm]{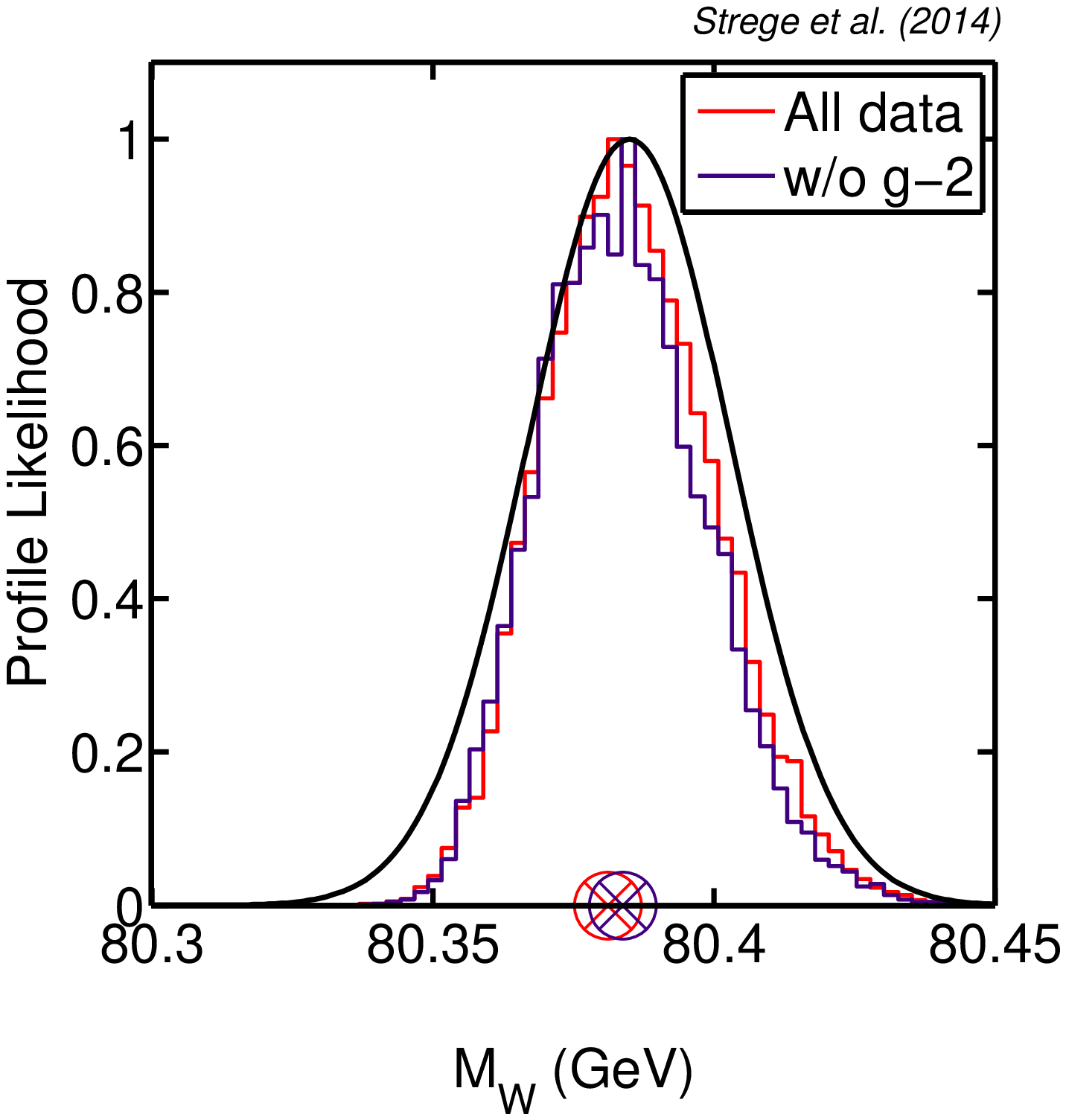}
\includegraphics[width=0.24\linewidth, trim = 0.7cm 0cm 0.7cm 0cm]{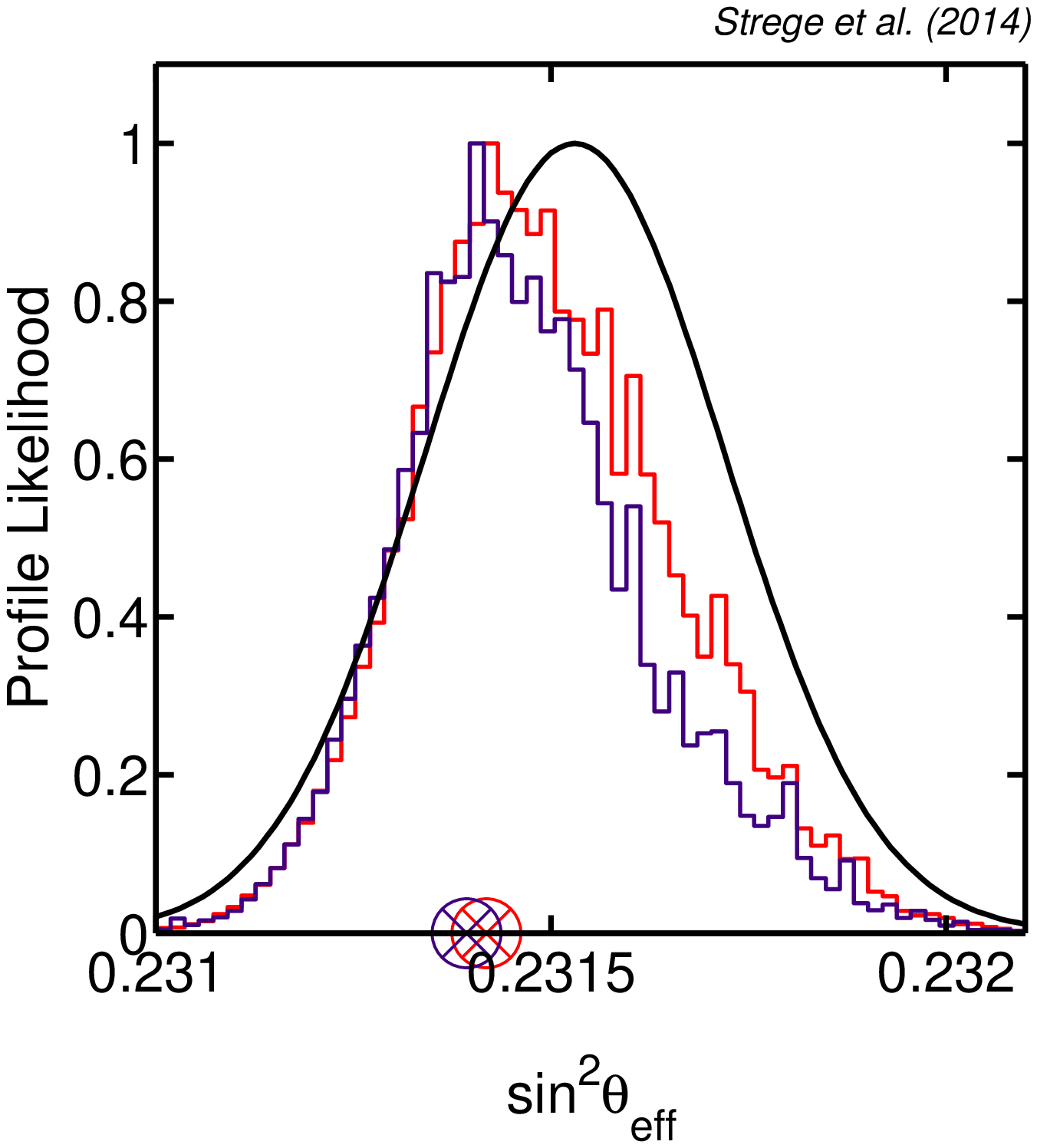}
\includegraphics[width=0.24\linewidth, trim = 0.7cm 0cm 0.7cm 0cm]{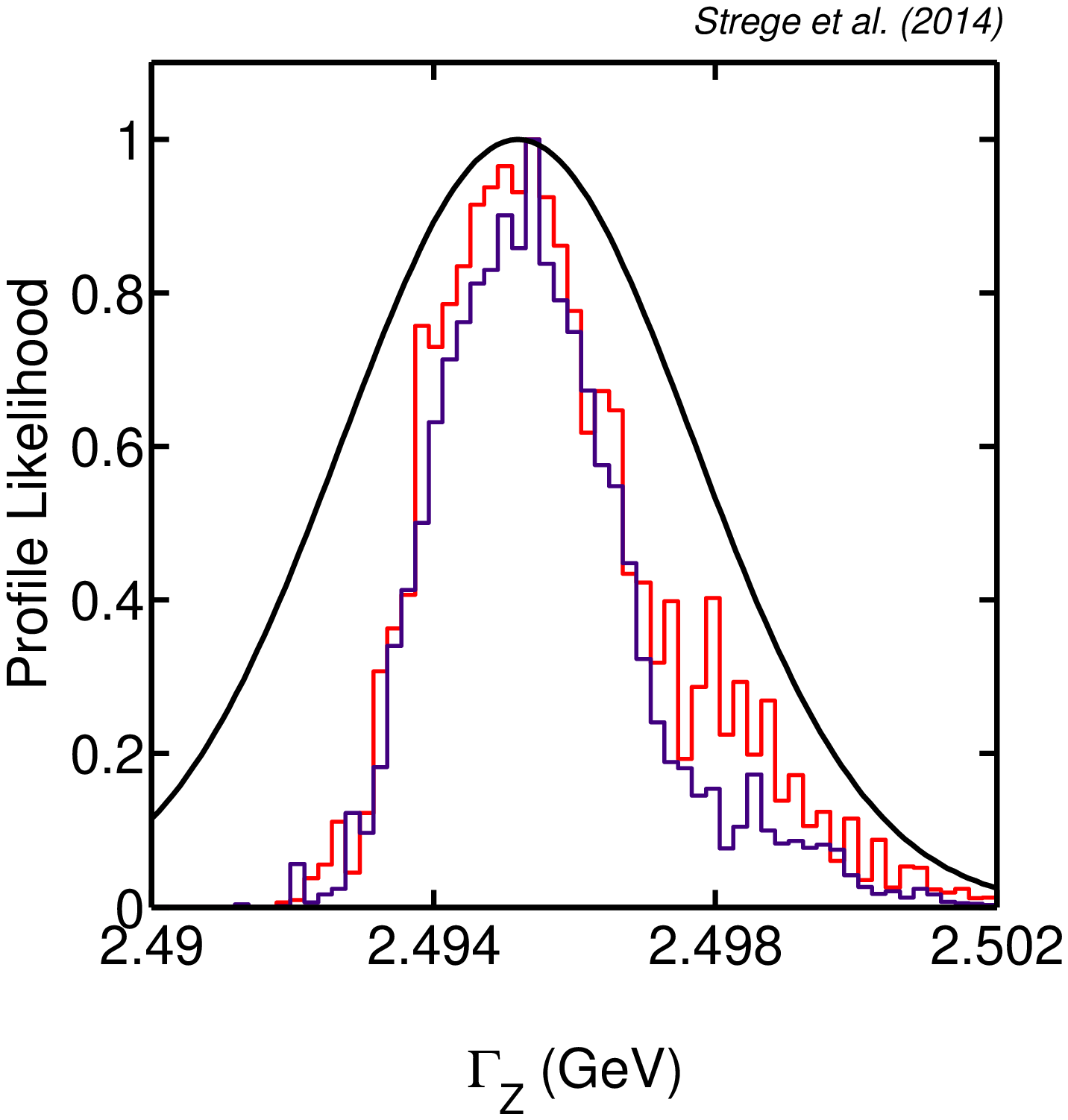}
\includegraphics[width=0.24\linewidth, trim = 0.7cm 0cm 0.7cm 0cm]{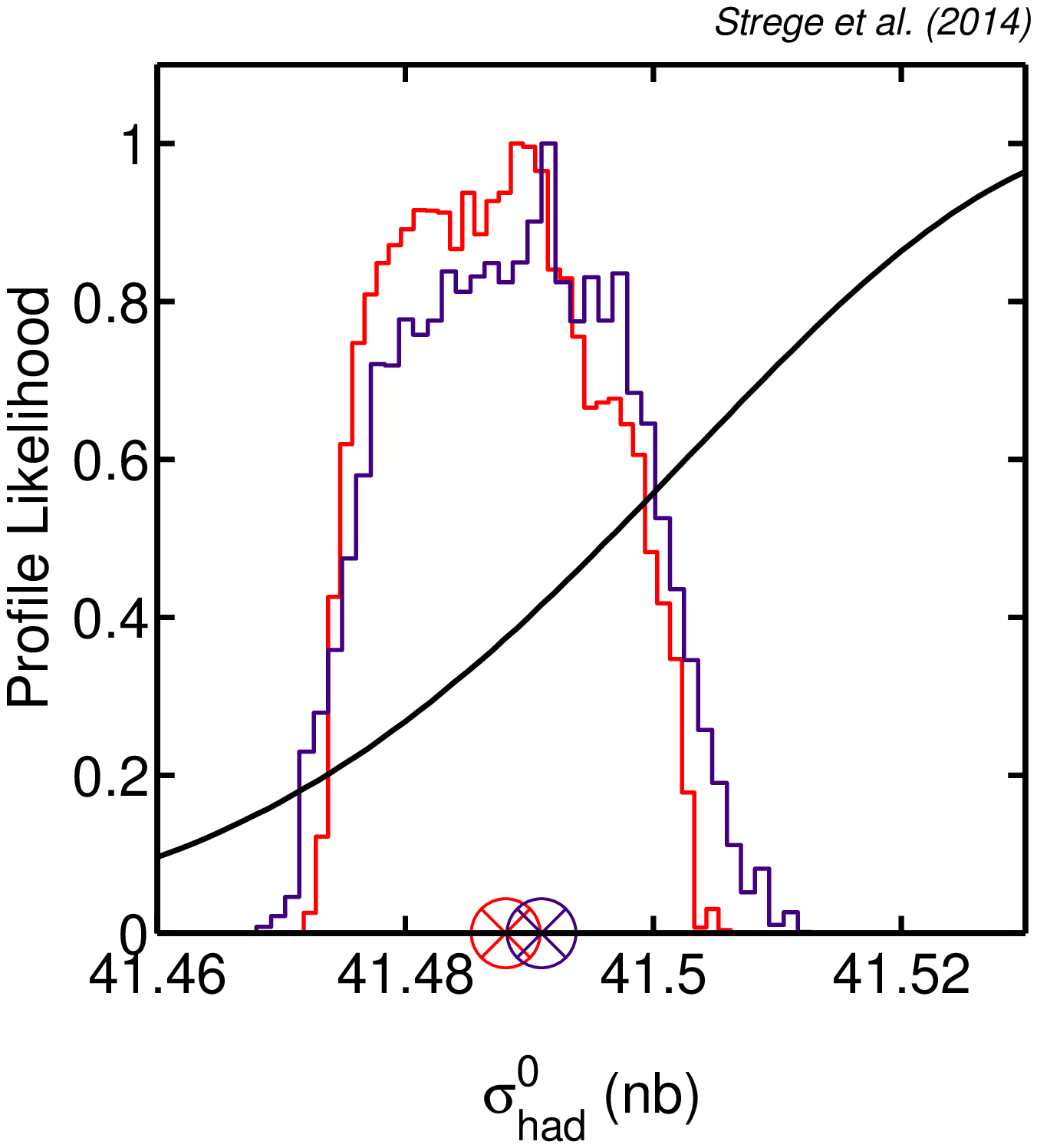} \\
\includegraphics[width=0.24\linewidth, trim = 0.7cm 0cm 0.7cm 0cm]{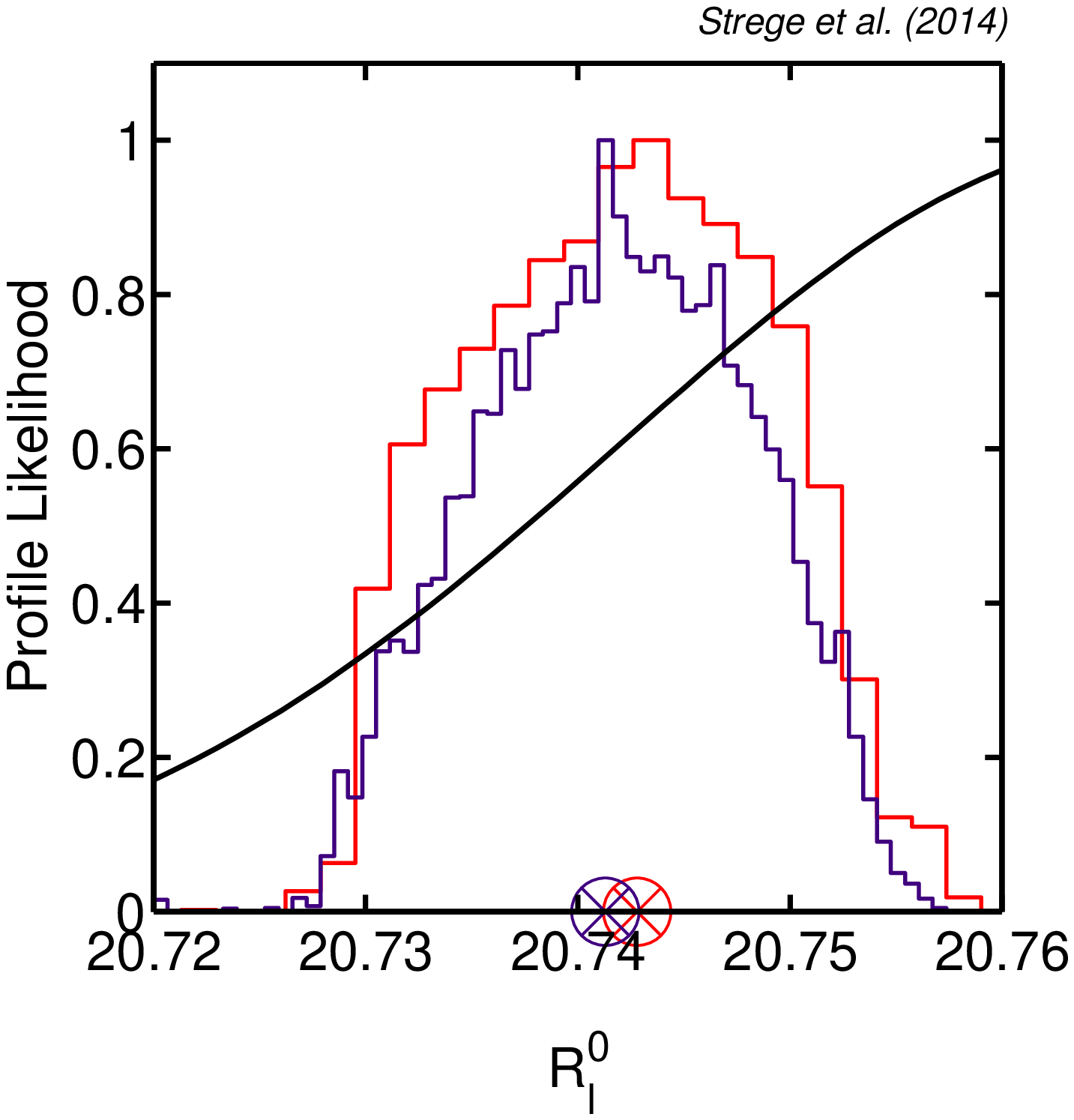}
\includegraphics[width=0.24\linewidth, trim = 0.7cm 0cm 0.7cm 0cm]{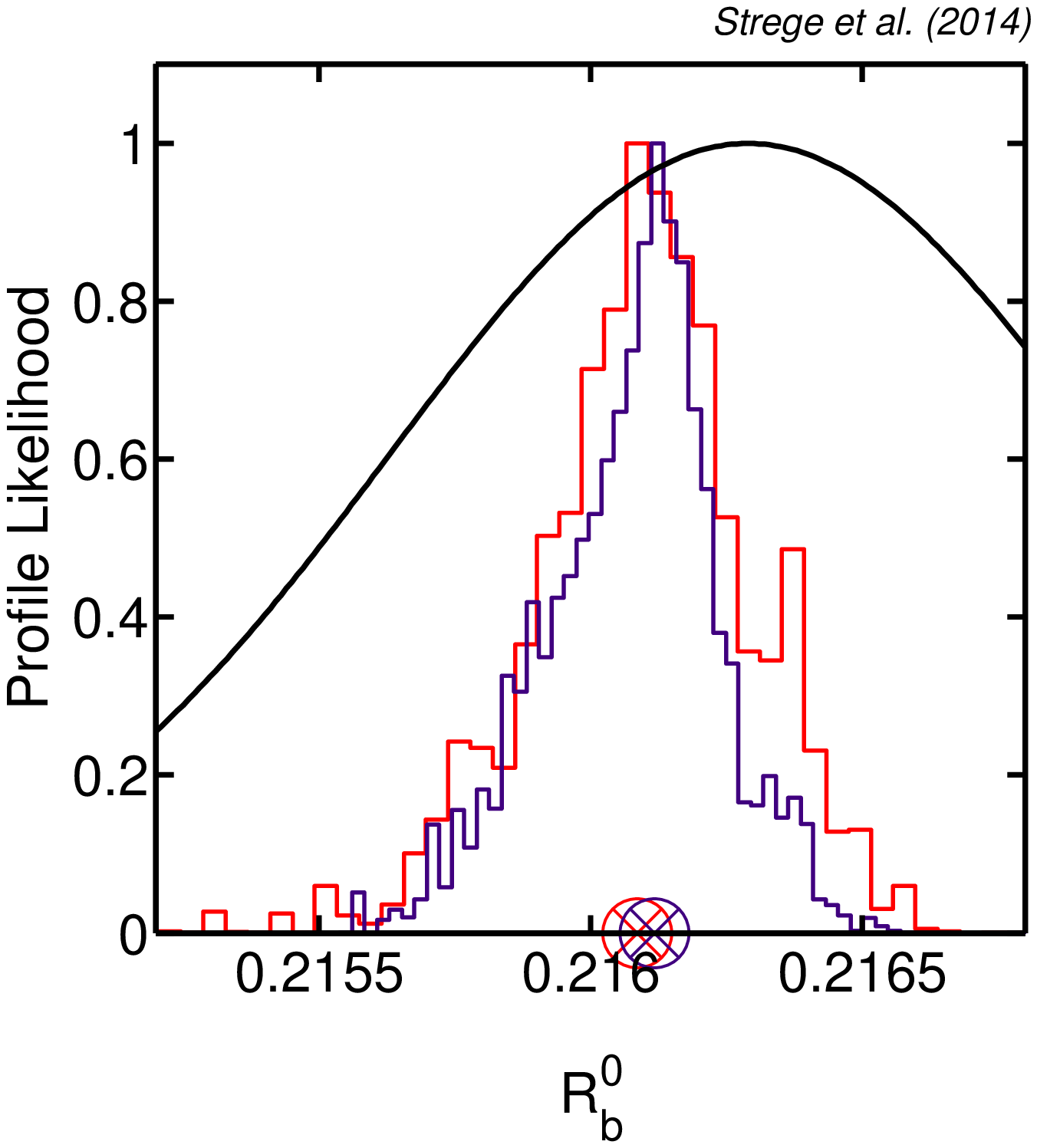}
\includegraphics[width=0.24\linewidth, trim = 0.7cm 0cm 0.7cm 0cm]{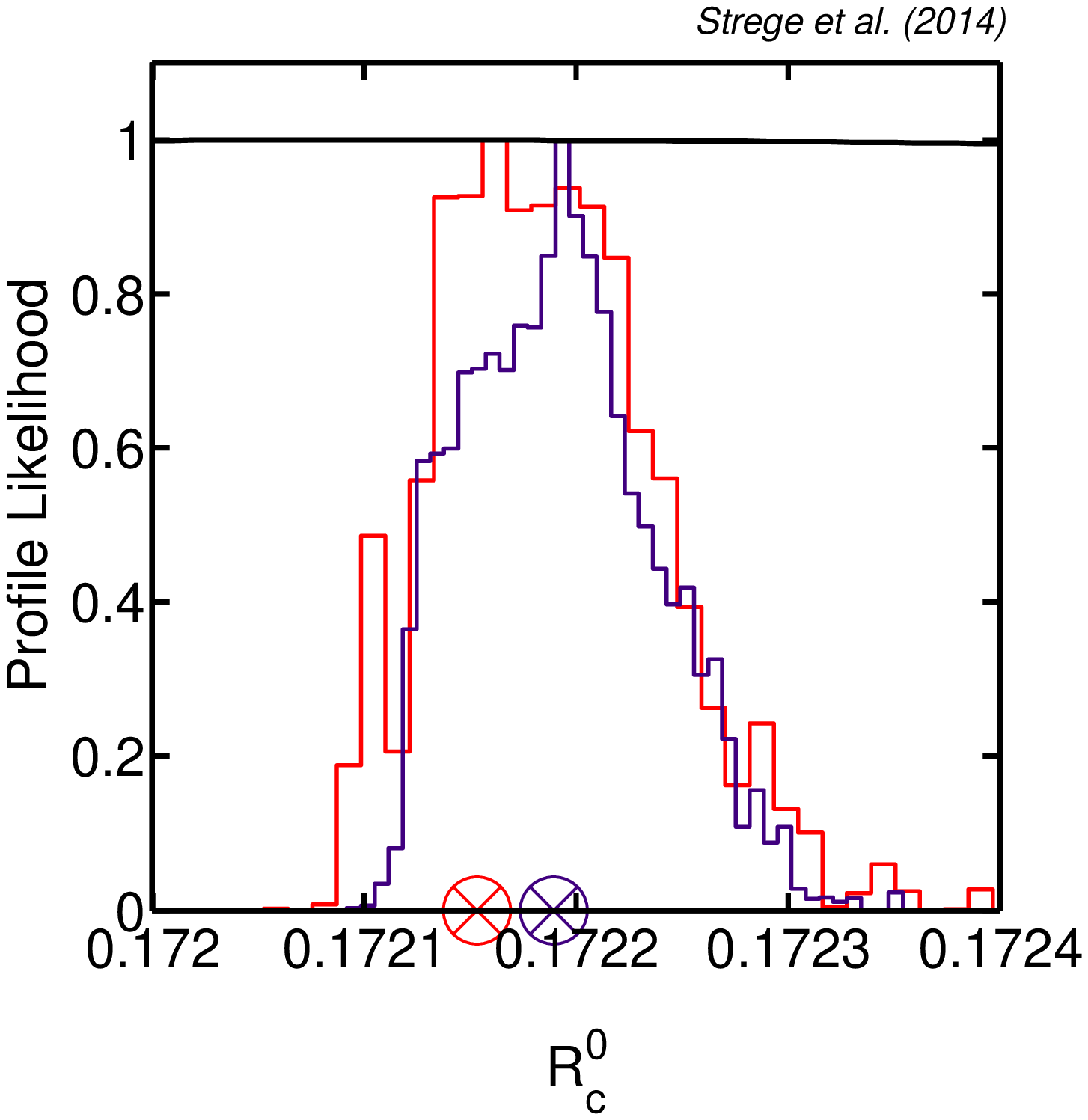}
\includegraphics[width=0.24\linewidth, trim = 0.7cm 0cm 0.7cm 0cm]{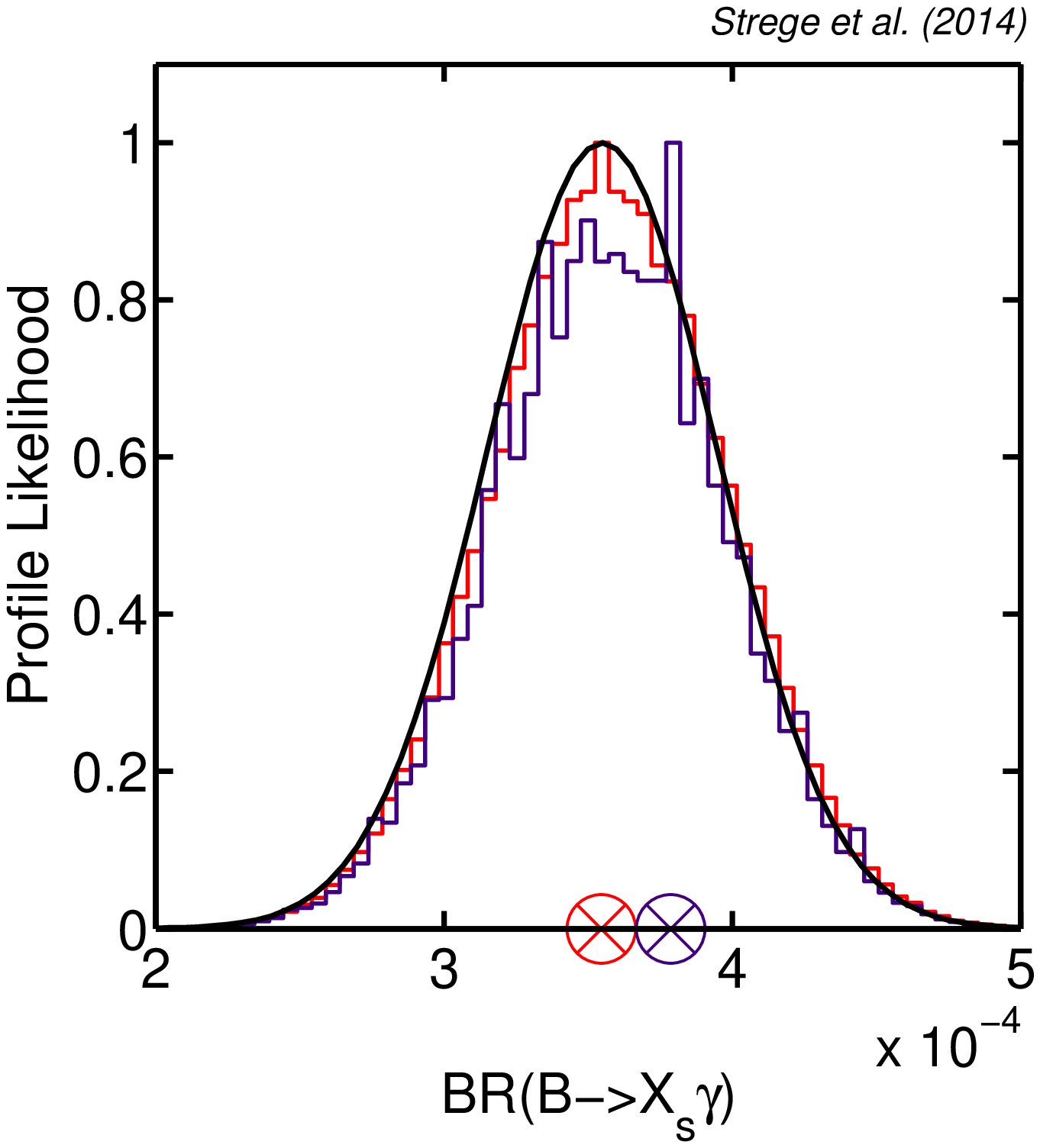}\\
\includegraphics[width=0.24\linewidth, trim = 0.7cm 0cm 0.7cm 0cm]{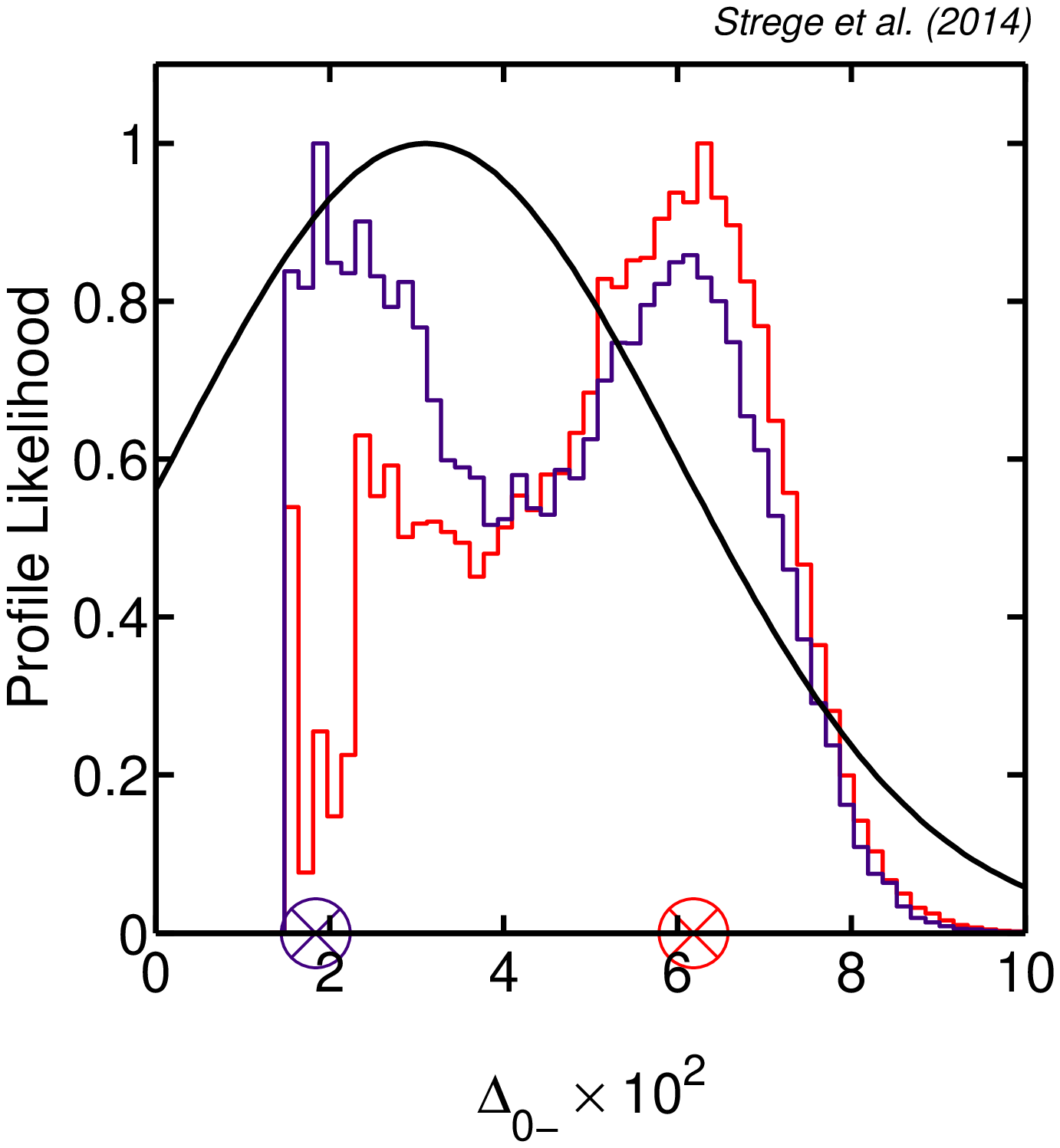}
\includegraphics[width=0.24\linewidth, trim = 0.7cm 0cm 0.7cm 0cm]{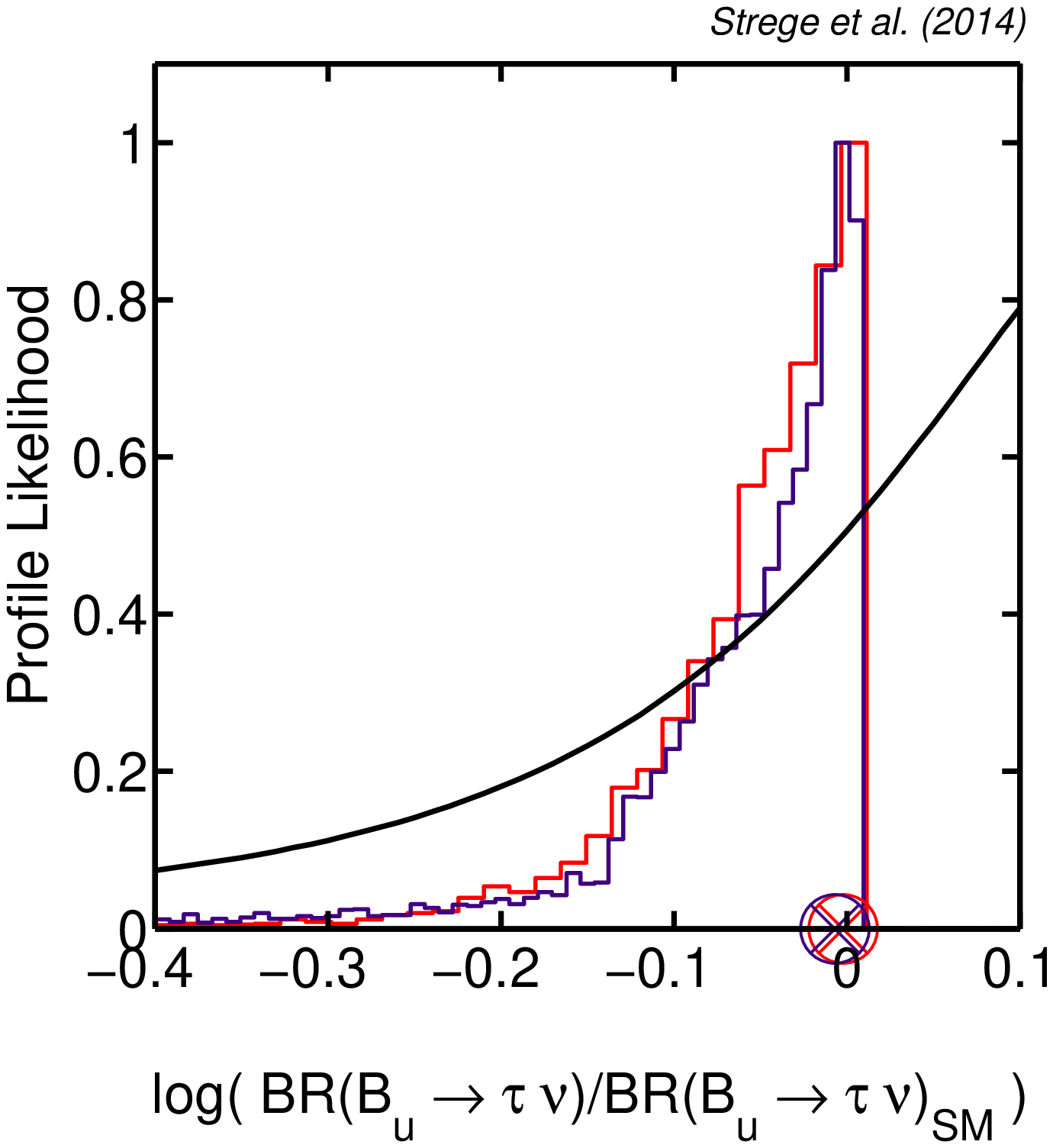}
\includegraphics[width=0.24\linewidth, trim = 0.7cm 0cm 0.7cm 0cm]{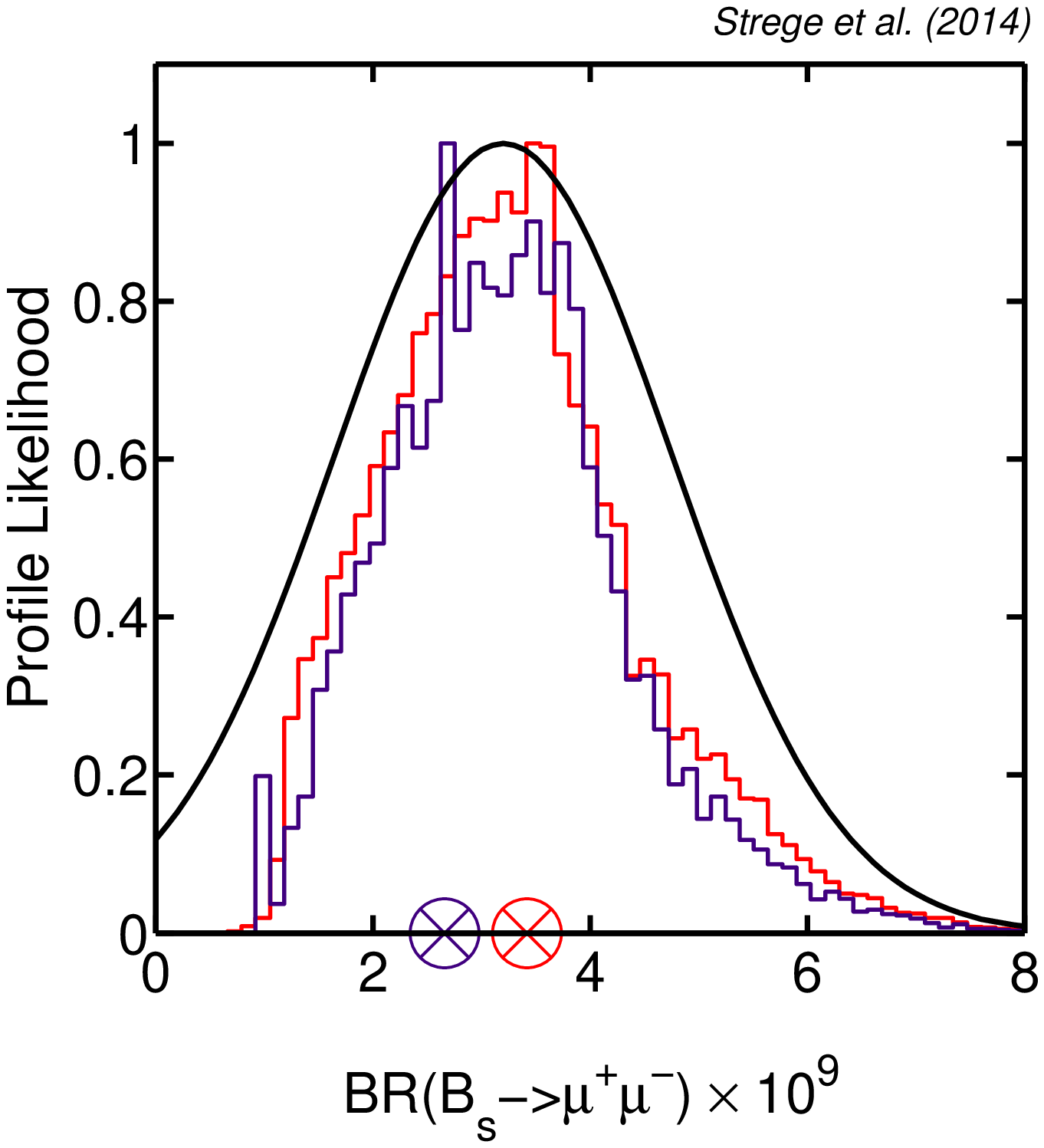}
\includegraphics[width=0.24\linewidth, trim = 0.7cm 0cm 0.7cm 0cm]{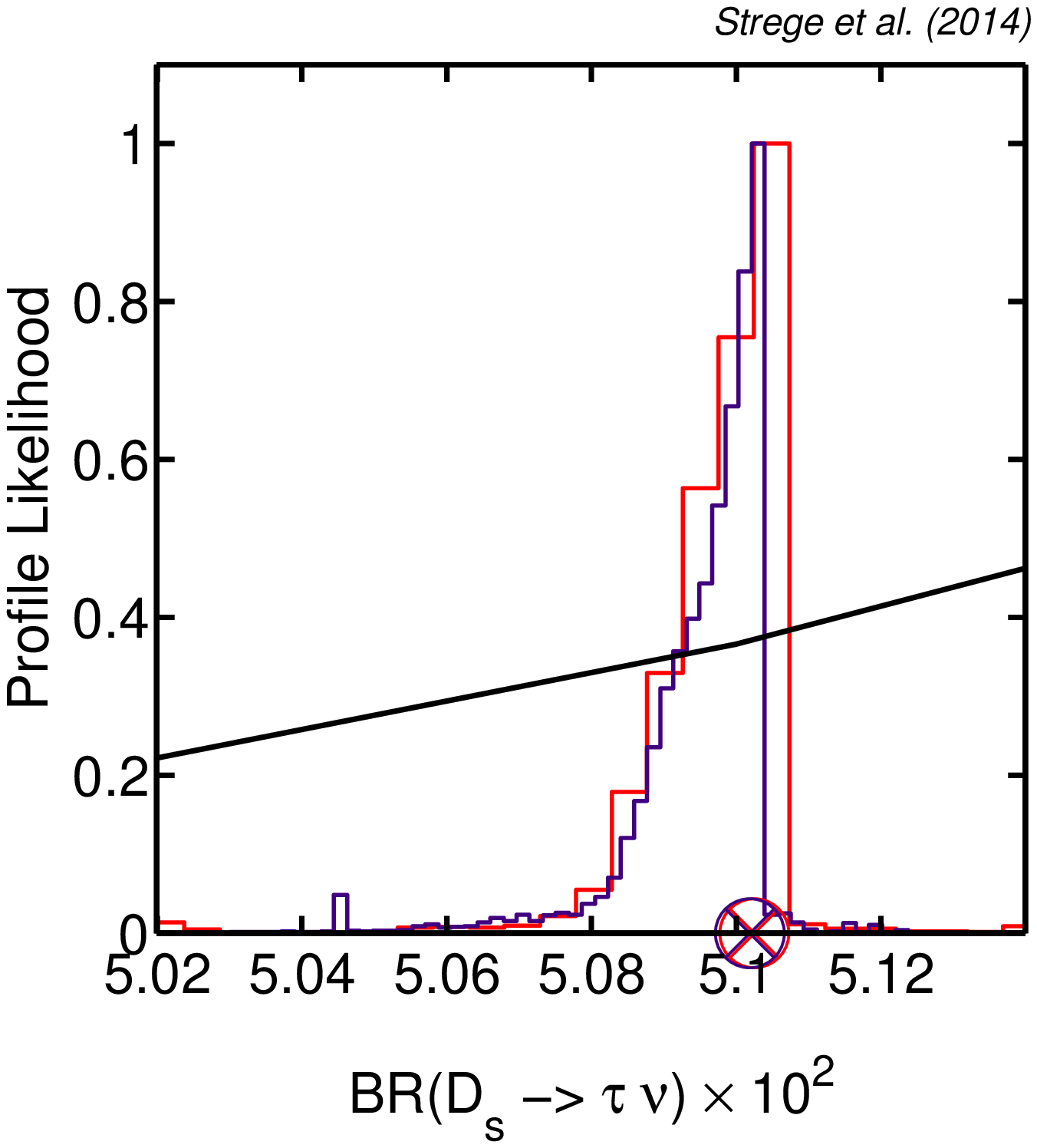}\\
\includegraphics[width=0.24\linewidth, trim = 0.7cm 0cm 0.7cm 0cm]{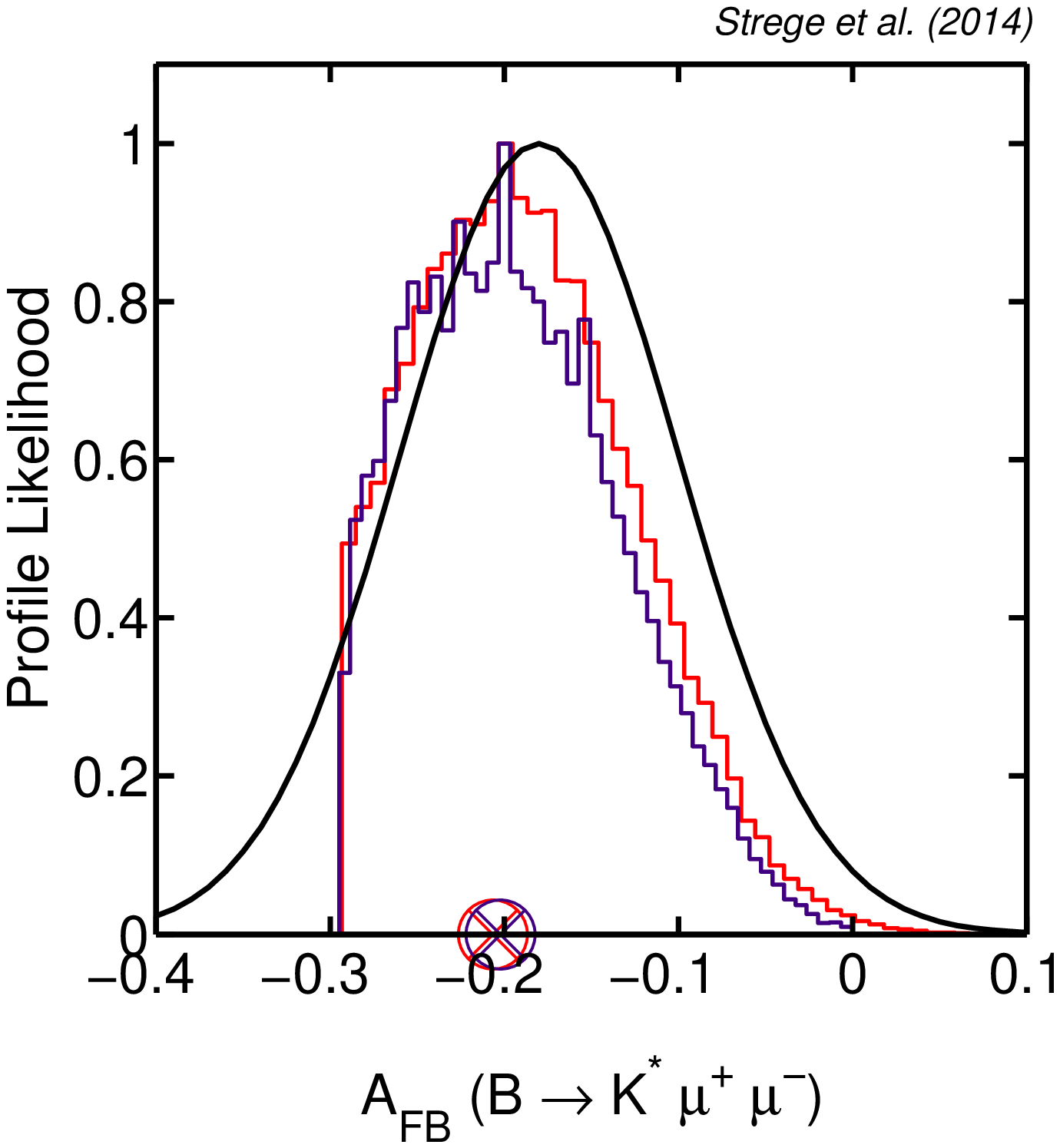}
\includegraphics[width=0.24\linewidth, trim = 0.7cm 0cm 0.7cm 0cm]{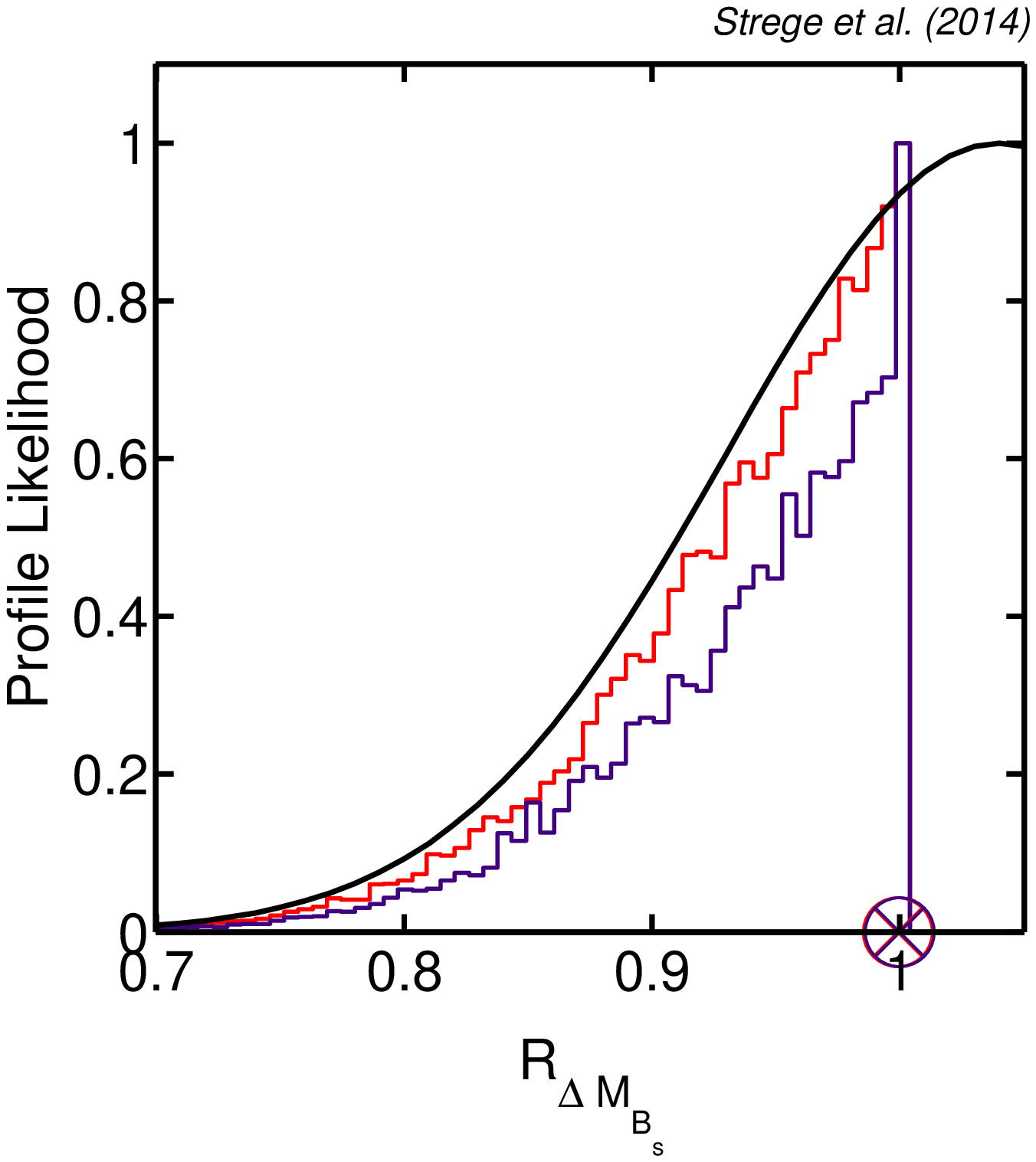}
\includegraphics[width=0.24\linewidth, trim = 0.7cm 0cm 0.7cm 0cm]{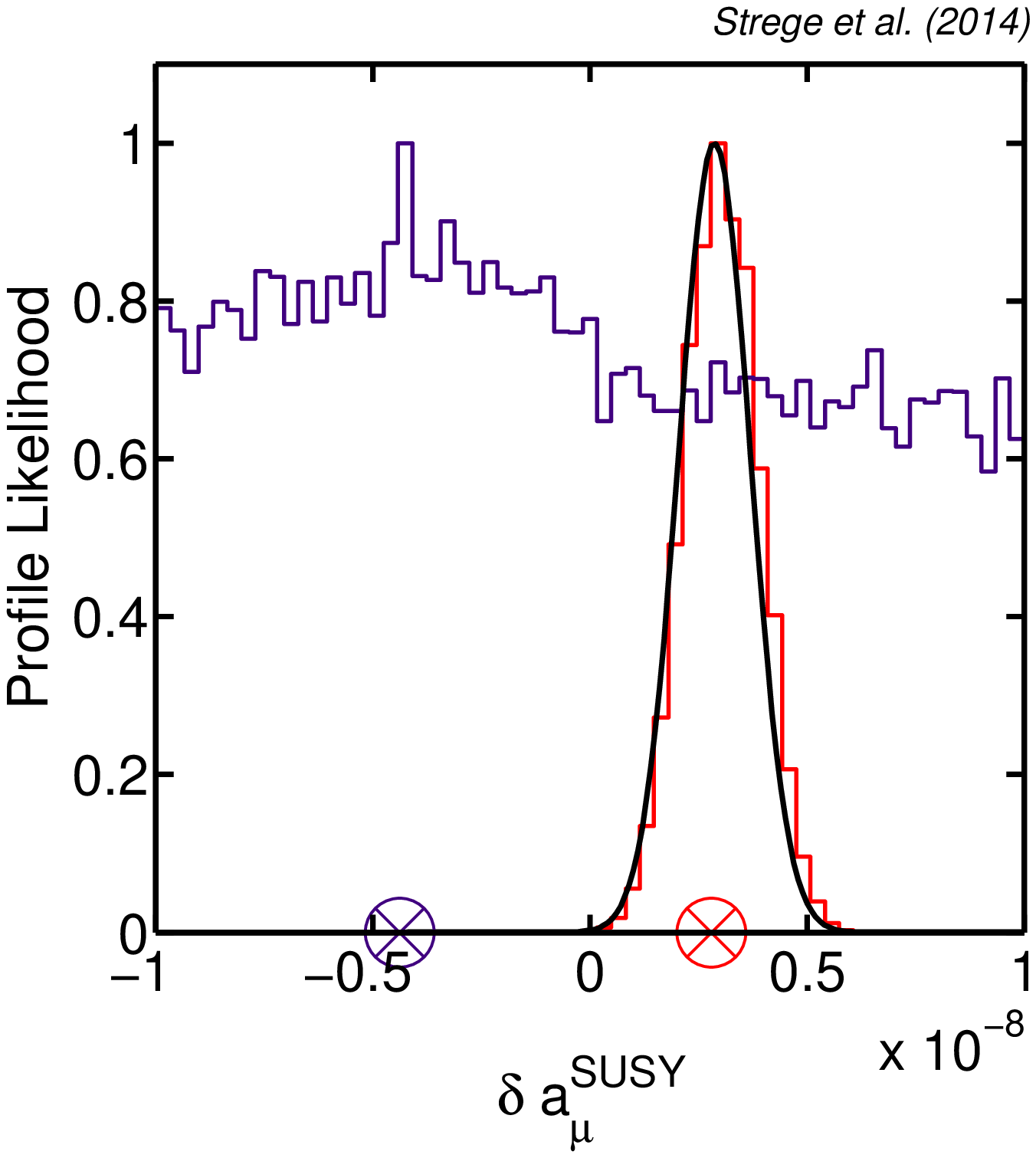}
\includegraphics[width=0.24\linewidth, trim = 0.7cm 0cm 0.7cm 0cm]{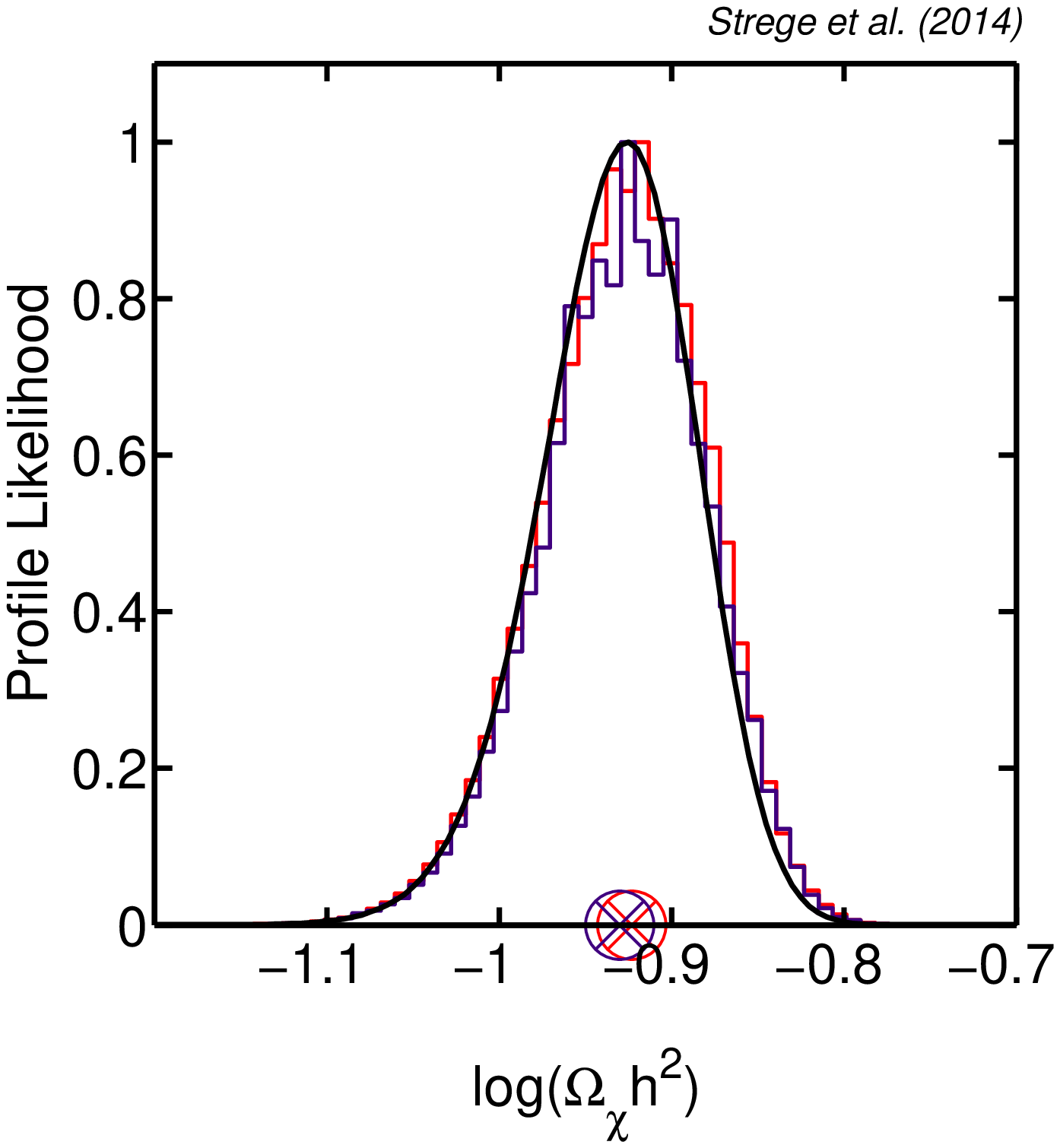}
\caption{1-D profile likelihood global fits results including all data except LHC SUSY searches and Higgs couplings (red) and further excluding the $g - 2$ constraint (purple), for observable quantities. Encircled crosses represent the best-fit points. Black lines are the likelihood function for the corresponding observable.}
\label{fig:1D_wog2_2}
\end{center}
\end{figure}

The 1D profile likelihood for the observables, shown in Fig.~\ref{fig:1D_wog2_2}, generally agree well with the likelihood functions imposed on these quantities (shown in black). In contrast to the tension that is observed in simpler SUSY models, such as the cMSSM and the NUHM, the experimentally measured values of $\delta a_\mu^{\mathrm{SUSY}}$ and $\brbsgamma$ can simultaneously be achieved in the \pMSSM. The Planck measurement of the relic density is also well fit. 

The EWPOs most sensitive to SUSY effects within the \pMSSM\ are $m_W$, $\sin^2\theta_\text{eff}$ and $\Gamma_Z$, with the most important role played by $\tilde{t}/\tilde{b}$ and --to a lesser extent-- by the chargino and neutralino sectors. Their dependence on the top mass is also significant~\cite{Heinemeyer:2007bw}. The SM prediction for $m_W$ and $\Gamma_Z$ is marginally (at $1 \sigma$ level) below the experimental value, assuming the current central value of the top mass. Since SUSY contributions are constructive, light squarks and/or light EWKinos are required to fit the experimental values of $m_W$ and $\Gamma_Z$, as also favoured by other experimental constraints. This leads to a good match between the 1D PL and the likelihood function for these quantities.  
The $\sin^2\theta_\text{eff}$ case is different since SUSY corrections are destructive for this observable and the SM prediction is compatible at $1 \sigma$ level with the measurements. Therefore a low SUSY spectrum pushes the predictions below the measurement which is what we observed in the PL.
The other EWPOs considered in our analysis, namely $\sigma_{had}^0$, $R_l$, $R_b$ and $R_c$, are mostly insensitive to SUSY effects, so that their PL peak at the values predicted in the SM.

The relevant flavour observables are generally well fit. A notable exception is the isospin asymmetry $\DeltaO$, which requires very large SUSY contributions (see Ref.~\cite{Strege:2012bt} for a discussion of the discrepancy between the experimental measurement of $\DeltaO$ and the values favoured in simple SUSY models). Note however that, for the reasons pointed out above, a good fit to this quantity can be obtained in the analysis excluding the \gmt\ constraint. As expected, the 1D PL for $\delta a_\mu^{\mathrm{SUSY}}$ becomes essentially flat when this constraint is dropped from the analysis.
 
%%%%%%%%%%%%%%%%%%%%%%%%%%%%%%%%%%%%%%%%%%%%%%%%%%%%%%%%%%%%%%%%%%%%%%%%%%%%%%%%%%%%%%%%%%
\subsubsection{Profile likelihood for the SUSY mass spectrum}
%%%%%%%%%%%%%%%%%%%%%%%%%%%%%%%%%%%%%%%%%%%%%%%%%%%%%%%%%%%%%%%%%%%%%%%%%%%%%%%%%%%%%%%%%

\begin{figure}
\begin{center}
\includegraphics[width=0.32\linewidth, trim = 0.7cm 0cm 0.7cm 0cm]{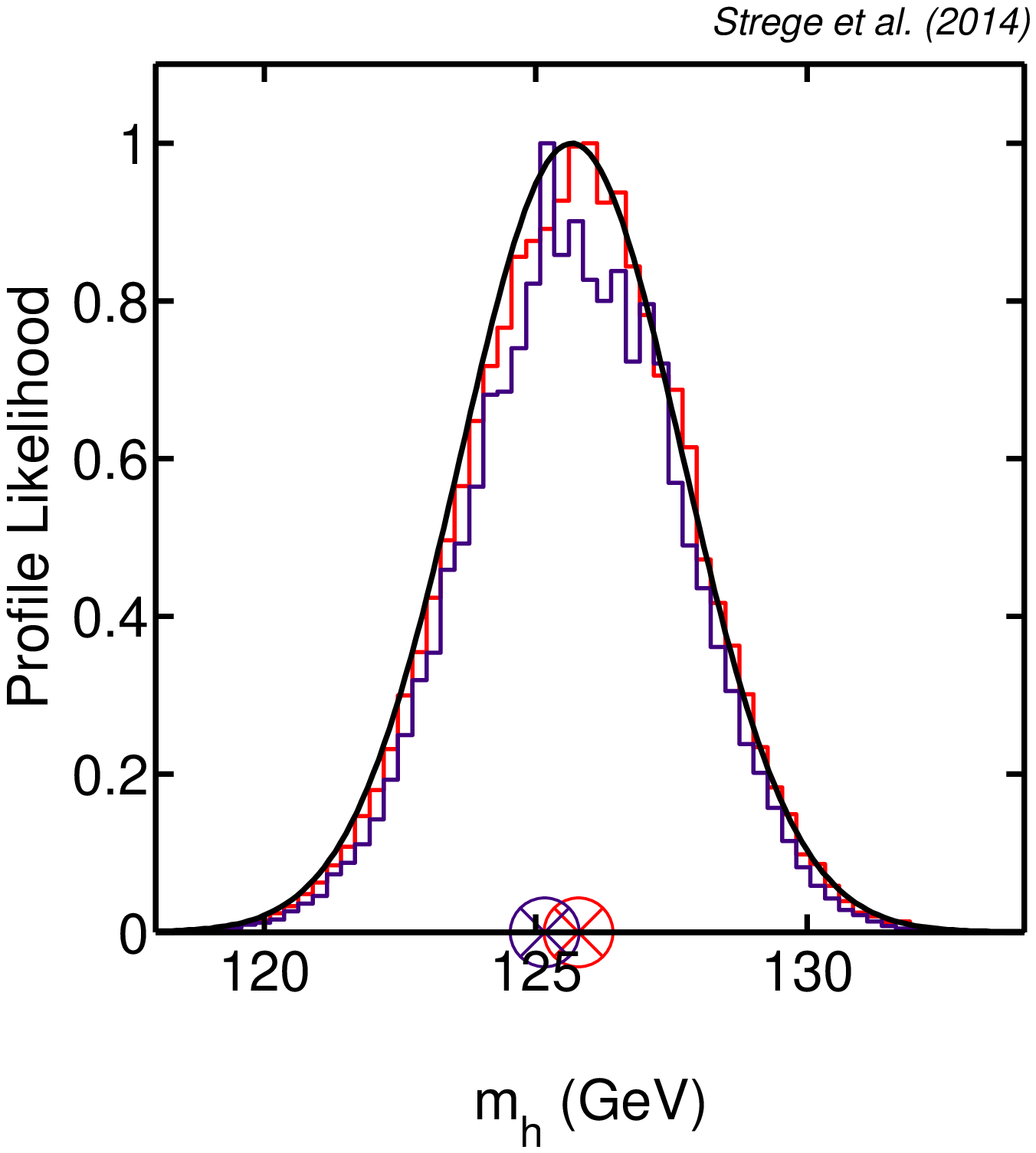}
\includegraphics[width=0.32\linewidth, trim = 0.7cm 0cm 0.7cm 0cm]{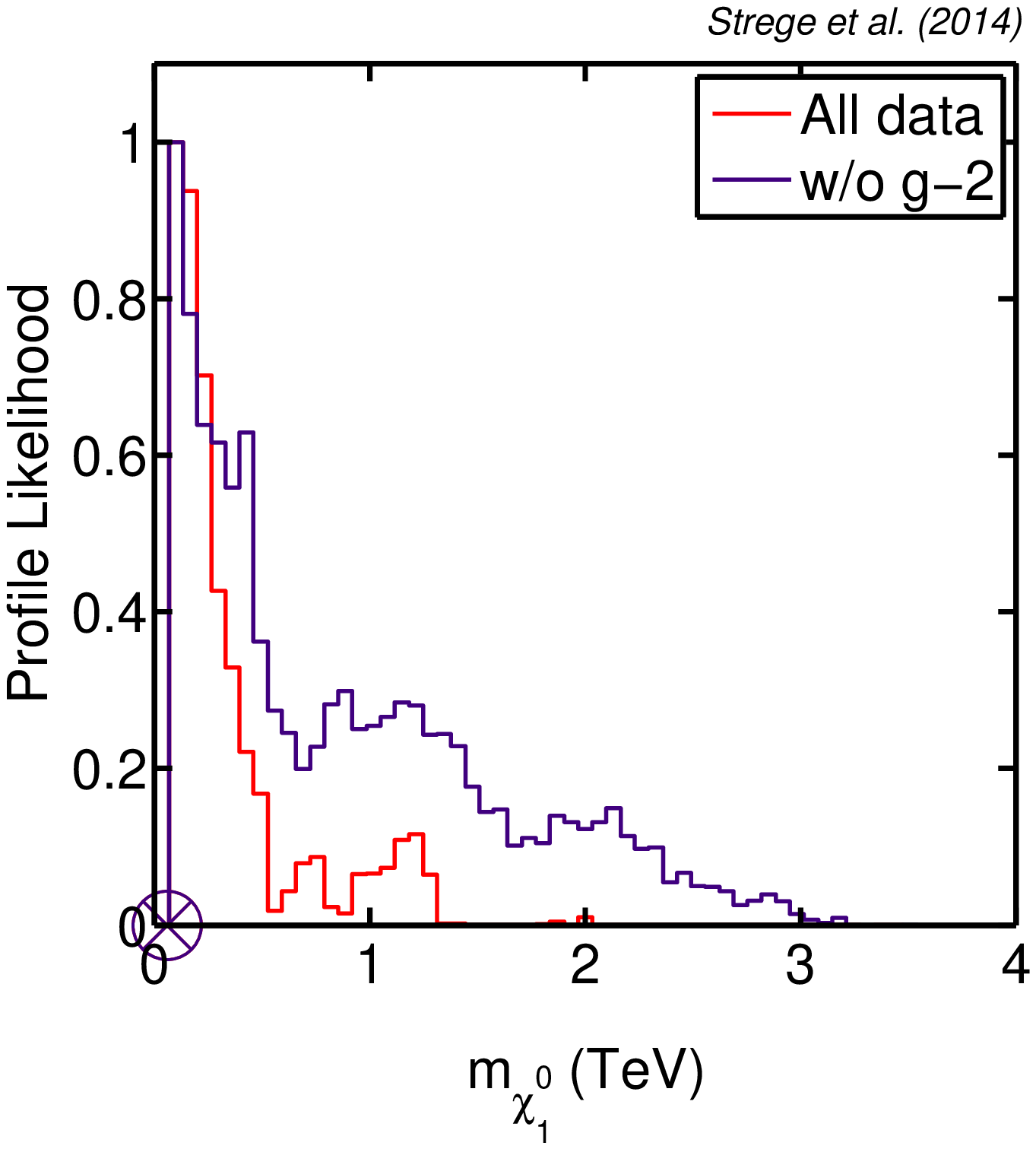}
\includegraphics[width=0.32\linewidth, trim = 0.7cm 0cm 0.7cm 0cm]{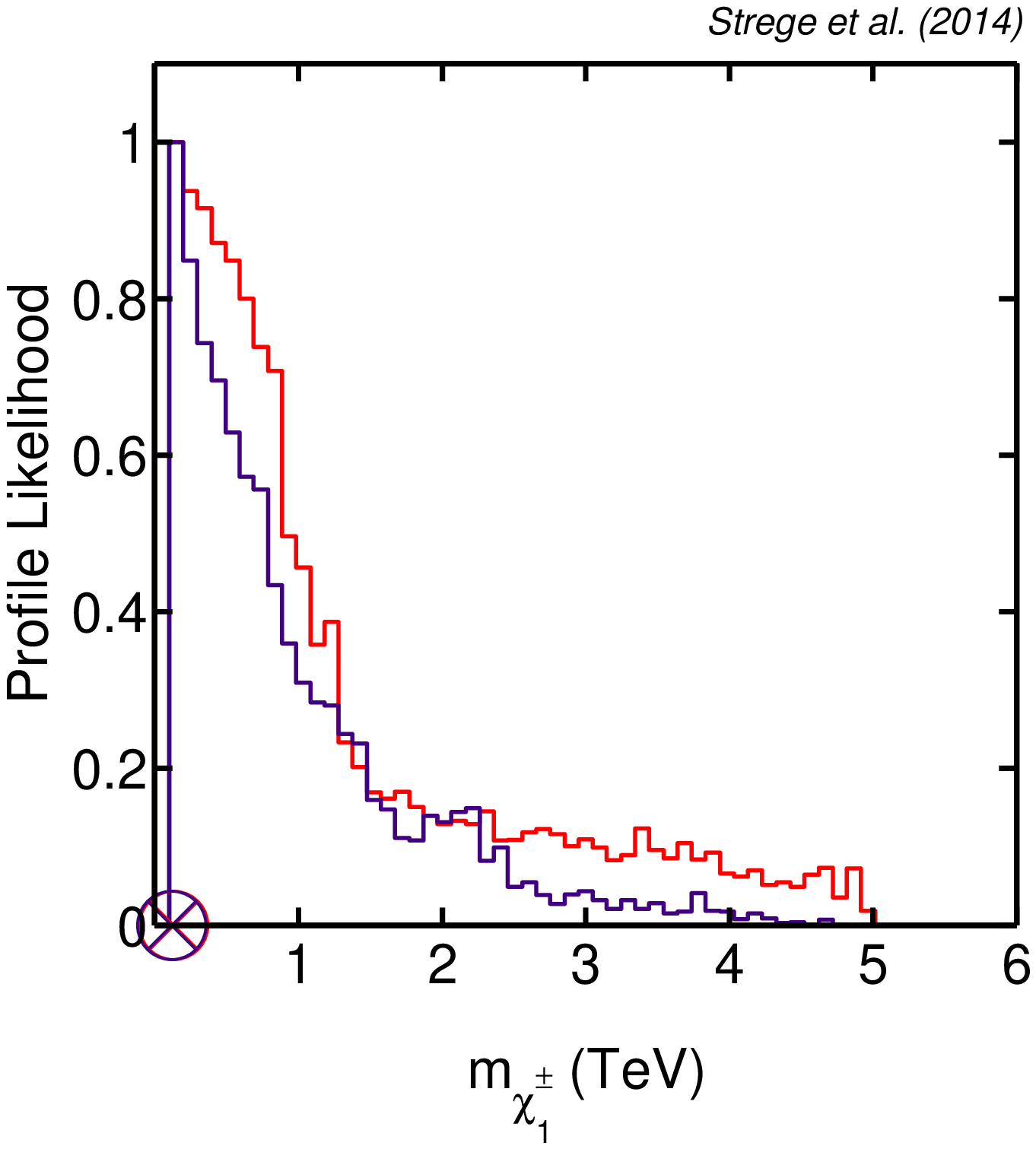}\\
\includegraphics[width=0.32\linewidth, trim = 0.7cm 0cm 0.7cm 0cm]{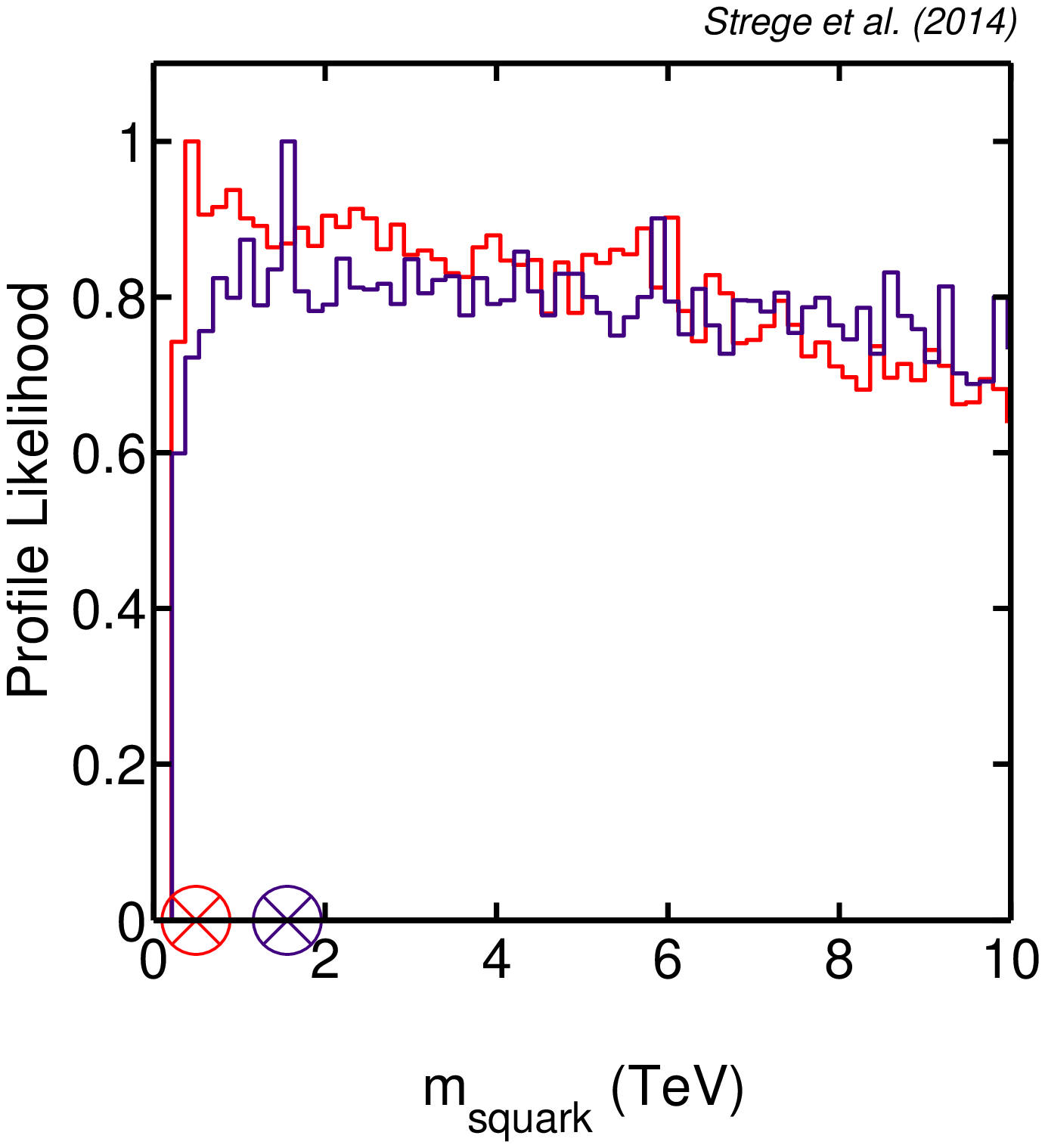}
\includegraphics[width=0.32\linewidth, trim = 0.7cm 0cm 0.7cm 0cm]{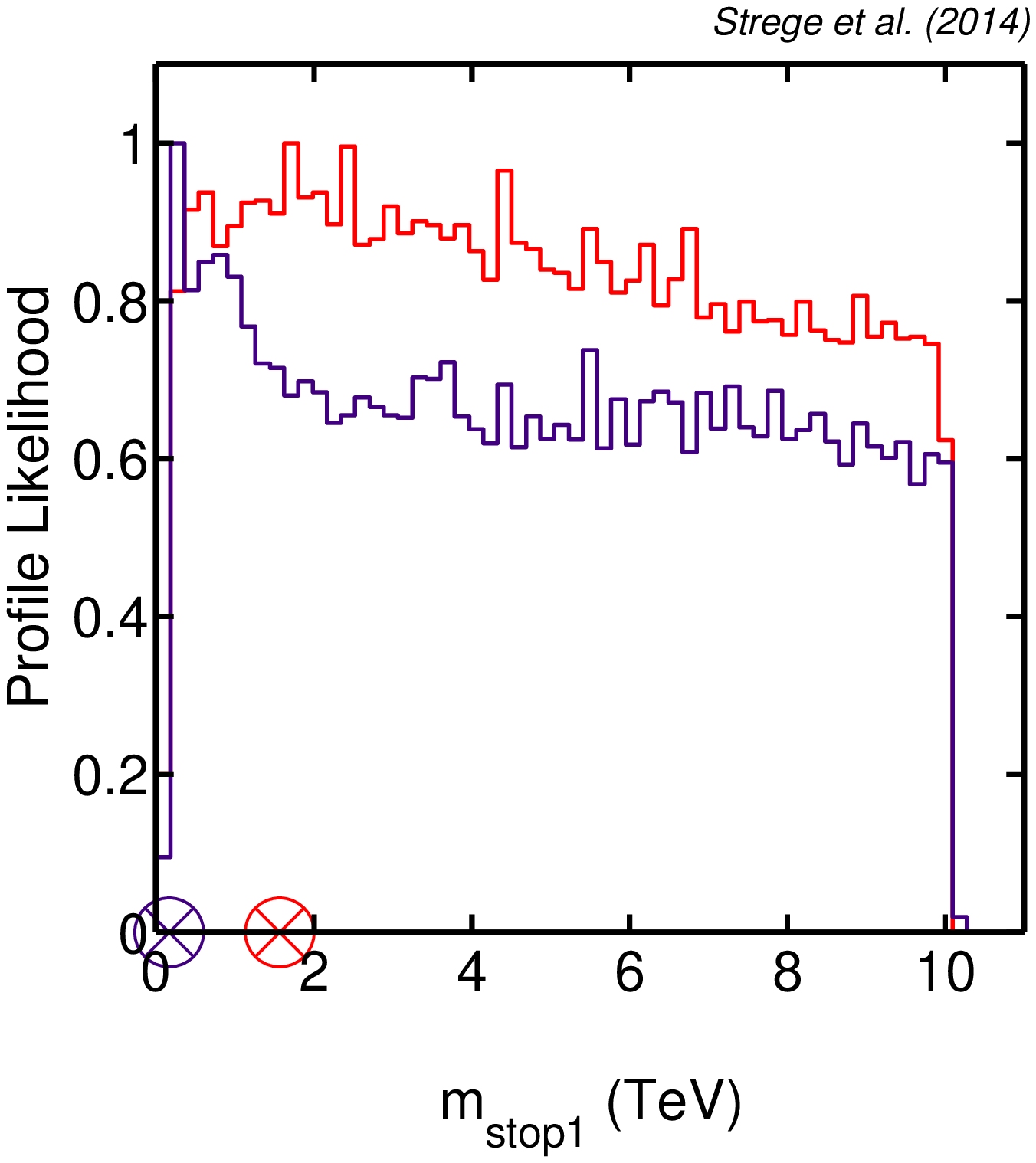}
\includegraphics[width=0.32\linewidth, trim = 0.7cm 0cm 0.7cm 0cm]{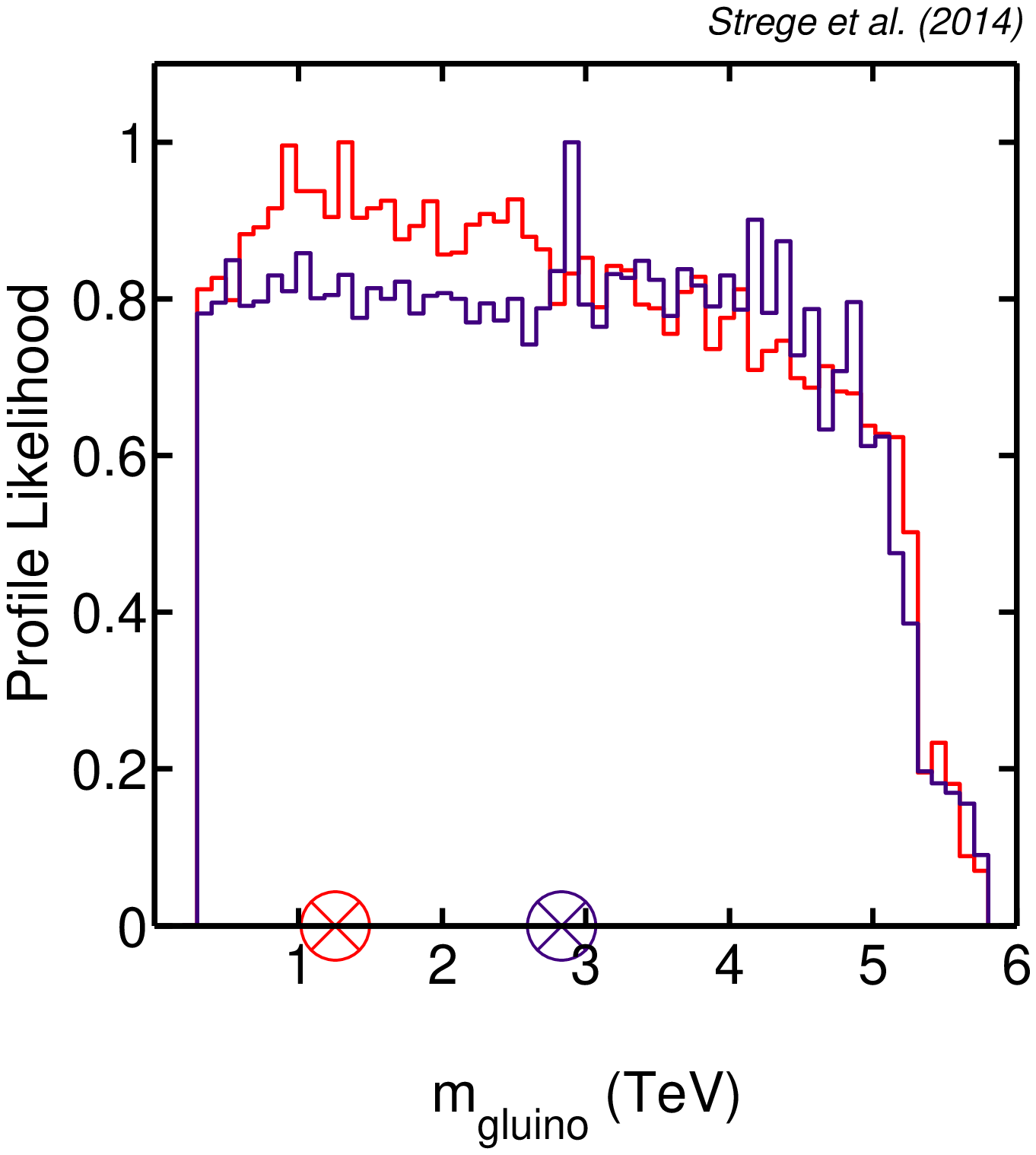}
\caption{1-D profile likelihood global fits results including all data except LHC SUSY searches and Higgs couplings (red) and further excluding the $g - 2$ constraint (purple) for some relevant SUSY quantities. Encircled crosses represent the best-fit points. For quantities constrained in the scan, the likelihood function applied is shown in black. Recall that these analyses does not include null SUSY searches at the LHC (see Section~\ref{sec:LHC_impact}).}
\label{fig:1D_wog2_3}
\end{center}
\end{figure}

The 1D PL for several SUSY masses are displayed in Fig.~\ref{fig:1D_wog2_3}. The mass of the lightest Higgs boson measured by the LHC can easily be satisfied in the \pMSSM. This is a reflection of the large number of degrees of freedom of the model, which allow to maximize the tree-level contribution to the Higgs mass by pushing \tanb\ to large values, while at the same time maximizing the leading 1-loop corrections either via heavy stops or maximal stop mixing.

The mass of the neutralino LSP is shown in the top-central panel. For the analysis including all data, the neutralino mass is constrained to $m_{\neut} < 1.5$ TeV at 99\% confidence level. In contrast, the 1D PL for the analysis excluding the \gmt\ constraint reaches significantly larger masses $m_{\neut} \leq 3.0$ TeV. In both cases, the PL peaks at low values, where the neutralino is bino-like, with an almost identical best fit at $m_\neut \approx 60$ GeV (see Table~\ref{tab:bestfit} below).
The bump in the neutralino PL around $\sim 1 \tev$ corresponds to a higgsino-like neutralino~(see Section~\ref{sec:Composition} below for further details), and it is more pronounced for the case without \gmt,  as expected from the above discussion.  In the latter case,  the small bump at $m_\neut \sim 2 \tev$ in the PL corresponds to a wino-like neutralino.

 The 1D PL for the mass of the lightest chargino stretches to large values, close to the prior boundary around $\sim 5$ TeV imposed by the prior on the input parameters. Nevertheless, similarly to what was observed for the neutralino mass, small chargino masses are favoured. In contrast, the 1D PL for the average squark mass, the lightest stop mass and the gluino mass remain almost unconstrained. The shape of the 1D PL for these quantities is a direct consequence of the 1D PL for the corresponding soft masses and $M_3$, respectively (discussed above).

%%%%%%%%%%%%%%%%%%%%%%%%%%%%%%%%%%%%%%%%%%%%%%%%%%%%%%%%%%%%%%%%%%%%%%%%%%%%%%%%%%%%%%%%%%%%
\subsection{Relic density as an upper limit}
%%%%%%%%%%%%%%%%%%%%%%%%%%%%%%%%%%%%%%%%%%%%%%%%%%%%%%%%%%%%%%%%%%%%%%%%%%%%%%%%%%%%%%%%%%%%

We now discuss the case where the Planck measurement of the relic density is applied as an upper limit, i.e. where the cosmological dark matter (DM) consists of multiple components (one of which is the neutralino LSP). While the 1D PL for the observables are slightly broader than for the analysis implementing the Planck measurement as a constraint, most of the 1D PL are qualitatively very similar for the two cases. Therefore, we focus on discussing the results for a few selected quantities that illustrate the phenomenological differences between these two analyses. 

\begin{figure}[tbh]
\begin{center}
\includegraphics[width=0.32\linewidth, trim = 0.7cm 0cm 0.7cm 0cm]{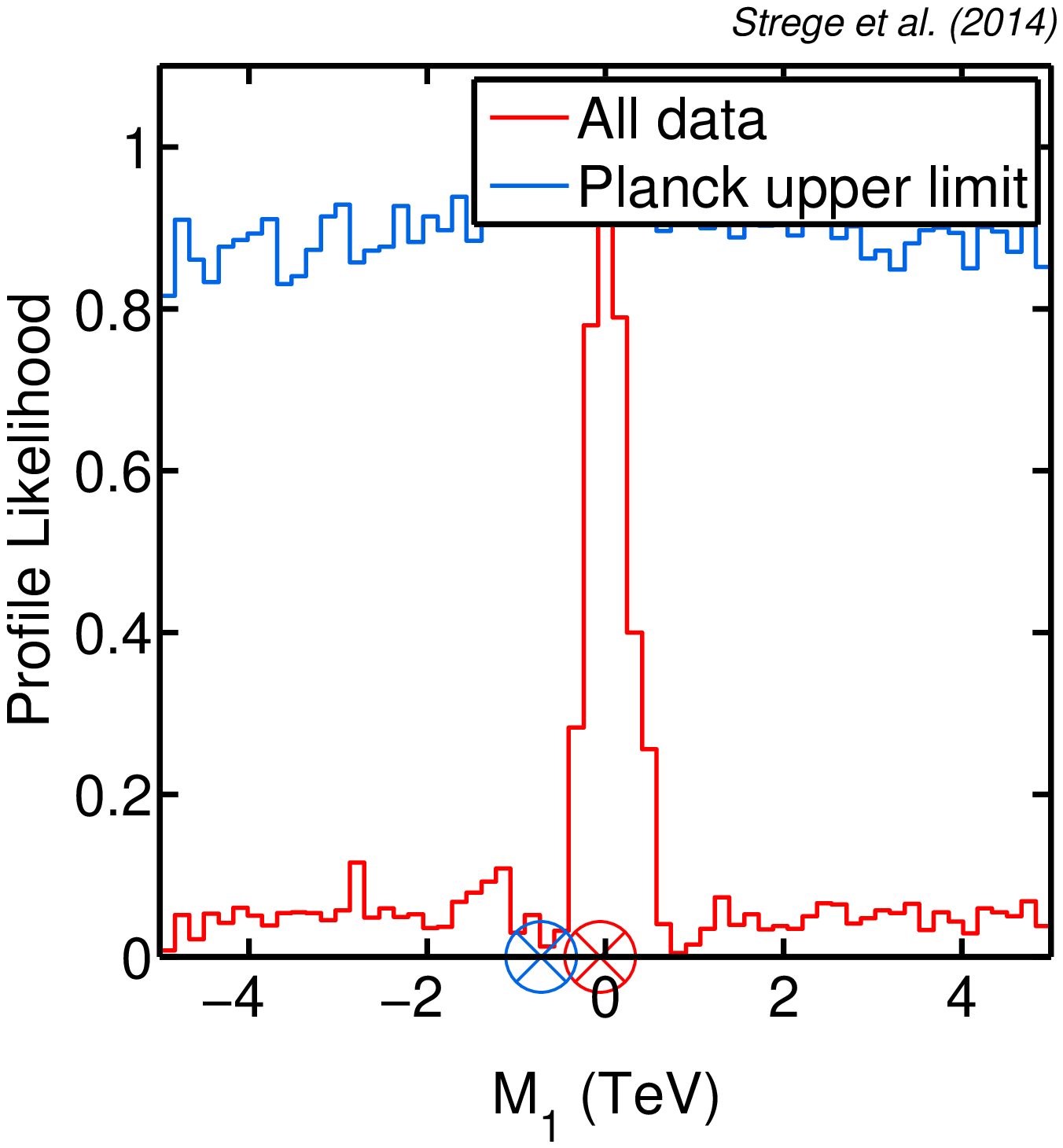}
\includegraphics[width=0.32\linewidth, trim = 0.7cm 0cm 0.7cm 0cm]{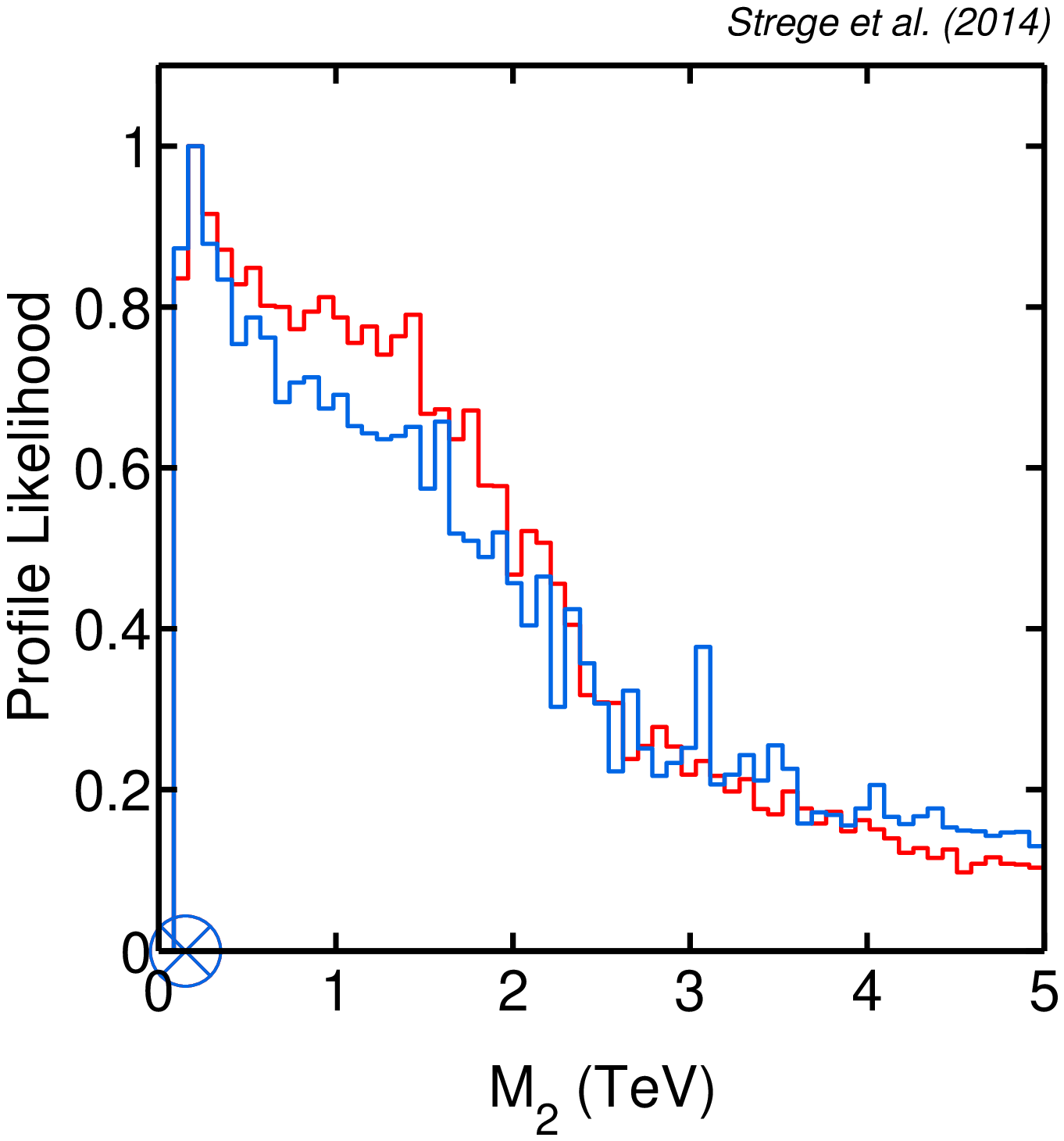}
\includegraphics[width=0.32\linewidth, trim = 0.7cm 0cm 0.7cm 0cm]{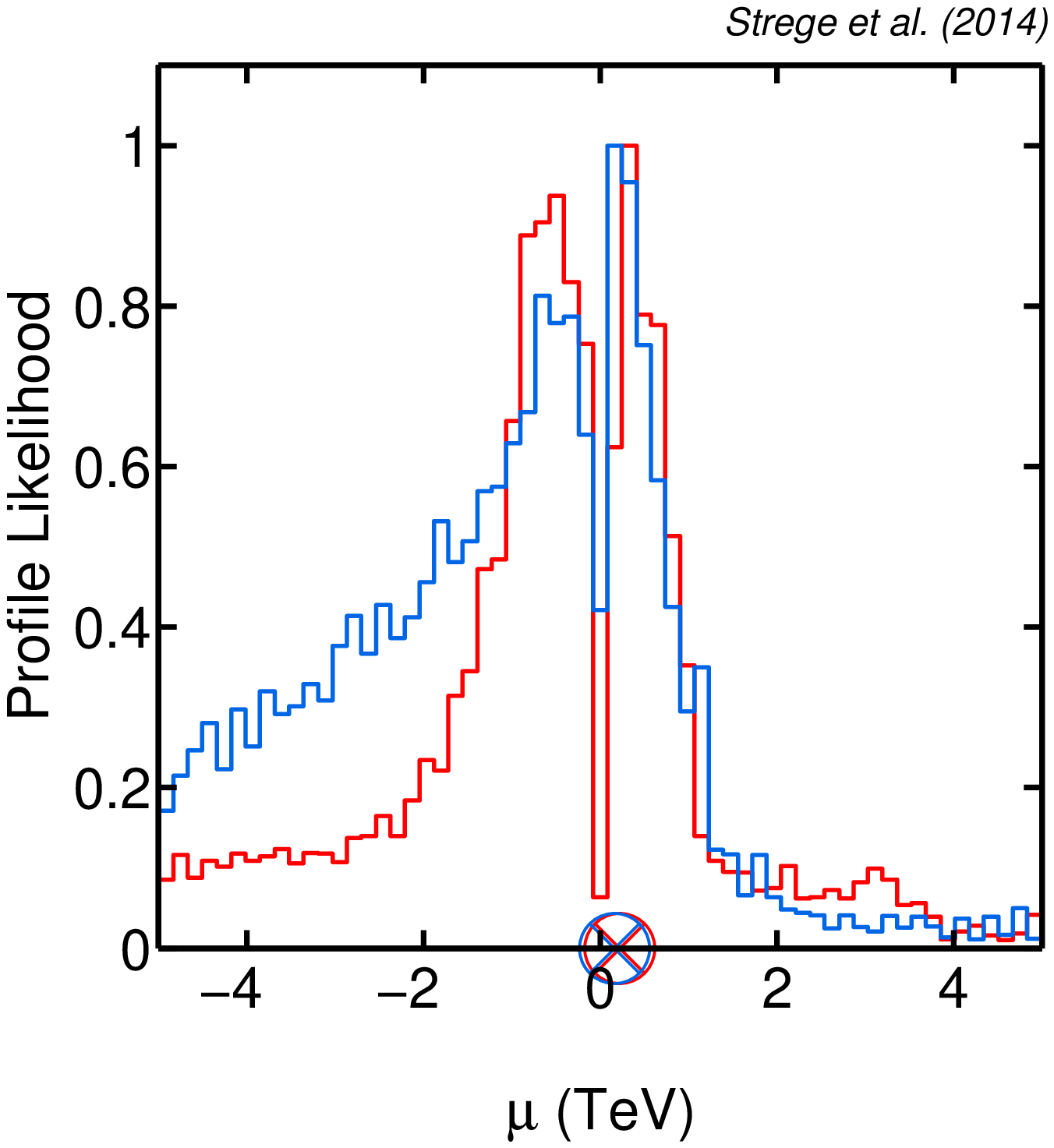} \\
\includegraphics[width=0.32\linewidth, trim = 0.7cm 0cm 0.7cm 0cm]{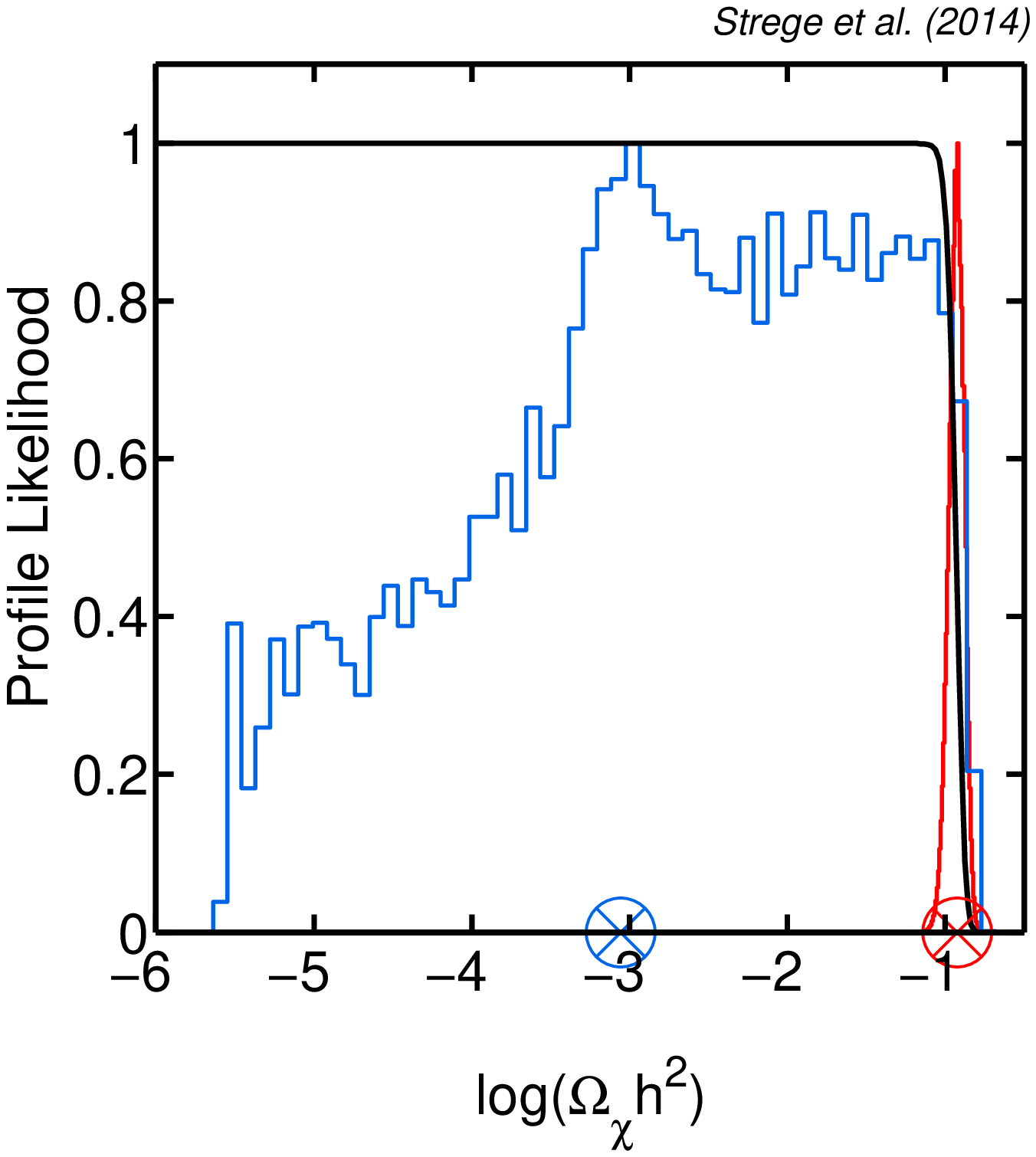}
\includegraphics[width=0.32\linewidth, trim = 0.7cm 0cm 0.7cm 0cm]{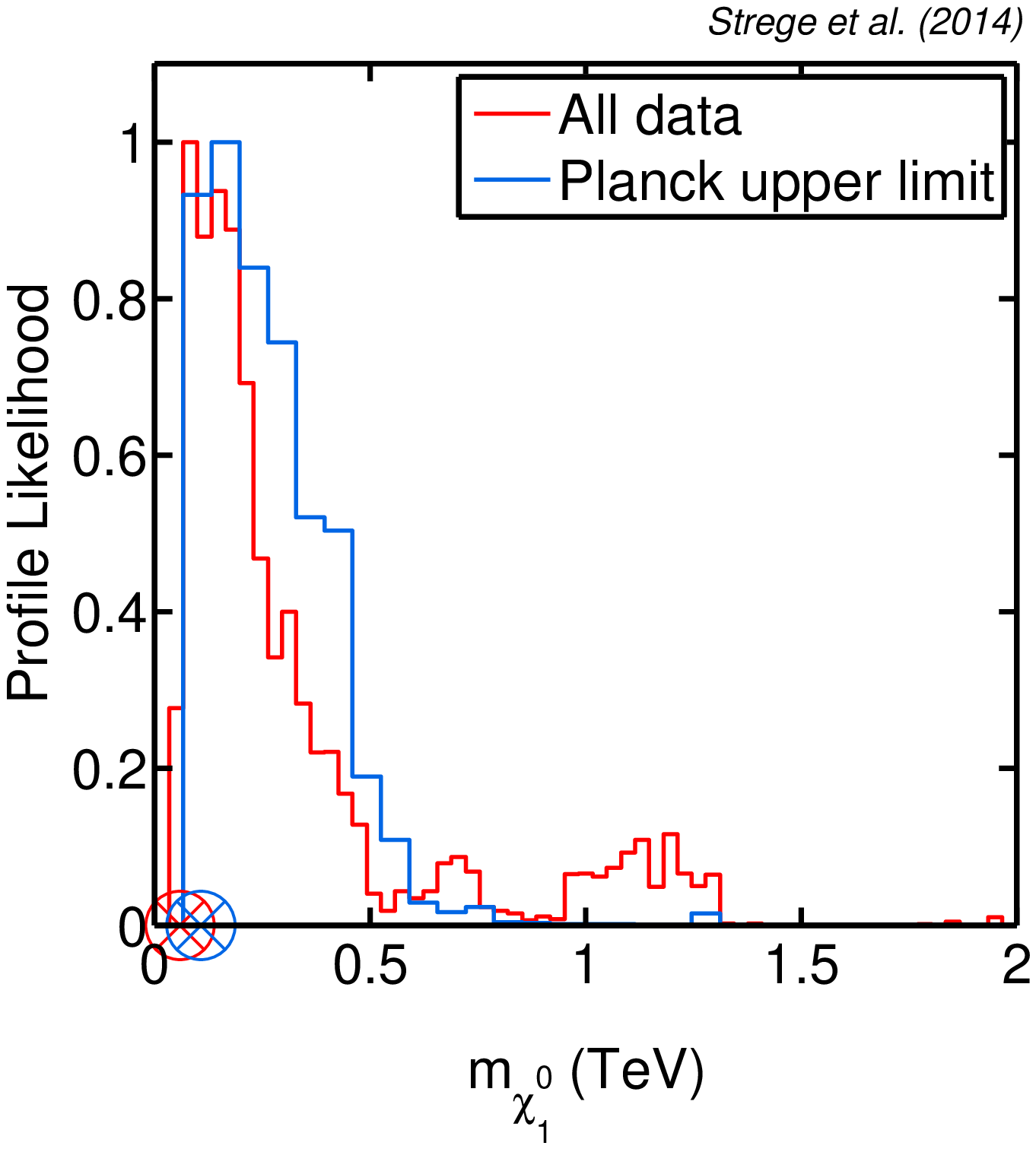}
\includegraphics[width=0.32\linewidth, trim = 0.7cm 0cm 0.7cm 0cm]{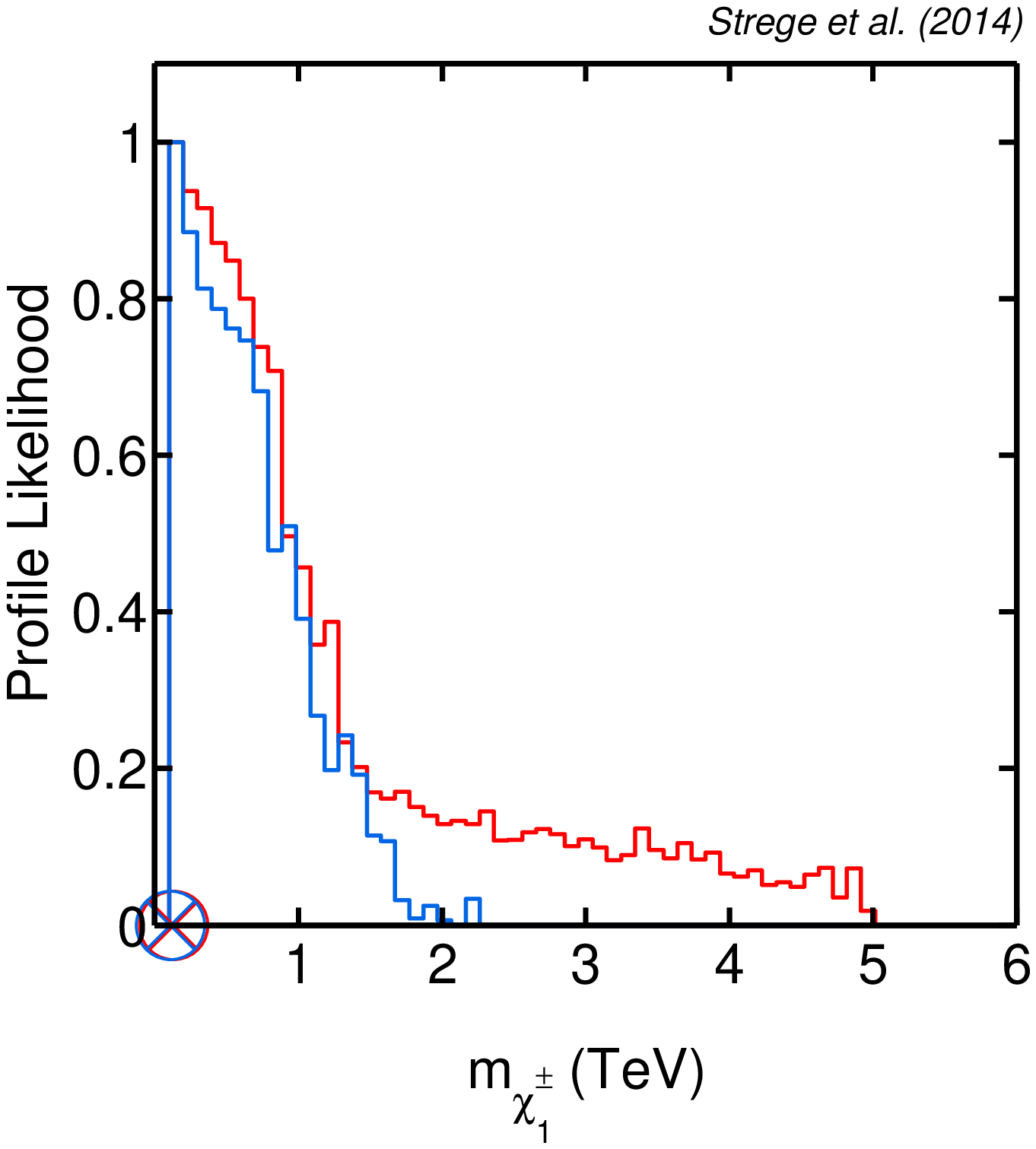}
\caption{Comparison of 1-D profile likelihood results between the case including all data (red) and using the Planck $\relic h^2$ measurement as an upper limit (blue).}
\label{fig:1D_DMlimit}
\end{center}
\end{figure}

In Fig.~\ref{fig:1D_DMlimit} we show the 1D PL for several quantities of interest, comparing the analysis in which the Planck constraint is applied as an upper limit (blue) and in which it is applied as a constraint (red). With the exception of the parameters related to the electroweakino sector, the differences with respect to the single-component DM scenario are small. The top row of Fig.~\ref{fig:1D_DMlimit} shows results for $M_1$, $M_2$ and $\mu$.
The bino mass $M_1$ now becomes essentially unconstrained over the entire prior range. The relaxation of the DM relic abundance constraint allows higher neutralino annihilation rates, so that light wino-like and higgsino-like neutralinos are now allowed, leading to heavier binos on average. Additionally, mixed neutralinos states (bino-higgsino, wino-higgsino and bino-wino-higgsino, so-called well-temped neutralinos~\cite{ArkaniHamed:2006mb}) are now allowed, as shown explicitly in Section~\ref{sec:Composition} below.

The 1D PL for the wino mass, $M_2$, is almost identical for the two cases shown. Differences in the 1D PL for the higgsino mass, $\mu$ are found in the negative branch, for which larger (more negative) values are now allowed. This is because, for larger bino masses, large $|\mu|$ help to fit the muon \gmt\ constraint (for $\text{sgn}(M_1 \mu) > 0$), as discussed in the previous section.

In the bottom row of Fig.~\ref{fig:1D_DMlimit} we show the 1D PL for the relic density, the lightest neutralino mass and the lightest chargino mass. 
As expected, the 1D PL for $\Omega_{\chi}h^2$ differs strongly for the two shown cases. When the Planck constraint is applied as an upper limit, the 1D PL stretches to very small values, almost five orders of magnitude below the measured dark matter relic density. While very small values $\Omega_{\chi}h^2 \lsim 10^{-3}$ are somewhat disfavoured, the PL peaks at $\Omega_{\chi}h^2 \sim 10^{-3}$ and is almost flat in the range $10^{-3} < \Omega_{\chi}h^2 < 10^{-1}$. 

The 1D PL for the mass of the neutralino LSP and the lightest chargino mass are now confined to significantly lower values than for the analysis requiring that $\Omega_{\chi} \sim \Omega_{\rm DM}$. The reason is that relatively light winos are allowed in the multi-component dark matter scenario, which makes it easier to fulfil the experimental constraints on a range of SM precision observables, most importantly \gmt, $\DeltaO$ and $\afb$. The experimentally measured values of these quantities are in disagreement with the SM predictions at $1 - 3\sigma$ level, and, upon relaxing the relic density constraint, play a dominant role in driving the profile likelihood results. In particular, low neutralino and chargino masses can lead to values of $\DeltaO$ and $\afb$ that are in reasonably good agreement with the observations, while at larger values of $m_\charg$ and $m_\neut$ these quantities approach their SM-like values, which are discrepant with the experimental constraints.  

%%%%%%%%%%%%%%%%%%%%%%%%%%%%%%%%%%%%%%%%%%%%%%%%%%%%%%%%%%%%%%%%%%%%%
\subsection{Best-fit points}
\label{sec:bf}
%%%%%%%%%%%%%%%%%%%%%%%%%%%%%%%%%%%%%%%%%%%%%%%%%%%%%%%%%%%%%%%%%%%%%

\begin{table*}
\begin{center}
%\begin{tabular}{|l|ll| l l}
\begin{tabular}{| l | c c | c | c c|}
\hline
& \multicolumn{2}{|c|}{All data} &  w/o g - 2 & \multicolumn{2}{|c|}{Planck upper limit} \\
\hline
\multicolumn{6}{|c|}{Input parameters}\\\hline
$M_1$ [GeV]  & -61.76 & -136.09 &  59.70 & -724.07 & -130.06 \\ 
$M_2$ [GeV]  & 150.23 & 149.98 &  123.96 & 147.96 & 814.37 \\ 
$M_3$ [GeV]  & 1191.2 & 2000.09 &  2967.70 & -1833.39 & 1294.62 \\ 
$m_{L}$ [GeV]  & 438.34 & 152.35 &  351.99 & 449.03 & 142.26 \\ 
$m_{L_3}$ [GeV]  & 286.68 & 1995.54 &  964.28 & 486.61 & 447.86 \\ 
$m_{E_3}$ [GeV]  & 389.88 & 1250.89 &  3850.93 & 1823.49 & 542.16 \\ 
$m_{Q}$ [GeV]  & 351.33 & 2234.41 &  1628.26 & 358.87 & 5860.04 \\ 
$m_{Q_3}$ [GeV]  & 2408.24 & 658.41 &  696.35 & 3573.49 & 396.24 \\ 
$m_{U_3}$ [GeV]  & 1579.95 & 1495.69 &  1341.55 & 804.81 & 1751.30 \\ 
$m_{D_3}$ [GeV]  & 503.38 & 332.04 &  920.19 & 262.12 & 141.28 \\ 
$A_t$ [GeV]  & 3025.88 & 2380.81 &  2219.57 & -3131.92 & 1962.58 \\ 
$A_0$ [GeV]  & -35.41 & 6396.91 &  1498.37 & -11.78 & 3827.41 \\ 
$\mu$ [GeV]  & 219.54 & -778.01 &  -224.60 & 158.52 & -582.89 \\ 
$m_A$ [GeV]  & 2297.46 & 1550.08 &  1298.28 & 3731.24 & 1676.59 \\ 
$\tan\beta$ & 21.82 & 17.82 &  21.85 & 20.75 & 14.93 \\ 
\hline 
$M_t$ [GeV]  & 173.34 & 173.30 &  173.19 & 173.11 & 173.06 \\ 
\hline
\multicolumn{6}{|c|}{Observables}\\\hline
$m_h$ [GeV] & 125.78 & 125.52 &  125.16 & 125.61 & 125.41 \\ 
$\delta a_\mu^{\mathrm{SUSY}} \times 10^{10}$ & 27.98 & 30.18 &  -43.91 & 28.63 & 27.87 \\ 
$m_{\rm squark}$ [GeV] & 489.57 & 2253.08 &  1554.61 & 497.96 & 5904.73 \\ 
$m_{\rm stop1}$ [GeV] & 1568.78 & 588.55 &  166.32 & 943.63 & 443.04 \\ 
$m_{\rm gluino}$ [GeV] & 1256.10 & 2050.19 &  2834.23 & 1883.16 & 1463.97 \\ 
$m_\neut$ [GeV] & 58.48 & 134.16 &  57.95 & 106.32 & 128.37 \\ 
$m_\charg$ [GeV] & 130.26 & 159.29 &  118.10 & 109.17 & 578.25 \\ 
$\sigmaSI $ [pb] & $3.56 \times 10^{-11}$ & $2.35 \times 10^{-10}$ &  $3.86  \times 10^{-11}$ & $4.40 \times 10^{-8}$ & $1.03 \times 10^{-9}$ \\ 
$\sigmaSDp$ [pb] & $2.34 \times 10^{-5}$ & $2.14 \times 10^{-7}$ &  $4.79 \times 10^{-5}$ & $9.78 \times 10^{-4}$ & $8.78 \times 10^{-7}$ \\ 
$\sigmaSDn$ [pb] & $3.48 \times 10^{-5}$ & $2.57 \times 10^{-7}$ &  $4.63 \times 10^{-5}$ & $1.02 \times 10^{-3}$ & $8.35 \times 10^{-7}$ \\ 
$\Omega_\chi h^2$ & 0.1194 & 0.1186 &  0.1174 & $8.84 \times 10^{-4}$ & $5.20 \times 10^{-2}$ \\ 
\hline
\multicolumn{6}{|c|}{$\chi^2$ values}\\\hline
Pre-LHC & 8.18 & 8.64 & 7.79 & 8.18 & 8.91 \\ 
Post-LHC & 1052.7 & 9.45 &  8.3 & 268.3 & 9.68 \\  
\hline 
\end{tabular}
\end{center}
\caption{Best-fit values of the MSSM-15 input parameters and several observables of interest. For the cases  ``All data'' and ``Planck upper limit'', we show both the overall pre-LHC best-fits in the second and fifth column (those points are ruled out by ATLAS data) and the best-fitting point surviving the addition of ATLAS constraints and Higgs boson properties data (third and sixth column). The bottom section gives the corresponding $\chi^2$ values. Notice that the ``Pre-LHC'' data do include the Higgs mass measurement. \label{tab:bestfit}}
\end{table*}

\begin{figure}
\begin{center}
\begin{center}
\includegraphics[width=0.8\linewidth]{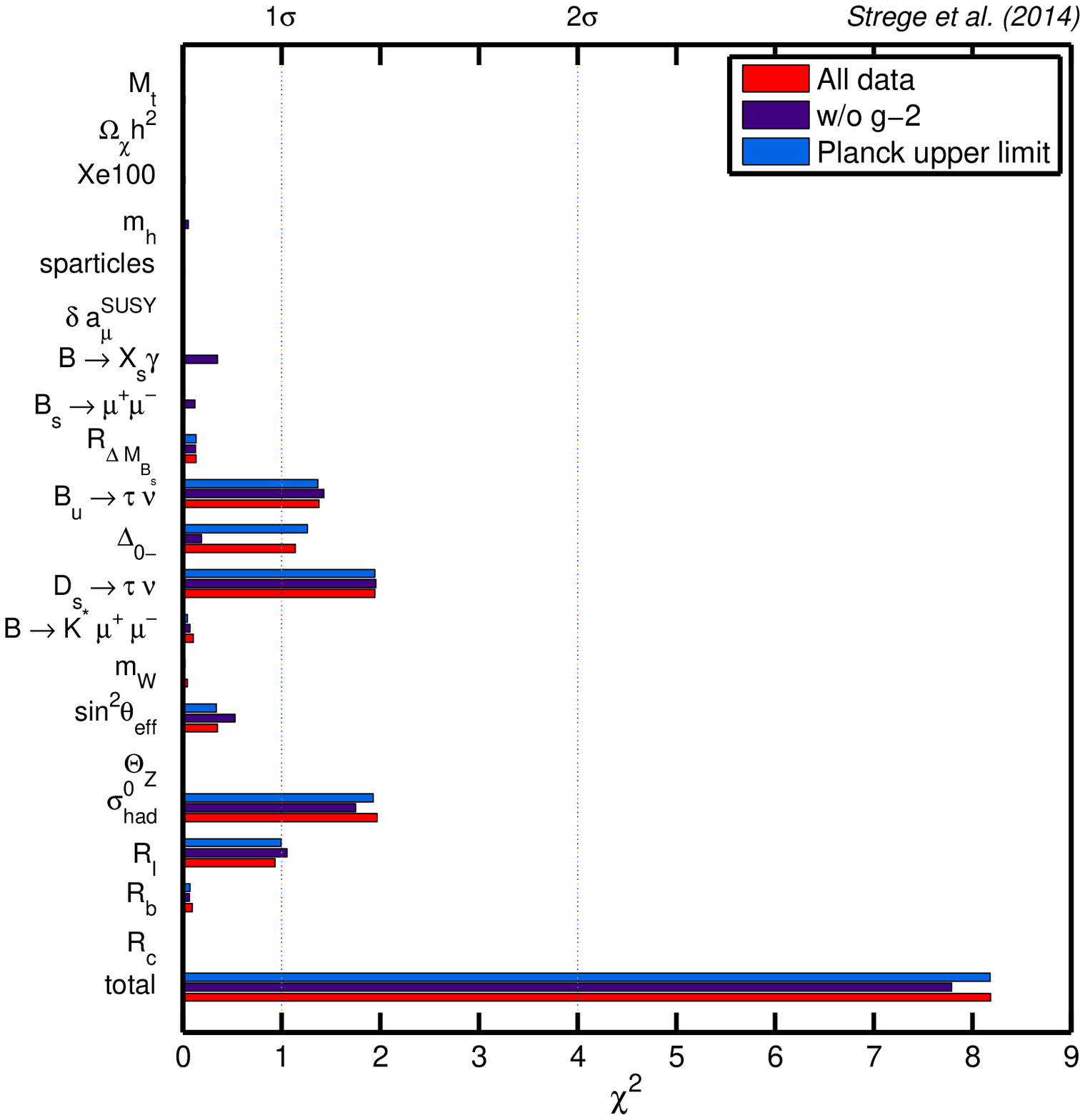}
%\vspace{-60pt}
\end{center}
\caption{Contribution to the best-fit $\chi^2$ from various observables, before including Higgs properties and LHC SUSY searches data. Once those data sets are added, the $\chi^2$ values become 1052.7 (all data), 8.3 (without \gmt) and 268.3 (Planck as upper limit). Thus the overall pre-LHC best-fit point becomes ruled out, while the one obtained without \gmt\ remains viable.}
\label{fig:BF}
\end{center}
\end{figure}

In Table~\ref{tab:bestfit} we show the coordinates of the best-fit points, as well as the best-fit values of several of the observables, for each of our three analyses. For the ``All data'' and ``Planck upper limit'' cases, the pre-LHC best-fit values become ruled out once the ATLAS null SUSY searches are added to the likelihood (see Section~\ref{sec:LHC_impact}), as a consequence of their low squark masses, which are excluded by the 0-lepton search. For those two cases, we also show the coordinates of the best-fitting points that survive the inclusion of ATLAS data at the post-processing stage.  

We do not provide an interpretation of the best-fit $\chi^2$ value in terms of goodness of fit. This is because our likelihood function receives contributions from experimental limits that are not Gaussian distributed, hence asymptotic distributions for the ensuing $\chi^2$ that assume Gaussian data do not apply. The  determination of the quantitative goodness of fit of our best-fit points would require detailed Monte Carlo realisations of the data sets.   

In Fig.~\ref{fig:BF} we display the contribution of each observable to the best-fit $\chi^2$, for the analysis including all data (red), excluding the $g - 2$ constraint (purple) and including the $\Omega_\chi h^2$ measurement as an upper limit (blue). In general, the largest contributions to the best-fit $\chi^2$ result from the same observables for each of the three analyses, namely $\sigma^0_{had}$, $\BR(B_u \to \tau \nu)/\BR(B_u \to \tau \nu)_{SM}$, $\Dstaunu$ and, to a lesser extent, $R^0_l$ (as already discussed in section~\ref{sec:1DPL_wog2}). Another large contribution to the best-fit $\chi^2$ for the analyses including the $g - 2$ constraint results from the isospin asymmetry $\DeltaO$. In contrast, for the analysis excluding the $g - 2$ constraint, the experimentally measured value of $\DeltaO$ can be reproduced (see the discussion in Section~\ref{sec:1DPL_wog2}), leading to a much smaller $\chi^2$ contribution. Largely as a consequence of this difference, the overall $\chi^2$ achieved by the analysis excluding $g - 2$ is slightly reduced compared to the other two analyses. 

Upon post-processing with the LHC data sets, the $\chi^2$ values of the pre-LHC best-fit points become 1052.7 (all data), 8.3 (without \gmt) and 268.3 (Planck as upper limit). Thus the overall pre-LHC best-fit point becomes ruled out, while the best-fit point obtained from the scans excluding \gmt\ remains viable. On one hand, this is a consequence of the larger best-fit values of the gluino (2.83 TeV) and squark (1.55 TeV) mass for this case, which are the main quantities constrained by the ATLAS 0-lepton search. On the other hand, even though the production cross-section of the lightest chargino and the second lightest neutralino  is of $\order(1 \pb)$, their branching ratios to leptons are only of a few percent, leading to a signal prediction for all the signal regions of the ATLAS 3-lepton search analysis compatible with the data at the $1\sigma$ level. The characteristics of the best-fit points surviving the inclusion of the LHC data sets are discussed in Section~\ref{sec:LHC_impact} below.

%%%%%%%%%%%%%%%%%%%%%%%%%%%%%%%%%%%%%
\subsection{Implications for direct detection}
\label{sec:2DPL}
%%%%%%%%%%%%%%%%%%%%%%%%%%%%%%%%%%%%%

Within the MSSM the dominant contribution to the spin-independent (SI) cross-section amplitude is generally the exchange of the two neutral Higgs bosons, although in some cases the contributions of the squark exchange and loop corrections are substantial.
When $m_H < m_h \sqrt{\tan \beta}$, the heavy Higgs is usually the dominant one. As we have seen in section~\ref{sec:1DPL_wog2}, values of $m_A \lesssim 1 \tev$ are disfavoured, which in turn implies that values of $m_H \gtrsim 1 \tev$ are preferred. Thus we expect that the light Higgs exchange dominates. The SI cross-section for $H/h$ exchange in the SUSY decoupling limit with moderate to large \tanb\ values is $\propto | (N_{12}-N_{11} \tan \theta_w)|^2 |N_{13/14}|^2/m^4_{H/h}  f^2_q$, where $\theta_w$ is the electroweak mixing angle, $N_{1i}$ represent the neutralino composition and $f_q$ are the quark-nucleon matrix elements. Therefore, a larger SI cross-section is expected in the ``well-tempered'' neutralino scenario, i.e.\ when the neutralino is a bino-higgsino,  wino-higgsino or bino-wino-higgsino mixture. In fact, a sizeable SI cross-section is obtained as long as the higgsino fraction is larger than $\order (0.1)$.
  
For the squark exchange, at tree level only the exchange of the $u$, $d$ and $s$ squarks contributes, though one can still consider heavy quarks in the effective field theory approach provided that $m_{\tilde{q}} \gg (m_{\chi^0} + m_q)$ . Otherwise, a one-loop treatment has to be considered to account for them~\cite{Drees:1993bu}. As the expression for the amplitude is lengthly we do not write it here explicitly (for details see for instance~\cite{Drees:1993bu}).  It consists of two parts, one coming from gaugino-higgsino mixing and a second one proportional to $\sin 2\theta_{\tilde{q}}$, in which pure binos or mixed bino-winos are involved, where $\theta_{\tilde{q}}$ is the squarks mixing angle. Of course, both are proportional to the propagator $1/(m^2_{\tilde{q}}-(m_{\chi^0} + m_q)^2)$. 

\begin{figure}
\begin{center}
\includegraphics[width=0.32\linewidth, trim = 0.7cm 0cm 0.7cm 0cm]{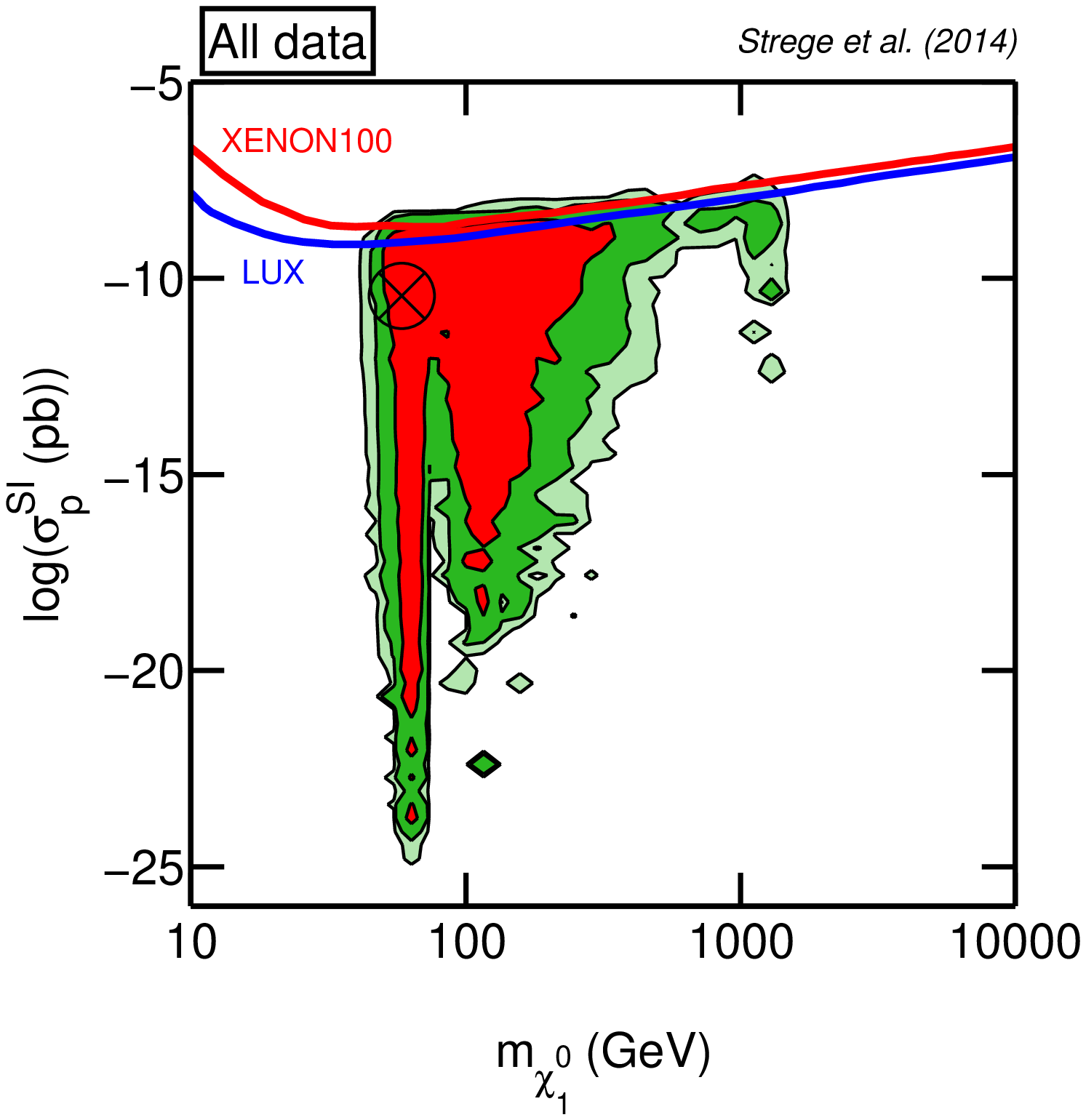} 
\includegraphics[width=0.32\linewidth, trim = 0.7cm 0cm 0.7cm 0cm]{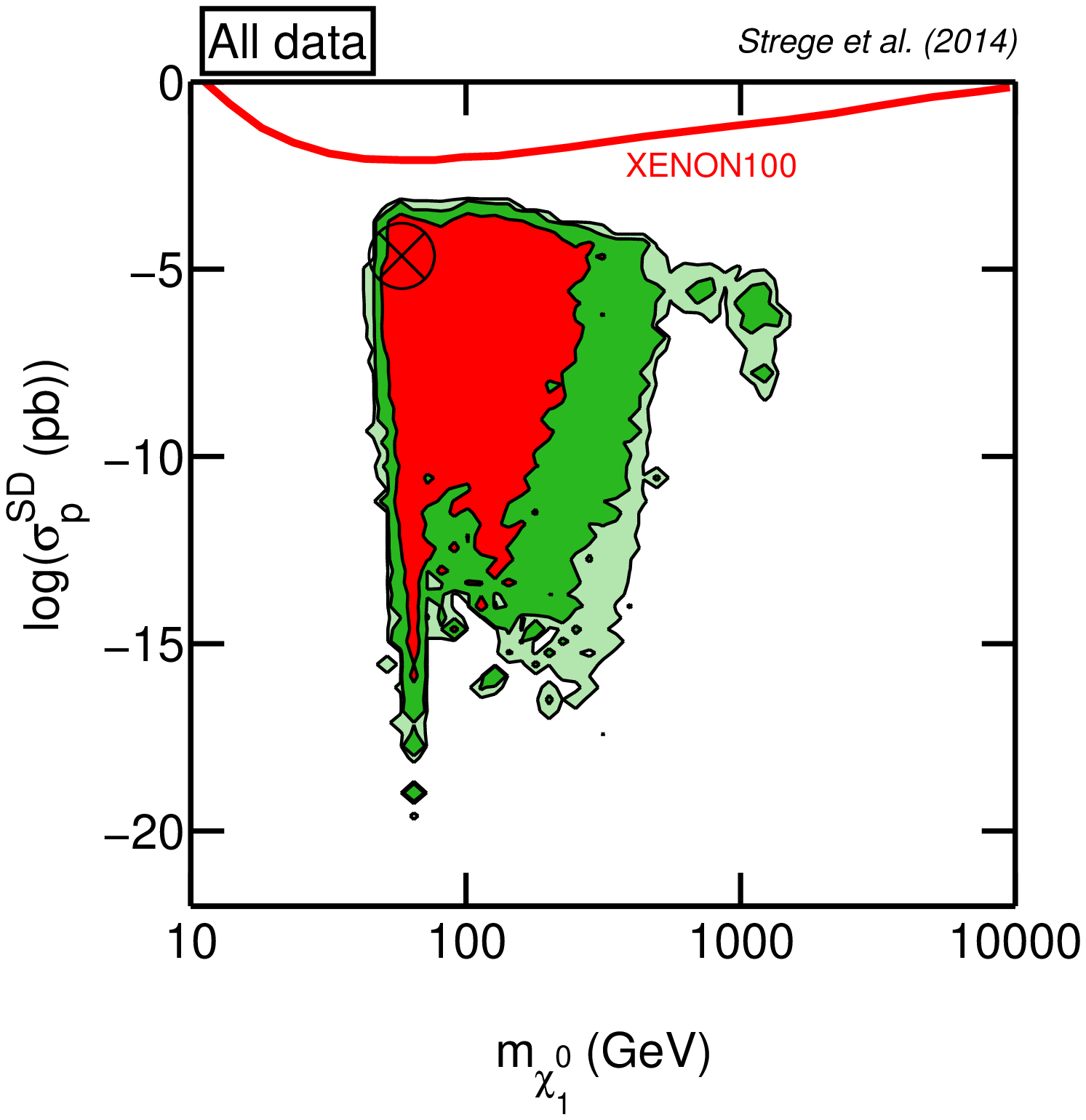}
\includegraphics[width=0.32\linewidth, trim = 0.7cm 0cm 0.7cm 0cm]{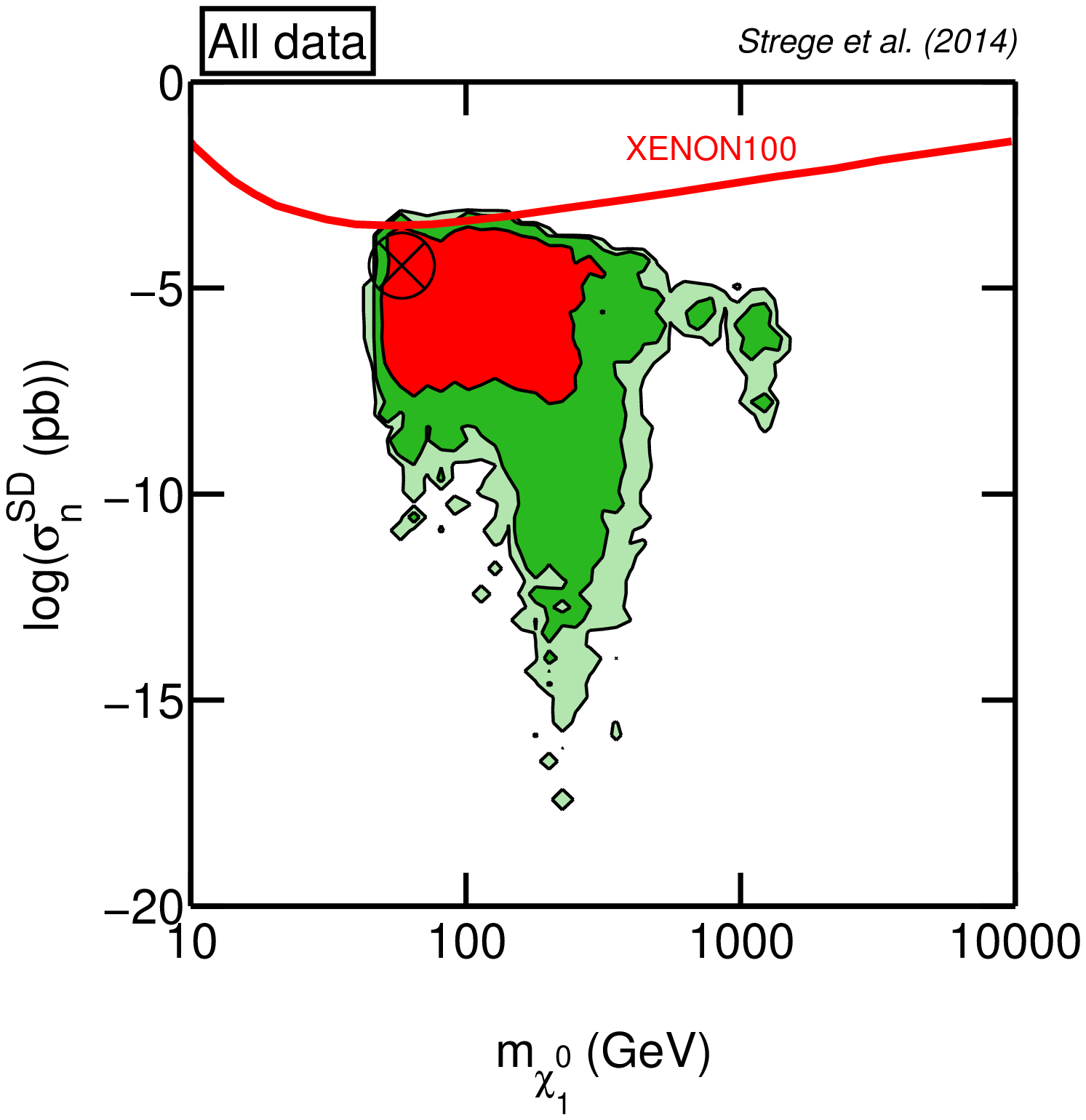} \\
\includegraphics[width=0.32\linewidth, trim = 0.7cm 0cm 0.7cm 0cm]{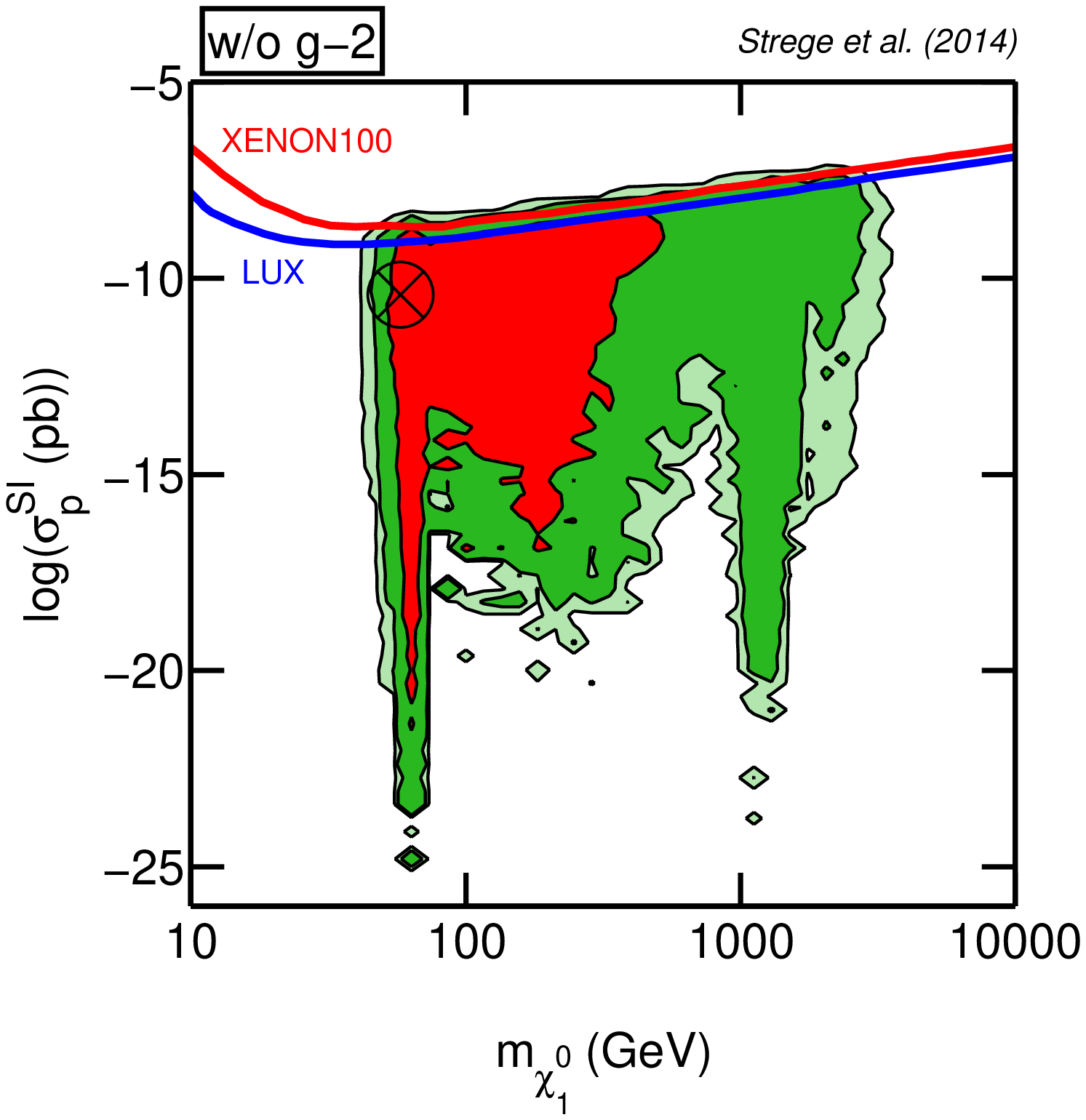}
\includegraphics[width=0.32\linewidth, trim = 0.7cm 0cm 0.7cm 0cm]{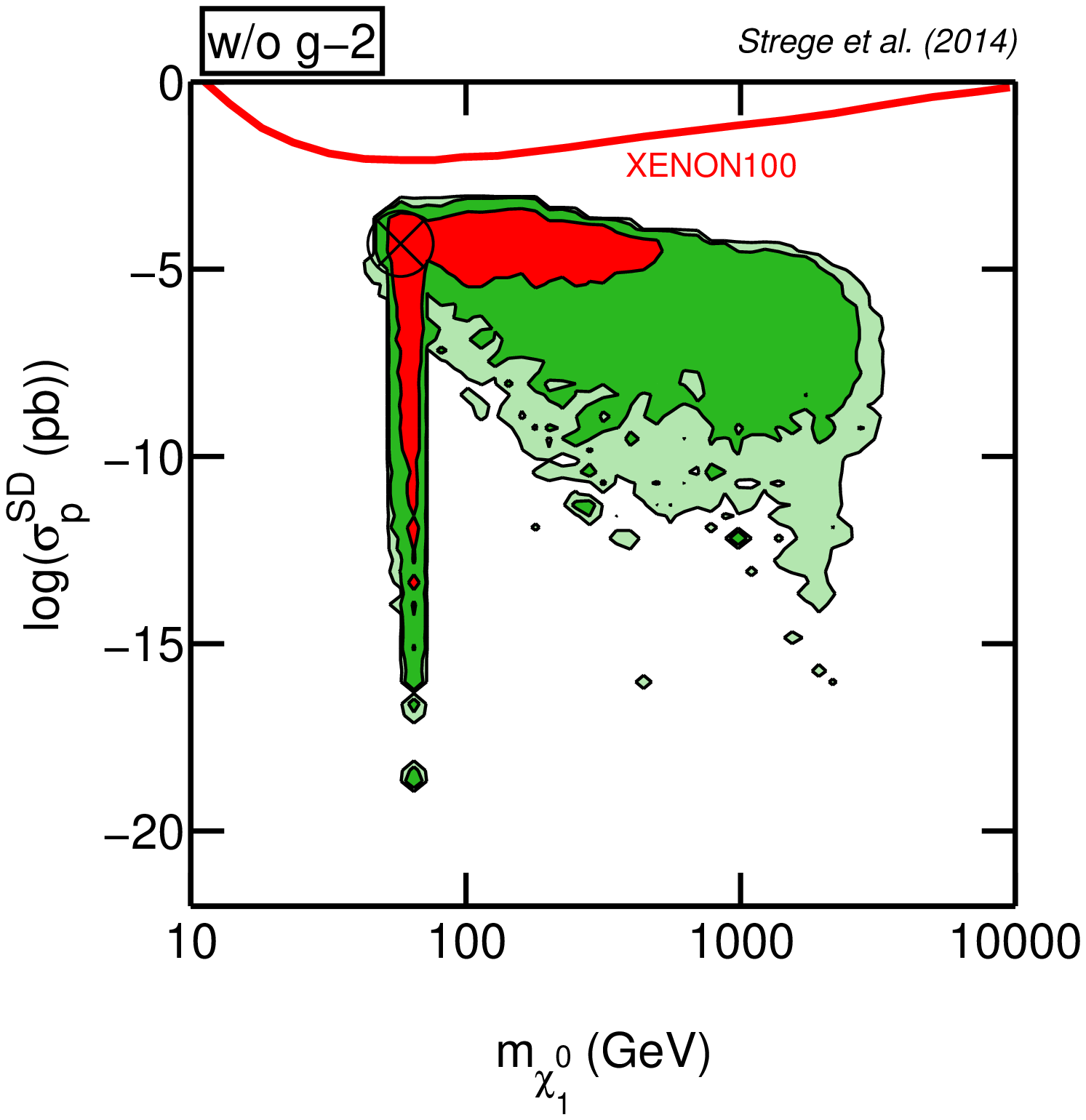}
\includegraphics[width=0.32\linewidth, trim = 0.7cm 0cm 0.7cm 0cm]{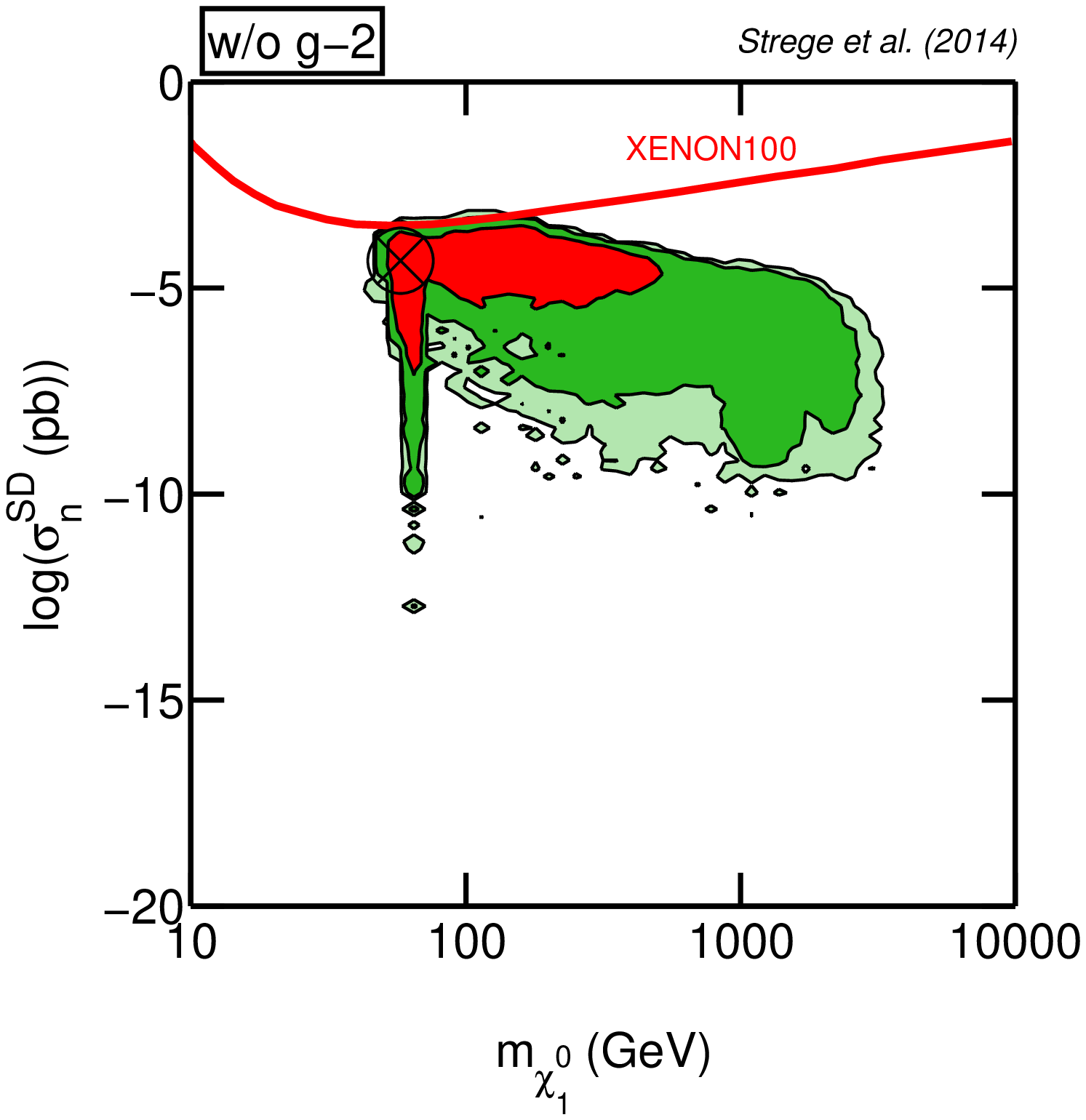} \\
\includegraphics[width=0.32\linewidth, trim = 0.7cm 0cm 0.7cm 0cm]{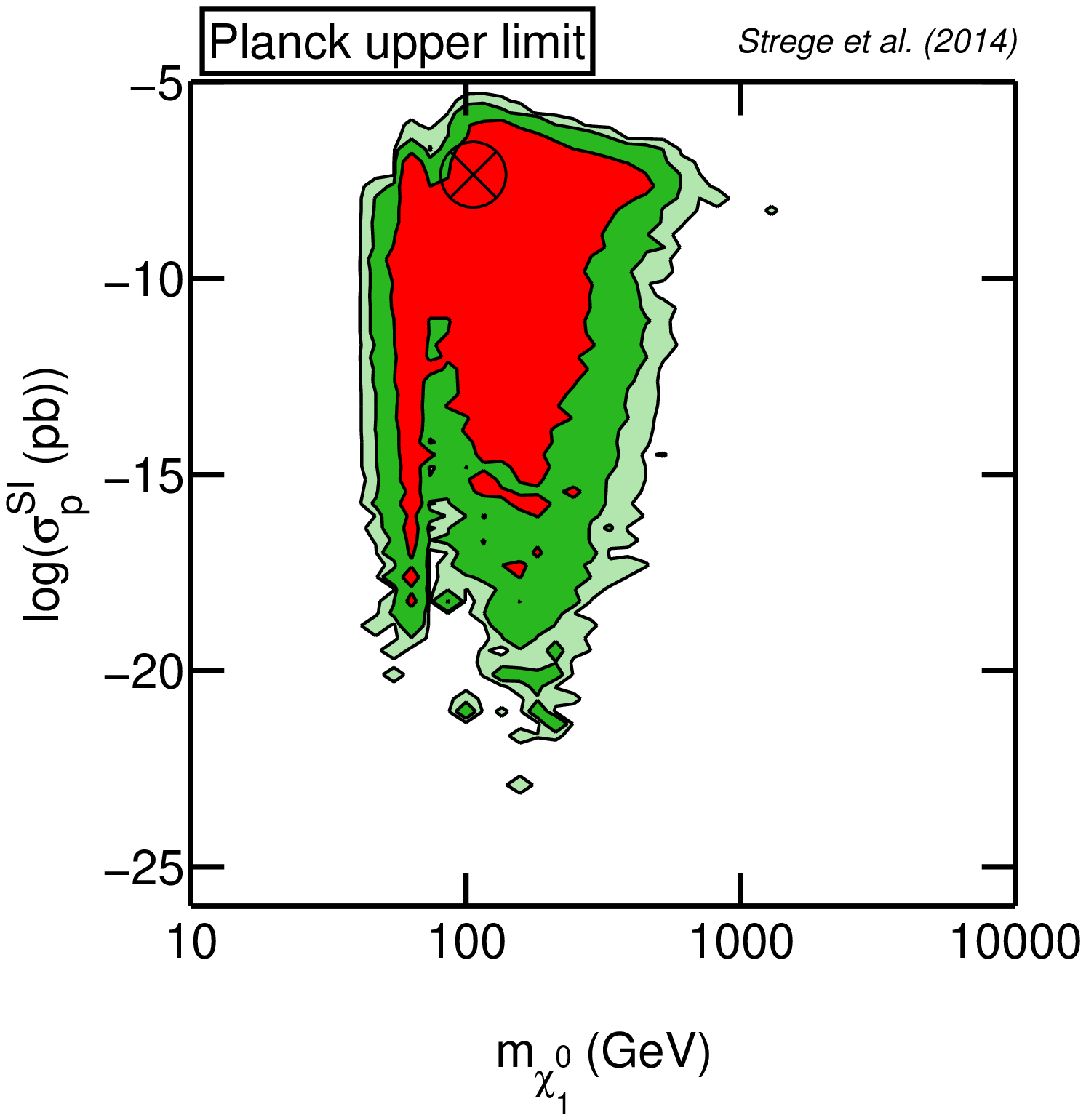}
\includegraphics[width=0.32\linewidth, trim = 0.7cm 0cm 0.7cm 0cm]{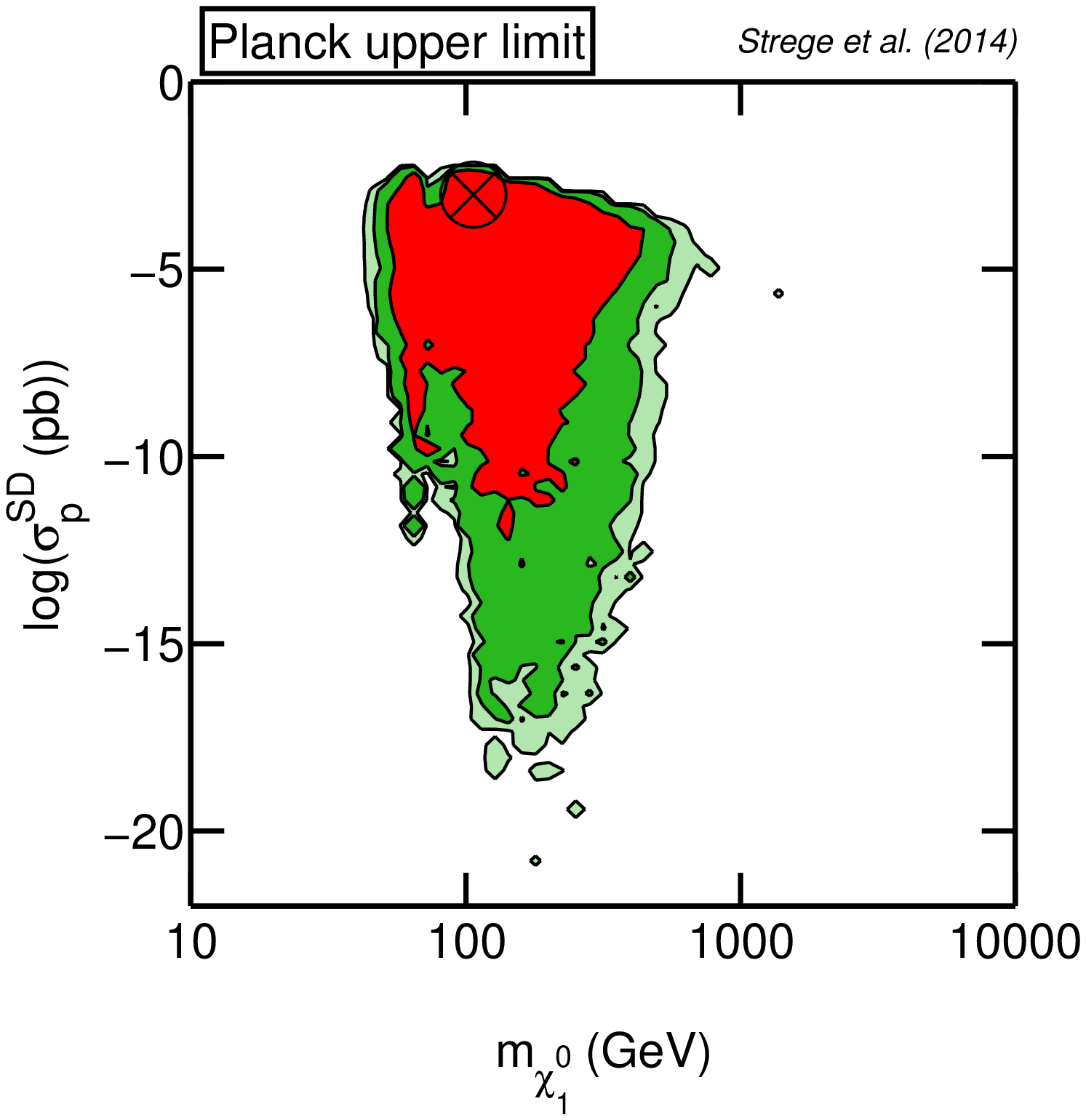}
\includegraphics[width=0.32\linewidth, trim = 0.7cm 0cm 0.7cm 0cm]{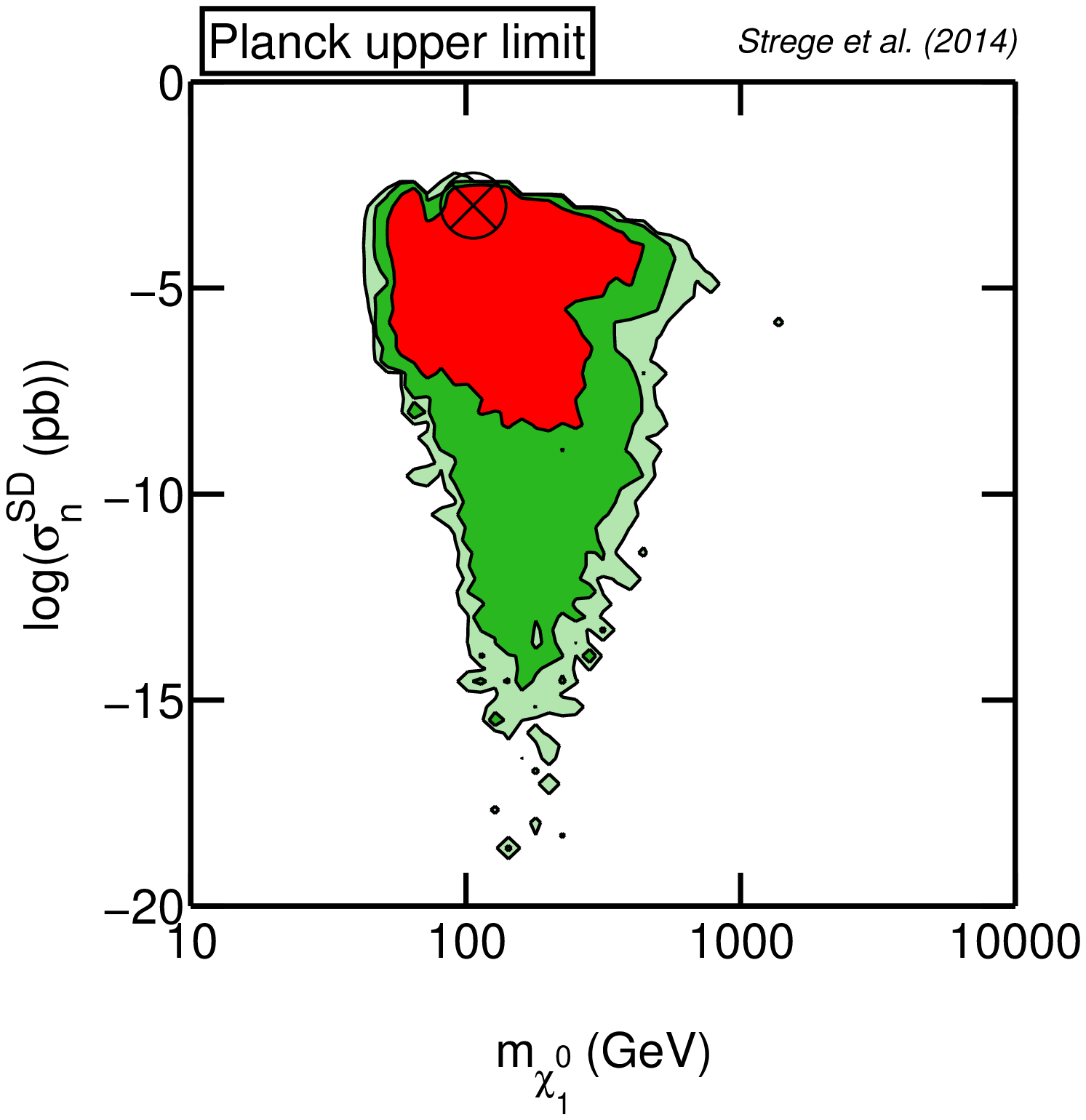}
\caption{2-D profile likelihood results in the variables relevant for direct detection experiments. From top to bottom: including all data,  excluding the $g - 2$ constraint, and applying the $\Omega_\chi h^2$ measurement as an upper limit. The encircled cross gives the location of the best fit. Recall that these analyses does not include null SUSY searches at the LHC (see Section~\ref{sec:LHC_impact}).}
\label{2D_plots}
\end{center}
\end{figure}

After this brief review of the anatomy of the contributions involved at tree-level in the SI cross section, we turn to the discussion of Fig.~\ref{2D_plots}, which shows the two-dimensional profile likelihood functions in the planes of neutralino mass vs. the cross section for spin-independent neutralino-proton (left panels), spin-dependent neutralino-proton (central panels) and spin-dependent neutralino-neutron (right panels) scattering. From top to bottom, the panels show the results for the analysis including all data, excluding the $g - 2$ constraint, and using the Planck relic density measurement as an upper limit. In each panel, the 68\%, 95\% and 99\% confidence regions are shown. In the top and central left-hand panels we also show the current 90\% exclusion limits from the XENON100 experiment (red) and the LUX collaboration (blue, not included in the analysis). As described in Section~\ref{sec:constraints}, when applying the Planck constraint on the relic density as an upper limit, the local dark matter density is rescaled with the scaling Ansatz of Eq.~\eqref{eq:scaling_ansatz}. Therefore, the XENON100 and LUX exclusion limits, that were computed for a fixed local density $\rho_{\neut} = 0.3$ GeV/cm$^{3}$, are not shown in the bottom left panel of Fig.~\ref{2D_plots}.

We start by discussing the 2D profile likelihood results for spin-independent neutralino-proton scattering (left-hand panels). Multiple modes of high likelihood can be identified. For each of the three analyses we observe a narrow area at $m_{\neut} \sim 50$ GeV spanning almost 15 orders of magnitude in $\sigmaSI$ that is favoured at $68\%$ C.L. A second region that is strongly favoured is found at WIMP masses of several hundred GeV, and stretches from cross-sections just below the XENON100 limit down to $\sigmaSI \sim 10^{-20}$ pb. Additionally, the $95\%$ region for the analysis excluding the $g - 2$ constraint also includes a sizeable region at larger neutralino masses $1$ TeV $\lsim m_{\neut} \lsim 1.5$ TeV, that spans a large cross-section range $10^{-20}$ pb $\lsim \sigmaSI \lsim 10^{-7}$ pb. This region in fact also appears in the top left panel (for the analysis including all data), albeit only at very large cross-sections. Finally, a small region favoured at $95\%$ C.L. is visible at very large $m_{\neut} \gsim 2$ TeV in the panel for the analysis excluding the $g - 2$ constraint. 

In the ($\sigmaSI$, $\mchi$) plane,  the 68\% C.L. region corresponds to a bino-like neutralino LSP.  Since the lightest Higgs mass is fixed to $\sim 126 \gev$ by the LHC measurement, and $m_A \gg m_Z$, one would expect $\sigmaSI$ to be $O(10^{-9} \pb)$, which is realised at the top of the narrow, vertical strip found at a neutralino mass $\sim 50 \gev$. Heavier bino-like neutralinos can acquire some mixing with higgsinos, which further enhances the SI cross-section. The degree of higgsino mixing is limited by the XENON100 constraint. On the other hand, cancellations among the different contributions might occur~\cite{Mandic:2000jz}, leading to values as low as $\sim 10^{-25}$ pb within the 95 \% C.L.  Such cancellations require unexpected relationships between the parameters in the Higgs and squark sectors, the parameters determining the neutralino composition and the nuclear matrix elements---something that in constrained SUSY models is quite unlikely to happen. We also observe a region favoured at the 95\% C.L. that corresponds to $\sim 1 \tev$  higgsino-like neutralinos. This region is disfavoured relative to the low-mass  regions because a heavy higgsino-like neutralino forces the EWKinos and sfermions to be heavy, which is in tension with the constraints on several observables, namely the muon \gmt\ constraint, $\DeltaO$ and $\afb$ (see the discussion in Section~\ref{sec:1DPL_wog2}).

Note that, for higgsino-like dark matter, the neutralino mass is strongly constrained to $m_{\neut} \sim 1$ TeV by the Planck constraint on the dark matter relic abundance. However, co-annihilations with the second lightest neutralino and the lightest chargino can further reduce the dark matter relic abundance, so that higgsino-like dark matter with $m_{\neut} \lsim 1.4$ TeV remains favoured at 95\% C.L. (see also Fig.~\ref{fig:1D_wog2_3}).

Excluding the muon \gmt\ data leads to a sizeable difference in the neutralino mass favoured at the 95\% C.L.
First, in the higgsino-like neutralino region ($\mchi \sim 1 \tev$), $\sigmaSI$ extends to significantly lower values, both because the neutralino becomes an increasingly pure Higgsino state, and due to cancellations between the Higgs sector contributions to $\sigmaSI$. Secondly, a region at large neutralino masses $m_{\neut} \sim 2 - 3 \tev$ is now favoured at 95\% level. In this region, which was previously disfavoured by the \gmt\ constraint, the neutralino is wino-like (see also Fig.~\ref{fig:Composition} below). 

The bottom-left panel depicts the SI cross-section for multi-component DM scenarios. Now larger cross-sections are not penalized because the scaling factor $\xi$ reduces the predicted number of recoil events, thus weakening the impact of the XENON100 constraint. As mentioned above, the lowest neutralino masses correspond to bino-like neutralinos.  Heavier neutralinos can be both well-tempered and almost pure wino-like or higgsino-like. Neutralinos with an admixture of wino and/or higgsino annihilate very efficiently via coannihilations, providing a relic abundance well below the Planck upper limit. This effect is largest for light neutralinos. Compared to the analysis assuming that neutralinos make up all dark matter in the universe, the contours are shifted towards smaller neutralino masses. As explained above, this is mainly a consequence of the flavour observables $\DeltaO$, $\afb$, and the \gmt\ constraint, which play a dominant role in driving the profile likelihood results when relaxing the Planck constraint on the neutralino relic density\footnote{In principle, the 95\% contours might also include the regions at high neutralino masses favoured at 95\% level in the analysis requiring that $\Omega_\chi \sim \Omega_{\rm DM}$. However, these points correspond to strong fine-tuning of the parameters in order to obtain an acceptable fit to observables such as $\DeltaO$, $\afb$, and \gmt. In the absence of the relic density constraint, which drives the scan towards these regions, the scan spends less time tuning the observables in this region. A dedicated
investigation of the profile likelihood coverage in this parameter space is the subject of future work.}.

Similar patterns can be observed for spin-dependent neutralino-proton and neutralino-neutron scattering (central and right-hand panels). The narrow region at low $m_{\neut}$ is clearly visible in the $m_{\neut} - \sigmaSDp$ plane for all three analyses, and also shows up in the $m_{\neut} - \sigmaSDn$ plane for the analysis excluding the $g - 2$ constraint. Likewise, the extended region at dark matter masses $m_{\neut} \sim \mathcal{O}(100)$ GeV can easily be identified in both planes, and spans a large range in $\sigmaSDp$ and $\sigmaSDn$ for both the analyses including all data and the analysis allowing for a relic density smaller than the value measured by Planck. Instead, for the analysis excluding the $g - 2$ constraint, this region only inhabits an area at large spin-dependent cross-sections, and is disfavoured at $99\%$ C.L. even at intermediate cross-section values $\sigmaSDp \sim 10^{-11}$ pb and $\sigmaSDn \sim 10^{-9}$ pb. Regions observed at large $m_{\neut} > 1$ TeV in the $m_{\neut} - \sigmaSI$ plane for the analyses including all data and excluding $g - 2$ also appear for spin-dependent scattering, although the two different high-mass regions observed in the spin-independent plane for the latter analysis are difficult to identify as separate regions.

In general, the dominant contribution to the spin-dependent (SD) cross-section is the $Z$ exchange contribution. Since the bino and wino are both SU(2) singlets, they do not couple to the $Z$ boson, so that the SD cross-section is largely determined by the higgsino content of the neutralino. The $Z$ exchange contribution (and hence the SD cross-section) is proportional to the higgsino asymmetry $(|N_{13}|^2-|N_{14}|^2)^2$. The squark exchange contributions has a similar structure to the SI case.

In the top central (right-most) panel the ($\sigma_{\neut-p(n)}^\text{SD}$, $\mchi$) plane is displayed. For the spin-dependent neutralino-proton interaction, the shape of the PL contours is similar to the results for the SI cross-section. This can be understood by the fact that the squark-exchange contribution follows a similar pattern and the $Z$ exchange contribution is non-negligible as long as the neutralino is well-tempered. In the absence of degeneracies between parameters in the neutralino mass matrix and if $m_Z$ can be treated as a perturbation, the asymmetry $|N_{13}|^2-|N_{14}|^2 \propto \cos 2 \beta/(\mu^2- M_i)$ for $|M_1|$, $|\mu|$, $|\mu|-|M_i| > m_Z$ and $M_i \rightarrow \infty$, with $i=1,2$. From this one recovers the limits of pure gaugino/Higgino in which the higgsino asymmetry vanishes. The asymmetry is maximized when either the binos and higgsinos or winos and higgsinos are close in mass, i.e.\ for well-tempered neutralinos. 
One expects a suppression when the $Z$ and squark exchange contributions cancel against one another, which requires fine-tuned relationships between the model parameters and the nuclear matrix elements. This is typically not the case for the scattering off both protons and neutrons simultaneously, which explains the differences between the results for the proton and neutron SD scattering cross-section.

In the middle central and right-most panel we display the ``w/o g-2" case. The most remarkable difference with respect to the ``All data" case (upper panels) occurs for neutralino masses $\order (100 \gev)$ where the SD cross-section is on average larger. The effect is more pronounced for masses $\sim 100 \gev$, where the neutralino, although bino-like, acquires a sizeable higgsino fraction, as required to fulfil the Planck measurement of the dark matter relic density. Recall that in the ``All data" case, in addition to acquiring a sizeable higgsino content, bino-like neutralinos may annihilate through both the exchange of light sleptons and through co-annihilations, while keeping their pure bino character. This suppresses the higgsino asymmetry factor and thus the SD cross-section. Additonally, at neutralino masses of a few hundred GeV, one can find the bulk region with relatively light sbottoms/stops, as outlined in Section~\ref{sec:1DPL_wog2}. As a result, the SD cross-section covers a large range of values. For masses $\sim 1 \tev$, the neutralino is higgsino-dominated and can exhibit either an enhancement or a suppression in the SD cross-section, depending  on its purity degree.

Finally the bottom central and right-most panels show the multi-component DM scenario. Here, the main difference with respect to the upper 
panels is that in this case the neutralino is mostly either wino-like or a wino-higgsino mixed state. A wino-like neutralino has an enhancement with respect to the bino-like state due to the larger SU(2) gauge coupling relative to the U(1) one. In this case the PL contours for the proton and neutron SD cross-sections are almost identical.

\begin{figure}
\begin{center}
\begin{center}
\includegraphics[width=0.5\linewidth]{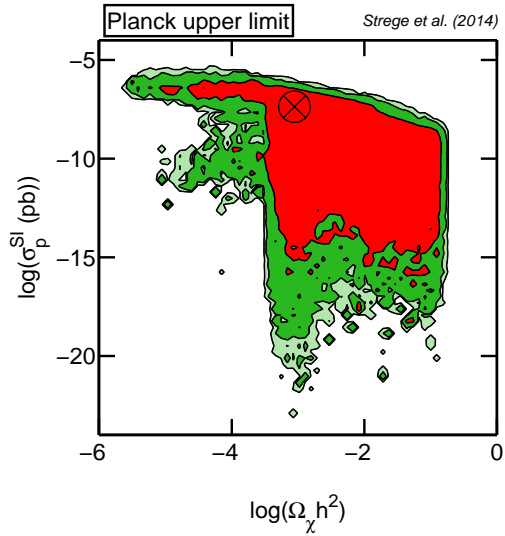}
\end{center}
\caption{2-D profile likelihood in the plane of spin-independent neutralino-proton cross-section vs dark matter relic density for the case where the relic density measurement is applied as an upper limit. The encircled black cross is the best-fit point.}
\label{2D_plot_DMlimit}
\end{center}
\end{figure}

In Fig.~\ref{2D_plot_DMlimit} we show the 2D profile likelihood in the plane of spin-independent neutralino-proton cross-section vs. the neutralino relic density. As mentioned above, we assume that the local neutralino density scales with the cosmological abundance. As a result, the XENON100 limit is shifted towards larger cross-section values for points in parameter space that lead to a relic density smaller than the value measured by Planck. This translates into a negative correlation visible in Fig.~\ref{2D_plot_DMlimit} for large values of the scattering cross-section. For small values of $\relic h^2$ the most favoured region of parameter space is a narrow band stretching along the currently largest allowed cross-section values, within reach of future direct detection searches. In this region, the neutralino is a mixed wino-higgsino state (see Section~\ref{sec:Composition} below), so that co-annihilation effects are maximized, and very small neutralino relic densities can be achieved. 

For $10^{-5} \lesssim \relic h^2 \lesssim 10^{-4}$ a second region, favoured at 95\% C.L., shows up at slightly lower cross-section values $\sigmaSI \sim 10^{-10}$ pb. This is a consequence of light pure-gaugino neutralinos 
with masses of $\order (100 \gev)$ still annihilating efficiently, but leading to a suppressed SI cross-section with respect to the wino-higgsino case.
For higgsino- (wino-)like neutralinos, annihilation remains efficient up to $\mchi \lesssim 1 (2) \tev$. This leads to a low relic abundance, a scenario that corresponds to the region $ 10^{-4} \lesssim \relic h^2 \lesssim 10^{-1}$. Finally when the neutralinos are either bino-like (with masses from $\sim 50 \gev$ to a few hundred GeV), higgsino-like (with $m_\neut \sim 1 \tev$), or wino-like (with $m_\neut \sim 2 \tev$) the relic density matches the Planck constraint. In this cases the SI cross-section reaches lower values because of the great purity of the neutralino. In Fig.~\ref{2D_plot_DMlimit} the largest values of $\relic h^2$ correspond to bino-like neutralinos, due to the preference for relatively small $m_\neut$ in the analysis allowing for multi-component dark matter (cf.\ Fig.~\ref{2D_plots}). 
The cutoff at large $\relic h^2$ is due to the Planck upper limit on the relic density. 

%%%%%%%%%%%%%%%%%%%%%%%%%%%%%%%%%%%%%
\subsection{Dark matter composition}
\label{sec:Composition}
%%%%%%%%%%%%%%%%%%%%%%%%%%%%%%%%%%%%%

\begin{figure}
%\begin{center}
\includegraphics[width=0.5\linewidth, trim=1cm 1.1cm 7cm 9.1cm, clip=true]{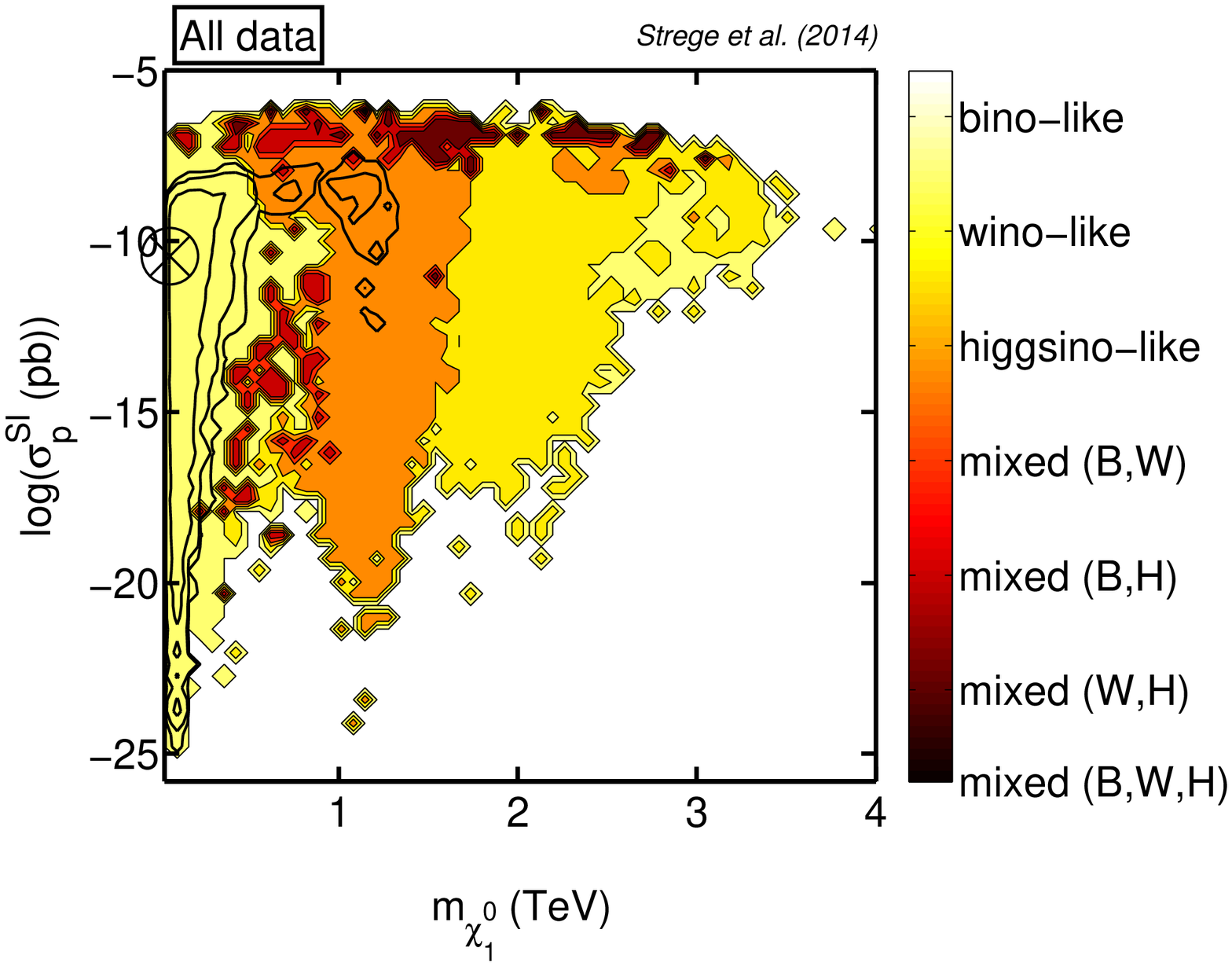}
\includegraphics[width=0.5\linewidth,  trim=1cm 1.1cm 7cm 9.1cm, clip=true]{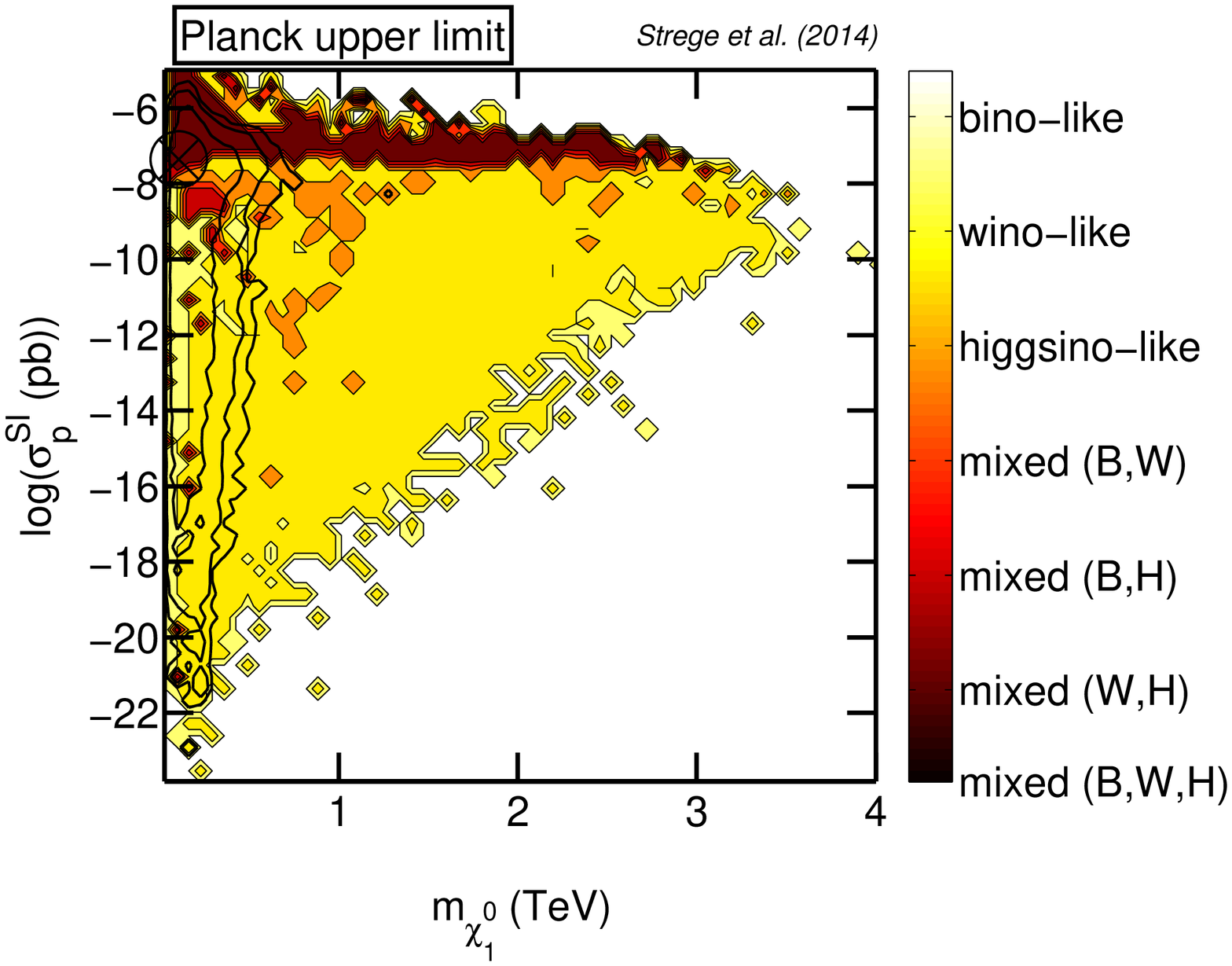} \hfill
\caption{Composition of neutralino dark matter in the mass vs. spin-independent scattering cross-section plane. For reference, the PL contours from Fig.~\ref{2D_plots} are shown in black (notice that the horizontal axis is on a linear scale to better show the region at larger neutralino masses).}
\label{fig:Composition}
%\end{center}
\end{figure}

We now discuss the neutralino compositions favoured in different regions of the MSSM-15 parameter space. The neutralino composition in the plane of neutralino mass vs. spin-independent cross-section is shown in Fig.~\ref{fig:Composition}, for the analysis including all data (left panel) and when using the Planck relic density constraint as an upper limit (right panel). The neutralino composition for the analysis excluding the $g - 2$ constraint is not shown, as it is qualitatively very similar to the ``All data'' case.

We define the neutralino to be bino-like if it has a bino fraction $b_f > 80\%$, wino-like  for a wino fraction $w_f > 80\%$ and higgsino-like  for a higgsino fraction $h_f > 80\%$. A mixed (B,W) neutralino has both a sizeable bino and wino fraction ($b_f, w_f > 20\%$), and similarly for mixed (B,H) and mixed (W,H) neutralinos. Neutralinos that do not fit into any of the above categories are considered mixed (B,W,H) states. For reference, we also show the $68\%, 95\%$ and $99\%$ 2D PL contours in this plane (black, empty), as well as the best-fit points (circled black crosses).

For the analysis including all data (left-hand panel) three dominant dark matter compositions can be identified. At low masses, $m_{\neut} \lsim 600$ GeV, the neutralino is bino-like. Pure bino dark matter tends to lead to relic densities that overclose the universe. However, for low and intermediate and neutralino masses, pole-resonances with Z/h or co-annihilation effects with light sleptons reduce the relic density sufficiently to achieve $\relic h^2 \sim \mathcal{O}(0.1)$. Additionally, in this mass range the neutralino can acquire a non-negligible higgsino fraction, leading to a relic density in agreement with the values measured by Planck. In the mass range $0.8$ TeV $\lsim m_{\neut} \lsim 1.6$ TeV the neutralino LSP is higgsino-like. For pure higgsino dark matter, annihilation in the early universe is very efficient, so that small values of $\mu \sim \mathcal{O}(100)$ GeV lead to relic densities much smaller than the Planck constraint. However, for large values $\mu \sim 1$ TeV (and thus $m_{\neut} \sim 1$ TeV) the correct dark matter density can be achieved, so that higgsino-like dark matter is favoured at neutralino masses $\sim 1$ TeV. At very large masses $m_{\neut} \gsim 1.6$ TeV the neutralino becomes predominantly wino-like.  Wino-like dark matter annihilates even more efficiently than higgsino-like states, so that very large wino masses $M_2 \gsim 2$ TeV are required to reproduce the Planck measurement of the dark matter density. As a result, wino-like dark matter is favoured at $m_{\neut} \gsim$ 2 TeV. Finally, we observe a small region of bino-like neutralinos at $m_{\neut} \sim 3$ TeV. In this region, the correct relic density is achieved via gluino co-annihilations, a phenomenological feature appearing in models without gaugino mass unification, as pointed out in~\cite{Profumo:2004wk}. Small islands of mixed $(B,H)$ dark matter show up in the transition region from bino-like to higgsino-like neutralinos. Additional small islands of mixed $(W,H)$ neutralinos can be found at large masses and large spin-independent cross-sections. Mixed $(B,W)$ and $(B,W,H)$ states are rare. 

The neutralino composition when the Planck relic density constraint is applied as an upper limit (right-hand panel) is largely driven by the SM precision observables, namely $\DeltaO$, $\afb$, and the \gmt.
The bulk of the parameter space correspond to wino-like neutralinos, with the exception of a narrow area at very low masses $m_{\neut} \sim 100$ GeV and a narrow diagonal area at the lowest allowed cross-sections (as a function of $m_{\neut}$) that correspond to bino-like dark matter. Higgsino-like neutralinos are now disfavoured, and only show up as isolated islands in different regions of parameter space. A second interesting feature is a pronounced region of mixed $(W,H)$ neutralinos that is found at large spin-independent cross-sections and stretches along almost the entire allowed neutralino mass range. Other mixed states ($(B,H)$,$(B,W)$,$(B,W,H)$) are rare. Note that, for low mass neutralinos, a large range of different neutralino compositions are possible. 

%%%%%%%%%%%%%%%%%%%%%%%%%%%%%%%%%%%
\subsection{Impact of LHC Higgs properties and ATLAS SUSY searches}
\label{sec:LHC_impact}
%%%%%%%%%%%%%%%%%%%%%%%%%%%%%%%%%%%

We now turn to the discussion of the impact of ATLAS null searches for SUSY and CMS measurements of the Higgs properties on the favoured regions of the \pMSSM\ parameter space.

The evaluation of the full LHC likelihood described in the Appendix is numerically very demanding. We estimate that post-processing of all the samples gathered for the above analysis would require approximately 400 CPU-years. This considerable task is the subject of a dedicated work~\cite{LHC_PP_Future}. 
For the more limited purpose of this paper, we adopt an intermediate approach, which gives an indication of the extra constraining power from LHC SUSY searches and Higgs signal strengths measurements. In what we call the ``mini-chains'' approach, we first produce profile likelihood maps from our full chains for several 2D planes of interest. Given typical binning sizes, this leads to approximately $10^4$ profile likelihood values for each 2D plane. For each of those values, we then compute the combined $\chi^2$ contribution from LHC constraints on the Higgs production cross-sections and LHC SUSY searches (0-lepton and 3-lepton), according to the procedure described and validated in the Appendix. We add the extra $\chi^2$ value to the pre-LHC $\chi^2$ obtained from all other experimental data sets.

We stress that this is not a fully consistent statistical approach, and that the ensuing maps cannot be interpreted probabilistically as PL maps (as the full likelihood has not been maximised out in the dimensions not shown). However, it does allow to draw some useful conclusions regarding the impact of LHC SUSY searches and measurements of the Higgs properties: mini-chain points that remain viable after inclusion of the LHC constraints would not be ruled out even under a full PL approach. In this sense, our approach gives an indication of the maximal possible constraints (in the plane under consideration) resulting from the included LHC data sets. Furthermore, this procedure allows us to investigate whether the best-fit points found in the above global fits analysis remain viable in the light of the LHC constraints. 

\begin{figure}
\begin{center}
\includegraphics[width=0.32\linewidth, trim = 0.7cm 0cm 0.7cm 0cm]{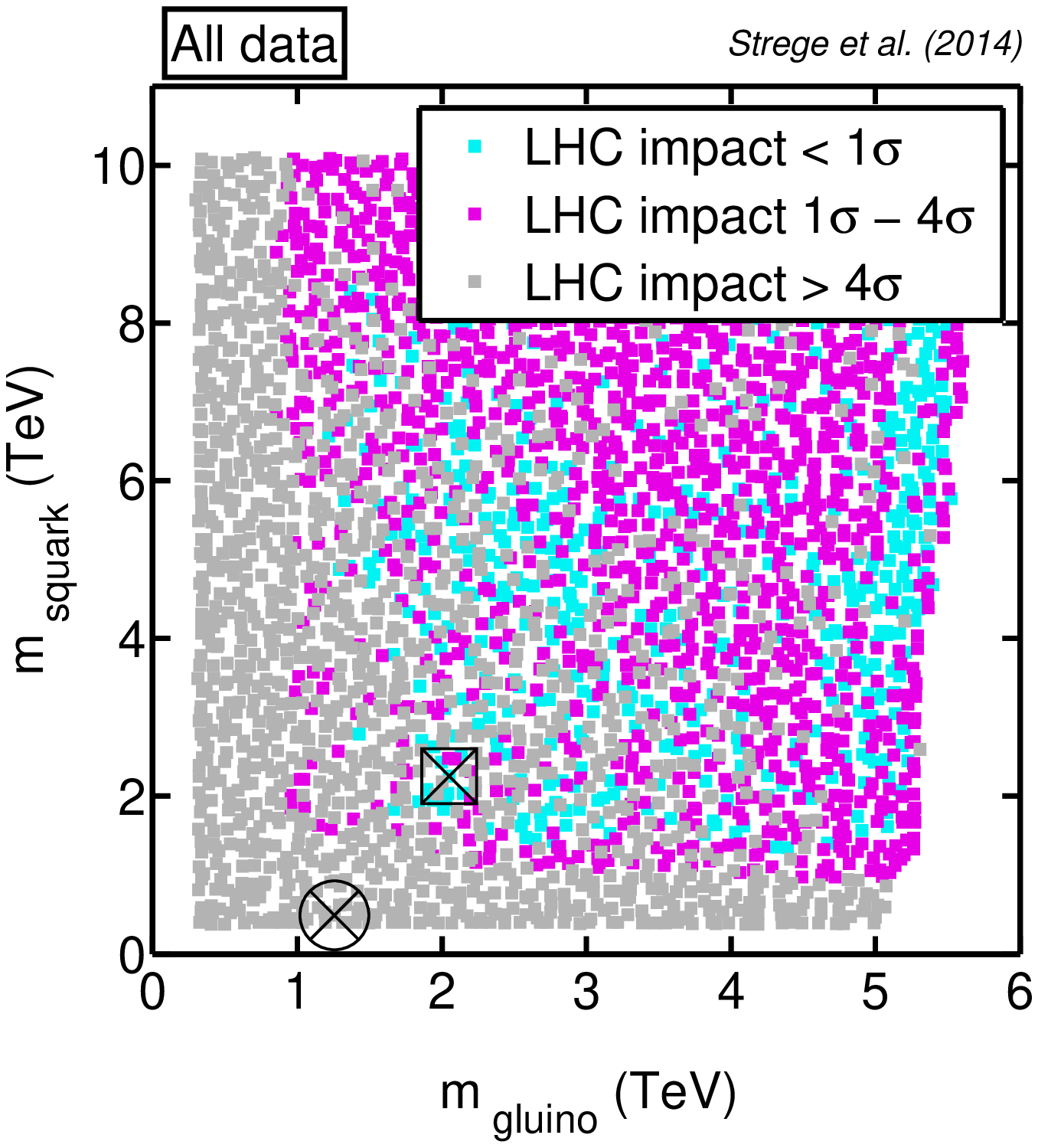}
\includegraphics[width=0.32\linewidth, trim = 0.7cm 0cm 0.7cm 0cm]{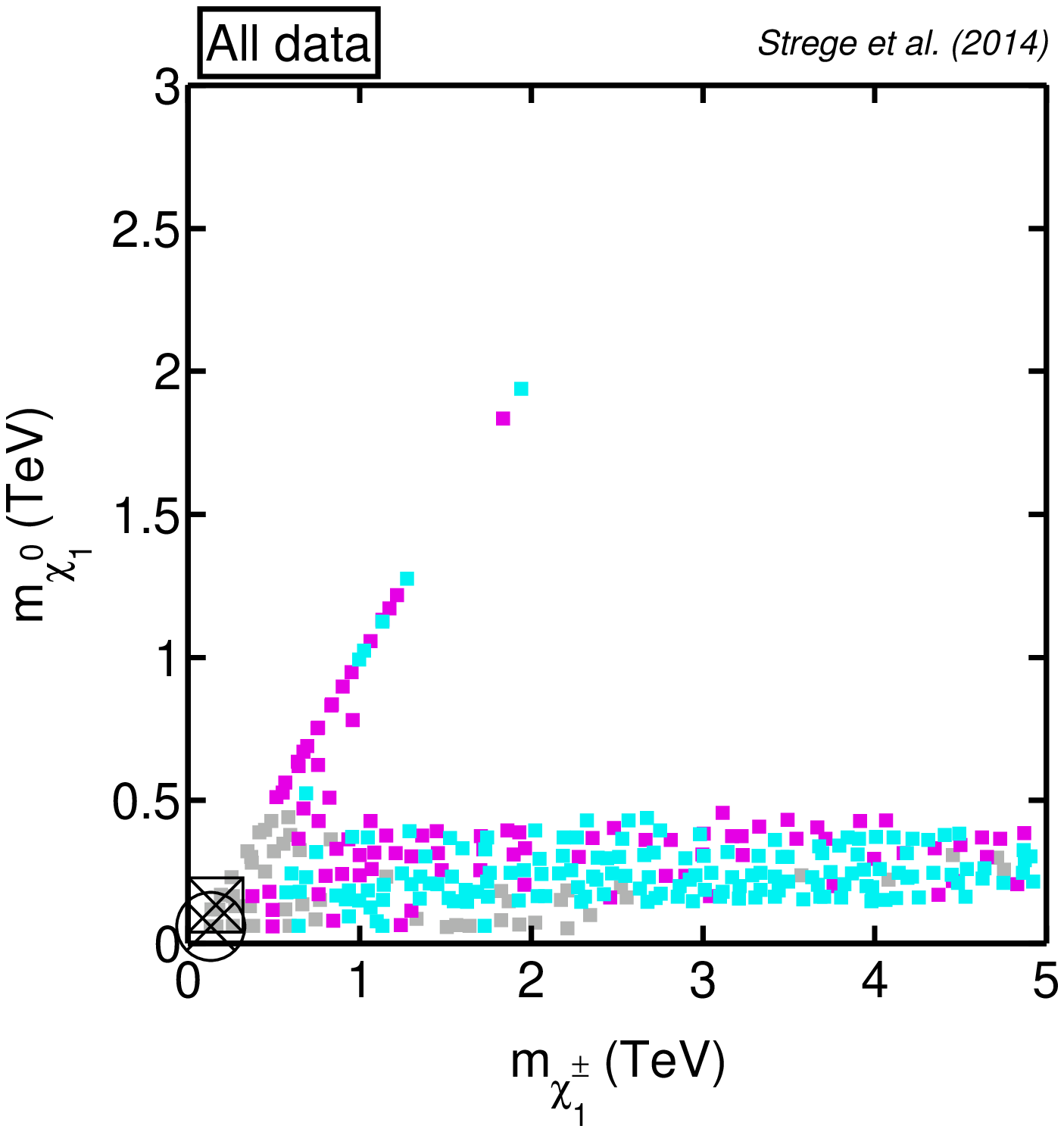}
\includegraphics[width=0.32\linewidth, trim = 0.7cm 0cm 0.7cm 0cm]{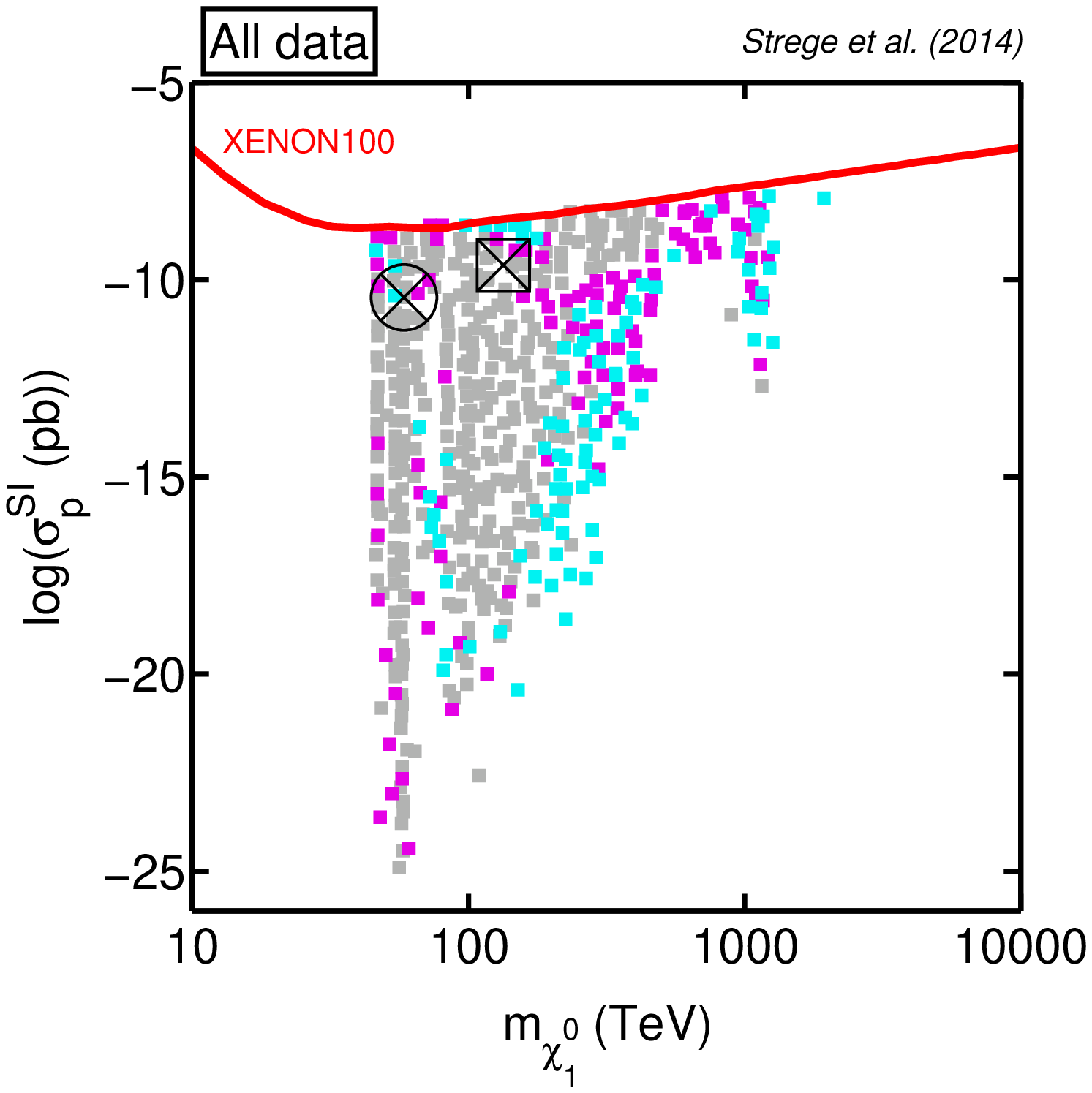} \\
\includegraphics[width=0.32\linewidth, trim = 0.7cm 0cm 0.7cm 0cm]{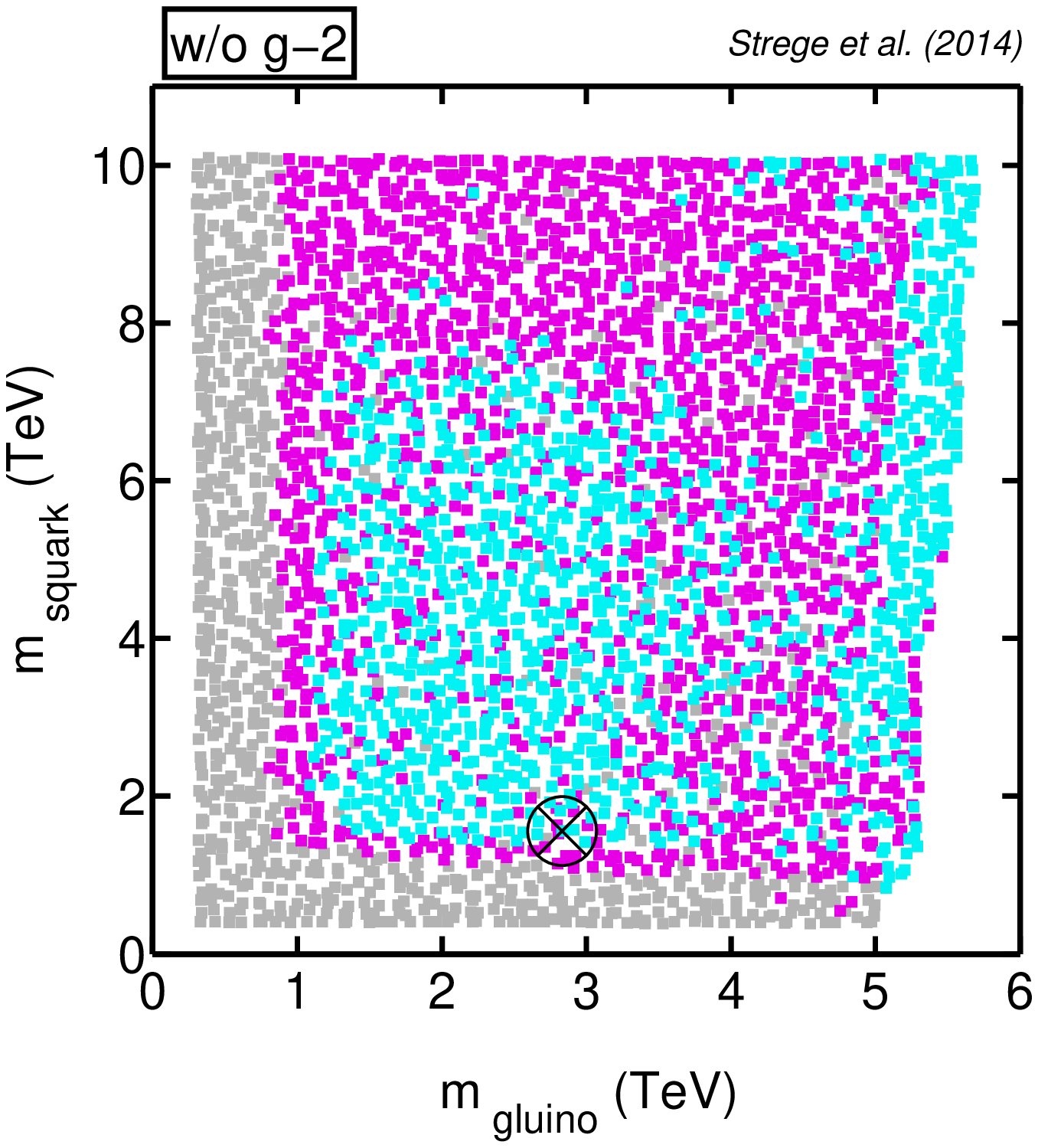}
\includegraphics[width=0.32\linewidth, trim = 0.7cm 0cm 0.7cm 0cm]{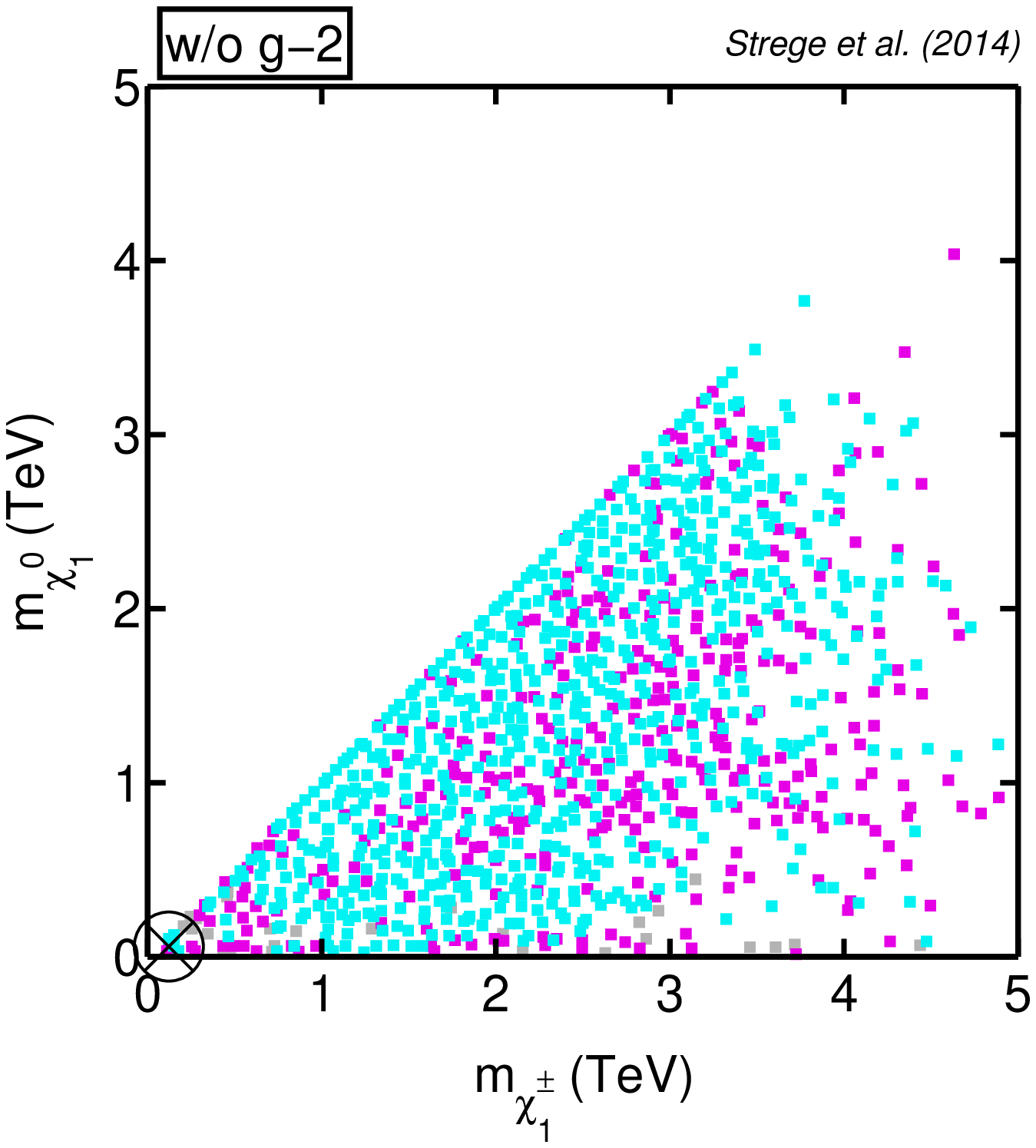}
\includegraphics[width=0.32\linewidth, trim = 0.7cm 0cm 0.7cm 0cm]{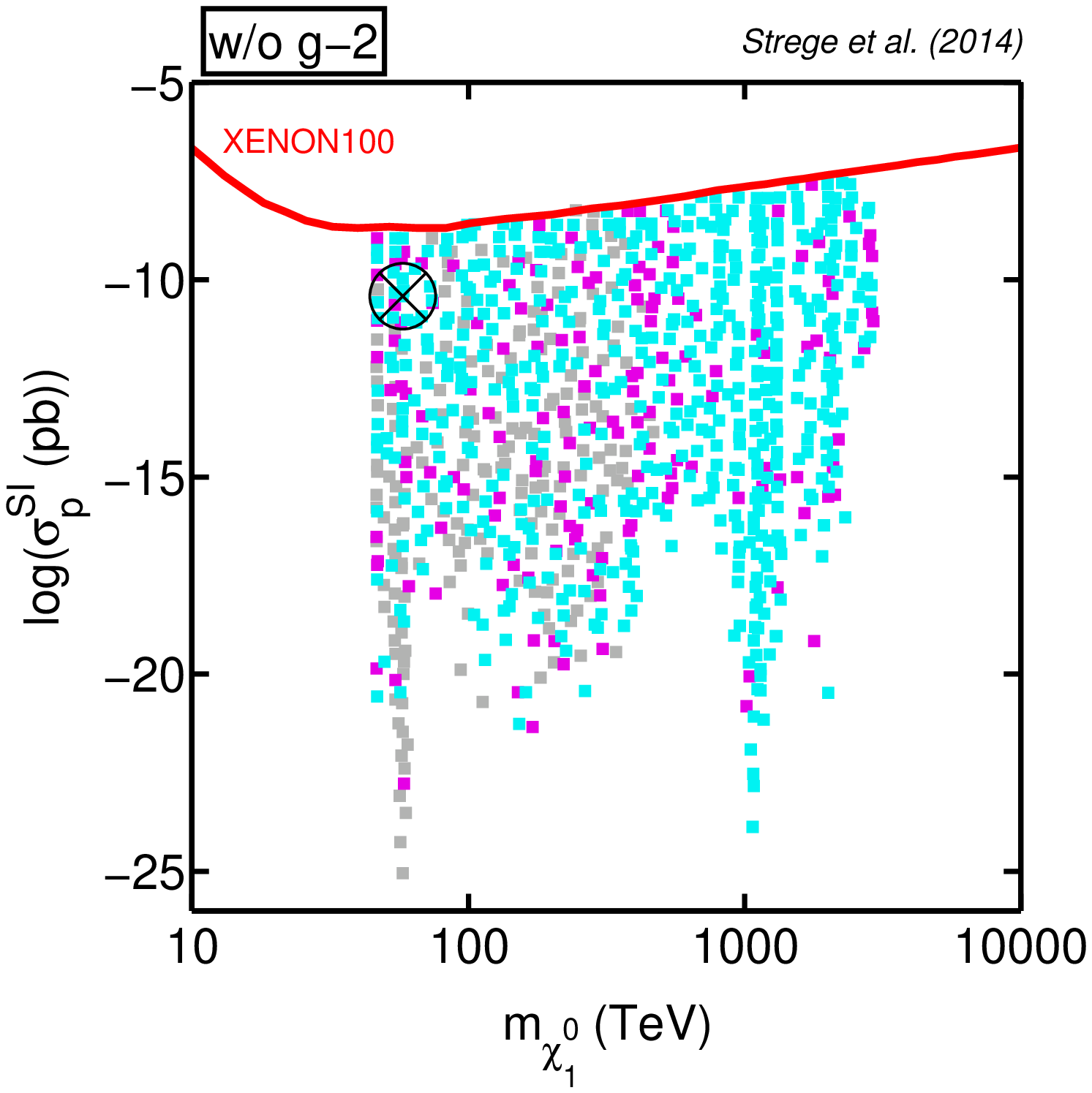}
\caption{Scatter plots from the 2D mini-chains, showing the impact of the LHC (SUSY searches and constraints on the Higgs decay cross-sections) on the chi-square of the best-fit point in each bin. The top (bottom) row shows results for the scans including all data (except the $g - 2$ constraint). The encircled black cross indicates the best-fit point prior to inclusion of the LHC constraints;  for the scans including all data, this point is ruled out by the LHC results, hence we also show the next best-fit point that survives the LHC constraints (cross inscribed in the square). The best-fit point for the analysis excluding $g - 2$ (bottom panels) remains viable after LHC data are included.}
\label{fig:LHC_impact}
\end{center}
\end{figure}

\begin{figure}
\begin{center}
\begin{center}
\includegraphics[width=0.5\linewidth]{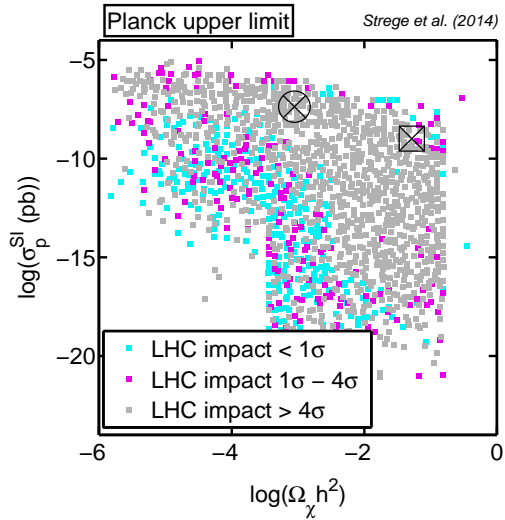}
\end{center}
\caption{Scatter plot from the 2D mini-chains, showing the impact of the LHC (SUSY searches and constraints on the Higgs decay cross-sections) on the chi-square of the best-fit point in each bin. The plot shows results for the scans including all data, with the Planck constraint applied as an upper limit. The encircled black cross indicates the best-fit point prior to inclusion of the LHC constraints (which is ruled out by the LHC),  while the cross inscribed in the square indicates the next best-fit point that survives the LHC constraints.}
\label{fig:LHC_impact_DMlimit}
\end{center}
\end{figure}

In Fig.~\ref{fig:LHC_impact} and Fig.~\ref{fig:LHC_impact_DMlimit} we show the impact of the ATLAS null searches for SUSY in the 0-lepton and 3-lepton channels, and of the CMS measurements of the Higgs boson properties. Bins that are almost unaffected by the LHC constraints (impact $< 1\sigma$) are shown in cyan, bins that are disfavoured with a significance $> 1\sigma$ and $< 4\sigma$ level are shown in pink, and bins that are ruled out by the LHC (impact $> 4\sigma$) are displayed in grey. Note that we only show bins that were included in the 99\% C.L. region before post-processing the mini-chains with the LHC constraints. 

Fig.~\ref{fig:LHC_impact} shows the LHC impact for the analysis including all data (top row) and the analysis excluding the $g - 2$ constraint (bottom row). Results for the ``Planck upper limit" analysis (not shown) are qualitatively very similar to the ``All data" case. From left to right the plots show the LHC impact in the planes of gluino mass vs. average squark mass, lightest chargino mass vs. lightest neutralino mass and neutralino mass vs. SI cross-section. As can be seen in the left-hand panels, the LHC 0-lepton search has a strong impact on the favoured regions of the MSSM-15, both for the analysis including and excluding the $g - 2$ constraint, ruling out gluino and squark masses $m_{\rm gluino}, m_{\rm squark} \lsim 1$ TeV. In addition, the measurements of the Higgs production cross-sections have a strong effect. In particular, in the regions most strongly affected by the Higgs signal strengths data we observe a suppression of the $b \bar{b}$ signal strength (and, to a lesser extent, the $\tau^+ \tau^-$ signal strength). As a consequence of this suppression, the other signal strengths are enhanced, in conflict with the experimental measurements. In particular, the constraint on $\muww$ leads to a significant contribution to the total $\chi^2$, as the central value is below the SM prediction at $\sim 1\sigma$ level, and the experimental error on this quantity is relatively small.

At tree-level, one would expect that the Higgs couplings are approximately SM-like, as $m_A \gtrsim 1 \tev$ for all points considered. However this argument breaks down when considering higher-order corrections. In fact, Ref.~\cite{Haber:2000kq} shows how SUSY QCD (SQCD) corrections to the $hb\bar{b}$ coupling can still be large in this limit, provided one or both of the sbottoms lie below the TeV scale (we have verified that this is the case for the regions most strongly affected by the Higgs couplings data). In this case, sizeable deviations from the SM prediction may arise. 

The effect of full decoupling can be seen in the narrow vertical band of cyan bins with $m_{\rm gluino}  \sim 5 \tev$, where all SUSY masses are large. For the analysis including all data, on the right-hand side of this band there is a narrow strip in which the full decoupling is not fulfilled. This particular region corresponds to very large values \tanb~$\sim 50$, for which the onset of decoupling is delayed~\cite{Haber:2000kq}. As a result, even though the gluino is heavy, the approach to decoupling is significantly slower. In general, the pink bins correspond to relatively larger values of \tanb\ than the cyan bins, for which full decoupling is not achieved.

Note that, for the analysis including all experimental constraints, there is a region at relatively low values of $m_{\rm squark}$ and $m_{\rm gluino}$ (but above the ATLAS 0-lepton limits), which is significantly disfavoured by the LHC constraints. In this region, \tanb $\sim 10$, so that the Higgs couplings data have a smaller impact. Instead, the ATLAS 3-lepton search (see below) impacts quite strongly on this region. The above discussion applies broadly also to the central and right-hand panels (for both the analysis including all data and excluding the \gmt\ constraint), in which the impact of the Higgs production cross-sections data follows a similar pattern.

In general, the impact of the 3-lepton channel search, which imposes constraints in the lightest chargino mass vs. lightest neutralino mass plane (central panels) is relatively weak compared to the 0-lepton channel. This is true in particular for the analysis excluding the muon \gmt\ constraint, for which larger neutralino masses are favoured (cf.\ Fig.~\ref{fig:1D_wog2_3} above). The impact of the constraint on the Higgs production cross-sections is again clearly visible, significantly disfavouring points that lead to strong deviations from the SM prediction for a large range of different values of $m_{\neut}$ and $m_{\charg}$.

The impact of the LHC SUSY and Higgs searches in the plane of neutralino mass vs.\ spin-independent scattering cross-section is shown in the right-hand panels of Fig.~\ref{fig:LHC_impact}. The main impact of the LHC in this plane is to rule out points at low/intermediate neutralino masses that were previously strongly favoured, mainly as a consequence of the 0-lepton channel search. Therefore, for small $m_{\neut} \lsim 300$ GeV, the LHC is extremely powerful, ruling out cross-sections orders of magnitudes below the reach of present and future direct detection experiments (and indeed below the ``ultimate'' limit represented by the solar neutrino background). For the analysis excluding the $g - 2$ constraint, a much smaller fraction of points is affected by the LHC, and several points at small $m_{\neut}$ are still allowed. 
This is largely a result of the 0-lepton search having less of an impact on the analysis excluding $g - 2$ (as very small squark masses are disfavoured for this analysis). Note that for $m_{\neut} \gsim 500$ GeV the \pMSSM\ parameter space is largely unaffected by constraints from LHC SUSY searches, but can be constrained by precise measurements of the Higgs production cross-sections. 

In Fig.~\ref{fig:LHC_impact_DMlimit} we show the impact of the LHC in the plane of spin-independent cross-section vs. neutralino relic density for the case when the Planck relic density measurement is taken as an upper limit (i.e., multi-component dark matter scenarios are allowed; the relic density is connected to the local density via the scaling Ansatz in Eq.~\eqref{eq:scaling_ansatz}). The LHC  has a strong impact in this plane, ruling out a large range of different relic densities and spin-independent cross-sections. Most of these points correspond to squark masses of $\order (100 \gev)$, and are thus ruled out at high significance by the ATLAS 0-lepton search. A narrow region at very large cross-sections, stretching along almost all allowed values of $\relic h^2$ is less affected by the LHC, especially at large relic density values. Likewise, a narrow horizontal region at large $\relic h^2 \sim \mathcal{O}(0.1)$ and intermediate cross-sections $10^{-15}$ pb $\lsim \sigmaSI \lsim 10^{-8}$ pb is unaffected by LHC SUSY searches, and only receives significant $\chi^2$ contributions from LHC constraints on the Higgs production cross-sections. Finally, a large selection of points that survive all LHC constraints is found at intermediate and small relic density and cross-section values.

The pre-LHC best-fit point from the ``w/o g - 2" analysis remains viable in light of the LHC data, while the best-fit points for the ``All data" and ``Planck upper limit" analyses are strongly disfavoured (see Section~\ref{sec:bf}). The best-fit points in the mini-chains after inclusion of the LHC constraints for the ``All data'' and ``Planck upper limit'' cases are given in Table~\ref{tab:bestfit}. Pre-LHC, those points have a $\chi^2$ value within 1$\sigma$ of the overall best-fit, and thus are perfectly viable. After adding the contributions from LHC SUSY null searches and constraints on the Higgs properties, their $\chi^2$ increases by $0.81$ (``All data'') and by $0.76$ (``Planck upper limit''). respectively. This indicates that they remain in good agreement with all experimental data sets considered in this analysis.

Compared to the pre-LHC best-fit points, we observed a shift of the squark mass to the multi-TeV region (2.3 TeV and 5.9 TeV, respectively), a slight increase in the neutralino mass (134 GeV and 128 GeV, respectively) and a gluino mass in the 1-2 TeV region. A squark mass of 2.3 TeV with gluinos in the 1-2 TeV range
will be accessible to the LHC searches in the upcoming high energy runs ~\cite{ATL-PHYS-PUB-2012-001}.
For the ``All data'' case, the best-fit SI cross-section shifts to a value of $2.3\times 10^{-10}$ pb, which is within the reach of the next generation of multi-ton scale direct detection experiments. 

Notice that in some panels in Figs.~\ref{fig:LHC_impact}--\ref{fig:LHC_impact_DMlimit}, the post-LHC best-fit points appear to be located in bins that are excluded according to the colour coding (grey). This is a consequence of the limitations of the mini-chains approach adopted here: as explained above, the mini-chain results have been obtained by post-processing the 2D PL values in each of the 2D planes separately. The post-LHC best-fit points, on the other hand, have been selected from the joint mini-chains encompassing PL values from all 4 2D planes shown in Fig.~\ref{fig:LHC_impact}--\ref{fig:LHC_impact_DMlimit} (in particular, in both cases the post-LHC best-fit point was taken from the $m_{\rm squark} - m_{\rm gluino}$ mini-chain). Therefore, the post-LHC best-fit points may appear to be excluded in some 2D projections since the latter have not been maximised over the entire parameter space with LHC data. 

The above results are suggestive of the constraints that a full analysis of our entire \pMSSM\ sample with our LHC likelihood would provide. As mentioned above, results from the mini-chains approach are indicative, but should not be interpreted probabilistically as PL maps. A thorough profile likelihood analysis of the LHC impact on the MSSM-15 will be presented in a future work~\cite{LHC_PP_Future}.

%%%%%%%%%%%%%%%%%%%%%%%%%%%%%%%%%%%%%%%%%%%%%%%%%%%%%%%%%%%%
\section{Conclusions}
\label{sec:conclusions}
%%%%%%%%%%%%%%%%%%%%%%%%%%%%%%%%%%%%%%%%%%%%%%%%%%%%%%%%%%%%

We have presented global fits of a phenomenological Minimal Supersymmetric Standard Model with 15 free parameters, including all available accelerator constraints, as well as constraints from cosmology and direct detection experiments. We have obtained high-resolution profile likelihood maps of the model parameter space, and discussed implications for the collider phenomenology and detection prospects in astro-particle physics experiments. We have discussed and compared the results for both the case in which the neutralino LSP is the only component of the dark matter in the universe, and the case in which it may be a subdominant dark matter component. We have also provided a detailed assessment of the impact of the \gmt\ constraint on the MSSM-15 profile likelihood maps. We summarise here the main results of our work:

\begin{itemize}
\item {\bf Constraints on input parameters:} Most of the input parameters remain almost unconstrained by current experimental results. However, relatively stringent constraints are placed on the parameters related to the dark matter phenomenology, $M_1$, $M_2$ and $\mu$, which are significantly affected by the relic density constraint, direct detection data, and several of the flavour observables, leading to a preference for small values of these quantities.

\item {\bf Constraints on SUSY mass spectrum:} In all considered cases, the profile likelihood function for the mass of the neutralino LSP peaks at very small values $m_\neut \lsim 100$ GeV. For single-component dark matter scenarios, a bino-like neutralino LSP with a mass of $\sim 60$ GeV is strongly favoured, although higgsino-like dark matter with $m_\neut \sim 1$ TeV is allowed at lower confidence. For the case excluding the \gmt\ constraint, the profile likelihood for the neutralino mass extends to significantly larger values, pushing the maximum value from $1.5$ TeV to about $3$ TeV. In this case, wino-like dark matter with $m_\neut \sim 2$ TeV is favoured at $95\%$ level. The profile likelihood functions for the squarks and gluinos are almost flat within the investigated parameter ranges.

\item {\bf Direct dark matter searches:} Direct detection constraints are found to be complementary to accelerator searches. Whereas upcoming experiments will allow to probe high neutralino scattering cross-sections, the very long tails in the parameter space extending to extraordinarily small cross-section values further strengthen the case for a combined analysis of astro-particle and accelerator data. Our current best-fit point, however, is within reach of the next generation of multi-ton scale direct detection experiments, exhibiting a spin-independent cross-section of $2.3\times 10^{-10}$ pb. 

\item {\bf Neutralino composition:} The rich phenomenology of the \pMSSM\ manifests itself in a broad range of neutralino compositions. We have provided a detailed discussion of the phenomenological consequences of the different compositions, and noticed in particular that in the case where the relic density constraint is applied as an upper limit, the favoured neutralino compositions are substantially different from the other cases, with the bulk of the favoured parameter space corresponding to wino-like (instead of bino-like) states. 

\item {\bf Impact of LHC searches:} We have demonstrated the strong impact of LHC SUSY searches, which provide stringent constraints in regions of the parameter space corresponding to very low values of $\sigmaSI$, which are not accessible with astro-particle physics experiments in the foreseeable future. Furthermore, we highlighted the significant impact of constraints on the Higgs signal strengths on the MSSM-15.

\end{itemize}

The full implementation of the LHC likelihood described in the Appendix is numerically very demanding: post-processing of all samples gathered for the above analysis would require approximately 400 CPU-years, even with our approximate likelihood based on fast simulations. We have adopted here an intermediate approach, which gives an indication of the extra constraining power from LHC searches and Higgs properties on the 2D profile likelihood maps. We will provide profile likelihood maps including the full LHC constraints in an upcoming dedicated work~\cite{LHC_PP_Future}. 

\vspace{\baselineskip} 
{\it Acknowledgements:} 
C.S. is partially supported by a scholarship of the ``Studienstiftung des deutschen Volkes". 
R. RdA, is supported by the Ram\'on y Cajal program of the Spanish MICINN and also thanks the support of the Spanish MICINN's Consolider-Ingenio 2010 Programme under the grant MULTIDARK CSD2209-00064, the Invisibles European ITN project (FP7-PEOPLE-2011-ITN, PITN-GA-2011-289442-INVISIBLES and the ``SOM Sabor y origen de la Materia" (FPA2011-29678) and the ``Fenomenologia y Cosmologia de la Fisica mas alla del Modelo Estandar e lmplicaciones Experimentales en la era del LHC" (FPA2010-17747) MEC projects. This research was supported in part by the National Science Foundation under Grant No. NSF PHY11-25915. GB acknowledges the support of the European Research Council through the ERC Starting Grant {\it WIMPs Kairos}.
  
We gratefully acknowledge the use of the Cartesius supercomputer (Amsterdam) and the High Performance Computing systems of Imperial College, IFT-UAM and IFIC-UV. R.Rda, C.S. and R.T. would like to thank GRAPPA and the University of Amsterdam for hospitality. G.B, R.Rda, C.S. and R.T. thank the Kavli Institute for Theoretical Physics at the University of California, Santa Barbara for hospitality during the programme ``Hunting for Dark Matter''. The authors would like to thank Maria Eugenia Cabrera,  Alberto Casas, Paul de Jong, Nazila Mahmoudi and Pat Scott for useful discussions, and Simon Burbidge at Imperial College London for help with the HPC resources.

%%%%%%%%%%%%%%%%%%%%%%%%%%%%%%%%%%%%%%%%%%%%%%%%%%%%%%%%%%%
\appendix
\label{sec:appendix}
%%%%%%%%%%%%%%%%%%%%%%%%%%%%%%%%%%%%%%%%%%%%%%%%%%%%%%%%%%%

%%%%%%%%%%%%%%%%%%%%%%%%%%%%%%%%%%%%%%%%%%%%%%%%%%%%%%%%%%%
\section{ATLAS Likelihood} \label{LHC_likelihood}
%%%%%%%%%%%%%%%%%%%%%%%%%%%%%%%%%%%%%%%%%%%%%%%%%%%%%%%%%%%
In this section we describe the construction adopted for the ATLAS likelihood. We adopt an approximate construction to exploit 0-lepton and 3-lepton inclusive searches from ATLAS data, as explained below. 

%%%%%%%%%%%%%%%%%%%%%%%%%%%%%%%%%%%%%%%%%%%%%%%%%%%%%%%%%%%
\subsection{ATLAS 0-lepton and 3-lepton signal regions}
%%%%%%%%%%%%%%%%%%%%%%%%%%%%%%%%%%%%%%%%%%%%%%%%%%%%%%%%%%%

The ATLAS 0-lepton analysis~\cite{ATLAS-CONF-2012-033-5fb} has 6 channels which are used to construct between one and three signal regions with ``tight", ``medium" and/or ``loose"  $m_\text{eff}(\text{incl.})$ selections, giving in total 11 signal regions. The different channels have been constructed for different SUSY particle production mechanisms. Signal region A is designed for squark-squark production, signal region A' especially for models with low mass splittings. Signal region B is designed for squark-gluino production whereas signal regions C-E are constructed for gluino-gluino production with high jet multiplicities.

The selection criteria for each signal region are shown in Table \ref{tab:sigreg47}. As the name implies there is a general veto on events containing leptons. The used selection variables are the minimum required number of jets and their respective transverse momentum, the missing transverse energy $E_{\text{T}}^{\text{miss}}$, the effective mass $m_{\text{eff}}$ calculated as the scalar sum of all transverse jet momenta larger than 40 GeV and the missing transverse energy, the ratio of $E_{\text{T}}^{\text{miss}}$ to $m_{\text{eff}}$ (where $m_{\text{eff}}$ only includes the required number of jets), the minimum angle between the required jets and the missing energy vector $\varDelta\phi(\text{jet}_i,\text{E}_{\text{T}}^{\text{miss}})_{min}$. For the signal region C-E an additional criterium is applied, a cut on $\varDelta\phi(\text{jet}_i,\text{E}_{\text{T}}^{\text{miss}})_{min}$ for all jets with transverse momenta larger than 40 GeV.\\\\

\begin{table}[tbh]
	\begin{center}
		\begin{footnotesize}
			\begin{tabular}{ccccccc}
				\toprule
				Signal region: & A & A' & B & C & D & E\\
				\midrule
				\etmis\ \gevklam $>$& \multicolumn{6}{c}{160}\\
				1$^{st}$ jet \pt\ \gevklam $>$& \multicolumn{6}{c}{130}\\
				2$^{nd}$ jet \pt\ \gevklam $>$&  \multicolumn{6}{c}{60}\\
				\cmidrule{2 - 7}
				
				3$^{rd}$ jet \pt\ \gevklam $>$&  -  &  -  &  60&  60&  60&  60\\
				4$^{th}$ jet \pt\ \gevklam $>$&  -  &  -  &  -  &  60&  60&  60\\
				5$^{th}$ jet \pt\ \gevklam $>$&  -  &  -  &  -  &  -  &  40&  40\\
				6$^{th}$ jet \pt\ \gevklam $>$&  -  &  -  &  -  &  -  &   -  & 40\\
				\cmidrule{2 - 7}
				
				\multirowbt[1]{2}{*}{$\varDelta\phi(\text{jet}_i,\text{E}_{\text{T}}^{\text{miss}})_{min}$  $ > $}& \multicolumn{3}{c}{ 0.4 (i=1,2,(3))}& \multicolumn{3}{l}{\hspace{9mm}0.4 (i=1,2,3)}
				\\
				& - & - &  - &\multicolumn{3}{l}{\hspace{9mm}0.2 (for all jets \pt $>40$ \gev )}\\
				\cmidrule{2 - 7}
				
				\etmis/\meff(Nj) $>$&   0.3 (2j)&   0.4 (2j)&  0.25 (3j)&  0.25 (4j)&  0.2 (5j)&  0.15 (6j)\\
				\cmidrule{2 - 7}
				
				\meff(\scriptsize incl.\footnotesize) [TeV]  $>$&  1.9/1.4/-&  -/1.2/-&  1.9/-/-&  1.5/1.2/0.9&  1.5/-/- &  1.4/1.2/0.9\\
				
				\bottomrule
			\end{tabular}
		\end{footnotesize}
	\end{center}
	\caption{\footnotesize Requirements for the inclusive channels A-E for the ATLAS 0-lepton analysis with an \lumitext\ of 4.7 \lumifb. For \meff (incl.) the limits are given in the order \textit{tight/medium/loose} (from~\cite{ATLAS-CONF-2012-033-5fb}).}
	\label{tab:sigreg47}
\end{table}

The ATLAS 3-lepton analysis~\cite{3-lepton-ana} consists of 3 signal regions. Signal regions 1a and 1b include a Z-veto, signal region 2 is designed for a on-shell Z boson. All signal regions require exactly three leptons, two of them form the same flavour opposite sign (SFOS) lepton pair. The selection criteria are shown in Table \ref{tab:sigreg3lep}. The transverse mass m$_{T}$ is calculated using the missing transverse energy and the third lepton.  

\begin{table}[tbh]
	\begin{center}
		\begin{footnotesize}
			\begin{tabular}{cccc}
				\toprule
				Signal region: & 1a & 1b & 2 \\
				\midrule
				lepton charge, flavour & \multicolumn{3}{c}{at least one SFOS pair with m$_{ll} > 20$ GeV}\\
				\etmis\ \gevklam $>$& \multicolumn{3}{c}{75}\\
				m$_{SFOS} $\gevklam &  \multicolumn{2}{c}{$< 81.2$ ~or $ > 101.2$} & $ 81.2 - 101.2$\\
				
				No. of b-jets &  0  &  0  & any  \\
				m$_{T}$ \gevklam &  any  &  $ >90$  &  $ >90$ \\
				\pt\ of all leptons \gevklam $>$&  10  &  30  & 10 \\
				\bottomrule
			\end{tabular}
		\end{footnotesize}
	\end{center}
	\caption{\footnotesize Requirements for the signal regions 1a, 1b and 2 for the 3-lepton ATLAS analysis with an \lumitext\ of 4.7 \lumifb.  In addition, the number of reconstructed leptons has to be three (from~\cite{3-lepton-ana}).}
	\label{tab:sigreg3lep}
\end{table}

Altogether, we thus have a total of 14 signal regions (11 from the 0-lepton analysis and 3 from the 3-lepton analysis).

%%%%%%%%%%%%%%%%%%%%%%%%%%%%%%%%%%%
\subsection{The likelihood function}
%%%%%%%%%%%%%%%%%%%%%%%%%%%%%%%%%%

The likelihood for each bin in a signal region $i$ ($i=1,\dots,14$) is given by
\begin{equation}
\like_i(n_i |s,b, {\boldsymbol \theta}) = \text{Poiss}(n_i| \lambda_s(s,b,\boldsymbol \theta)) \times \like_C(\boldsymbol \theta),  \label{eq:LHC_likelihood}    
\end{equation}
where the first factor reflects the Poisson probability of 
observing a number of events $n$ in the signal region given the signal (background) expected value $s$ ($b$). The Poisson expectation value 
$\lambda_s$ also depends on 
the nuisance parameters $\boldsymbol \theta$ that parameterize systematic 
uncertainties, such as luminosity or jet energy scale. Those uncertainties are constrained via the likelihood term $\like_C(\boldsymbol \theta)$, which is taken to be a multivariate Gaussian distribution around the nominal value $\boldsymbol \theta = 0$, with diagonal covariance matrix entries given by the quoted nominal uncertainties in each of the systematic factors. Then we write the Poisson expectation value as 
\begin{equation} 
\lambda_s = s (1+\Delta_s \theta_s) + b (1+\Delta_b \theta_b), 
\end{equation}
where $s$ and $b$ are the nominal values of the signal and background, 
$\Delta_{s}$ and $\Delta_b$ are their relative uncertainties and $\theta_{s}$ and $\theta_b$ are 
nuisance parameters, so that $\boldsymbol \theta= \{ \theta_s, \theta_b \}$. 

Experimental analyses provide the overall uncertainty in the 
background expectation in the signal region, $\Delta_b$. For the systematic uncertainty on $s$ we 
can use the fact that
\begin{equation} 
s = L \sigma \epsilon, \label{eq:LHC_signal}
\end{equation}
where $L$ is the integrated luminosity, $\sigma$ is the SUSY cross-section 
and $\epsilon$ is the acceptance times the detector efficiency. 
The errors on each of the terms above can then be propagated linearly to obtain 
\begin{equation} \label{eq:like_syst_err}
\frac{\Delta_ s}{s} = \sqrt{\left(\frac{\Delta L}{L}\right)^2 + \left(\frac{\Delta \sigma}{\sigma}\right)^2 + \left(\frac{\Delta \epsilon }{\epsilon}\right)^2}. 
\end{equation}
The theoretical uncertainties involved in the 
SUSY cross-sections determination, $\Delta \sigma$, are computed at each point of 
the parameter space of the model under consideration for the 0-lepton analysis via the  NLL-fast 1.2 package \cite{nllfast}, while they are neglected for the 3-lepton analysis. The relative  
error in the efficiency can be determined by comparing the official efficiencies 
maps from the ATLAS collaboration with ours (see below). The value of $\Delta L$ is subdominant compared with the other uncertainties, and hence can be neglected.
We further neglect uncertainties that are subdominant compared to the ones affecting the efficiencies, such as the jet energy scale.

We then obtain an effective likelihood,  $\like_{\text{eff}, i}$, by eliminating the above nuisance parameters $\boldsymbol \theta$ via marginalisation as follows:
\begin{equation}
\like_{\text{eff}, i}(n_i|s,b) = \int \like_i(n_i |s,b,\boldsymbol \theta) p(\boldsymbol \theta) d \boldsymbol \theta, 
\end{equation}
where the prior over $\boldsymbol \theta$ is uniform around $\boldsymbol \theta = 0$ and of length 6 standard deviations on either side. 

%%%%%%%%%%%%%%%%%%%%%%%%%%%%%%%%%%%%%%%%%%%%%%%%%%%%%%%%%%%
\subsection{Approximate joint likelihood for inclusive searches}
%%%%%%%%%%%%%%%%%%%%%%%%%%%%%%%%%%%%%%%%%%%%%%%%%%%%%%%%%%%

The method for combining different SUSY analyses depends on whether the 
analyses are exclusive (i.e., without overlapping data samples), or inclusive (i.e., 
with overlapping data samples). 

For exclusive analyses the corresponding data samples are statistically independent, whether they are signal regions or control samples to constrain the background prediction. However, the combined likelihood of two exclusive analyses cannot be constructed as the simple
product of the two individual likelihoods, as the systematics term $\like_C$ is in general correlated between the two searches. 
Every (fully) correlated 
systematic uncertainty must use the same nuisance parameter in both analyses, 
and only one constraint on this single parameter should be used in $\like_C$.

For analyses with statistically overlapping data samples or signal regions 
that are not exclusive (i.e., ``inclusive'' analyses), the likelihoods for different signal regions are not statistically independent, hence a joint likelihood is difficult to construct. In this case, for each value of $s$ we want to test, we select the best signal region based on the median (expected) value of the likelihood $P(q_s|s+b)$, where $q_s$ denotes the test statistics (as appropriate for setting upper limits)
% TODO: use the cases environment here
\begin{equation}
q_s = \left\{
\begin{array}{c l}
  -2 \ln \lambda(s)  & \mbox{ if } \hat{s}<s,  \\
  0 &\mbox{ if } \hat{s} > s.
\end{array}
\right.
\end{equation}
In the above test statistics, we have defined the profile likelihood ratio

\begin{equation}
\lambda(s) \equiv \frac{\like(s, \hat{\hat{\boldsymbol \theta}})}{\like(\hat{s}, \hat{\boldsymbol \theta})},
\end{equation}
where $\hat{\hat{\boldsymbol \theta}}$ denotes the conditional ML estimator for the nuisance parameters, ${\boldsymbol \theta}$, $\hat{s}$ is the unconditional MLE for $s$ and $\hat{\boldsymbol\theta}$ the unconditional MLE for $\boldsymbol \theta$. 
The distribution of the likelihood is obtained from Monte Carlo simulations (assuming the alternative $s+b$). The best signal region is the one leading to a median likelihood with the smallest p-value.  This procedure thus selects the signal region that is expected to give the strongest upper limits for each value of $s$. 

We then evaluate the likelihood using the observed number of events in that optimal signal region,  $P(n_{\mathrm{obs}}|s+b_{\mathrm{fit}})$, where $n_\mathrm{obs}$ is the observed number of events and `\rm{fit}' refers to the data-constrained background value.

This approach however would in general lead to a discontinuity in the value of the likelihood whenever one crosses regions in parameter space where the best signal region changes. This is because there is no reason why absolute values of the likelihood function for different signal regions should be continuous across optimal signal regions boundaries. We solve this problem by defining the full likelihood as
\begin{equation} \label{eq:non-exclusive}
\like = \like^{\rm{obs}}_{i} \prod_{j \neq i} E[\like_{j}],
\end{equation}
where $\like^{\rm{obs}}_{i}$ is the observed likelihood for the signal region selected by the above procedure, while  $E[\like_{j}]$ is the expected 
likelihood in signal region $j \neq i$ (i.e., in all the other signal regions that are less optimal for the given $s$ being tested). 

%%%%%%%%%%%%%%%%%%%%%%%%%%%%%%%%%
\section{Validation via simulations}
 \label{LHC_validation}
%%%%%%%%%%%%%%%%%%%%%%%%%%%%%%%%%

%%%%%%%%%%%%%%%%%%%%%%%%%%%%%%%%%
\subsection{Likelihood validation}
%%%%%%%%%%%%%%%%%%%%%%%%%%%%%%%%%

We validate the likelihood \eqref{eq:LHC_likelihood} and the approach of Eq.~\eqref{eq:non-exclusive} in the case of non-overlapping signal regions as follows. For every signal region $i$ ($i=1,\dots,14$) in the analyses used, the number of expected events in the signal region under the null hypothesis ($s=0$) is given by the number of expected background events $b_i\pm \sigma_{b_i}$. In order to not bias ourselves towards any particular SUSY model or particular number of expected signal events $s_i$ in general, we generate 10,000 toy events around the background expectation only.

To take into account systematic and statistical fluctuations, the number of toy observed events is generated according to
\begin{equation}
n_{{\rm toy}_i} = \mathrm{Poisson} ( \mathrm{Normal} (b_i, \sigma_{b_i})),
\end{equation}
where the extra Gaussian smearing approximately accounts for systematic effects. For $\sigma_{b_i}$ we adopt the uncertainty on the background prediction as given by ATLAS. We then compute the joint likelihood of Eq.~\eqref{eq:non-exclusive} for the simulated events for each non-exclusive signal region $i$. 
\begin{figure}
%\begin{center}
\includegraphics[width=0.32\linewidth]{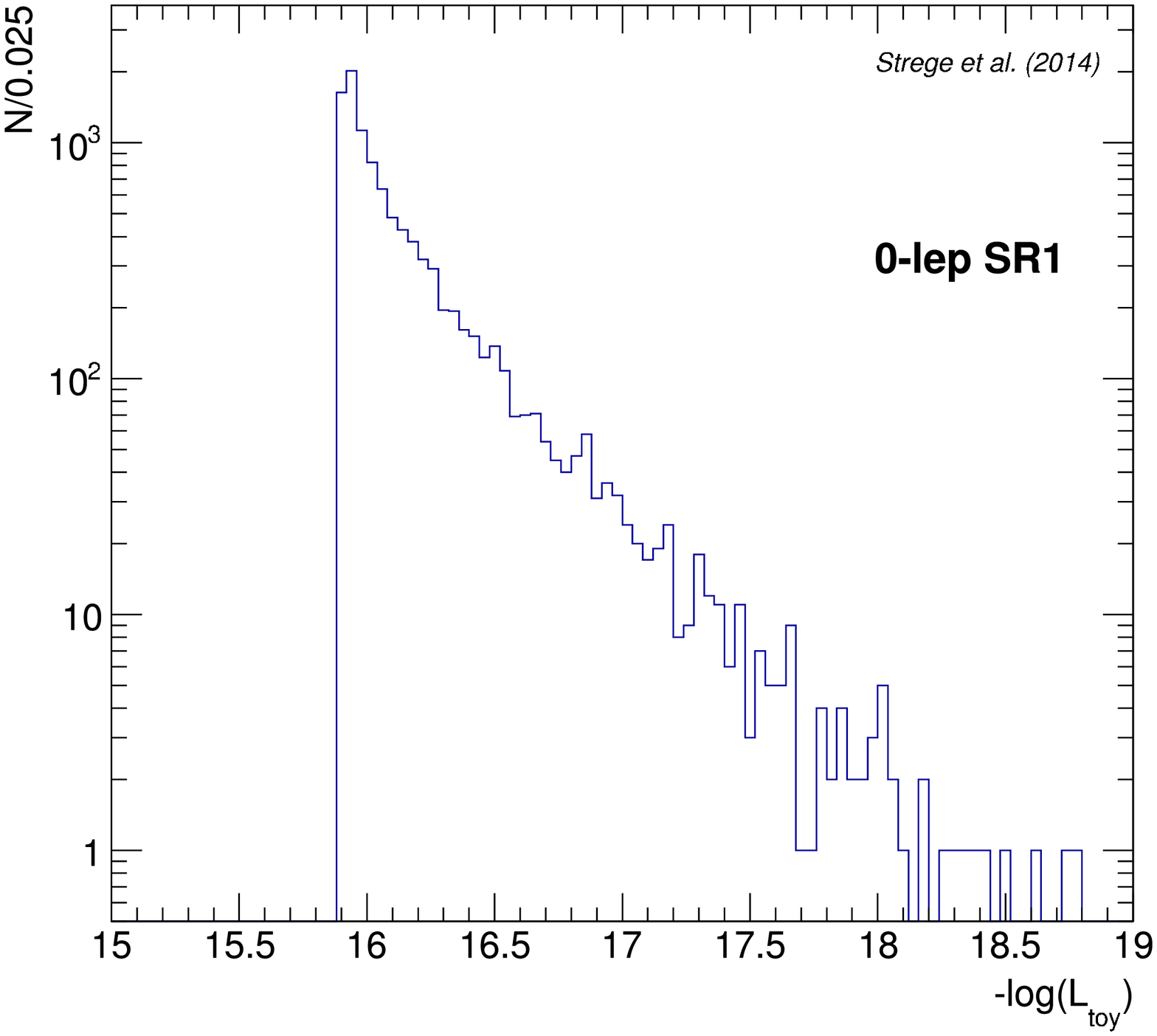}
\includegraphics[width=0.32\linewidth]{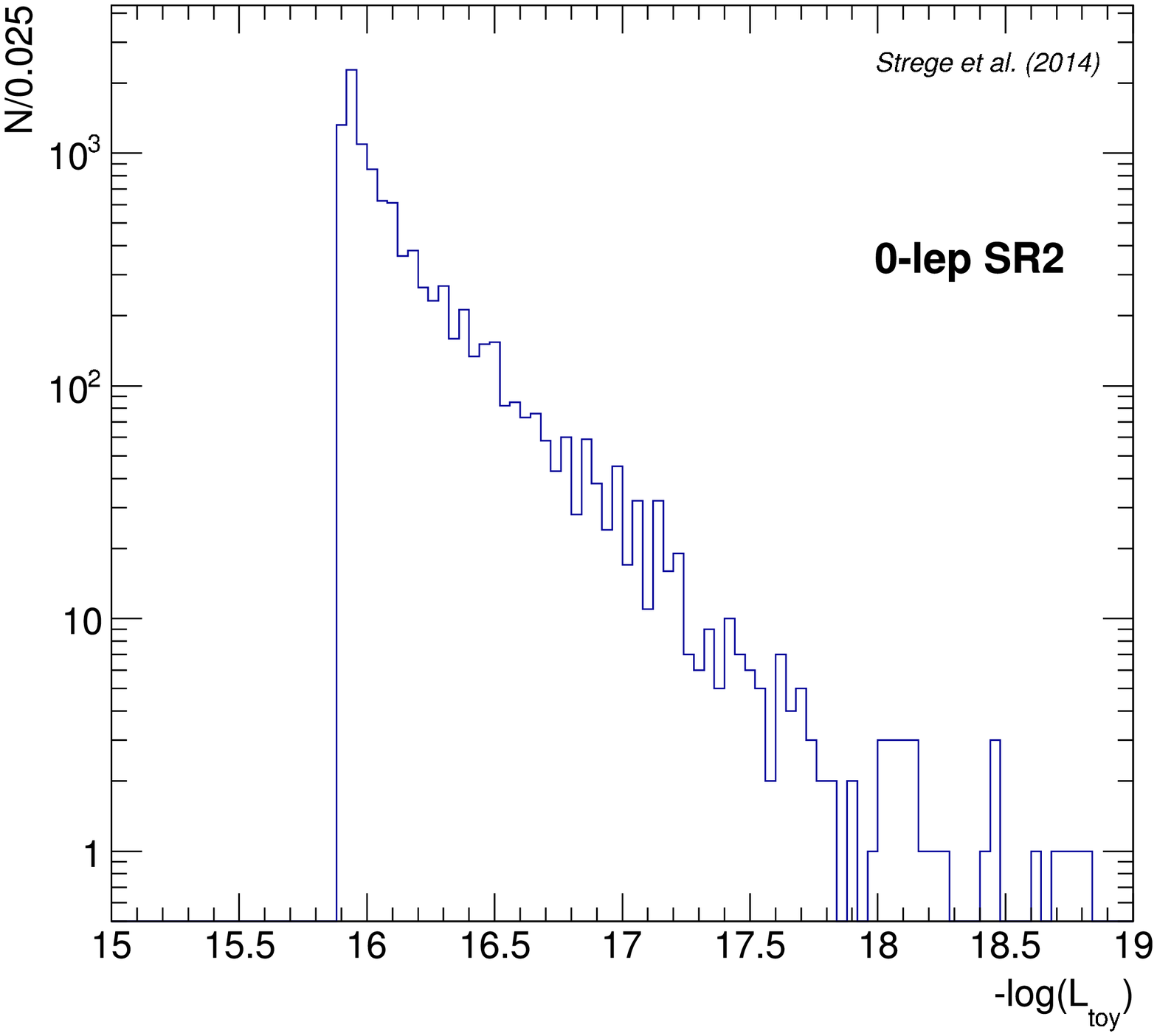}
\includegraphics[width=0.32\linewidth]{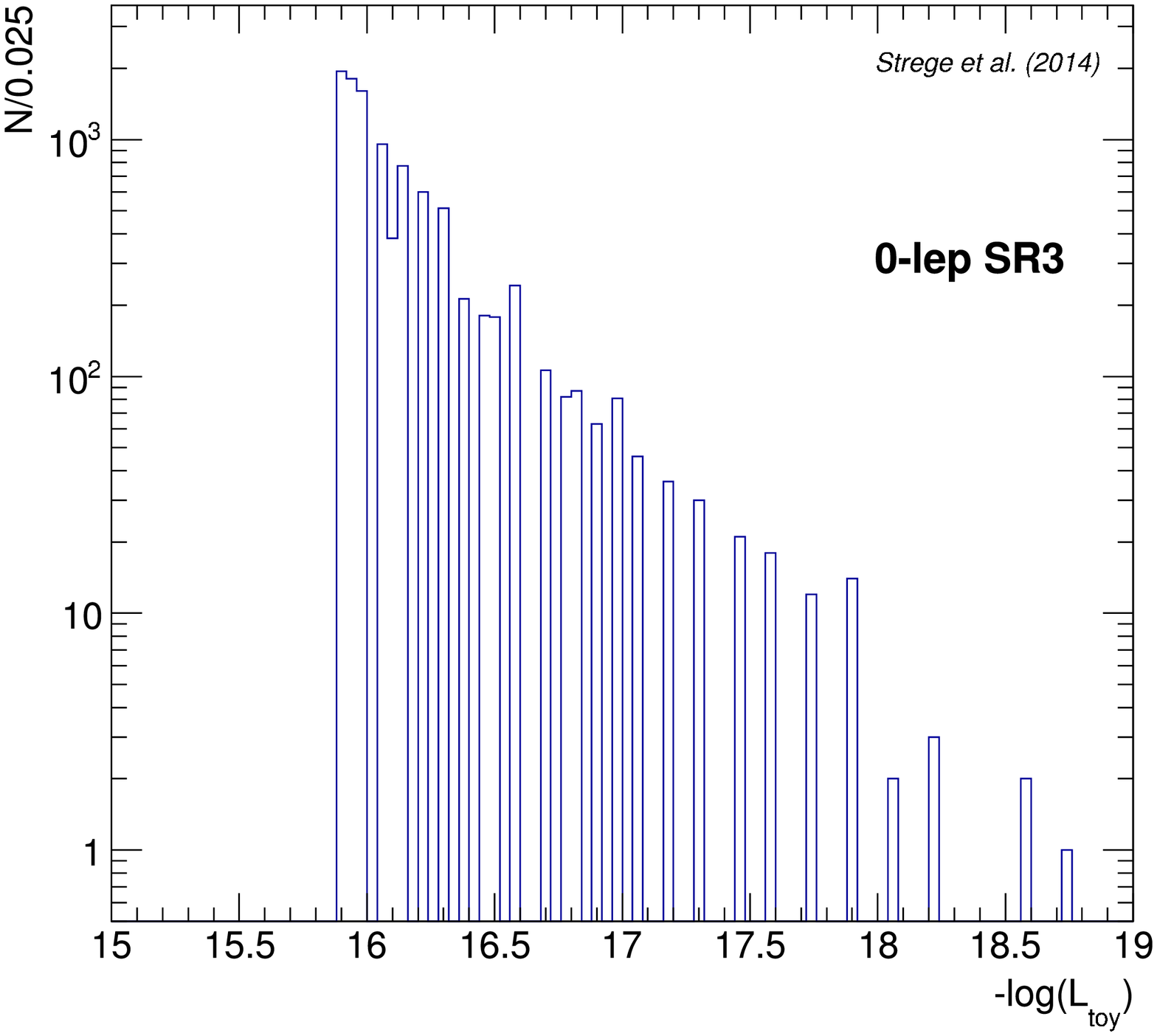}\\
\includegraphics[width=0.32\linewidth]{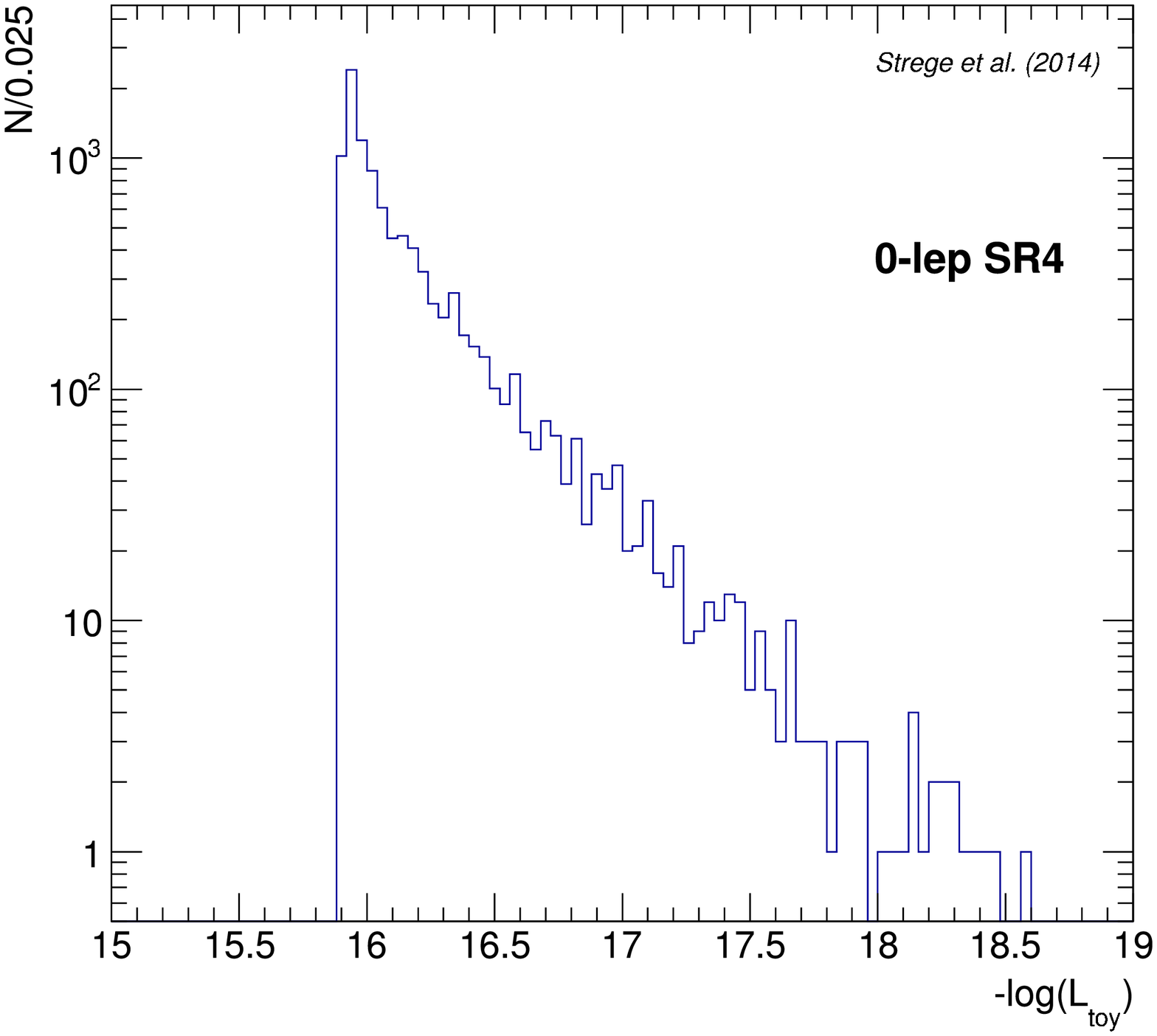}
\includegraphics[width=0.32\linewidth]{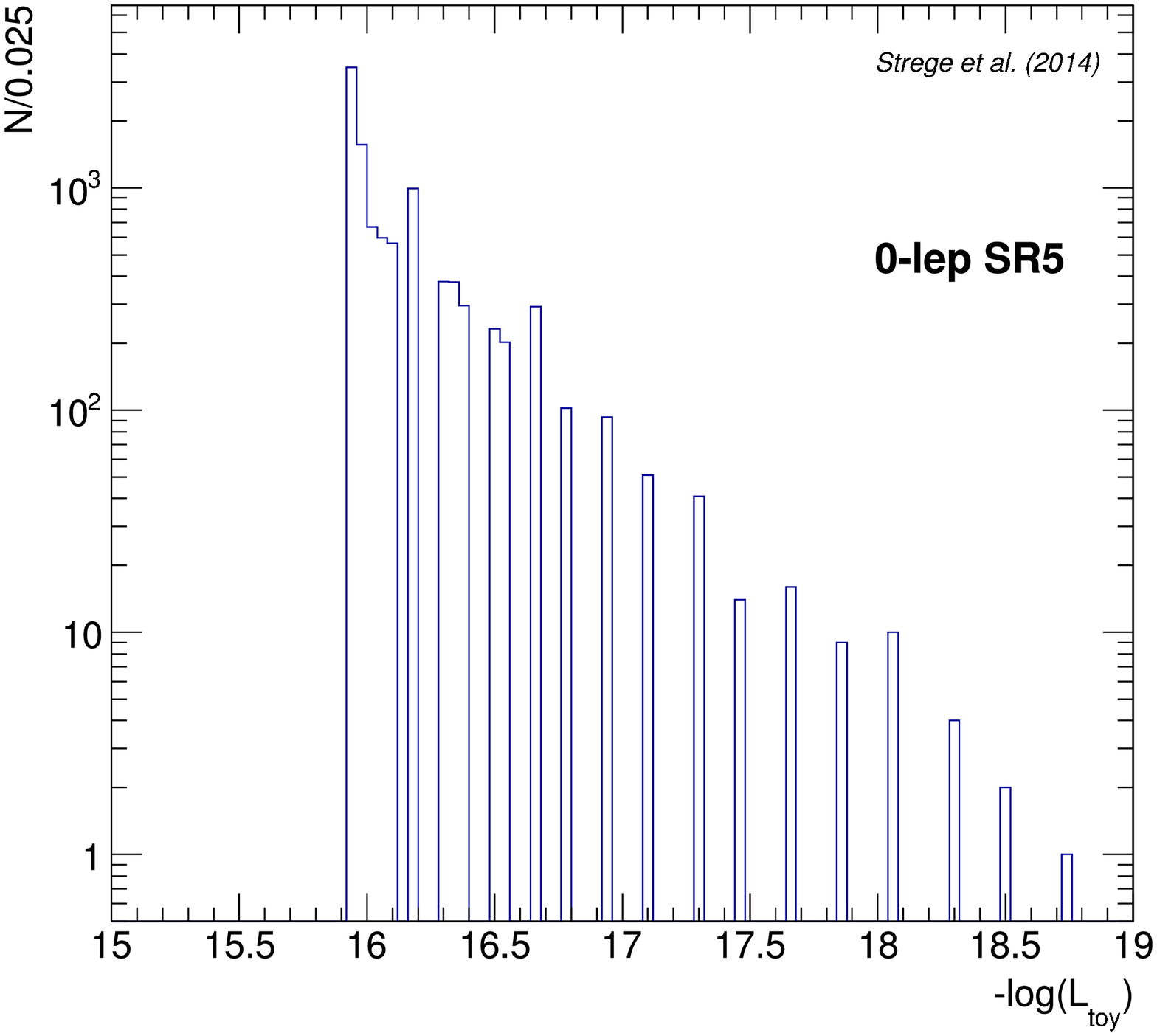}
\includegraphics[width=0.32\linewidth]{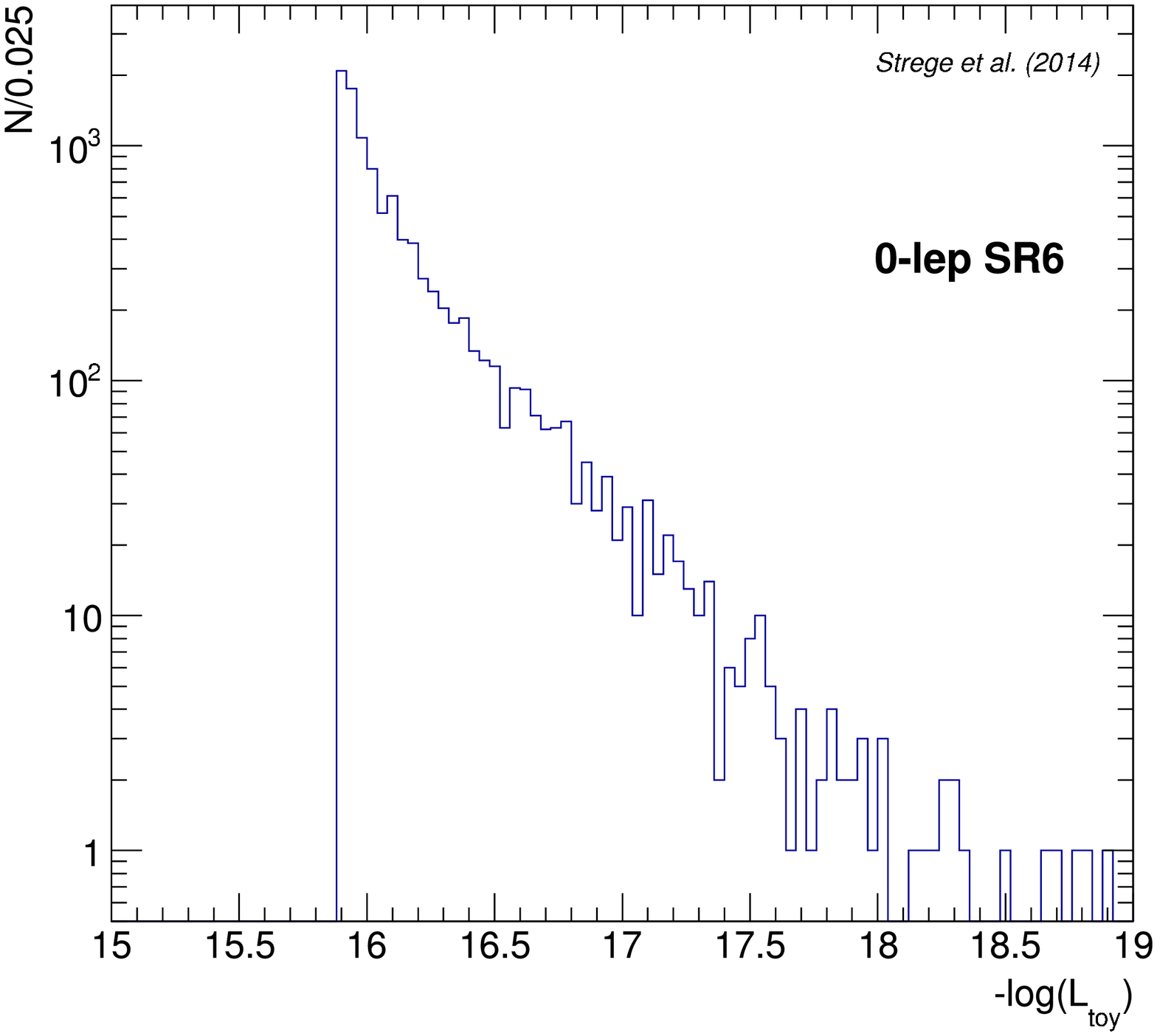}\\
\includegraphics[width=0.32\linewidth]{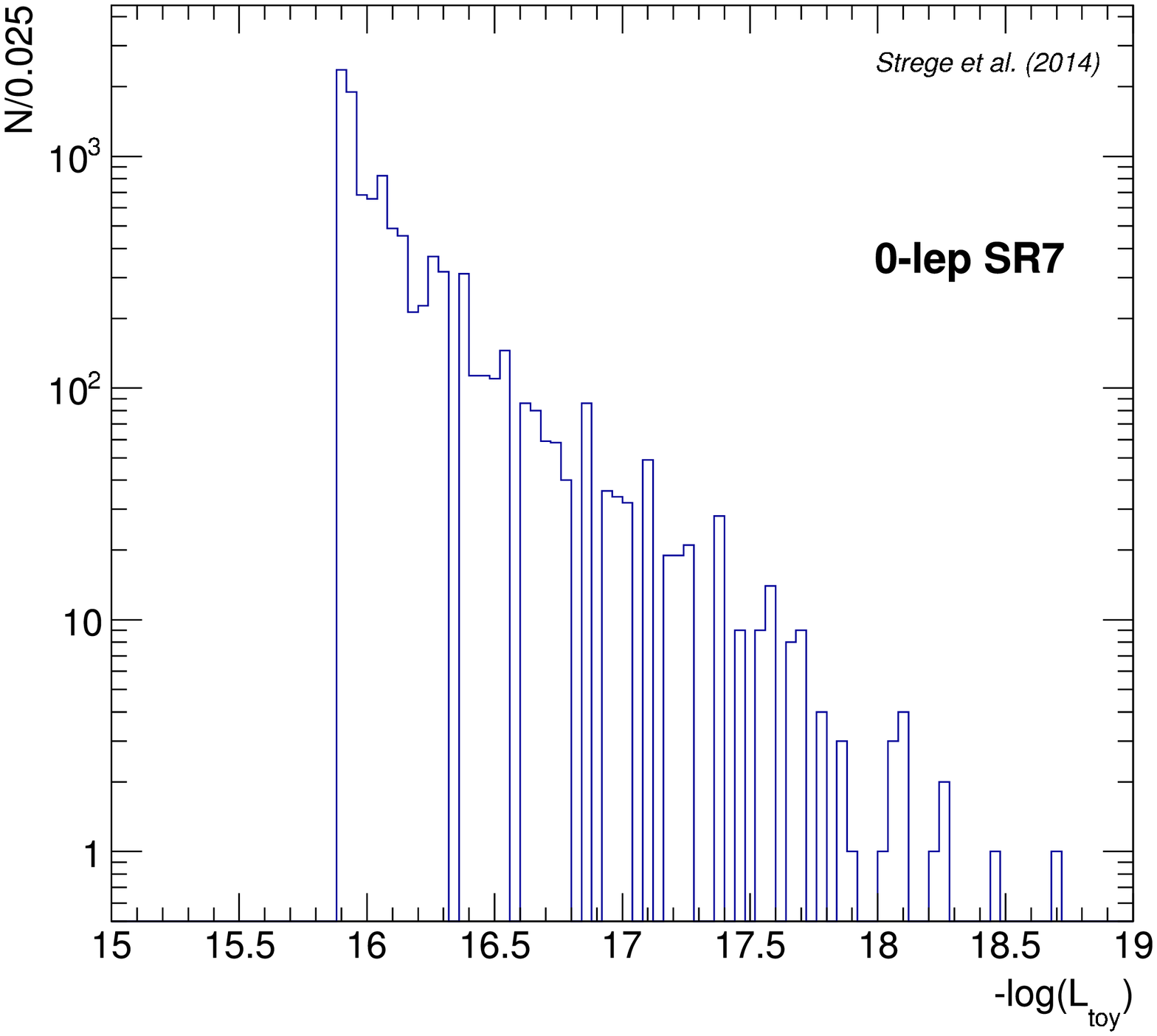}
\includegraphics[width=0.32\linewidth]{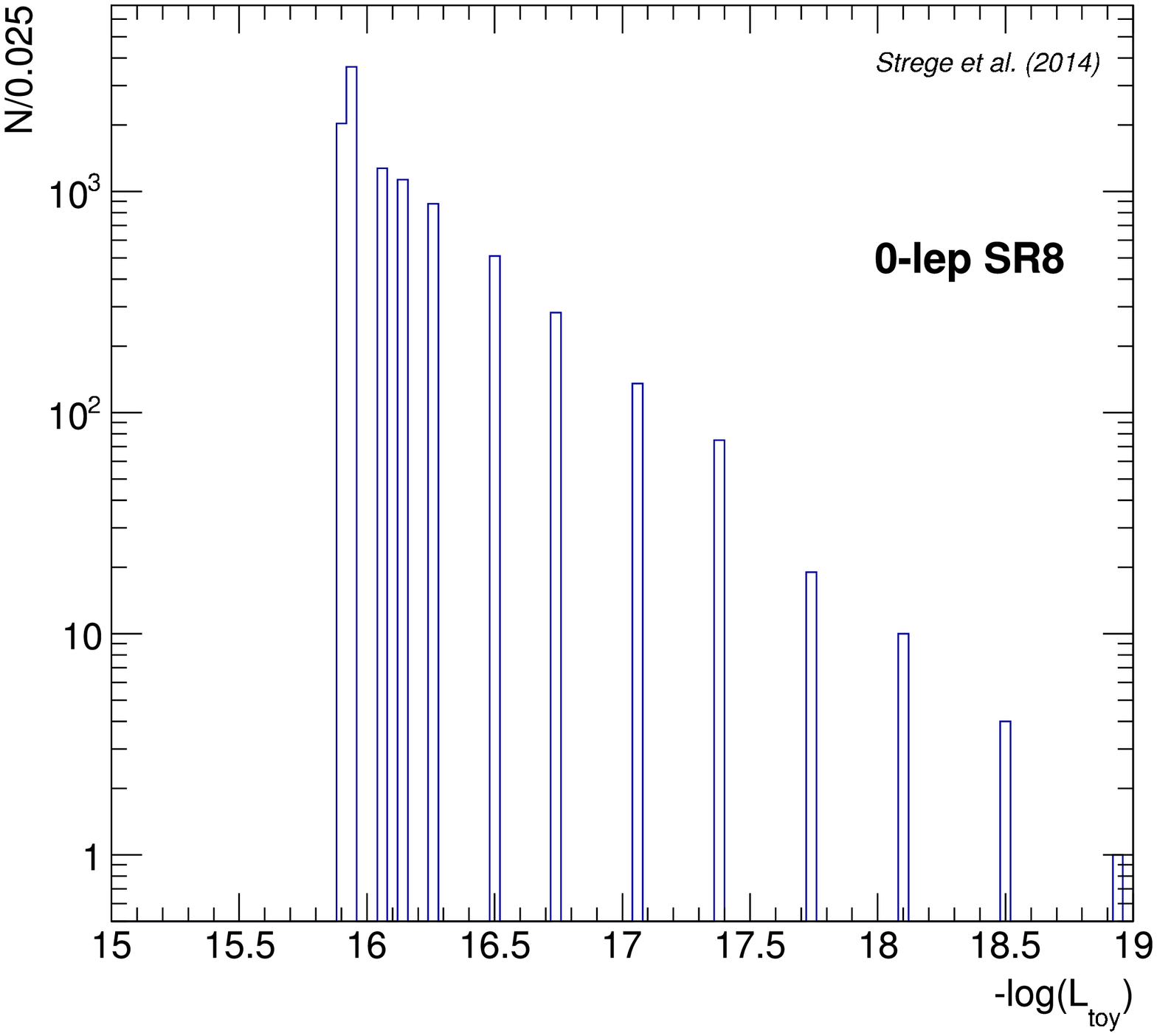}
\includegraphics[width=0.32\linewidth]{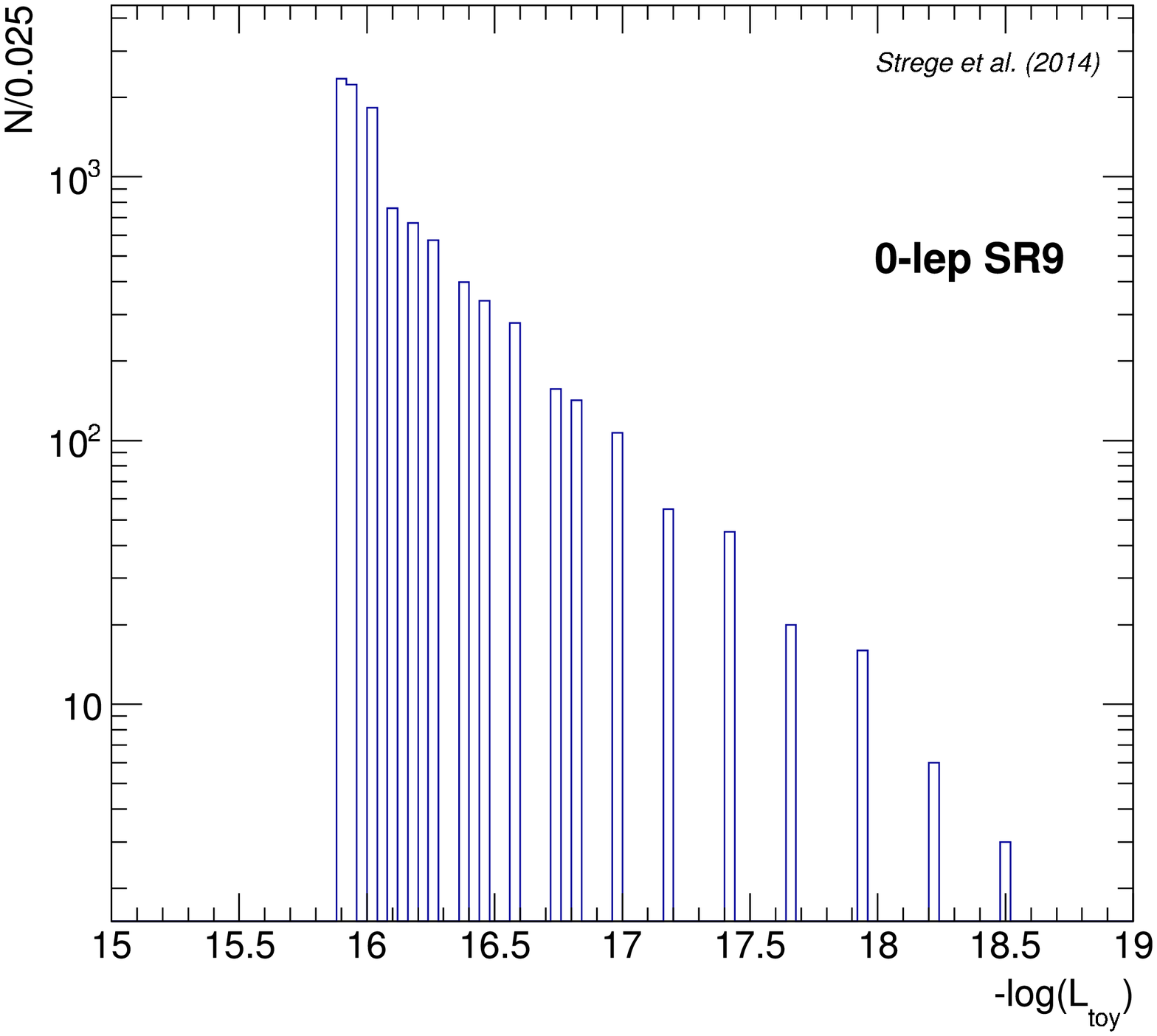}\\
\includegraphics[width=0.32\linewidth]{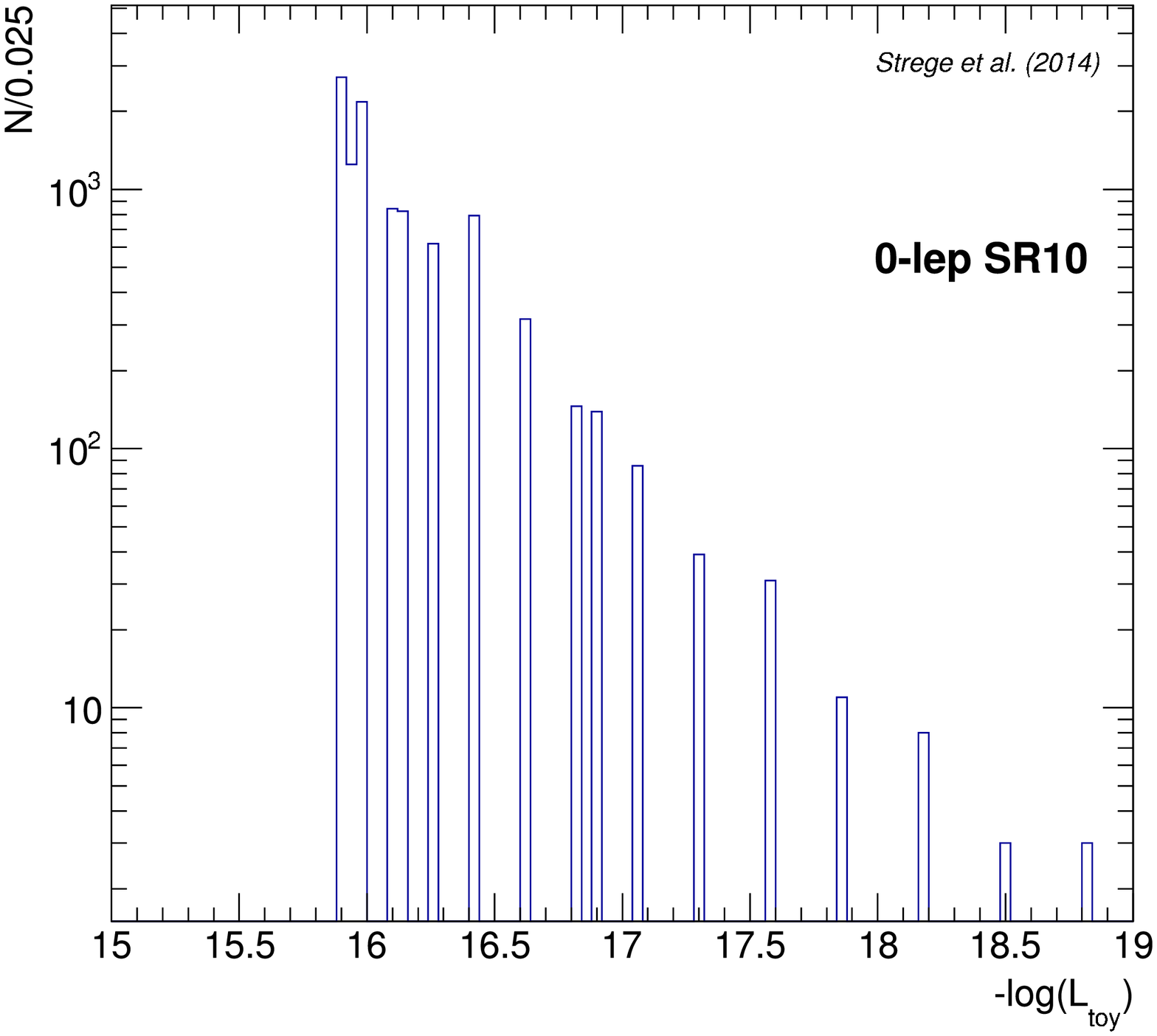}
\includegraphics[width=0.32\linewidth]{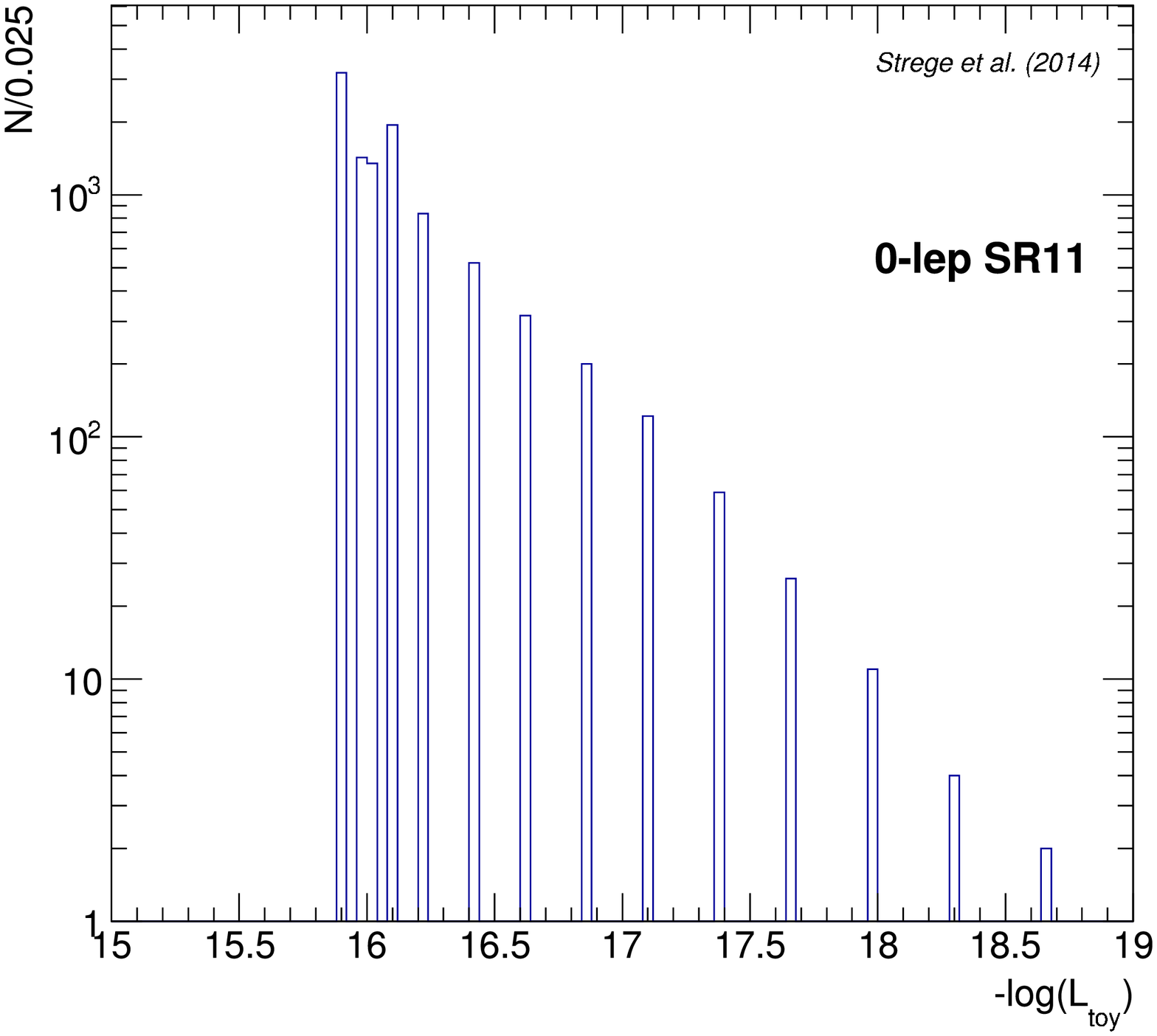}
\caption{Log likelihood results for toy data in each signal region in the ATLAS 0-lepton inclusive analysis. All curves can be seen to follow the same distribution, which validates our approach.} 
\label{fig:LHC_likelihood_validation}
%\end{center}
\end{figure}

The results of the validation studies are shown in Fig. \ref{fig:LHC_likelihood_validation}, for the 11 signal regions of the 0-lepton analysis (the case of the 3-lepton analysis is similar). We can see that the distribution's shapes are identical, up to an irrelevant normalisation factor.  The spikes are a normalisation issue, due to the fact that for regions with a small number of events the number of possible values for $\like_{\mathrm{obs_i}}(n = n_{\mathrm{toy}_i})$ is smaller than it is for regions with high $n_{b_j}$. 

We can conclude from this validation study that our procedure to use the most powerful signal region for each sampled value of the \pMSSM\ parameter space while normalising the likelihood via the expected value of the other signal regions leads to no large bias. 

%%%%%%%%%%%%%%%%%%%%%%%%%%%%%%%%%%%%%%%%%%%&
\subsection{Signal simulation validation}
%%%%%%%%%%%%%%%%%%%%%%%%%%%%%%%%%%%%%%%%%%%

For the validation of the event and detector simulations we adopted two different ATLAS analyses. For both analyses we cross-checked the resulting event selection efficiencies of our simulation done with PYTHIA 6.4~\cite{pythia} and DELPHES3.1~\cite{delphes} against the corresponding ATLAS acceptance times efficiency values.

To cover a broad spectrum of signals the ATLAS SUSY searches with zero leptons~\cite{ATLAS-CONF-2012-033-5fb} and with 3 leptons~\cite{3-lepton-ana} were chosen. Both analyses use a total integrated luminosity of $4.7$ fb$^{-1}$ of data taken at $\sqrt{s}=7$ TeV.

%%%%%%%%%%%%%%%%%%%%%%%%%%%%%%%%%%%%%%%%%%
\subsection{Comparison of efficiencies}
%%%%%%%%%%%%%%%%%%%%%%%%%%%%%%%%%%%%%%%%%

To validate our simulation setup the relative efficiency difference 
\begin{equation}
\frac {\Delta \varepsilon} {\varepsilon} = \frac{(A\varepsilon)_{\rm{ATLAS}} - (A\varepsilon)_{\rm{Sim}}} {(A\varepsilon)_{\rm{ATLAS}}} 
\end{equation}
between our setup and the official ATLAS analyses was determined. Here $(A\varepsilon)_{\rm{ATLAS}} $ is the acceptance times efficiency of the ATLAS analyses. A negative value of $\frac {\Delta \varepsilon} {\varepsilon} $ corresponds to an overestimation of the efficiency by the simulation, a positive value to an underestimation. 

For the validation the default ATLAS detector card supplied with DELPHES 3.1 was modified. For both analyses the value of the jet cone parameter $R$ of the anti-$k_t$ jet algorithm was set to 0.4. For the 3-lepton analysis the lepton efficiencies were increased and the lepton isolation value set to 0.7. 

\begin{figure}
\begin{center}
	\includegraphics[width=0.32\linewidth, trim=0.5cm 0cm 1.8cm 0cm, clip=true]{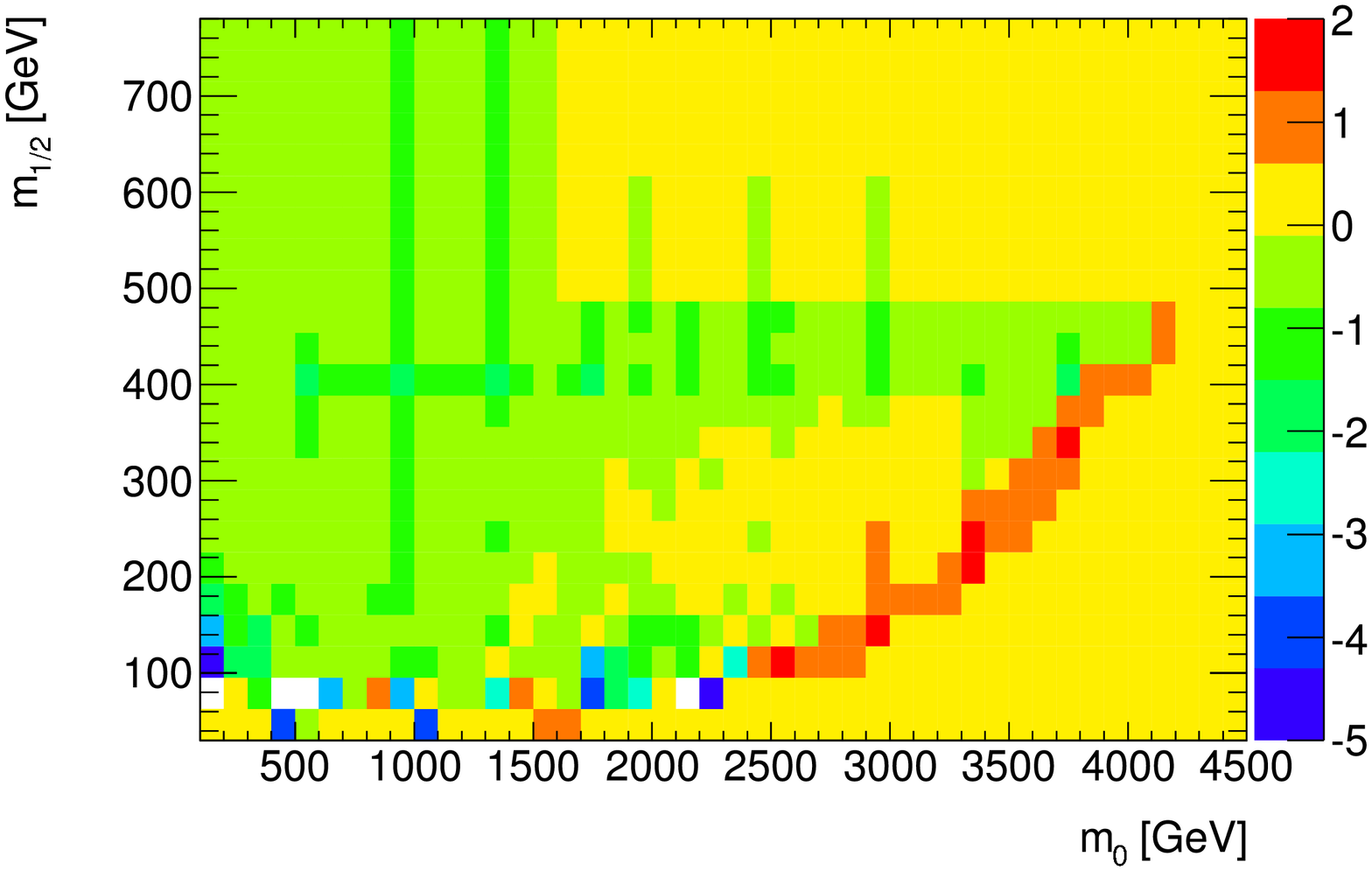}
	\includegraphics[width=0.32\linewidth, trim=0.5cm 0cm 1.8cm 0cm, clip=true]{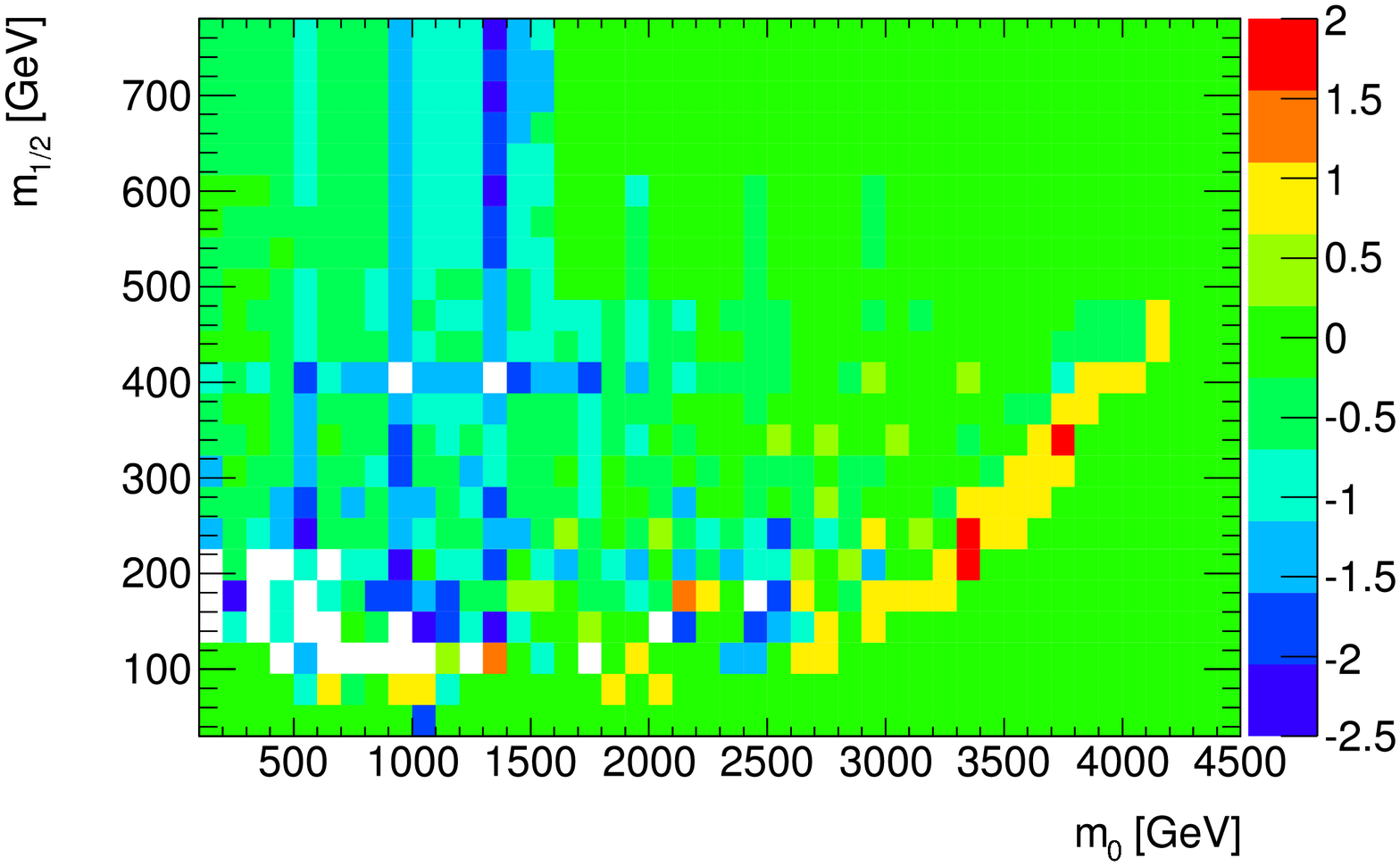}
	\includegraphics[width=0.32\linewidth, trim=0.5cm 0cm 1.8cm 0cm, clip=true]{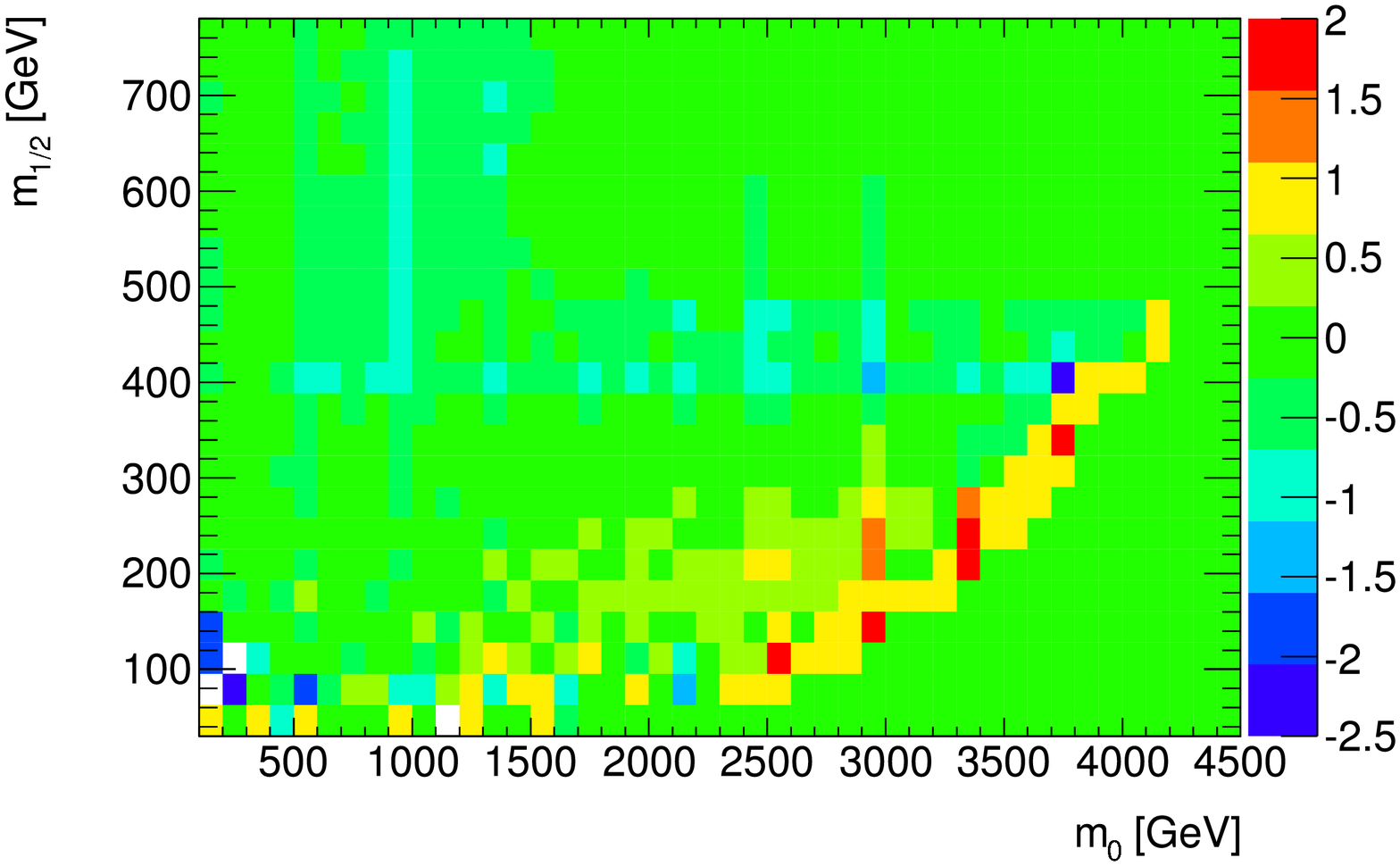}
	\caption{\effbruch\ of the ATLAS and simulation setups for signal region A \textit{medium/loose} and A' of the 0-lepton analysis.}
	\label{fig:0lep-A}
\end{center}
\end{figure}

\begin{figure}
\begin{center}
	\includegraphics[width=0.49\linewidth, trim=0.5cm 0cm 1.8cm 0cm, clip=true]{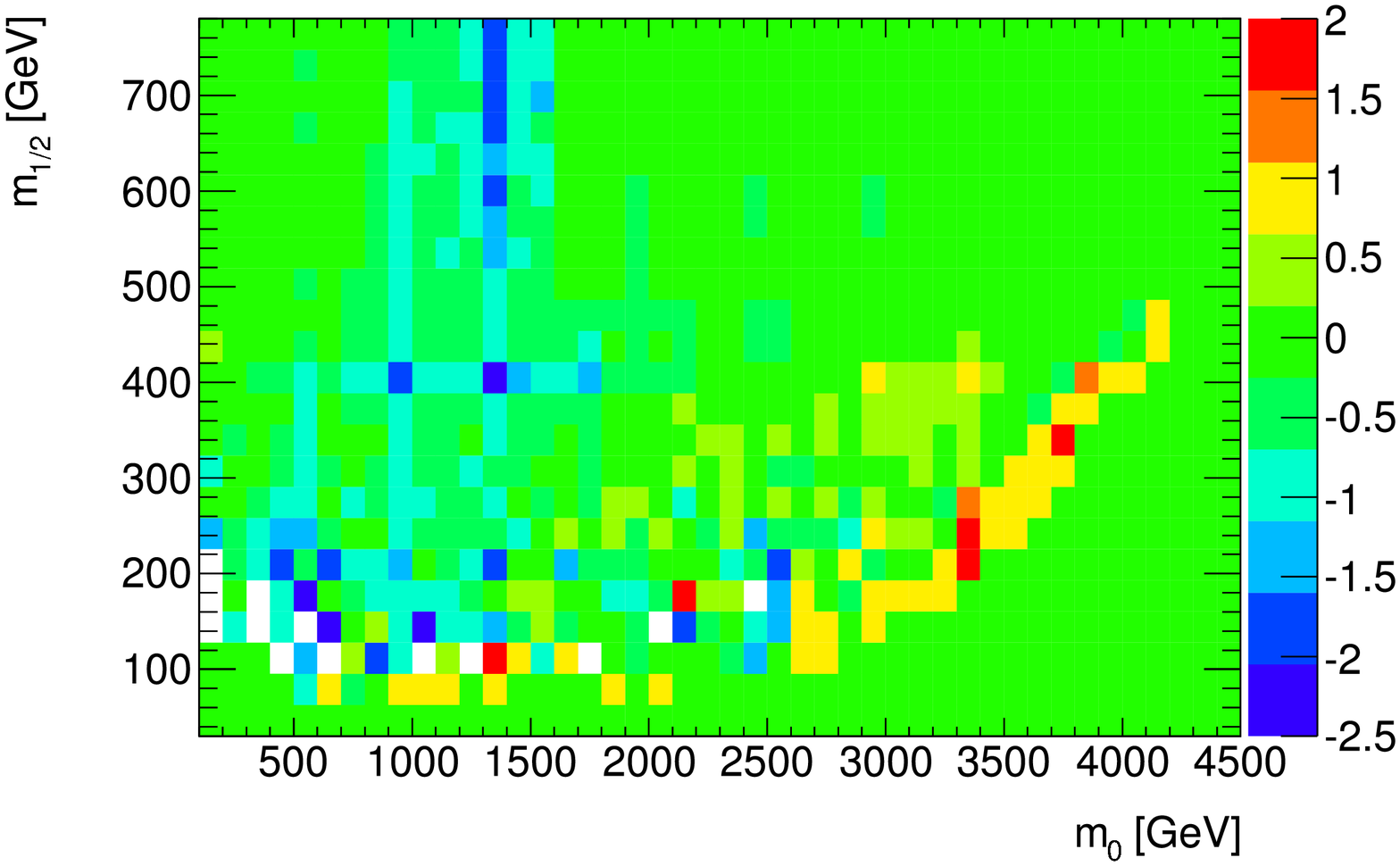}
	\includegraphics[width=0.49\linewidth, trim=0.5cm 0cm 1.8cm 0cm, clip=true]{figs/histEFF_DELPHESproATLASColour_regionBtight_Delphes310_offGridR04_0LEPtanb10.ps.pdf.ps}\\
	\includegraphics[width=0.49\linewidth, trim=0.5cm 0cm 1.8cm 0cm, clip=true]{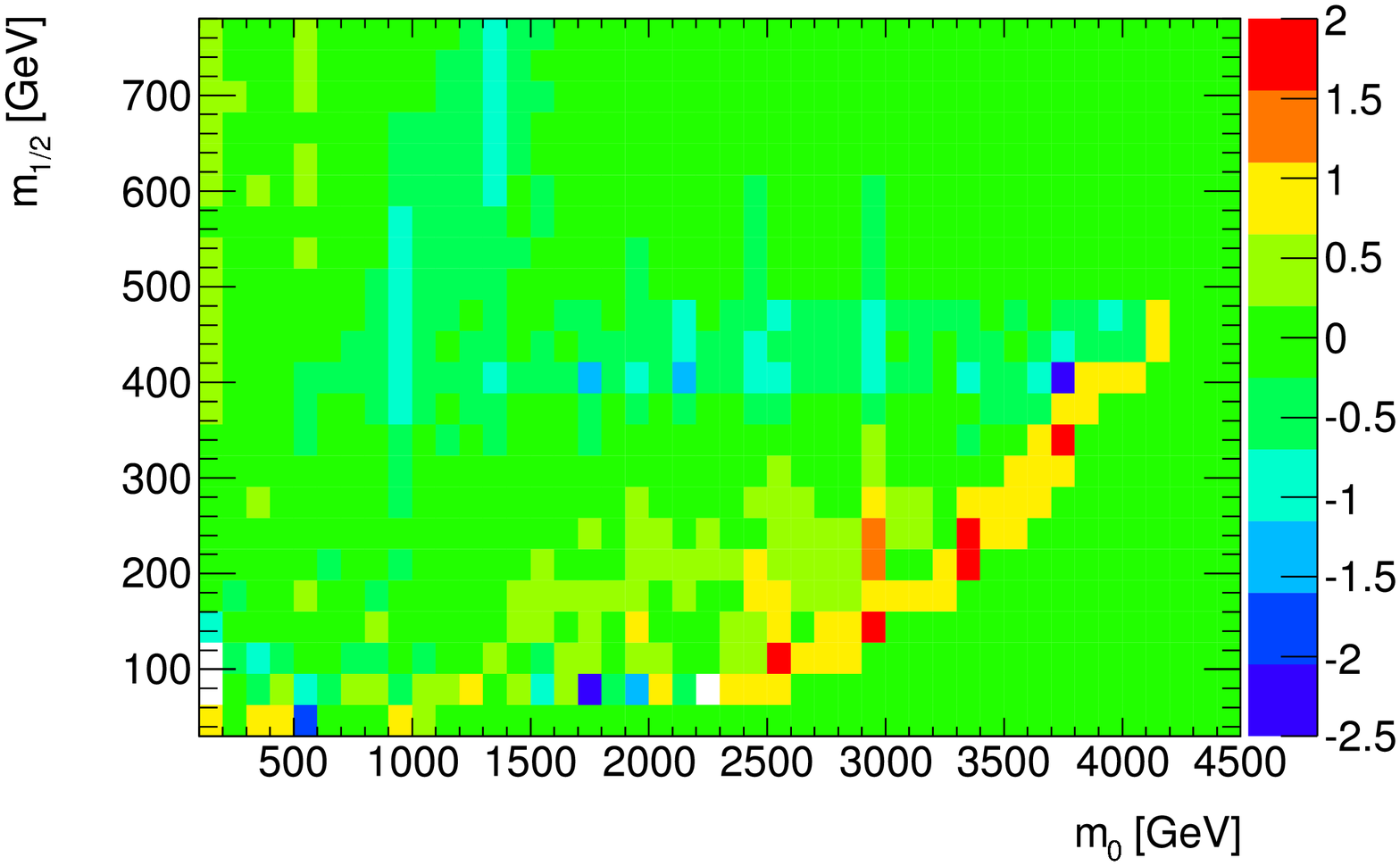}
	\includegraphics[width=0.49\linewidth, trim=0.5cm 0cm 1.8cm 0cm, clip=true]{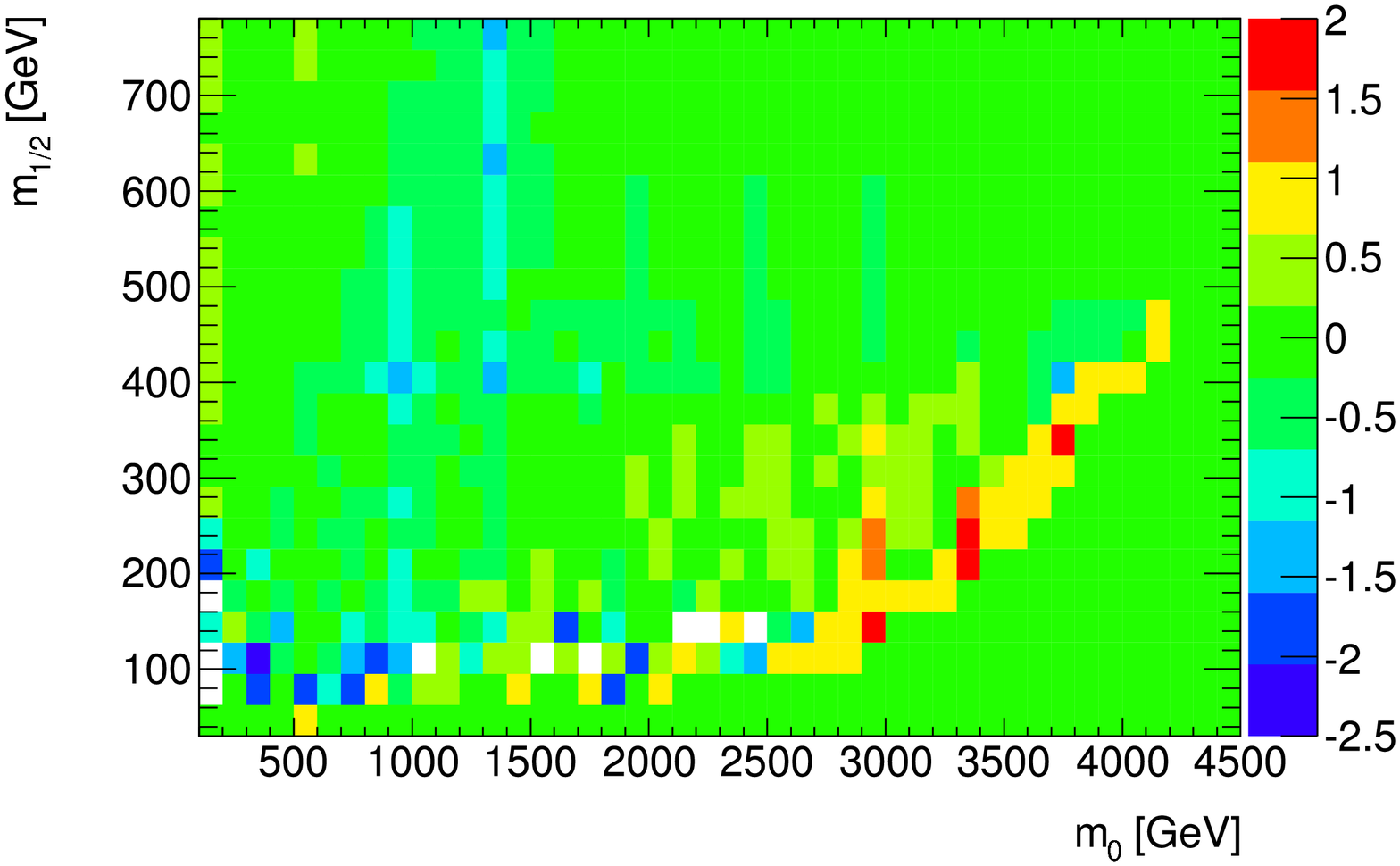}
	\caption{\effbruch\ of the ATLAS and simulation setups for signal region B \textit{tight} and C \textit{loose/medium/tight} of the 0-lepton analysis.}
	\label{fig:0lep-BC}
\end{center}
\end{figure}

\begin{figure}
\begin{center}
	\includegraphics[width=0.49\linewidth, trim=0.5cm 0cm 1.8cm 0cm, clip=true]{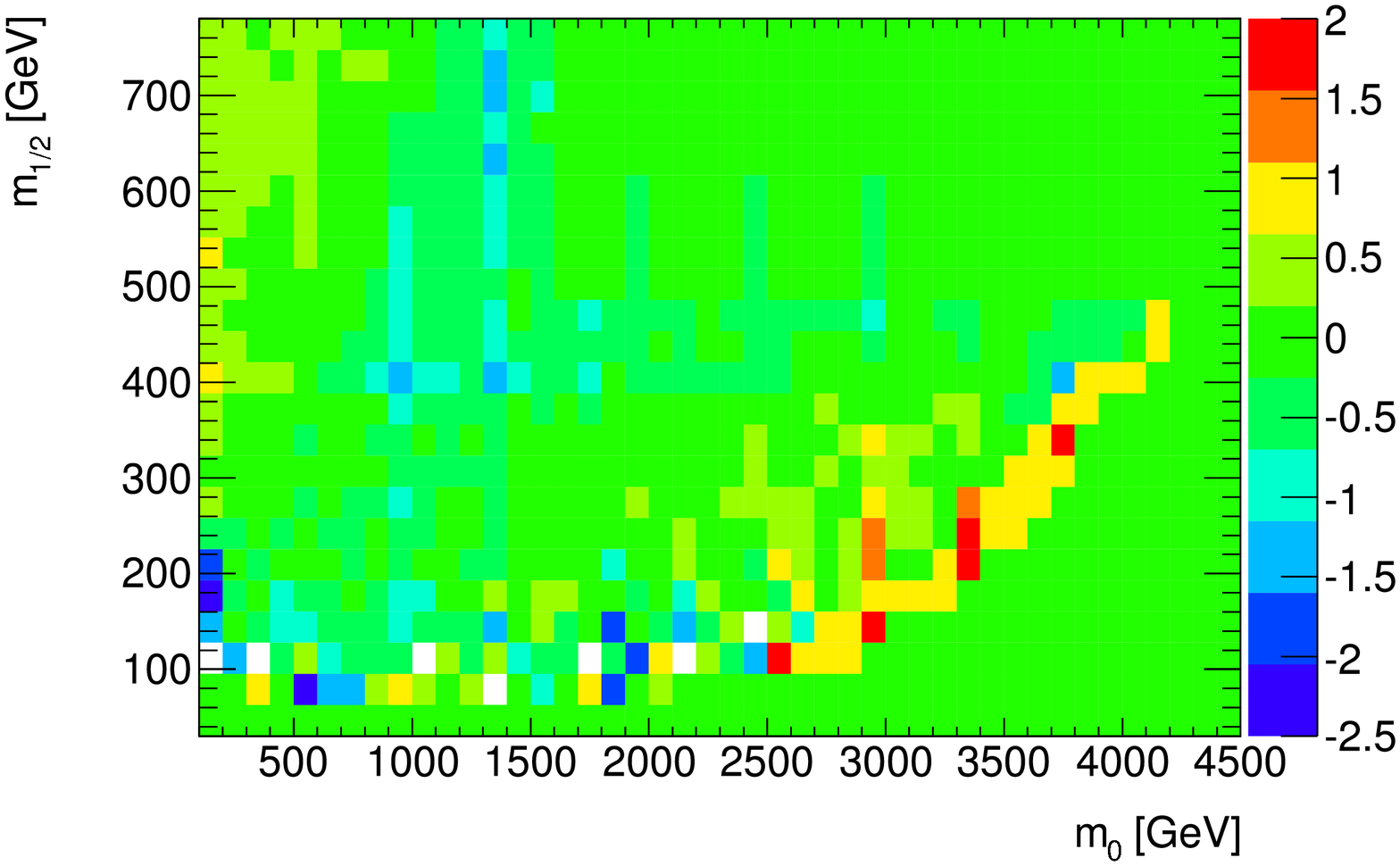}
	\includegraphics[width=0.49\linewidth, trim=0.5cm 0cm 1.8cm 0cm, clip=true]{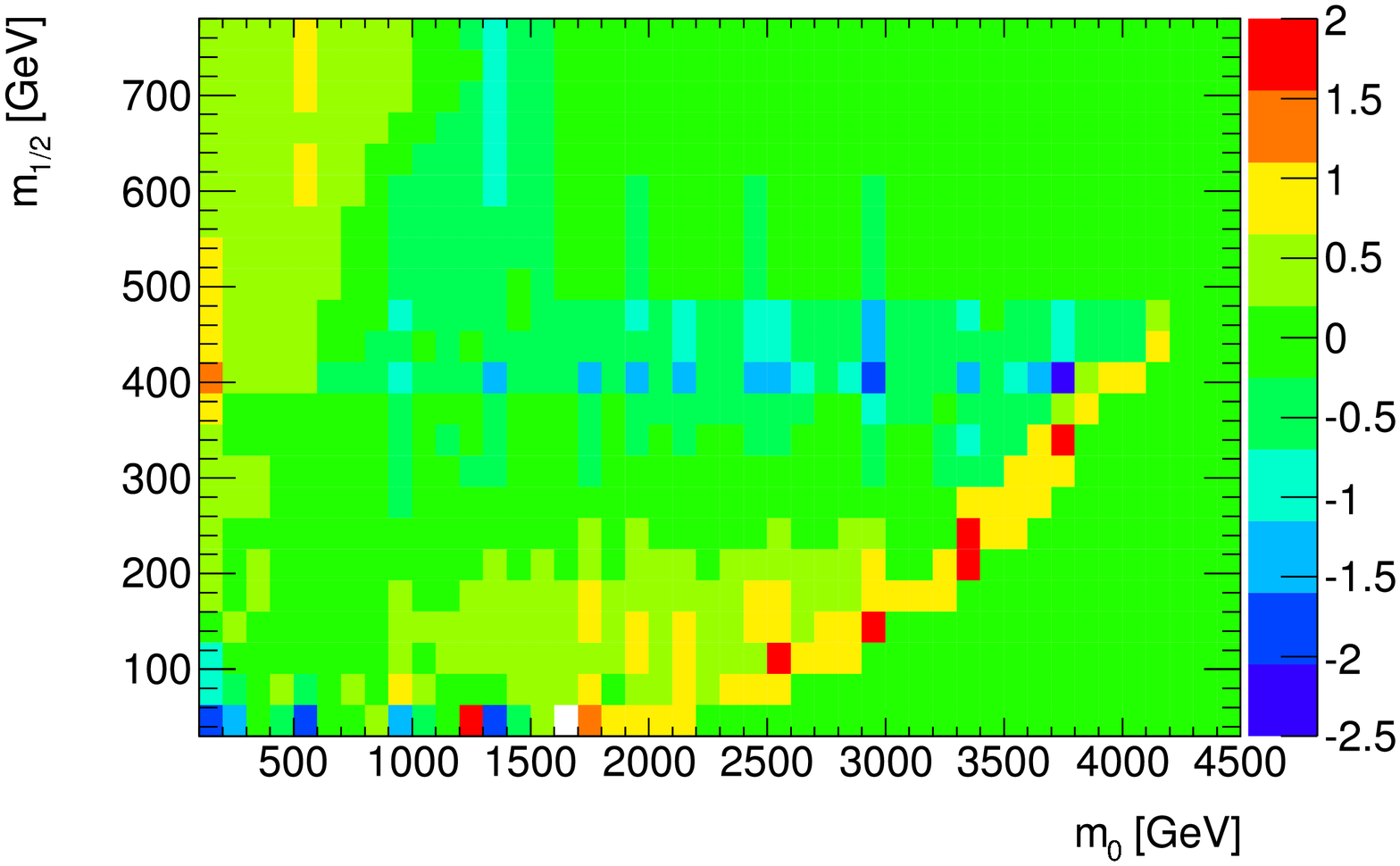}\\
	\includegraphics[width=0.49\linewidth, trim=0.5cm 0cm 1.8cm 0cm, clip=true]{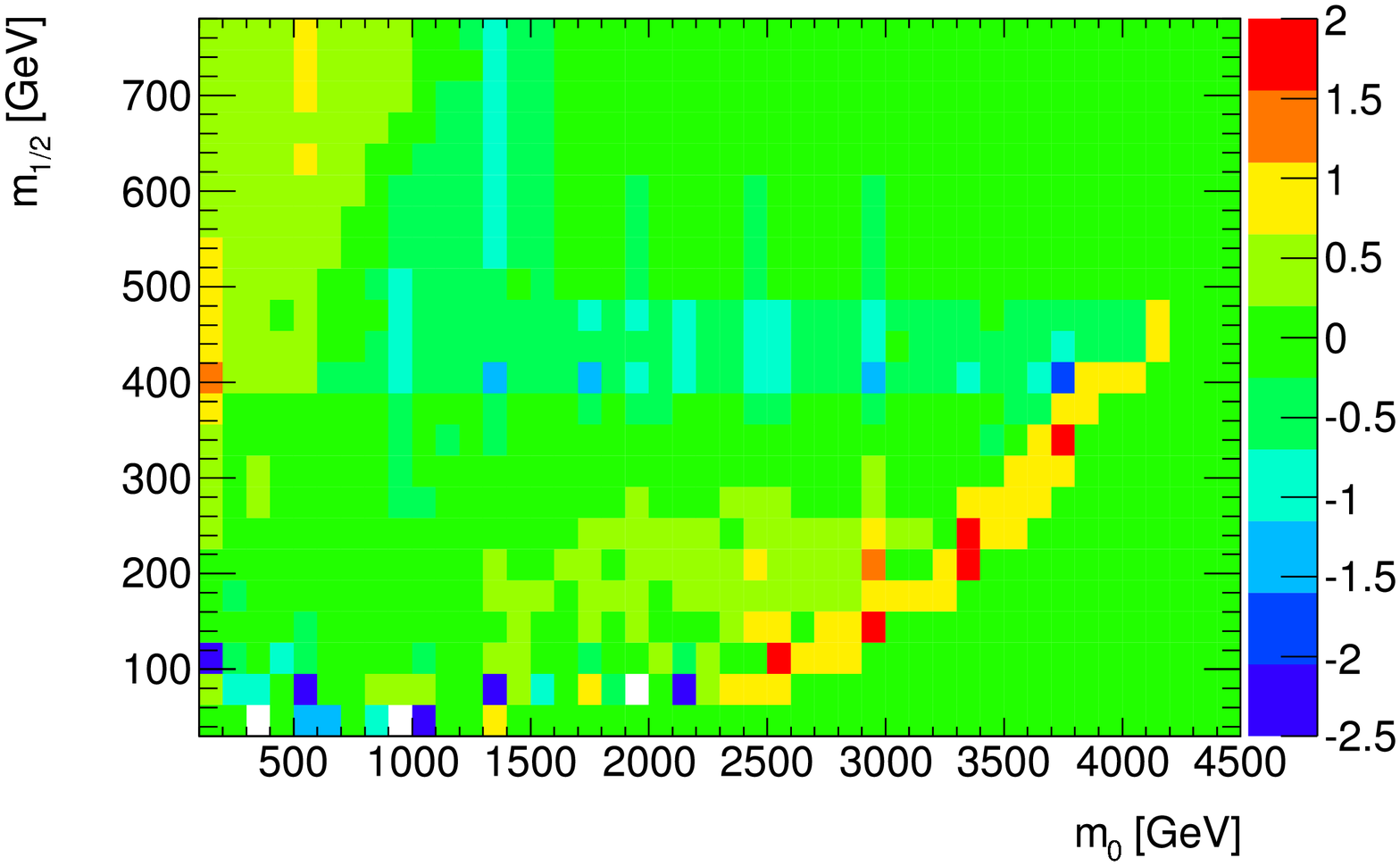}
	\includegraphics[width=0.49\linewidth, trim=0.5cm 0cm 1.8cm 0cm, clip=true]{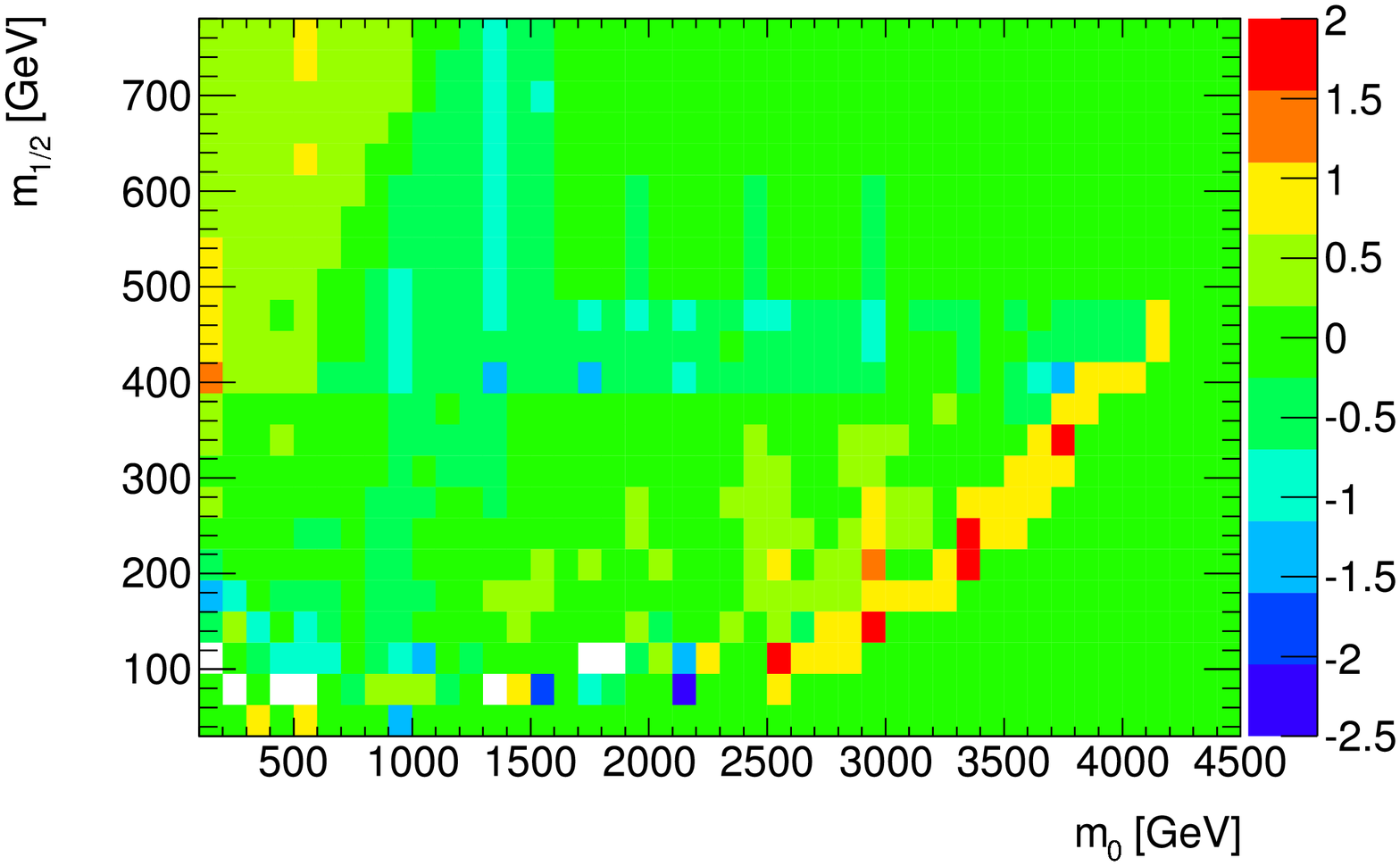}
	\caption{\effbruch\ of the ATLAS and simulation setups for signal region D and E \textit{loose/medium/tight} of the 0-lepton analysis.}
	\label{fig:0lep-DE}
\end{center}
\end{figure}

For the ATLAS 0-lepton analysis the validation was done in a cMSSM-grid with $\tan \beta = 10$, while $m_0$ runs from 100 GeV to 4180 GeV, $m_{1,2}$ from 60 GeV to 750 GeV.  The comparison was done for each signal region individually.  The results are shown in Figs.~\ref{fig:0lep-A}--\ref{fig:0lep-DE}. The minimum value was fixed for all plots due to some large deviations in regions with efficiencies close to zero, as indicated by the color scale. Grid points with values below the minimum are shown in white. For the large areas with value zero in the upper and lower right corner no data points were given by the ATLAS analyses.

\begin{figure}
\begin{center}
	\includegraphics[width=0.32\linewidth, trim=0.5cm 0cm 1.8cm 0cm, clip=true]{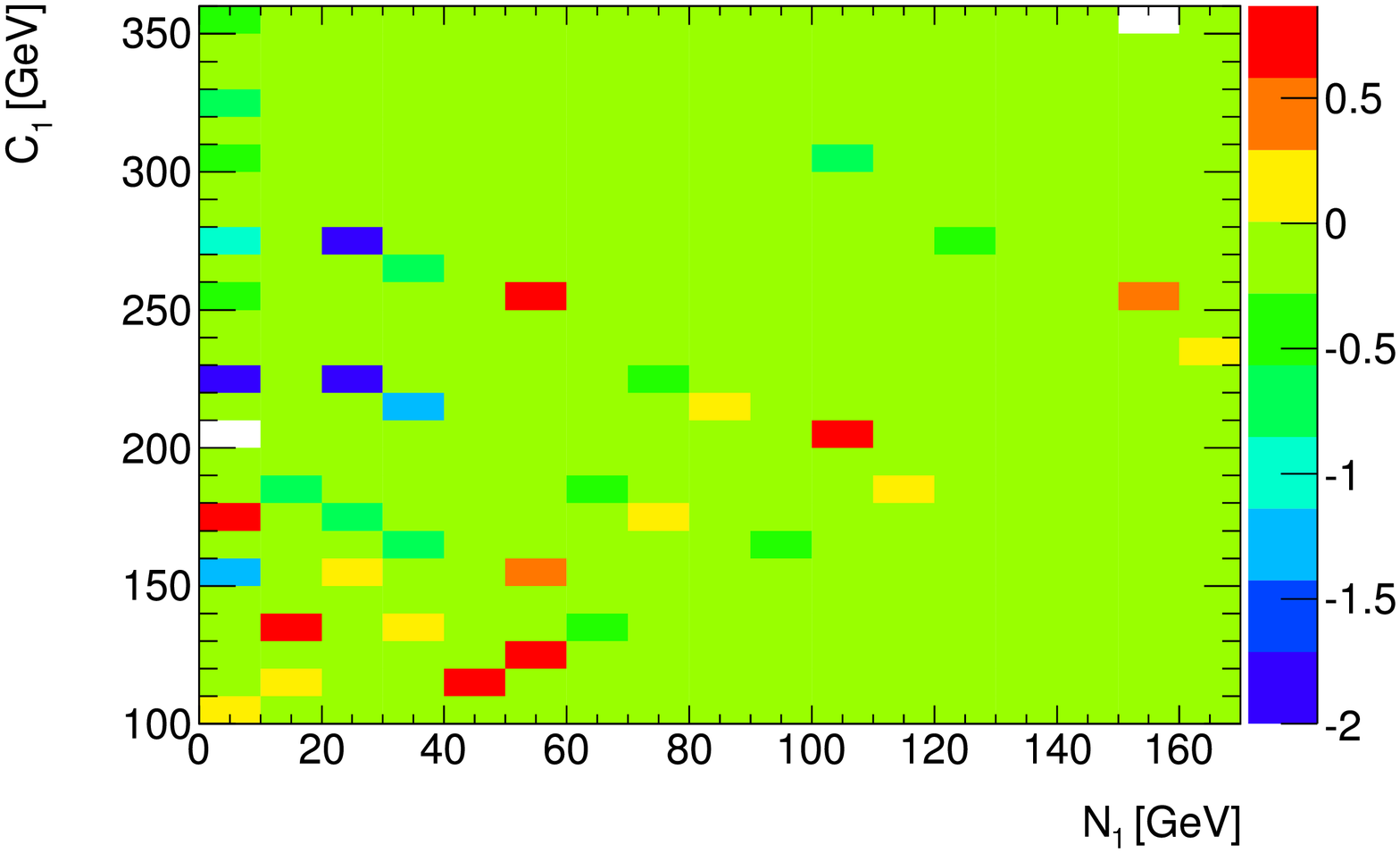}
	\includegraphics[width=0.32\linewidth, trim=0.5cm 0cm 1.8cm 0cm, clip=true]{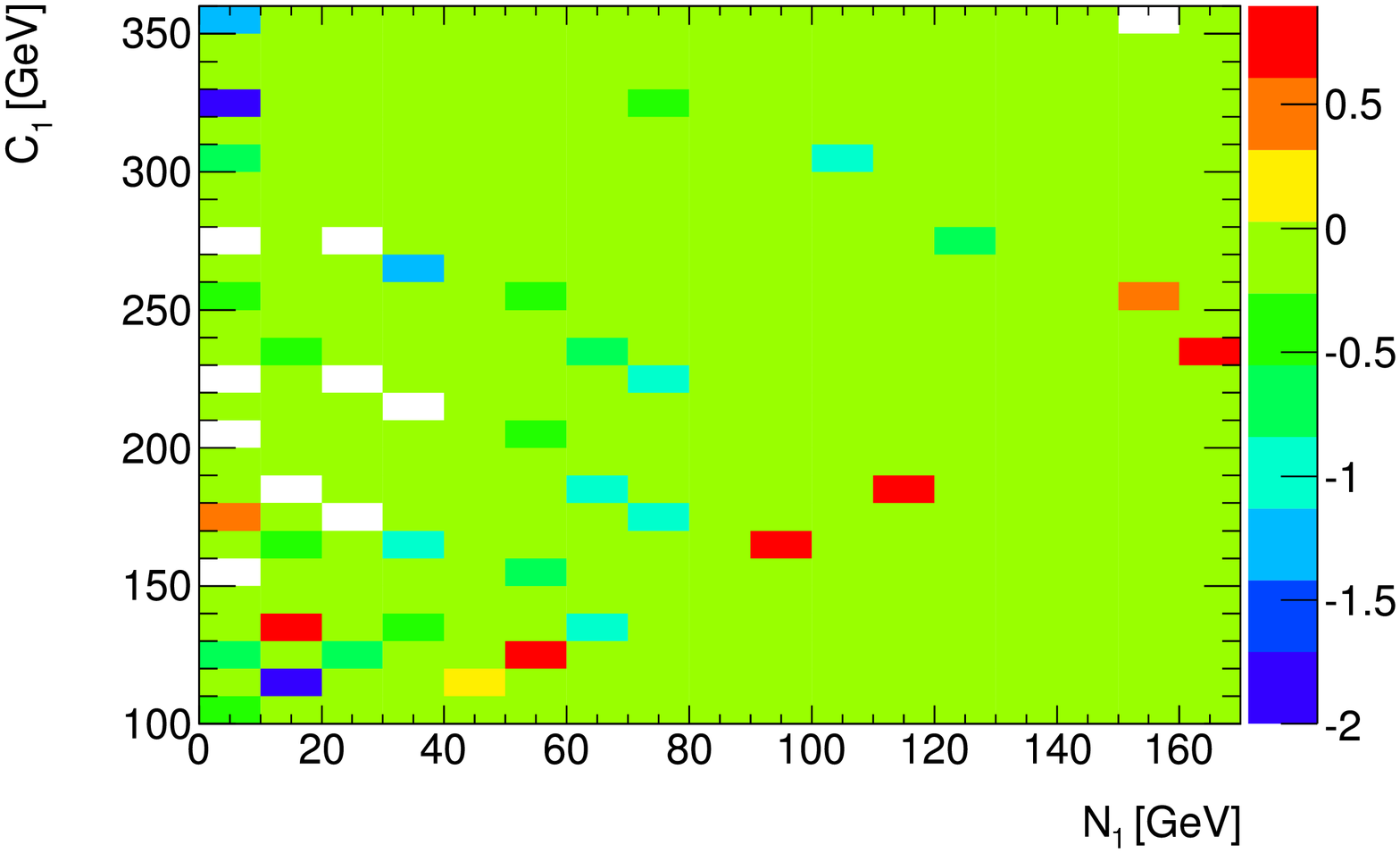}
	\includegraphics[width=0.32\linewidth, trim=0.5cm 0cm 1.8cm 0cm, clip=true]{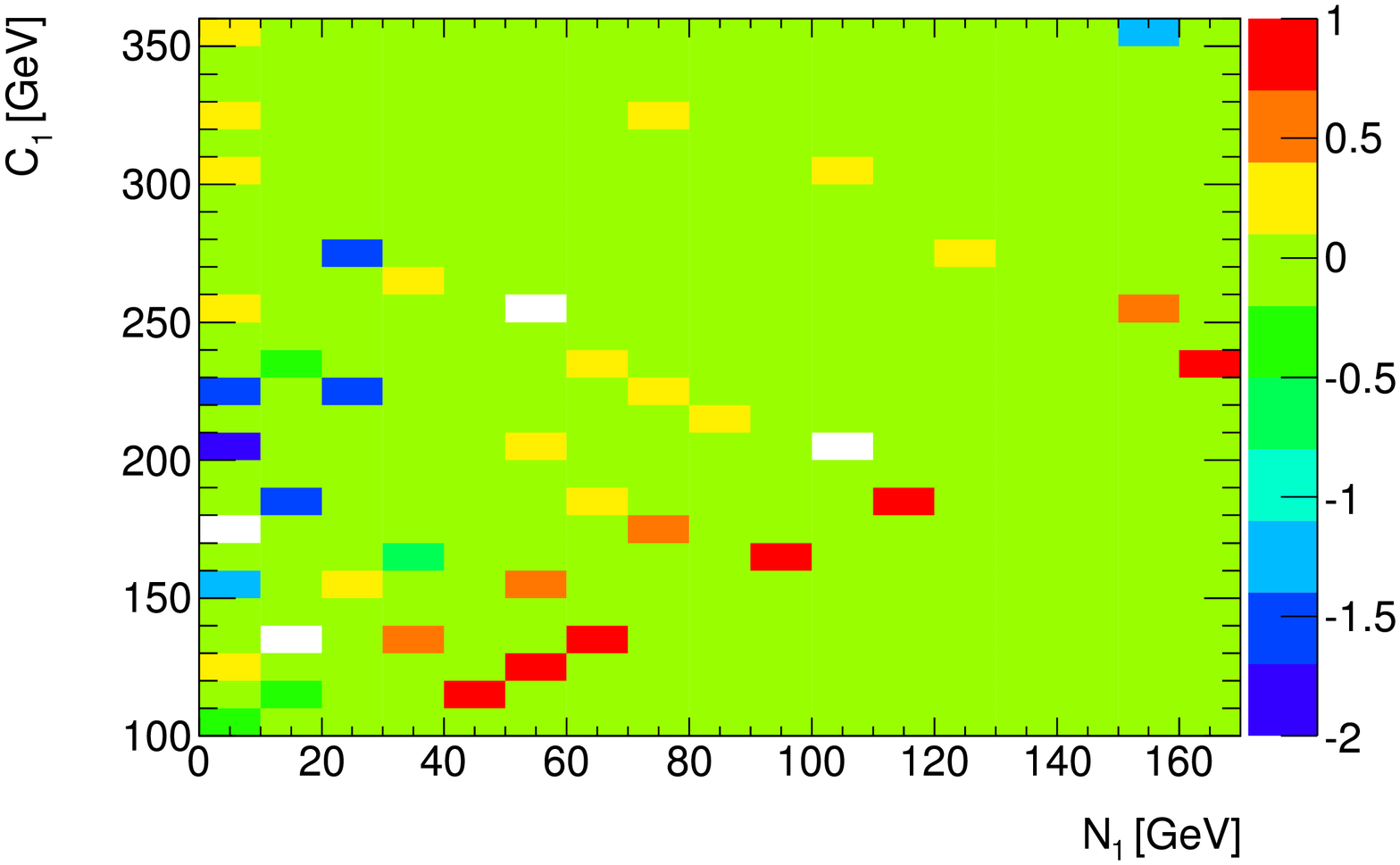}
	\caption{\effbruch\ of the ATLAS and simulation setups for signal region 1a, 1b and 2 of the 3-lepton analysis.}
	\label{fig:3lep}
\end{center}
\end{figure}

\begin{figure}
\begin{center}
	\includegraphics[width=0.32\linewidth, trim=0.5cm 0cm 1.8cm 0cm, clip=true]{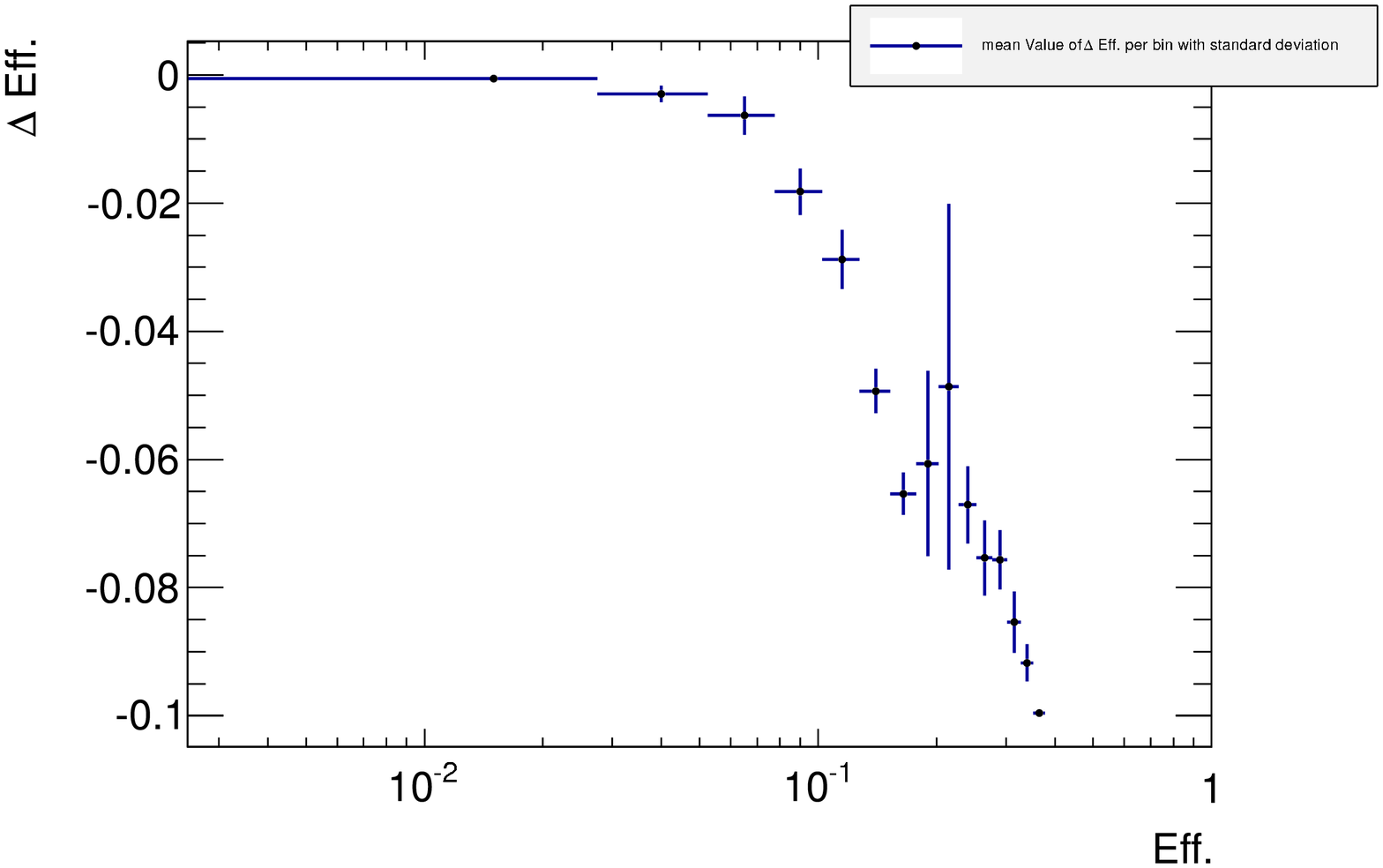}
	\includegraphics[width=0.32\linewidth, trim=0.5cm 0cm 1.8cm 0cm, clip=true]{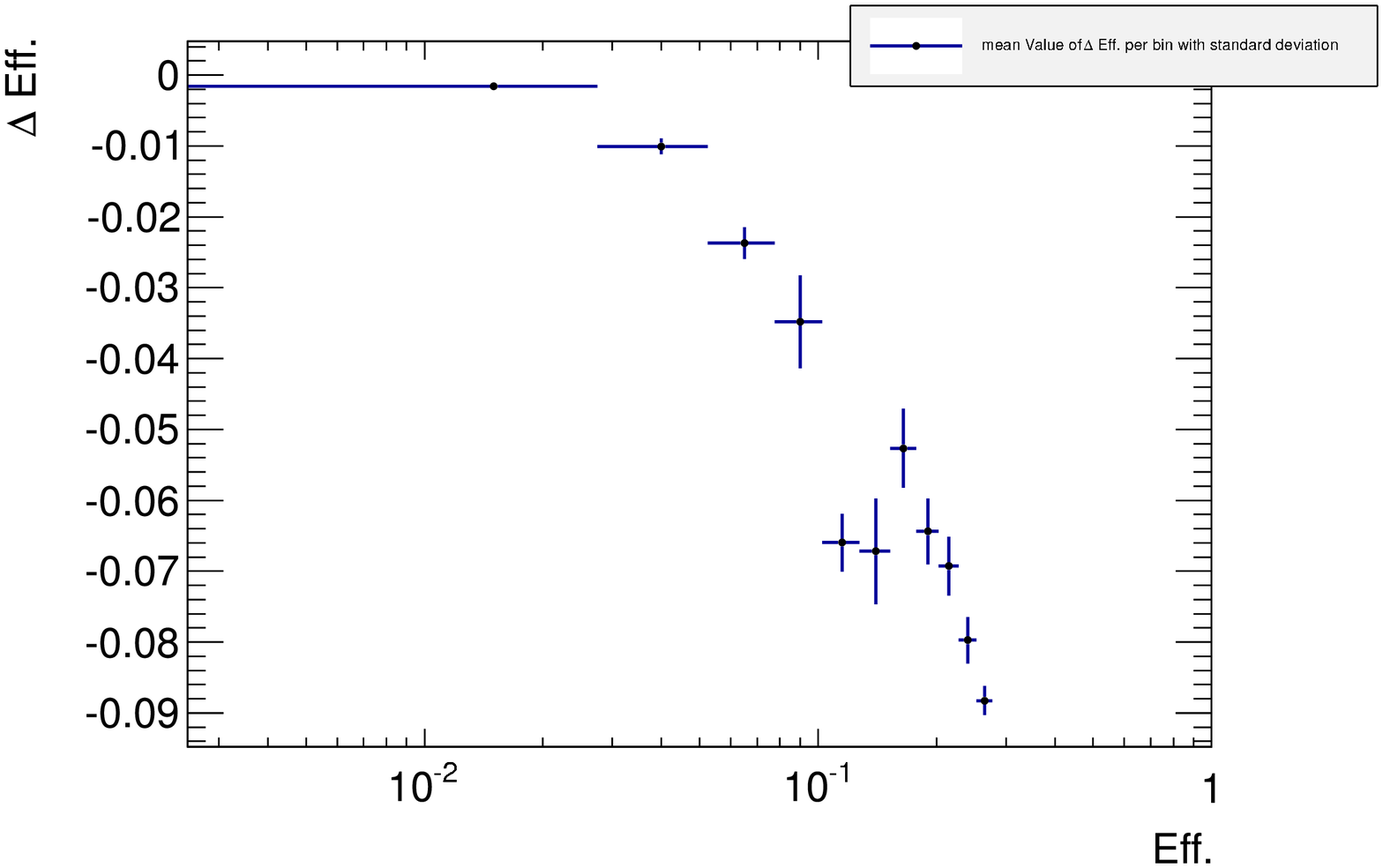}
	\includegraphics[width=0.32\linewidth, trim=0.5cm 0cm 1.8cm 0cm, clip=true]{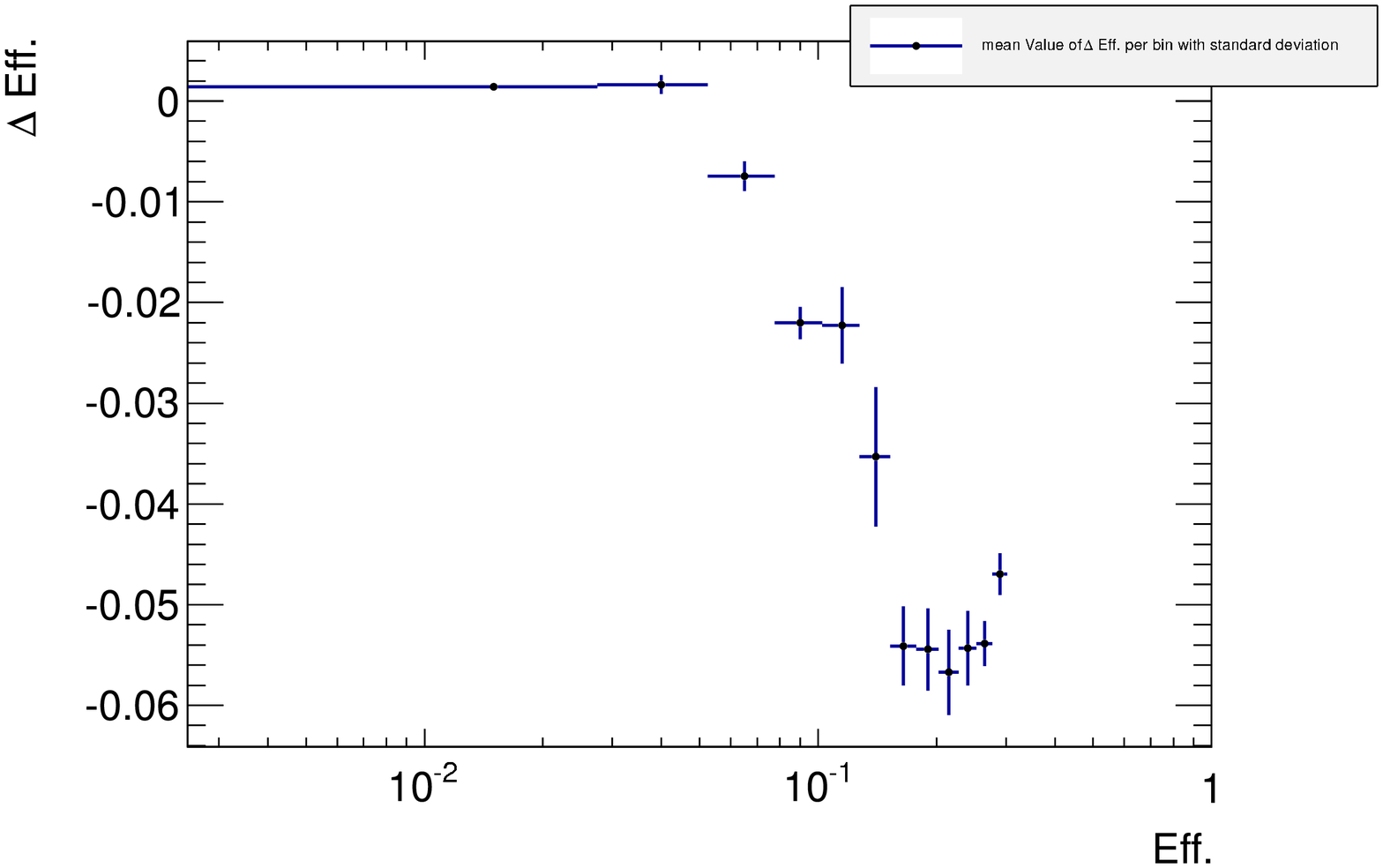}
	\caption{Mean efficiency value with estimated standard deviation for signal region A \textit{medium/loose} and A' of the 0-lepton analysis.}
	\label{fig:0lep-A-eff}
\end{center}
\end{figure}

\begin{figure}
\begin{center}
	\includegraphics[width=0.49\linewidth, trim=0.5cm 0cm 1.8cm 0cm, clip=true]{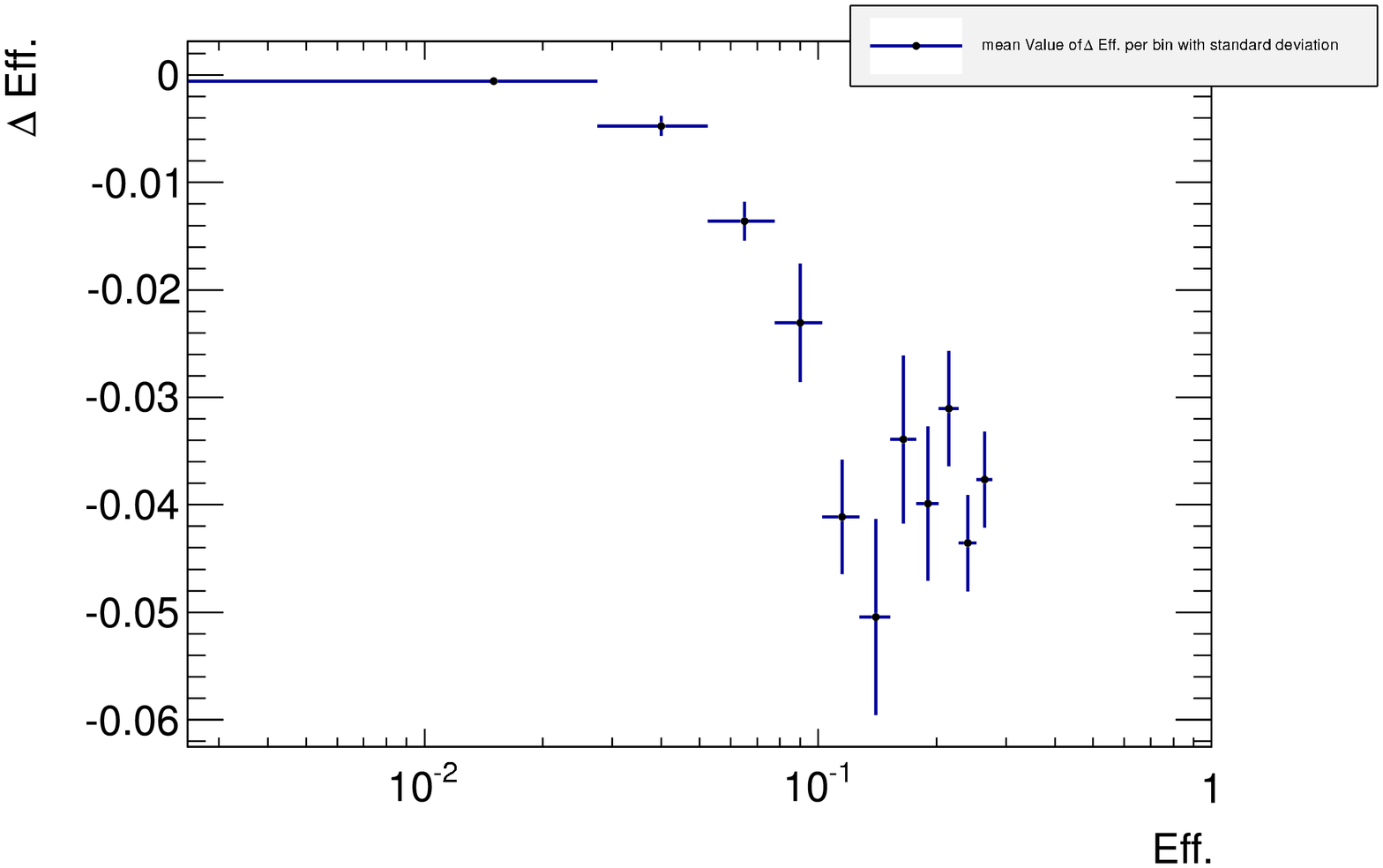}
	\includegraphics[width=0.49\linewidth, trim=0.5cm 0cm 1.8cm 0cm, clip=true]{figs/histEFF_meanEffProfile_regionBtight_Delphes310_offGridR04_0LEPtanb10.ps.pdf.ps}\\
	\includegraphics[width=0.49\linewidth, trim=0.5cm 0cm 1.8cm 0cm, clip=true]{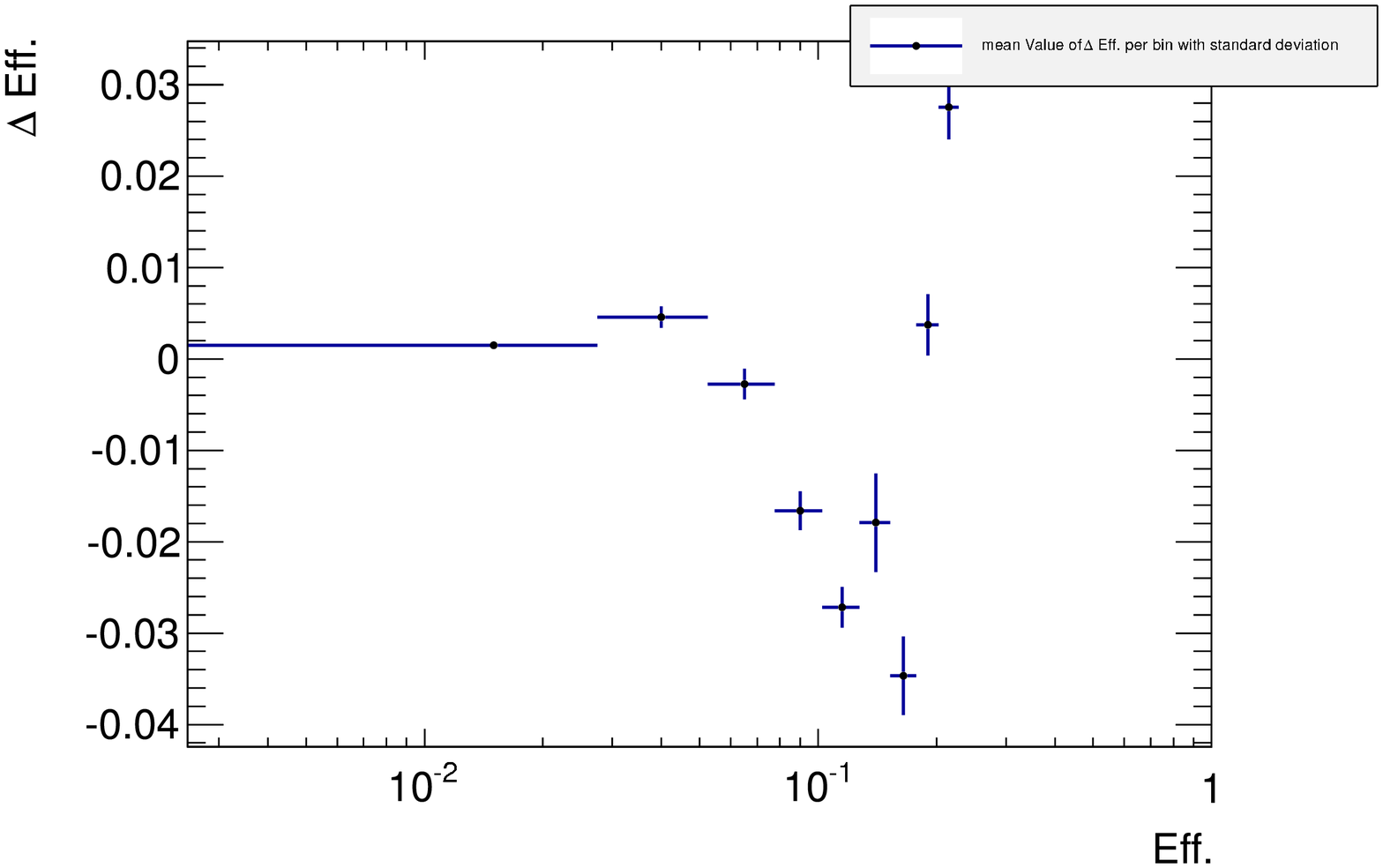}
	\includegraphics[width=0.49\linewidth, trim=0.5cm 0cm 1.8cm 0cm, clip=true]{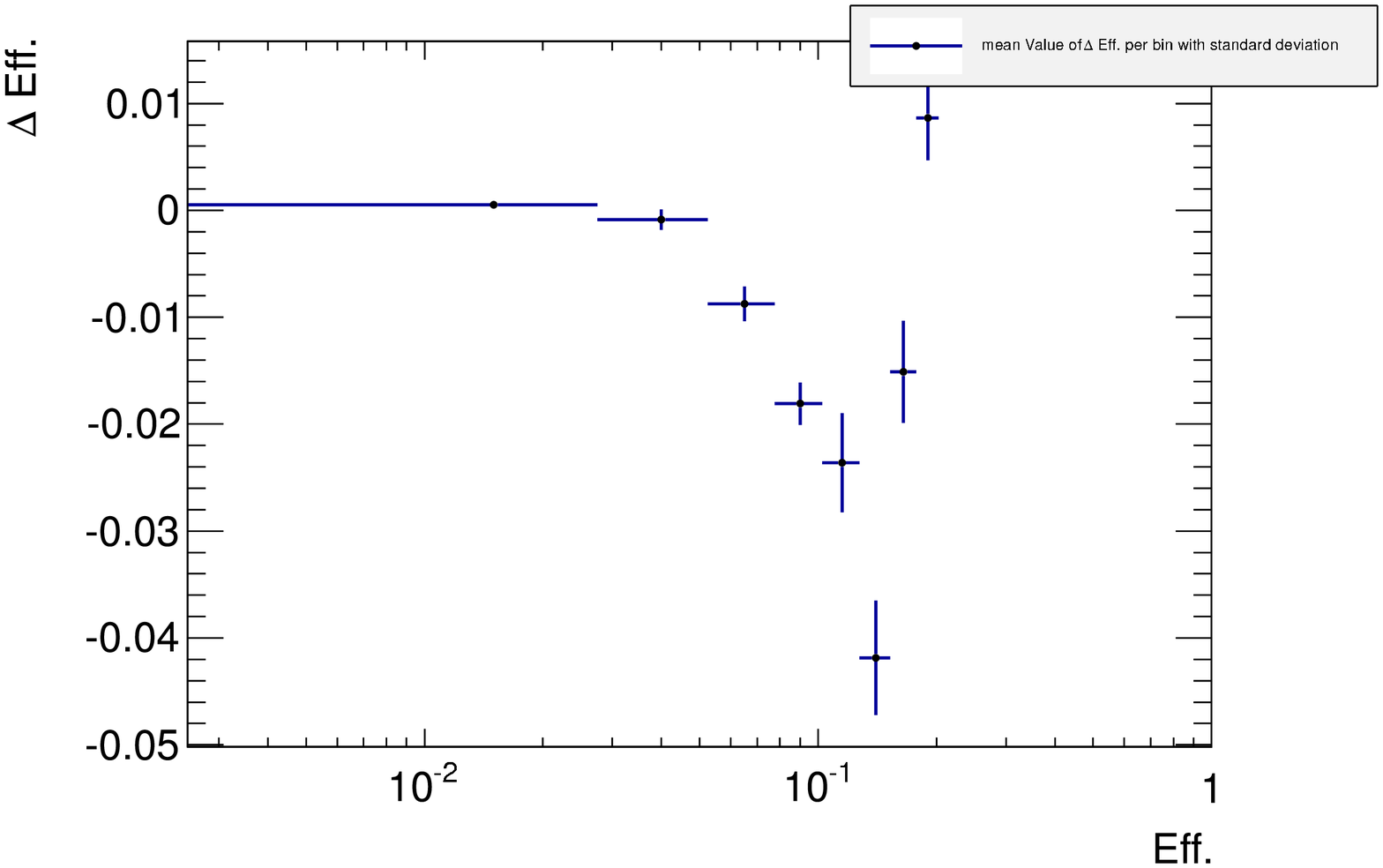}
	\caption{Mean efficiency value with standard deviation for signal region B \textit{tight} and C \textit{loose/medium/tight} of the 0-lepton analysis.}
	\label{fig:0lep-BC-eff}
\end{center}
\end{figure}

\begin{figure}
\begin{center}
	\includegraphics[width=0.49\linewidth, trim=0.5cm 0cm 1.8cm 0cm, clip=true]{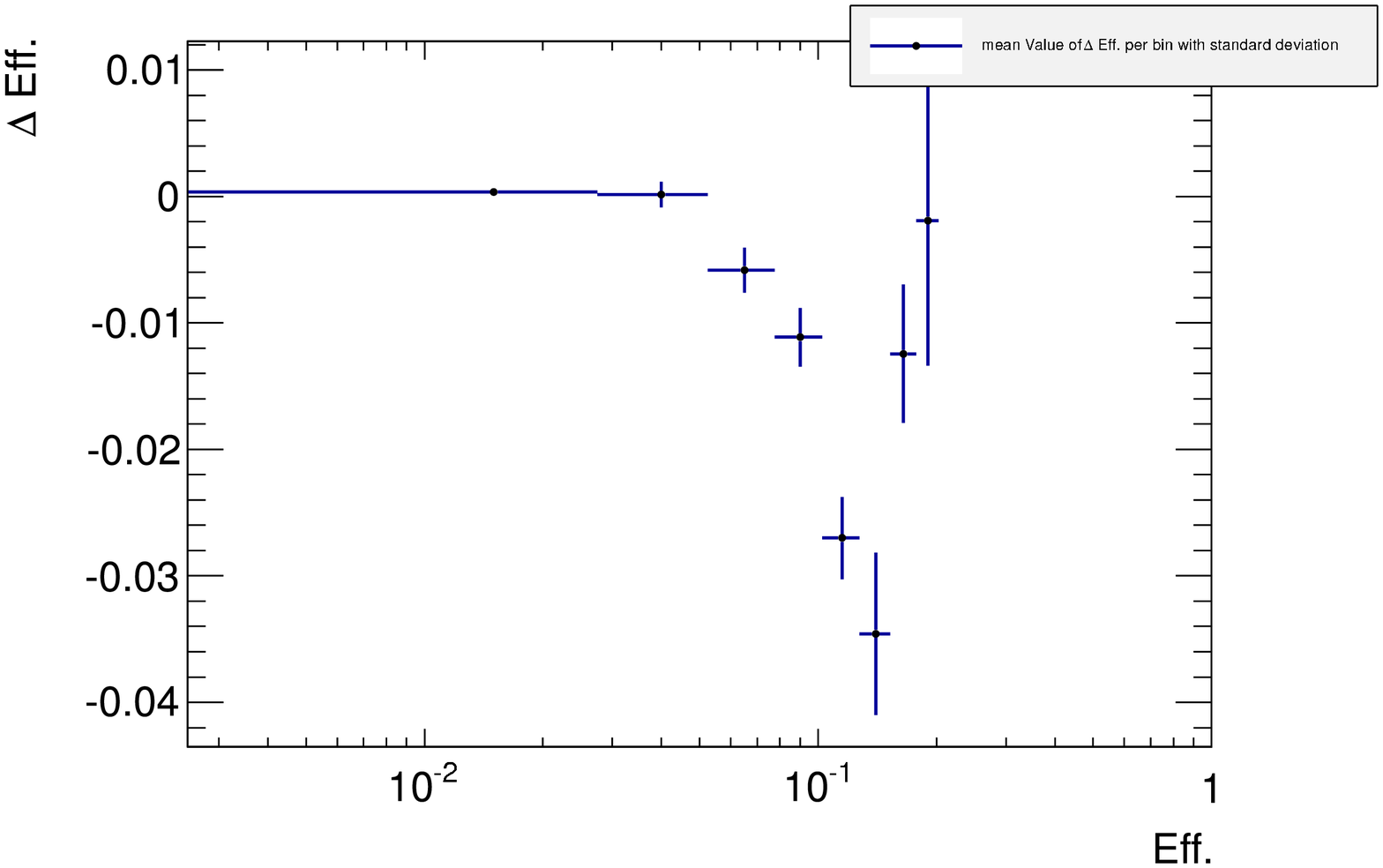}
	\includegraphics[width=0.49\linewidth, trim=0.5cm 0cm 1.8cm 0cm, clip=true]{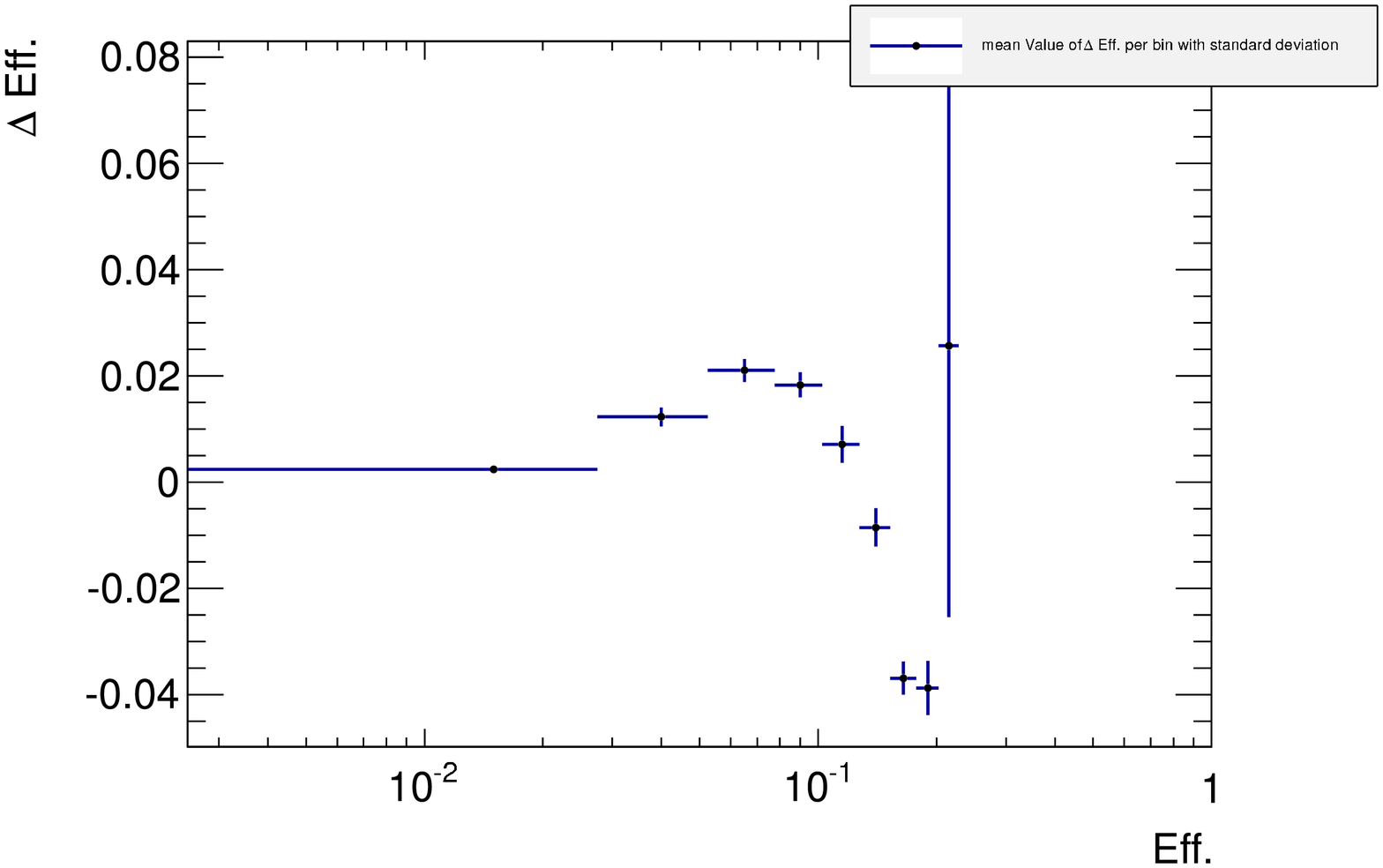} \\
	\includegraphics[width=0.49\linewidth, trim=0.5cm 0cm 1.8cm 0cm, clip=true]{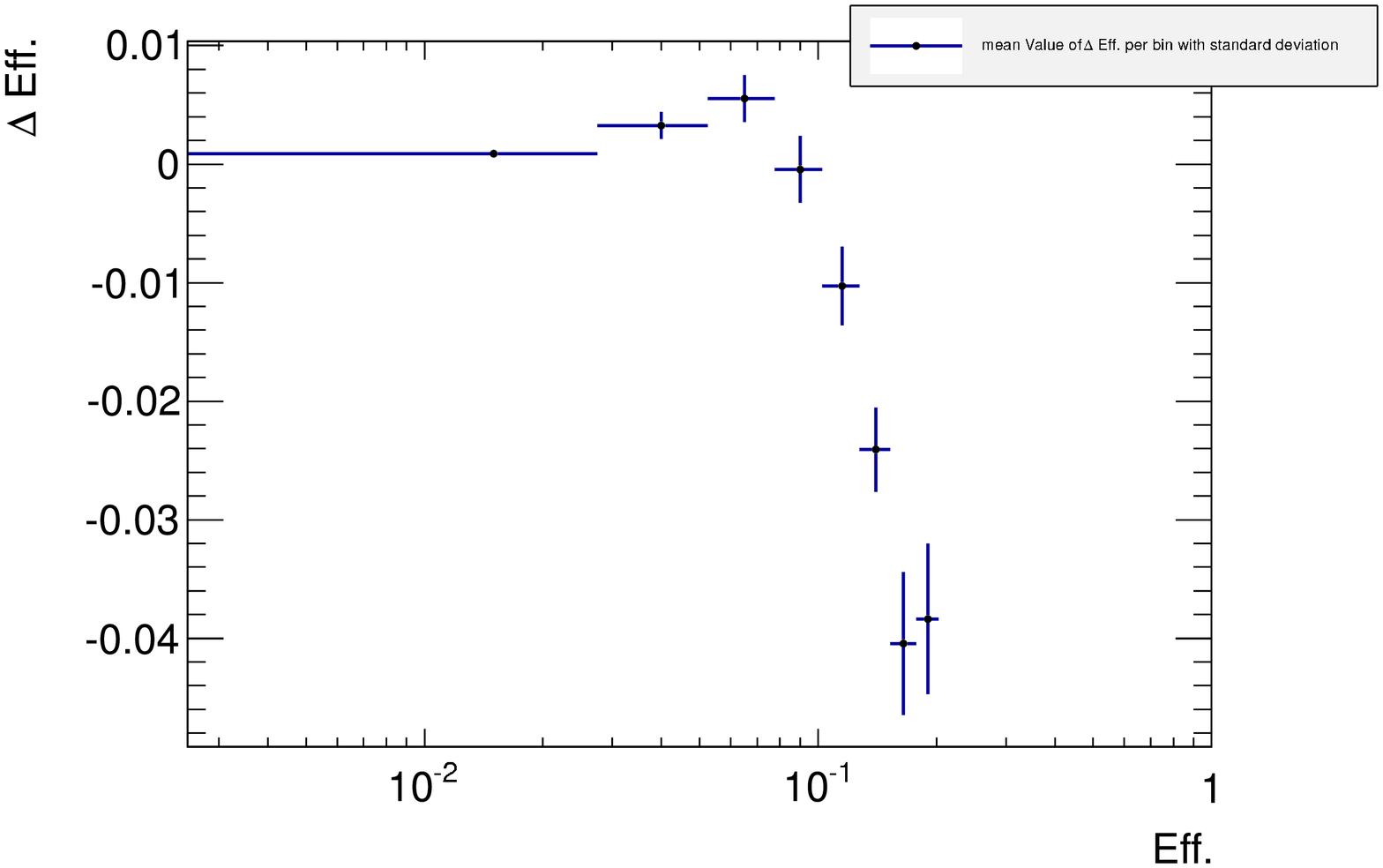}
	\caption{Mean efficiency value with standard deviation for signal region D and E \textit{loose/medium/tight} of the 0-lepton analysis.}
	\label{fig:0lep-DE-eff}
\end{center}
\end{figure}

\begin{figure}
\begin{center}
	\includegraphics[width=0.32\linewidth, trim=0.5cm 0cm 1.8cm 0cm, clip=true]{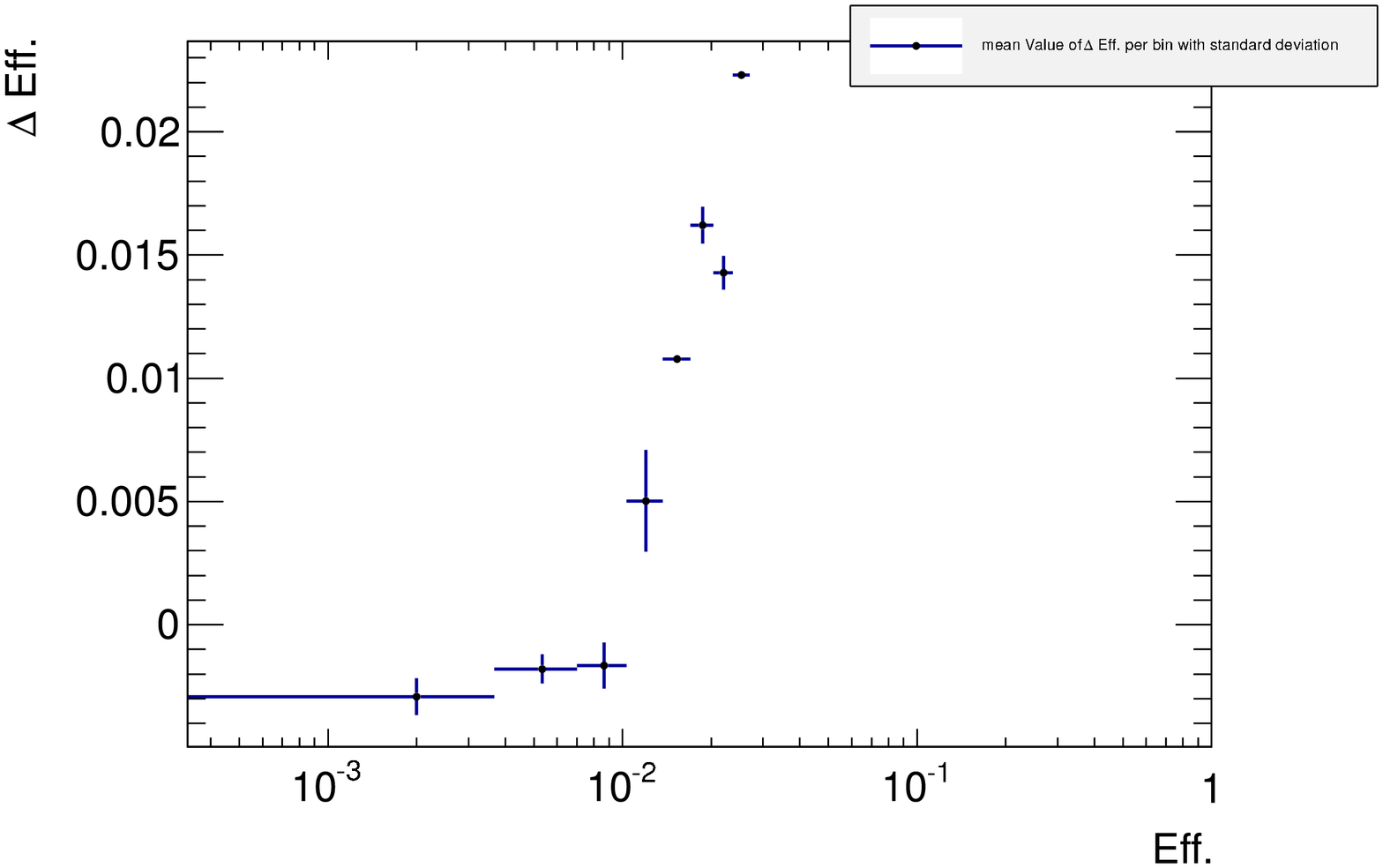}
	\includegraphics[width=0.32\linewidth, trim=0.5cm 0cm 1.8cm 0cm, clip=true]{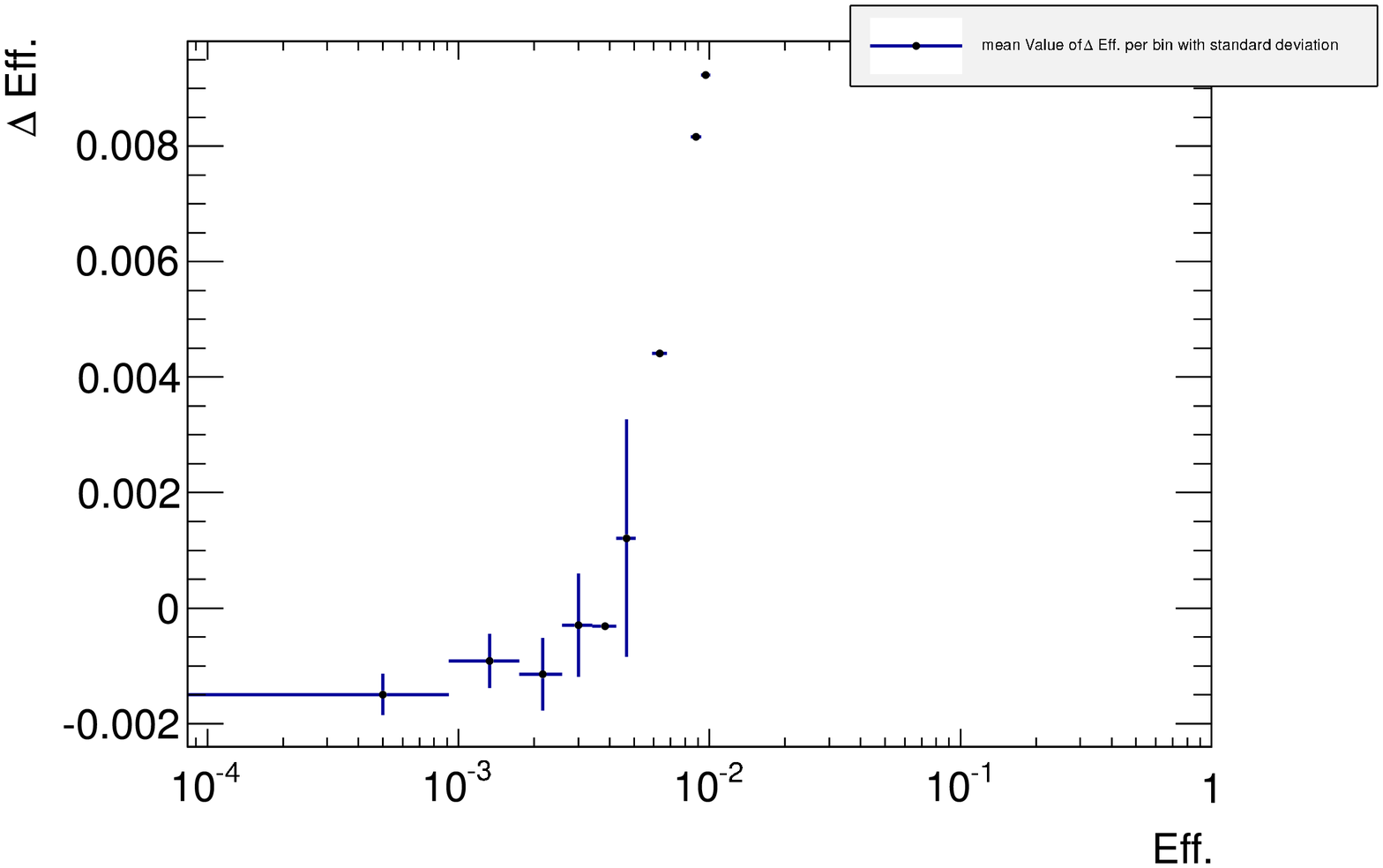}
	\includegraphics[width=0.32\linewidth, trim=0.5cm 0cm 1.8cm 0cm, clip=true]{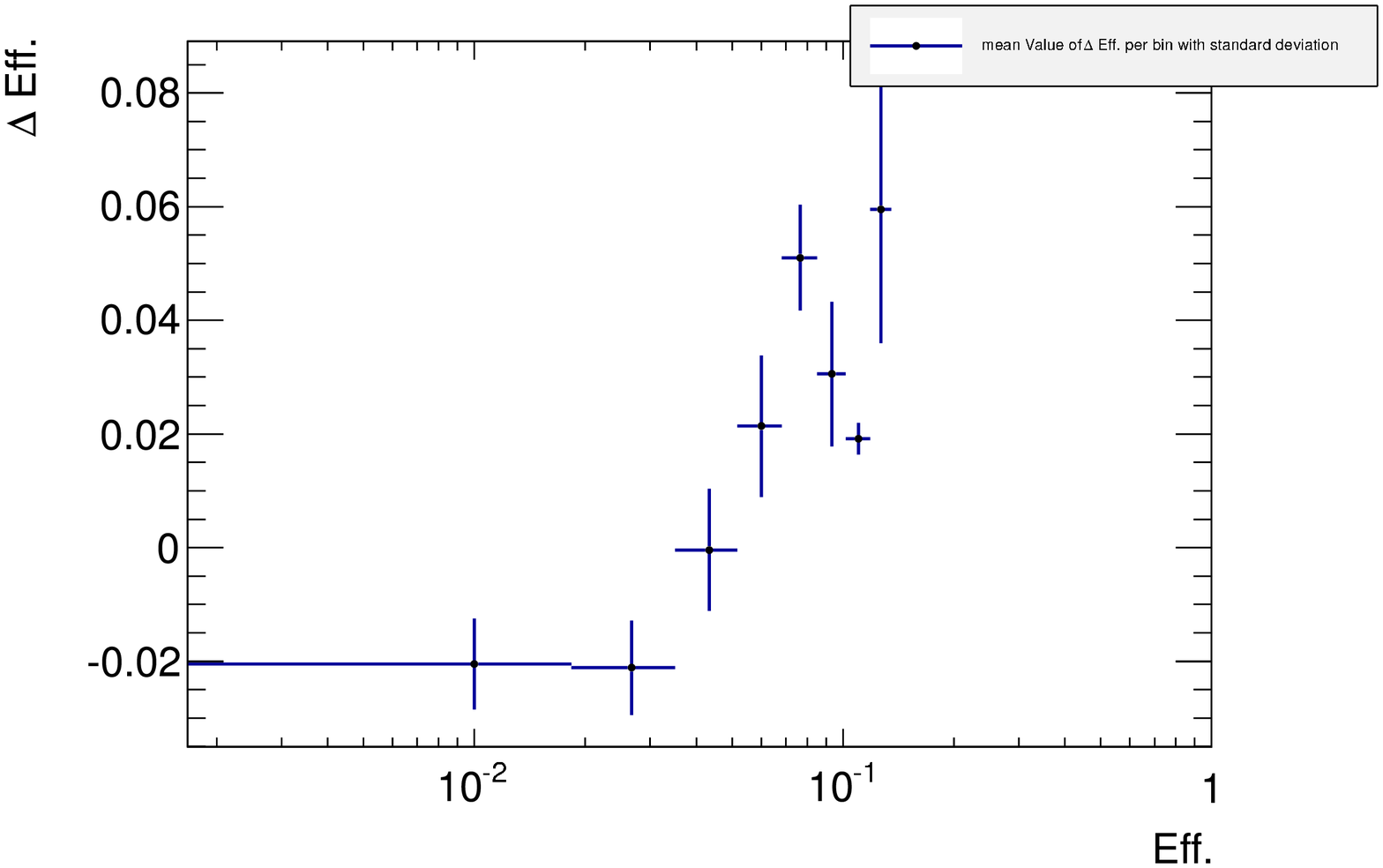}
	\caption{Mean efficiency value with standard deviation for signal region 1a, 1b and 2 of the 3-lepton analysis.}
	\label{fig:3lep-eff}
\end{center}
\end{figure}

For the 3-lepton analysis the validation was done for a simplified model where only the masses of the neutralinos, charginos and sleptons are free parameters and the $\tilde{\chi}_1^{\pm} $ and $\tilde{\chi}_2^{0} $ decay to W and Z bosons. The employed grid has values of 70 GeV to 350 GeV for $m_{\tilde{\chi}_1^{\pm}}$ and 0 GeV to 200 GeV for $m_{\tilde{\chi}_1^{0}}$.  The results are presented in Figure~\ref{fig:3lep}.
As error estimation the mean value of $\frac{\Delta \varepsilon}{\varepsilon} $ and its standard deviation were computed for each efficiency bin. The results are presented in Figs.~\ref{fig:0lep-A-eff}--\ref{fig:3lep-eff}.  Those values have then been used as estimates for \effbruch in the likelihood, Eq.~\eqref{eq:like_syst_err} assuming that the \effbruch can be parameterized as a function of the efficiency. 

%%%%%%%%%%%%%%%%%%%%%%%%%%%%%%%%%%%%%%%%%%%%%%%%%%%%%%%%%%%%%%%%%
\subsection{Comparison with official ATLAS result for the cMSSM case}
%%%%%%%%%%%%%%%%%%%%%%%%%%%%%%%%%%%%%%%%%%%%%%%%%%%%%%%%%%%%%%%%

For completeness, we also validated our likelihood construction in the cMSSM framework and compared it with the results of the ATLAS Collaboration.

\begin{figure}[tbh]
\begin{center}
\includegraphics[width=0.49\linewidth]{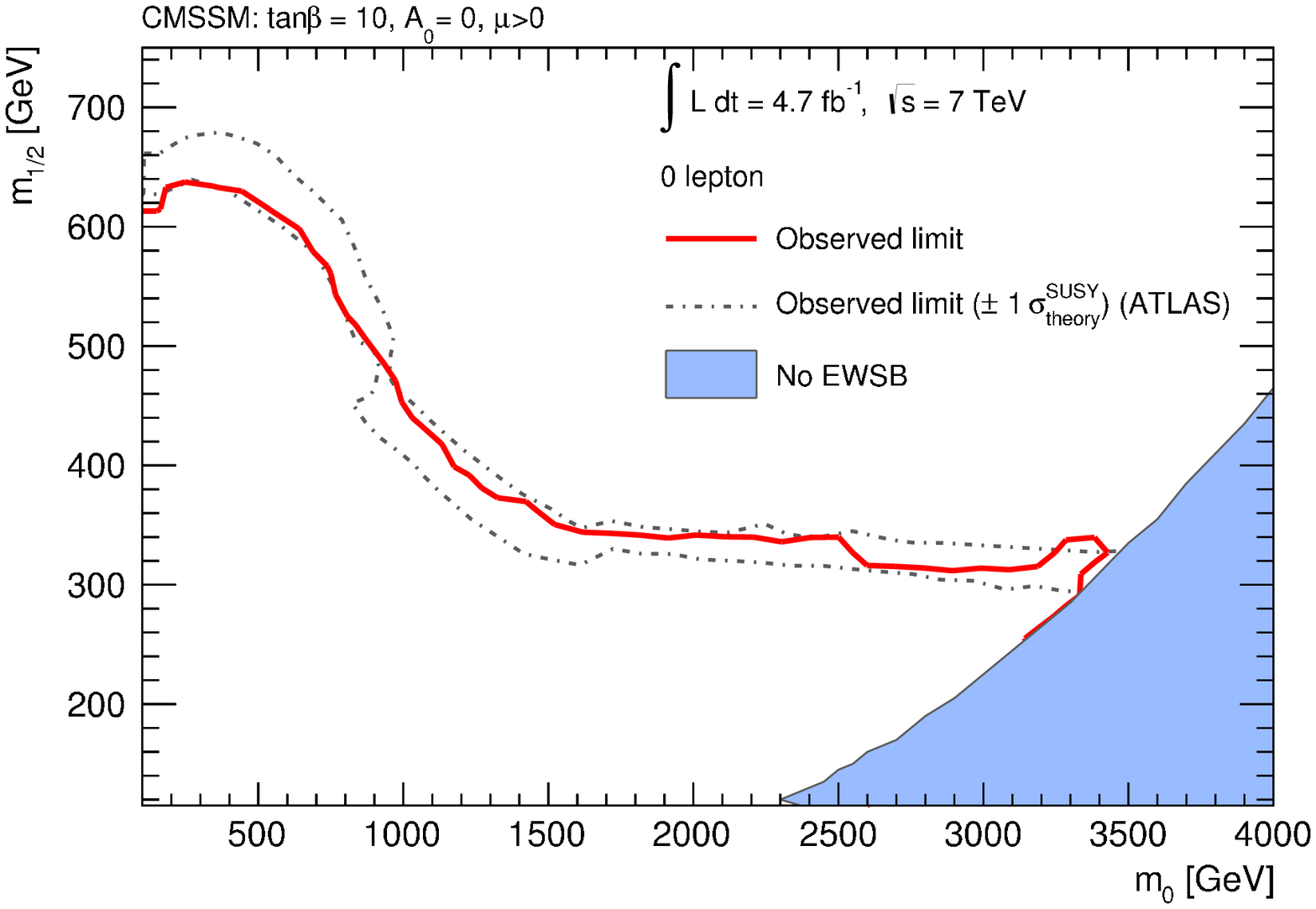}
\includegraphics[width=0.49\linewidth]{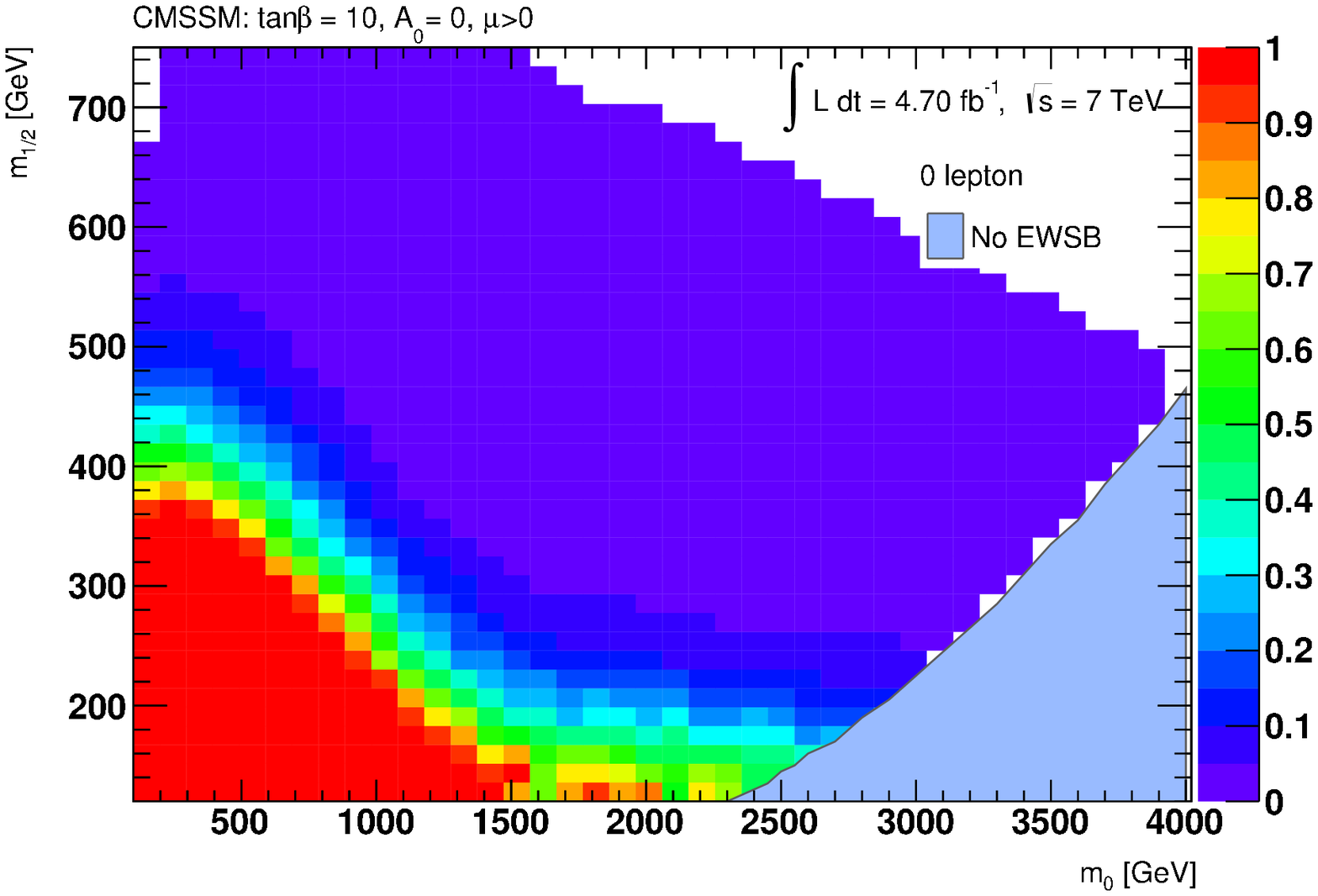}
 \caption{In the left panel, we show the 95\% C.L. observed exclusion limit for the 0-lepton analysis in the cMSSM from our likelihood construction (red line) and the C.L.$_{s}$ method, which is in remarkably good agreement with the ATLAS result~\cite{ATLAS-CONF-2012-033-5fb}. The band limited by the gray dash-dotted lines is the exclusion limit by the ATLAS collaboration, accounting for uncertainties in the SUSY production cross-section.
On the right, we show the normalized full log-likelihood.}
\label{fig:cmssm}
\end{center}
\end{figure}

We computed the C.L.$_{s}$ to estimate the observed exclusion limits using the prescription outlined in Ref.~\cite{Cowan:2010js} which uses the concept of Asimov data and Wilk's theorem for its efficient evaluation. The results are shown in the left-panel of Fig. \ref{fig:cmssm}. The continuous-red line represents our estimated exclusion limits at 95\% \cl, whereas the region between the dash-dotted gray lines gives the ATLAS Collaboration exclusion limit, accounting for uncertainties in the determination of the SUSY production cross section. The agreement is very satisfactory, indicating that both the signal prediction procedure and the likelihood construction we adopted work remarkably well. 

In the right-panel of Fig. \ref{fig:cmssm} we display the shape of the full-log likelihood function in our setup. 

\begin{thebibliography}{199}

%\cite{Cabrera:2012vu}
\bibitem{Cabrera:2012vu}
  M.~E.~Cabrera, J.~A.~Casas and R.~R.~de Austri,
  %``The health of SUSY after the Higgs discovery and the XENON100 data,''
  JHEP {\bf 1307} (2013) 182
  [arXiv:1212.4821 [hep-ph]].
  %%CITATION = ARXIV:1212.4821;%%
  %20 citations counted in INSPIRE as of 19 Apr 2014

%\cite{Buchmueller:2013rsa}
\bibitem{Buchmueller:2013rsa}
  O.~Buchmueller, R.~Cavanaugh, A.~De Roeck, M.~J.~Dolan, J.~R.~Ellis, H.~Flacher, S.~Heinemeyer 
  and G.~Isidori {\it et al.},
  %``The CMSSM and NUHM1 after LHC Run 1,''
  arXiv:1312.5250 [hep-ph].
  %%CITATION = ARXIV:1312.5250;%%
  %11 citations counted in INSPIRE as of 19 Apr 2014

%\cite{Chakrabortty:2013voa}
\bibitem{Chakrabortty:2013voa}
  J.~Chakrabortty, S.~Mohanty and S.~Rao,
  %``Non-universal gaugino mass GUT models in the light of dark matter and LHC constraints,''
  JHEP {\bf 1402} (2014) 074
  [arXiv:1310.3620 [hep-ph]].
  %%CITATION = ARXIV:1310.3620;%%
  %2 citations counted in INSPIRE as of 19 Apr 2014

%\cite{Cabrera:2013jya}
\bibitem{Cabrera:2013jya}
  M.~E.~Cabrera, A.~Casas, R.~R.~de Austri and G.~Bertone,
  %``LHC and dark matter phenomenology of the NUGHM,''
  arXiv:1311.7152 [hep-ph].
  %%CITATION = ARXIV:1311.7152;%%
  %1 citations counted in INSPIRE as of 19 Apr 2014

%\cite{Djouadi:1998di}
\bibitem{Djouadi:1998di} 
  A.~Djouadi {\it et al.}  [MSSM Working Group Collaboration],
  %``The Minimal supersymmetric standard model: Group summary report,''
  hep-ph/9901246.
  %%CITATION = HEP-PH/9901246;%%
  %145 citations counted in INSPIRE as of 01 May 2014

%\cite{Berger:2008cq}
\bibitem{Berger:2008cq}
  C.~F.~Berger, J.~S.~Gainer, J.~L.~Hewett and T.~G.~Rizzo,
  %``Supersymmetry Without Prejudice,''
  JHEP {\bf 0902} (2009) 023
  [arXiv:0812.0980 [hep-ph]].
  %%CITATION = ARXIV:0812.0980;%%
  %126 citations counted in INSPIRE as of 17 Apr 2014

%\cite{Arbey:2011un}
\bibitem{Arbey:2011un}
  A.~Arbey, M.~Battaglia and F.~Mahmoudi,
  %``Implications of LHC Searches on SUSY Particle Spectra: The pMSSM Parameter Space with Neutralino Dark Matter,''
  Eur.\ Phys.\ J.\ C {\bf 72} (2012) 1847
  [arXiv:1110.3726 [hep-ph]].
  %%CITATION = ARXIV:1110.3726;%%
  %59 citations counted in INSPIRE as of 17 Apr 2014
  
\bibitem{Cahill-Rowley:2013dpa} 
  M.~Cahill-Rowley, R.~Cotta, A.~Drlica-Wagner, S.~Funk, J.~Hewett, A.~Ismail, T.~Rizzo and M.~Wood,
  %``Complementarity and Searches for Dark Matter in the pMSSM,''
  arXiv:1305.6921 [hep-ph].
  %%CITATION = ARXIV:1305.6921;%%
  %18 citations counted in INSPIRE as of 01 May 2014

%\cite{Baltz:2006fm}
\bibitem{Baltz:2006fm}
  E.~A.~Baltz, M.~Battaglia, M.~E.~Peskin and T.~Wizansky,
  %``Determination of dark matter properties at high-energy colliders,''
  Phys.\ Rev.\ D {\bf 74} (2006) 103521
  [hep-ph/0602187].
  %%CITATION = HEP-PH/0602187;%%
  %229 citations counted in INSPIRE as of 17 Apr 2014

%\cite{AbdusSalam:2009qd}
\bibitem{AbdusSalam:2009qd}
  S.~S.~AbdusSalam, B.~C.~Allanach, F.~Quevedo, F.~Feroz and M.~Hobson,
  %``Fitting the Phenomenological MSSM,''
  Phys.\ Rev.\ D {\bf 81} (2010) 095012
  [arXiv:0904.2548 [hep-ph]].
  %%CITATION = ARXIV:0904.2548;%%
  %81 citations counted in INSPIRE as of 17 Apr 2014

%\cite{AbdusSalam:2012ir}
\bibitem{AbdusSalam:2012ir}
  S.~S.~AbdusSalam,
  %``LHC-7 supersymmetry search interpretation within the phenomenological MSSM,''
  Phys.\ Rev.\ D {\bf 87} (2013) 11,  115012
  [arXiv:1211.0999 [hep-ph]].
  %%CITATION = ARXIV:1211.0999;%%
  %9 citations counted in INSPIRE as of 17 Apr 2014

%\cite{Boehm:2013qva}
\bibitem{Boehm:2013qva}
  C.~Boehm, P.~S.~B.~Dev, A.~Mazumdar and E.~Pukartas,
  %``Naturalness of Light Neutralino Dark Matter in pMSSM after LHC, XENON100 and Planck Data,''
  JHEP {\bf 1306} (2013) 113
  [arXiv:1303.5386 [hep-ph]].
  %%CITATION = ARXIV:1303.5386;%%
  %32 citations counted in INSPIRE as of 17 Apr 2014

\bibitem{Jungman:1995df}
  G.~Jungman, M.~Kamionkowski and K.~Griest, 
  %`Supersymmetric dark matter', 
  {\it Phys.\ Rept.\ } {\bf 267} (1996) 195 [arXiv:hep-ph/9506380].

\bibitem{Munoz:2003gx}
  C. Mu\~noz,
 % `Dark matter detection in the light of recent experimental
 %  results',
  {\it Int. J. Mod. Phys.} {\bf A19} (2004) 2093 [arXiv:hep-ph/0309346].

\bibitem{Bertone:2004pz}
  G.~Bertone, D.~Hooper and J.~Silk,
  %``Particle dark matter: Evidence, candidates and constraints,''
  Phys.\ Rept.\  {\bf 405} (2005) 279 [arXiv:hep-ph/0404175].

\bibitem{su5}
 For recent reviews, see R.~Mohapatra, hep-ph/9911272 (1999) and 
 S.~Raby, in Rept.\ Prog.\ Phys.\ {\bf 67} (2004) 755.

 %\cite{CDF:2013jga}
\bibitem{CDF:2013jga} 
  M.~Muether {\it et al.}  [Tevatron Electroweak Working Group and CDF and D0 Collaborations],
  %``Combination of CDF and DO results on the mass of the top quark using up to 8.7 fb^{-1} at the Tevatron,''
  arXiv:1305.3929 [hep-ex].
  %%CITATION = ARXIV:1305.3929;%%
  %48 citations counted in INSPIRE as of 23 Mar 2014
  
\bibitem{deAustri:2006pe}
  R.~Ruiz~de Austri, R.~Trotta and L.~Roszkowski,
  %``A Markov chain Monte Carlo analysis of the cMSSM,''
  JHEP {\bf 0605}, 002 (2006)
  [arXiv:hep-ph/0602028].
  %%CITATION = JHEPA,0605,002;%%

\bibitem{Roszkowski:2007fd} 
  L.~Roszkowski, R.~Ruiz de Austri and R.~Trotta,
  %``Implications for the Constrained MSSM from a new prediction for $b \to s \gamma$,''
  JHEP {\bf 0707}, 075 (2007)
  [arXiv:0705.2012 [hep-ph]].
  %%CITATION = ARXIV:0705.2012;%%
  %126 citations counted in INSPIRE as of 09 Jul 2013
  
 \bibitem{Trotta:2008bp}
 R.~Trotta, F.~Feroz, M.~P.~Hobson, L.~Roszkowski and R.~Ruiz de Austri,
 %``The Impact of priors and observables on parameter inferences in the
 %Constrained MSSM,''
 JHEP {\bf 0812} (2008) 024 [arXiv:0809.3792 [hep-ph]].
 
 %\cite{Bertone:2011nj}
\bibitem{Bertone:2011nj} 
  G.~Bertone, D.~G.~Cerdeno, M.~Fornasa, R.~Ruiz de Austri, C.~Strege and R.~Trotta,
  %``Global fits of the cMSSM including the first LHC and XENON100 data,''
  JCAP {\bf 1201}, 015 (2012)
  [arXiv:1107.1715 [hep-ph]].
  %%CITATION = ARXIV:1107.1715;%%
  %43 citations counted in INSPIRE as of 12 Jul 2013
  
\bibitem{SoftSUSY}
  \texttt{http://projects.hepforge.org/softsusy/}

\bibitem{Allanach:2001kg}
  B.~C.~Allanach,
  %``SOFTSUSY: a program for calculating supersymmetric spectra,''
  Comput.\ Phys.\ Commun.\  {\bf 143} (2002) 305
  [arXiv:hep-ph/0104145].
  %%CITATION = CPHCB,143,305;%%

\bibitem{MicrOMEGAs}
  \texttt{http://lapth.in2p3.fr/micromegas/}

\bibitem{Belanger:2006is}
  G.~Belanger, F.~Boudjema, A.~Pukhov and A.~Semenov,
  %``micrOMEGAs2.0: A program to calculate the relic density of dark matter  in
  %a generic model,''
  Comput.\ Phys.\ Commun.\  {\bf 176} (2007) 367
  [arXiv:hep-ph/0607059].
  %%CITATION = CPHCB,176,367;%%

\bibitem{DarkSUSY}
  P. Gondolo, J. Edsj\"o, P. Ullio, L. Bergstr\"m, M. Schelke, E.A. Baltz, T. Bringmann and G. Duda, \texttt{http://www.darksusy.org/}
  
\bibitem{Gondolo:2004sc}
  P.~Gondolo, J.~Edsjo, P.~Ullio, L.~Bergstrom, M.~Schelke and E.~A.~Baltz,
  %``DarkSUSY: Computing supersymmetric dark matter properties numerically,''
  JCAP {\bf 0407} (2004) 008
  [arXiv:astro-ph/0406204].
  %%CITATION = JCAPA,0407,008;%%

\bibitem{SuperIso}
  \texttt{http://superiso.in2p3.fr/}

\bibitem{Mahmoudi:2008tp}
 F.~Mahmoudi, 
 %``SuperIso v2.3: A Program for calculating flavor physics observables in
 %Supersymmetry,''
 Comput.\ Phys.\ Commun.\  {\bf 180}, 1579 (2009)
 [arXiv:0808.3144 [hep-ph]].
 %%CITATION = CPHCB,180,1579;%%
  
\bibitem{SusyBSG}
  \texttt{http://slavich.web.cern.ch/slavich/susybsg/}
  
\bibitem{Degrassi:2007kj}
  G.~Degrassi, P.~Gambino and P.~Slavich,
  %``SusyBSG: a fortran code for BR[B -> Xs gamma] in the MSSM with Minimal
  %Flavor Violation,''
  Comput.\ Phys.\ Commun.\  {\bf 179} (2008) 759
  %[arXiv:0712.3265 [hep-ph]].
  %%CITATION = CPHCB,179,759;%%
  
\bibitem{feynhiggs}
  \texttt{http://www.feynhiggs.de/}
 
%\cite{Heinemeyer:2007bw}
\bibitem{Heinemeyer:2007bw}
  S.~Heinemeyer, W.~Hollik, A.~M.~Weber and G.~Weiglein,
  %``$Z$ Pole Observables in the MSSM,''
  JHEP {\bf 0804} (2008) 039
  [arXiv:0710.2972 [hep-ph]].
  %%CITATION = ARXIV:0710.2972;%%
  %68 citations counted in INSPIRE as of 08 Apr 2014
  
\bibitem{Feroz:2007kg}
  F.~Feroz and M.~P. Hobson, 
 %{\it {Multimodal nested sampling: an efficient and 
 %  robust alternative to MCMC methods for astronomical data analysis}},  {\em
 {Mon. Not. Roy. Astron. Soc.} {\bf 384} (2008) 449--463.
 
\bibitem{Feroz:2008xx}
 F.~Feroz, M.~P. Hobson, and M.~Bridges, 
%{\it {MultiNest: an efficient and robust Bayesian inference tool for cosmology and particle physics}}, 
  {Mon. Not. Roy. Astron. Soc.} {\bf 398} (2009) 1601--1614.
  
\bibitem{skilling}
 J.~Skilling,  
 ``Nested Sampling for General Bayesian Computation". Bayesian Analysis 1 (4): 833–860.

  %\cite{Feroz:2011bj}
\bibitem{Feroz:2011bj} 
  F.~Feroz, K.~Cranmer, M.~Hobson, R.~Ruiz de Austri and R.~Trotta,
  %``Challenges of Profile Likelihood Evaluation in Multi-Dimensional SUSY Scans,''
  JHEP {\bf 1106}, 042 (2011)
  [arXiv:1101.3296 [hep-ph]].
  %%CITATION = ARXIV:1101.3296;%%
  %29 citations counted in INSPIRE as of 09 Jul 2013
  
%\bibitem{Higgs}
 % http://moriond.in2p3.fr/QCD/2013/MorQCD13Prog.html

  %\cite{Strege:2012bt}
\bibitem{Strege:2012bt} 
  C.~Strege, G.~Bertone, F.~Feroz, M.~Fornasa, R.~Ruiz de Austri and R.~Trotta,
  %``Global Fits of the cMSSM and NUHM including the LHC Higgs discovery and new XENON100 constraints,''
  JCAP {\bf 1304}, 013 (2013)
  [arXiv:1212.2636 [hep-ph]].
  %%CITATION = ARXIV:1212.2636;%%
  %22 citations counted in INSPIRE as of 09 Jul 2013
  
\bibitem{MTopCombo}
 The ATLAS, CDF, CMS, D0 Collaborations,  [arXiv:1403.4427].

\bibitem{SMnuis}
 http://pdg.lbl.gov/

%\cite{Strege:2011pk}
\bibitem{Strege:2011pk} 
  C.~Strege, G.~Bertone, D.~G.~Cerdeno, M.~Fornasa, R.~Ruiz de Austri and R.~Trotta,
  %``Updated global fits of the cMSSM including the latest LHC SUSY and Higgs searches and 
  %XENON100 data,''
  JCAP {\bf 1203}, 030 (2012)
  [arXiv:1112.4192 [hep-ph]].
  %%CITATION = ARXIV:1112.4192;%%
  %44 citations counted in INSPIRE as of 09 Jul 2013
  
    %\cite{Pato:2010zk}
\bibitem{Pato:2010zk}
  M.~Pato, L.~Baudis, G.~Bertone, R.~Ruiz de Austri, L.~E.~Strigari and R.~Trotta,
  %``Complementarity of Dark Matter Direct Detection Targets,''
  Phys.\ Rev.\ D {\bf 83} (2011) 083505
  [arXiv:1012.3458 [astro-ph.CO]].
  %%CITATION = ARXIV:1012.3458;%%
  %38 citations counted in INSPIRE as of 19 Jun 2013
  
  %\cite{Ren:2012aj}
\bibitem{Ren:2012aj}
  X.~-L.~Ren, L.~S.~Geng, J.~M.~Camalich, J.~Meng and H.~Toki,
  %``Octet baryon masses in next-to-next-to-next-to-leading order covariant baryon chiral perturbation theory,''
  J.\  High Energy Phys.\  {\bf 12} (2012) 073
  [arXiv:1209.3641 [nucl-th]].
  %%CITATION = ARXIV:1209.3641;%%
  %11 citations counted in INSPIRE as of 19 Jun 2013
  
  %\cite{Stahov:2012ca}
\bibitem{Stahov:2012ca} 
  J.~Stahov, H.~Clement and G.~J.~Wagner,
  %``Evaluation of the Pion-Nucleon Sigma Term from CHAOS data,''
  arXiv:1211.1148 [nucl-th].
  %%CITATION = ARXIV:1211.1148;%%
  %4 citations counted in INSPIRE as of 22 Jul 2013
  
      %\cite{Junnarkar:2013ac}
\bibitem{Junnarkar:2013ac}
  P.~Junnarkar and A.~Walker-Loud,
  %``The Scalar Strange Content of the Nucleon from Lattice QCD,''
  arXiv:1301.1114 [hep-lat].
  %%CITATION = ARXIV:1301.1114;%%
  %12 citations counted in INSPIRE as of 19 Jun 2013
  
    %\cite{QCDSF:2011aa}
\bibitem{QCDSF:2011aa}
  G.~S.~Bali {\it et al.}  [QCDSF Collaboration],
  %``Strangeness Contribution to the Proton Spin from Lattice QCD,''
  Phys.\ Rev.\ Lett.\  {\bf 108} (2012) 222001
  [arXiv:1112.3354 [hep-lat]].
  %%CITATION = ARXIV:1112.3354;%%
  %15 citations counted in INSPIRE as of 19 Jun 2013

%\cite{deAustri:2013saa}
\bibitem{deAustri:2013saa} 
  R.~R.~de Austri and C.~Pér.~d.~l.~Heros,
  %``Impact of nucleon matrix element uncertainties on the interpretation of direct and indirect dark matter search results,''
  arXiv:1307.6668 [hep-ph].
  %%CITATION = ARXIV:1307.6668;%%
  
\bibitem{lepwwg}
  \texttt{http://lepewwg.web.cern.ch/LEPEWWG}.

\bibitem{hfag} Heavy Flavor Averaging
  D. Asner {\it et al.}, 
  %"Averages of b-hadron, c-hadron, and tau-lepton properties as of early 2012," 
  arXiv:1207.1158.
%

\bibitem{deltambs}
   R.~Aaij {\it et al.} [LHCb Collaboration], Phys. Lett. B 709 (2012) 177 
   [arXiv:1112.4311 [hep-ex]];
   A. Abulencia {\it et al.} [CDF Collaboration], Phys. Rev. Lett. 97 (2006) 242003 
   [hep-ex/0609040];
%  CKMfitter Group, http://ckmfitter.in2p3.fr
%  The CDF Collaboration,

%\cite{Mahmoudi:2012un}
\bibitem{Mahmoudi:2012un}
  F.~Mahmoudi, S.~Neshatpour and J.~Orloff,
  %``Supersymmetric constraints from $B_s -> \mu^+\mu^-$ and $B -> K* \mu^+\mu^-$ observables,''
  JHEP {\bf 1208} (2012) 092
  [arXiv:1205.1845 [hep-ph]].
  %%CITATION = ARXIV:1205.1845;%%
  %36 citations counted in INSPIRE as of 18 Jun 2013
  
 %\cite{:2012ct}
\bibitem{:2012ct} 
  RAaij {\it et al.}  [LHCb Collaboration],
  %``First evidence for the decay Bs -> mu+ mu-,''
  arXiv:1211.2674
  %%CITATION = ARXIV:1211.2674;%%
  
      %\cite{Chatrchyan:2013bka}
\bibitem{Chatrchyan:2013bka} 
  S.~Chatrchyan {\it et al.}  [CMS Collaboration],
  %``Measurement of the B(s) to mu+ mu- branching fraction and search for B0 to mu+ mu- with the CMS Experiment,''
  Phys.\ Rev.\ Lett.\  {\bf 111}, 101804 (2013)
  [arXiv:1307.5025 [hep-ex]].
  %%CITATION = ARXIV:1307.5025;%%
  %82 citations counted in INSPIRE as of 30 Apr 2014

%\cite{Arbey:2012ax}
\bibitem{Arbey:2012ax}
  A.~Arbey, M.~Battaglia, F.~Mahmoudi and D.~Martinez Santos,
  %``Supersymmetry confronts Bs -> mu+mu-: Present and future status,''
  Phys.\ Rev.\ D {\bf 87} (2013) 035026
  [arXiv:1212.4887 [hep-ph]].
  %%CITATION = ARXIV:1212.4887;%%
  %19 citations counted in INSPIRE as of 14 Feb 2014

\bibitem{delta0}
 Value obtained combining the Babar measurement 
 B. Aubert et al. [BABAR Collaboration], 
 %%\Measurement of Branching Fractions and
 %%CP and Isospin Asymmetries in B ! K
 arXiv:0808.1915 [hep-ex]
 with the results of 
 K. Nakamura et al. [Particle Data Group], 
 %%\Review of particle physics", 
 J. Phys. G37, 075021 (2010) and 
 M. Nakao et al. [BELLE Collaboration], 
 %%\Measurement of the B ! K
 % %branching 
 %%fractions and asymmetries", 
 Phys. Rev. D69, 112001 (2004) [hep-ex/0402042].

\bibitem{Ade:2013zuv} 
  P.~A.~R.~Ade {\it et al.}  [Planck Collaboration],
  %``Planck 2013 results. XVI. Cosmological parameters,''
  arXiv:1303.5076 [astro-ph.CO].
  %%CITATION = ARXIV:1303.5076;%%
  %1446 citations counted in INSPIRE as of 23 Mar 2014

 \bibitem{Bertone:2010ww} 
  G.~Bertone, K.~Kong, R.~R.~de Austri and R.~Trotta,
  %``Global fits of the Minimal Universal Extra Dimensions scenario,''
  Phys.\ Rev.\ D {\bf 83}, 036008 (2011)
  [arXiv:1010.2023 [hep-ph]].
  %%CITATION = ARXIV:1010.2023;%%
  %13 citations counted in INSPIRE as of 20 Mar 2014

\bibitem{Bertone:2010rv}
  G.~Bertone, D.~G.~Cerdeno, M.~Fornasa, R.~R.~de Austri and R.~Trotta,
  %``Identification of Dark Matter particles with LHC and direct detection data,''
  Phys.\ Rev.\ D {\bf 82} (2010) 055008
  [arXiv:1005.4280 [hep-ph]].
  %%CITATION = ARXIV:1005.4280;%%
  %41 citations counted in INSPIRE as of 20 Mar 2014
  
  %\cite{Aprile:2012nq}
\bibitem{Aprile:2012nq} 
  E.~Aprile {\it et al.}  [XENON100 Collaboration],
  %``Dark Matter Results from 225 Live Days of XENON100 Data,''
  arXiv:1207.5988 [astro-ph.CO].
  %%CITATION = ARXIV:1207.5988;%%
  
  %\cite{Aprile:2013doa}
\bibitem{Aprile:2013doa} 
  E.~Aprile {\it et al.}  [XENON100 Collaboration],
  %``Limits on spin-dependent WIMP-nucleon cross sections from 225 live days of XENON100 data,''
  arXiv:1301.6620 [astro-ph.CO].
  %%CITATION = ARXIV:1301.6620;%%
  %15 citations counted in INSPIRE as of 17 Jun 2013

%\cite{Menendez:2012tm}
\bibitem{Menendez:2012tm}
  J.~Menendez, D.~Gazit and A.~Schwenk,
  %``Spin-dependent WIMP scattering off nuclei,''
  Phys.\ Rev.\ D {\bf 86} (2012) 103511
  [arXiv:1208.1094 [astro-ph.CO]].
  %%CITATION = ARXIV:1208.1094;%%

%\cite{Akerib:2013tjd}
\bibitem{Akerib:2013tjd} 
  D.~S.~Akerib {\it et al.}  [LUX Collaboration],
  %``First results from the LUX dark matter experiment at the Sanford Underground Research Facility,''
  arXiv:1310.8214 [astro-ph.CO].
  %%CITATION = ARXIV:1310.8214;%%
  %155 citations counted in INSPIRE as of 30 Mar 2014

\bibitem{Davier:2010nc}
  M.~Davier, A.~Hoecker, B.~Malaescu and Z.~Zhang,
  %``Reevaluation of the Hadronic Contributions to the Muon g-2 and to alpha(MZ),''
  Eur.\ Phys.\ J.\ C {\bf 71} (2011) 1515
  [arXiv:1010.4180 [hep-ph]].
  %%CITATION = ARXIV:1010.4180;%%

\bibitem{Aubert:2007dsa}
  B.~Aubert {\it et al.}  [BABAR Collaboration],
  %``Observation of the semileptonic decays $B \to D^{*} \tau^{-} \bar{\nu}$(
  %$\tau^{)}$ and evidence for $B \to D \tau^{-} \bar{\nu}$( $\tau^{)}$,''
 Phys.\ Rev.\ Lett.\  {\bf 100} (2008) 021801
  [arXiv:0709.1698 [hep-ex]].
  %%CITATION = PRLTA,100,021801;%%

\bibitem{Antonelli:2008jg} 
  M.~Antonelli {\it et al.}, 
  %\An evaluation of jVusj and precise tests of the Standard Model
  %from world data on leptonic and semileptonic kaon decays", 
  Eur. Phys. J. C 69, 399 (2010) [arXiv:1005.2323].

\bibitem{CMS_Higgs}   
 CMS collaboration,  
 CMS-PAS-HIG-12-045.

\bibitem{ATLAS_Higgs}   
 ATLAS collaboration,  
 ATL-PHYS-PUB-2012-001.

%\cite{Allanach:2004rh}
\bibitem{Allanach:2004rh}
  B.~C.~Allanach, A.~Djouadi, J.~L.~Kneur, W.~Porod and P.~Slavich,
  %``Precise determination of the neutral Higgs boson masses in the MSSM,''
  JHEP {\bf 0409} (2004) 044
  [hep-ph/0406166].
  %%CITATION = HEP-PH/0406166;%%
  %212 citations counted in INSPIRE as of 08 Apr 2014

\bibitem{Higgsgg}   
 CMS collaboration, 
 CMS-PAS-HIG-13-001.

\bibitem{Higgsww}
 CMS collaboration, 
 CMS-PAS-HIG-13-003.

\bibitem{Higgszz}
 CMS collaboration, 
 CMS-PAS-HIG-13-002.

\bibitem{Higgstautau}
 CMS collaboration,
 CMS-PAS-HIG-13-004.     

\bibitem{Higgsbb}
 CMS collaboration, 
 CMS-PAS-HIG-12-044.

 %\cite {ATLAS-CONF-2012-033-5fb} 
\bibitem{ATLAS-CONF-2012-033-5fb}
ATLAS Collaboration,
%"{Search for squarks and gluinos using final states with jets
%	and missing transverse momentum with the ATLAS detector in 
%	sqrt(s) = 7 TeV proton-proton collisions}",
%institution  = "CERN",
%number       = "CERN-PH-EP-2012-195",
%year         = "2012",  
arXiv:1208.0949v3 [hep-ex].

%\cite{3-lepton-ana}
\bibitem{3-lepton-ana}
 ATLAS Collaboration,
 % number      = "CERN-PH-EP-2012-217"
 arXiv:1208.3144v2 [hep-ex].

%\cite{pythia}
\bibitem{pythia}
  T.~Sjostrand, S.~Mrenna and P.~Z.~Skands,
  %``PYTHIA 6.4 Physics and Manual,''
  JHEP {\bf 0605} (2006) 026
  [hep-ph/0603175].
  %%CITATION = HEP-PH/0603175;%%
  %4745 citations counted in INSPIRE as of 07 Jan 2014
  
%\cite{ATLAS:2010zyu}
\bibitem{ATLAS:2010zyu}
  [ATLAS Collaboration],
  %``ATLAS Monte Carlo tunes for MC09,''
  ATL-PHYS-PUB-2010-002.
  %%CITATION = ATL-PHYS-PUB-2010-002;%%
  %135 citations counted in INSPIRE as of 08 Apr 2014

%\cite{Pumplin:2002vw}
\bibitem{Pumplin:2002vw}
  J.~Pumplin, D.~R.~Stump, J.~Huston, H.~L.~Lai, P.~M.~Nadolsky and W.~K.~Tung,
  %``New generation of parton distributions with uncertainties from global QCD analysis,''
  JHEP {\bf 0207} (2002) 012
  [hep-ph/0201195].
  %%CITATION = HEP-PH/0201195;%%
  %3786 citations counted in INSPIRE as of 08 Apr 2014

\bibitem{nllfast}
 %Squark and Gluino Production at Hadron Colliders, 
 W.~Beenakker, R.~Höpker, M.~Spira, and P.~M.~Zerwas, Nucl. Phys. B492 (1997) 51-103;
 %Threshold resummation for squark-antisquark and gluino-pair production at the LHC, A. Kulesza, L. Motyka, Phys. Rev. Lett. 102 (2009) 111802
 %Soft gluon resummation for the production of gluino-gluino and squark-antisquark pairs at the LHC, 
 A.~Kulesza and L.~Motyka, Phys. Rev. D80 (2009) 095004;
 %Soft-gluon resummation for squark and gluino hadroproduction, 
 W.~ Beenakker, S.~Brensing, M.~Krämer, A.~Kulesza, E.~Laenen and I.~ Niessen, JHEP {\bf 0912} (2009) 041;
 %Squark and gluino hadroproduction, 
 W.~Beenakker, S.~Brensing, M.~Krämer, A.~Kulesza, E.~Laenen, L.~Motyka and  I.~Niessen, Int. J. Mod. Phys. A26 (2011) 2637-2664.

\bibitem{prospino}
  W.~Beenakker, R.~Hopker, M.~Spira and P.~M.~Zerwas,
  %``Squark and gluino production at hadron colliders,''
  Nucl.\ Phys.\ B {\bf 492} (1997) 51
  [hep-ph/9610490]; 
 W.~Beenakker, M.~Klasen, M.~Kramer, T.~Plehn, M.~Spira and P.~M.~Zerwas,
  %``The Production of charginos / neutralinos and sleptons at hadron colliders,''
  Phys.\ Rev.\ Lett.\  {\bf 83} (1999) 3780
   [Erratum-ibid.\  {\bf 100} (2008) 029901]
  [hep-ph/9906298].

 %\cite{delphes}
\bibitem{delphes}
  J.~de Favereau, C.~Delaere, P.~Demin, A.~Giammanco, V.~Lemaître, A.~Mertens and M.~Selvaggi,
  %``DELPHES 3, A modular framework for fast simulation of a generic collider experiment,''
  arXiv:1307.6346 [hep-ex].
  %%CITATION = ARXIV:1307.6346;%%
  %50 citations counted in INSPIRE as of 07 Jan 2014

\bibitem{LHC_PP_Future} R.~Ruiz de Austri et al., In preparation. 

\bibitem{LEP_SUSY}
http://lepsusy.web.cern.ch/lepsusy/

%\cite{Endo:2013bba}
\bibitem{Endo:2013bba}
  M.~Endo, K.~Hamaguchi, S.~Iwamoto and T.~Yoshinaga,
  %``Muon $g-2$ vs LHC in Supersymmetric Models,''
  JHEP {\bf 1401} (2014) 123
  [arXiv:1303.4256 [hep-ph]].
  %%CITATION = ARXIV:1303.4256;%%
  %21 citations counted in INSPIRE as of 08 Apr 2014

%\cite{DescotesGenon:2011yn}
\bibitem{DescotesGenon:2011yn}
  S.~Descotes-Genon, D.~Ghosh, J.~Matias and M.~Ramon,
  %``Exploring New Physics in the C7-C7' plane,''
  JHEP {\bf 1106} (2011) 099
  [arXiv:1104.3342 [hep-ph]].
  %%CITATION = ARXIV:1104.3342;%%
  %59 citations counted in INSPIRE as of 12 Mar 2014

%\cite{Altmannshofer:2013foa}
\bibitem{Altmannshofer:2013foa}
  W.~Altmannshofer and D.~M.~Straub,
  %``New physics in $B \to K^*\mu\mu$?,''
  Eur.\ Phys.\ J.\ C {\bf 73} (2013) 2646
  [arXiv:1308.1501 [hep-ph]].
  %%CITATION = ARXIV:1308.1501;%%
  %29 citations counted in INSPIRE as of 10 Apr 2014

%\cite{Gunion:2002zf}
\bibitem{Gunion:2002zf}
  J.~F.~Gunion and H.~E.~Haber,
  %``The CP conserving two Higgs doublet model: The Approach to the decoupling limit,''
  Phys.\ Rev.\ D {\bf 67} (2003) 075019
  [hep-ph/0207010].
  %%CITATION = HEP-PH/0207010;%%
  %229 citations counted in INSPIRE as of 29 Apr 2014

%\cite{Shamim:2007yy}
\bibitem{Shamim:2007yy} 
  M.~Shamim [D0 Collaboration],
  %``Searches for Squarks and Gluinos with D0 Detector,''
  arXiv:0710.2897 [hep-ex].
  %%CITATION = ARXIV:0710.2897;%%
  %4 citations counted in INSPIRE as of 30 Mar 2014

%\cite{Altmannshofer:2012ks}
\bibitem{Altmannshofer:2012ks} 
  W.~Altmannshofer, M.~Carena, N.~R.~Shah and F.~Yu,
  %``Indirect Probes of the MSSM after the Higgs Discovery,''
  JHEP {\bf 1301}, 160 (2013)
  [arXiv:1211.1976 [hep-ph]].
  %%CITATION = ARXIV:1211.1976;%%
  %37 citations counted in INSPIRE as of 28 Apr 2014

%\cite{Degrassi:2006eh}
\bibitem{Degrassi:2006eh}
  G.~Degrassi, P.~Gambino and P.~Slavich,
  %``QCD corrections to radiative B decays in the MSSM with minimal flavor violation,''
  Phys.\ Lett.\ B {\bf 635} (2006) 335
  [hep-ph/0601135].
  %%CITATION = HEP-PH/0601135;%%
  %68 citations counted in INSPIRE as of 10 Apr 2014

%\cite{Bobeth:2004jz}
\bibitem{Bobeth:2004jz}
  C.~Bobeth, A.~J.~Buras and T.~Ewerth,
  %``Anti-B ---> X(s) l+ l- in the MSSM at NNLO,''
  Nucl.\ Phys.\ B {\bf 713} (2005) 522
  [hep-ph/0409293].
  %%CITATION = HEP-PH/0409293;%%
  %29 citations counted in INSPIRE as of 22 Apr 2014

%\cite{ArkaniHamed:2006mb}
\bibitem{ArkaniHamed:2006mb}
  N.~Arkani-Hamed, A.~Delgado and G.~F.~Giudice,
  %``The Well-tempered neutralino,''
  Nucl.\ Phys.\ B {\bf 741} (2006) 108
  [hep-ph/0601041].
  %%CITATION = HEP-PH/0601041;%%
  %185 citations counted in INSPIRE as of 09 Apr 2014

%\cite{Drees:1993bu}
\bibitem{Drees:1993bu} 
  M.~Drees and M.~Nojiri,
  %``Neutralino - nucleon scattering revisited,''
  Phys.\ Rev.\ D {\bf 48}, 3483 (1993)
  [hep-ph/9307208].
  %%CITATION = HEP-PH/9307208;%%
  %218 citations counted in INSPIRE as of 09 Apr 2014

%\cite{Mandic:2000jz}
\bibitem{Mandic:2000jz}
  V.~Mandic, A.~Pierce, P.~Gondolo and H.~Murayama,
  %``The Lower bound on the neutralino nucleon cross-section,''
  hep-ph/0008022.
  %%CITATION = HEP-PH/0008022;%%
  %41 citations counted in INSPIRE as of 10 Apr 2014

\bibitem{Profumo:2004wk} 
  S.~Profumo and C.~E.~Yaguna,
  %``Gluino coannihilations and heavy bino dark matter,''
  Phys.\ Rev.\ D {\bf 69}, 115009 (2004)
  [hep-ph/0402208].
  %%CITATION = HEP-PH/0402208;%%
  %33 citations counted in INSPIRE as of 14 Apr 2014

%\cite{Haber:2000kq}
\bibitem{Haber:2000kq}
  H.~E.~Haber, M.~J.~Herrero, H.~E.~Logan, S.~Penaranda, S.~Rigolin and D.~Temes,
  %``SUSY QCD corrections to the MSSM h0 $b \bar{b}$ vertex in the decoupling limit,''
  Phys.\ Rev.\ D {\bf 63} (2001) 055004
  [hep-ph/0007006].
  %%CITATION = HEP-PH/0007006;%%
  %93 citations counted in INSPIRE as of 15 Apr 2014

%\cite{ATL-PHYS-PUB-2012-001}
\bibitem{ATL-PHYS-PUB-2012-001}
  ATLAS collaboration,
  ATL-PHYS-PUB-2012-001.

%\cite{Cowan:2010js}
\bibitem{Cowan:2010js}
  G.~Cowan, K.~Cranmer, E.~Gross and O.~Vitells,
  %``Asymptotic formulae for likelihood-based tests of new physics,''
  Eur.\ Phys.\ J.\ C {\bf 71} (2011) 1554
  [arXiv:1007.1727 [physics.data-an]].
  %%CITATION = ARXIV:1007.1727;%%


%%%%%%%%%%%%%%%%%%%%%%%%%%%%%%%%%%%%%%%%%%%%%%%%%%%%%%%%%%
%The following citations are not used


%\bibitem{Bernabei:2008yi}
%  R.~Bernabei {\it et al.}  [DAMA Collaboration],
  %``First results from DAMA/LIBRA and the combined results with DAMA/NaI,''
 % Eur.\ Phys.\ J.\  C {\bf 56} (2008) 333 [arXiv:0804.2741 [astro-ph]].

%\bibitem{Aalseth:2008rx}
 % C.~E.~Aalseth {\it et al.}  [CoGeNT Collaboration],
  %``Experimental constraints on a dark matter origin for the DAMA annual
  %modulation effect,''
  %Phys.\ Rev.\ Lett.\  {\bf 101} (2008) 251301 [Erratum-ibid.\  {\bf 102} (2009) 109903] [arXiv:0807.0879 [astro-ph]].

%\bibitem{Aalseth:2010vx}
 % C.~E.~Aalseth {\it et al.}  [CoGeNT collaboration],
  %``Results from a Search for Light-Mass Dark Matter with a P-type Point
  %Contact Germanium Detector,''
 % arXiv:1002.4703 [astro-ph.CO].

%\bibitem{Ahmed:2008eu}
 % Z.~Ahmed {\it et al.}  [CDMS Collaboration],
  %``Search for Weakly Interacting Massive Particles with the First Five-Tower
  %Data from the Cryogenic Dark Matter Search at the Soudan Underground
  %Laboratory,''
  %Phys.\ Rev.\ Lett.\  {\bf 102} (2009) 011301 [arXiv:0802.3530 [astro-ph]].

%\bibitem{Ahmed:2009zw}
 % Z.~Ahmed {\it et al.}  [The CDMS-II Collaboration and CDMS-II Collaboration],
  %``Results from the Final Exposure of the CDMS II Experiment,''
  %arXiv:0912.3592 [astro-ph.CO].

%\cite{Roszkowski:2009sm}
%\bibitem{Roszkowski:2009sm} 
 % L.~Roszkowski, R.~Ruiz de Austri, R.~Trotta, Y.~-L.~S.~Tsai and T.~A.~Varley,
  %``Global fits of the Non-Universal Higgs Model,''
  %Phys.\ Rev.\ D {\bf 83}, 015014 (2011)
  %[arXiv:0903.1279 [hep-ph]].
  %%CITATION = ARXIV:0903.1279;%%
  %32 citations counted in INSPIRE as of 09 Jul 2013

%\bibitem{Angle:2007uj}
 % J.~Angle {\it et al.}  [XENON Collaboration],
  %``First Results from the XENON10 Dark Matter Experiment at the Gran Sasso
  %National Laboratory,''
  %Phys.\ Rev.\ Lett.\  {\bf 100} (2008) 021303 [arXiv:0706.0039 [astro-ph]].

%\bibitem{Scott:2009jn}
 % P.~Scott, J.~Conrad, J.~Edsjo, L.~Bergstrom, C.~Farnier and Y.~Akrami,
  %``Direct Constraints on Minimal Supersymmetry from Fermi-LAT Observations of
  %the Dwarf Galaxy Segue 1,''
  %arXiv:0909.3300 [astro-ph.CO].

%\bibitem{Green:2007rb}
 % A.~M.~Green,
  %``Determining the WIMP mass using direct detection experiments,''
  %JCAP {\bf 0708} (2007) 022 [arXiv:hep-ph/0703217].

%\bibitem{Green:2008rd}
 % A.~M.~Green,
  %``Determining the WIMP mass from a single direct detection experiment, a more
  %detailed study,''
  %JCAP {\bf 0807} (2008) 005 [arXiv:0805.1704 [hep-ph]].

%\bibitem{Drees:2008bv}
 % M.~Drees and C.~L.~Shan,
  %``Model-Independent Determination of the WIMP Mass from Direct Dark Matter
  %Detection Data,''
  %JCAP {\bf 0806} (2008) 012 [arXiv:0803.4477 [hep-ph]].

%\bibitem{Bertone:2007xj}
 % G.~Bertone, D.~G.~Cerde\~no, J.~I.~Collar and B.~C.~Odom,
  %``WIMP identification through a combined measurement of axial and scalar
  %couplings,''
  %Phys.\ Rev.\ Lett.\  {\bf 99} (2007) 151301 [arXiv:0705.2502 [astro-ph]].

%\bibitem{Cho:2007qv}
 % W.~S.~Cho, K.~Choi, Y.~G.~Kim and C.~B.~Park,
  %``Gluino Stransverse Mass,''
  %Phys.\ Rev.\ Lett.\  {\bf 100} (2008) 17180 [arXiv:0709.0288 [hep-ph]].

%\bibitem{Cho:2008tj}
 % W.~S.~Cho, K.~Choi, Y.~G.~Kim and C.~B.~Park,
  %``M_T2-assisted on-shell reconstruction of missing momenta and its
  %application to spin measurement at the LHC,''
  %Phys.\ Rev.\  D {\bf 79} (2009) 031701 [arXiv:0810.4853 [hep-ph]].

%\bibitem{Baltz:2006fm}
 % E.~A.~Baltz, M.~Battaglia, M.~E.~Peskin and T.~Wizansky,
  %``Determination of dark matter properties at high-energy colliders,''
  %Phys.\ Rev.\  D {\bf 74} (2006) 103521 [arXiv:hep-ph/0602187].

%\bibitem{SuperBayeS}
%  http://www.ft.uam.es/personal/rruiz/superbayes/

%\bibitem{Khotilovich:2005gb}
 %V.~Khotilovich, R.~L.~Arnowitt, B.~Dutta and T.~Kamon,
 %``The stau neutralino co-annihilation region at an international linear
 %collider,''
 %Phys.\ Lett.\  B {\bf 618} (2005) 182 [arXiv:hep-ph/0503165].

%\bibitem{Bernal:2008zk}
 % N.~Bernal, A.~Goudelis, Y.~Mambrini and C.~Munoz,
  %``Determining the WIMP mass using the complementarity between direct and
  %indirect searches and the ILC,''
 % JCAP {\bf 0901} (2009) 046 [arXiv:0804.1976 [hep-ph]].

%\bibitem{Lewin:1995rx}
 % J.~D.~Lewin and P.~F.~Smith,
  %``Review of mathematics, numerical factors, and corrections for dark  matter
  %experiments based on elastic nuclear recoil,''
  %Astropart.\ Phys.\  {\bf 6} (1996) 87.

%\bibitem{Hagiwara:2006jt}
 %K.~Hagiwara, A.~D.~Martin, D.~Nomura and T.~Teubner,
 %{\it Improved predictions for g-2 of the muon and $\alpha_{\rm QED}(M_Z^2)$},
 %Phys. Lett. B {\bf 649}, 173 (2007).
 % [hep-ph/0611102].

%\cite{hepdata-0lep}
%\bibitem{hepdata-0lep}
%The Durham HepData Project,
%http://hepdata.cedar.ac.uk/view/ins1125961  %hepdata-0lep

%\cite{hepdata-3lep}
%\bibitem{hepdata-3lep}
%The Durham HepData Project,
%http://hepdata.cedar.ac.uk/view/ins1127601  %hepdata-3lep



\end{thebibliography}
\end{document}